%% file: kappa_v2.tex
\documentclass{amsarte}

\usepackage{latexsym,pifont}
\usepackage{epsfig}
\usepackage{rotating}

\usepackage{ifpdf}
\ifpdf
\usepackage{epstopdf}
\usepackage{hyperref}
\else
\usepackage[hypertex]{hyperref}
\fi

\usepackage{amsfonts,amsmath,amssymb,mathrsfs,extarrows,MnSymbol}
\usepackage[all]{xy}

\usepackage{tikz}
\usetikzlibrary{matrix,arrows}

\usepackage[makeroom]{cancel}

\include{1armagedef}

\newtheorem{Thm}{Theorem}
\newtheorem{Prop}[Thm]{Proposition}
\newtheorem{Lem}[Thm]{Lemma}
\newtheorem{Cor}[Thm]{Corollary}
\theoremstyle{definition}
\newtheorem{Rem}[Thm]{Remark}
\newtheorem{Def}[Thm]{Definition}
\newtheorem{Eg}[Thm]{Example}
\newtheorem{Conv}[Thm]{Convention}
\newtheorem{Fact}[Thm]{Fact}
\newtheorem{Quest}[Thm]{Question}
\newtheorem{Prob}[Thm]{Problem}

\numberwithin{equation}{section} \numberwithin{Thm}{section}

\newcount\hour\newcount\minute
        \hour=\time \divide\hour by60 \minute=\time
        {\multiply\hour by60 \global\advance\minute by-\hour}
        \edef\militarytime{\number\hour:\ifnum\minute<10 0\fi\number\minute}

\begin{document}

\title{The square root of the vacuum \\ I. Equivariance for the $\k$-symmetry superdistribution}

\author{Rafa\l ~R. ~Suszek}
\address{R.R.S.:\ Katedra Metod Matematycznych Fizyki, Wydzia\l ~Fizyki
Uniwersytetu Warszawskiego, ul.\ Pasteura 5, PL-02-093 Warszawa,
Poland} \email{suszek@fuw.edu.pl}

\begin{abstract}
A complete and natural geometric and physical interpretation of the tangential gauge supersymmetry, also known as $\k$-symmetry, of a large class of Green--Schwarz(-type) super-$\si$-models for the super-$p$-brane in a homogeneous space of a (supersymmetry) Lie supergroup is established in the convenient setting of the topological Hughes--Polchinski formulation of the super-$\si$-model and illustrated on a number of physical examples. The supersymmetry is identified as an odd superdistribution in the tangent sheaf of the supertarget of the super-$\si$-model, generating -- through its weak derived flag -- the vacuum foliation of the supertarget. It is also demonstrated to canonically lift to the vacuum restriction of the extended Hughes--Polchinski $p$-gerbe associated with the superbackground of the field theory, and that in the form of a canonical linearised equivariant structure thereon, canonically compatible with the residual global supersymmetry of the vacuum.
\end{abstract}


\maketitle

\tableofcontents

\section{Introduction}

Symmetry, a polyseme with a broad semantic field in the physical parlance stretching between -- on one hand -- the concept of `correspondence' between \emph{independent} entities that implies `balance' in their quantity\footnote{In this denotation, the `symmetry' is customarily qualified as `global' or `rigid'.}, and -- on the other hand -- the concept of `ambiguity' and `unphysical redundancy' of \emph{equivalent} representations of a given entity that calls for `reduction'\footnote{Here, the usual qualifiers are `local' or `gauge'.}, is widely recognised as one of the most robust organising principles in the mathematical modelling of physical phenomena. From admissible interaction vertices and superselection rules, through conserved charges and constraints imposed upon quantum-mechanical correlators, all the way to the exclusion of theories exhibiting anomalies and the removal of pure-gauge degrees of freedom, it helps to organise the physical content of the theory and its dynamics in terms of the representation theory of the symmetry structure, typically given as a group or an algebra. In extreme cases, a symmetry can be employed towards a complete resolution of the theory, as exemplified by the BPZ approach of \Rcite{Belavin:1984vu} to the derivation of correlators in a conformal field theory in two dimensions (in which the conformal group becomes infinite-dimensional), especially in the presence of an extension of the Virasoro algebra of symmetries modelled on a Ka\v{c}--Moody algebra, {\it cp} \Rcite{Knizhnik:1984nr}, and by the localisation of the theory on a Cauchy hypersurface (an equitemporal slice carrying complete data of a state) through identification of `propagation' of a state with a gauge transformation in the three-dimensional Chern--Simons topological gauge field theory on a cylinder over a (punctured) Riemann surface, {\it cp} \Rcite{Freed:1992vw}. Therefore, it is of utmost importance to have a good understanding of the symmetries of a field theory under study, and of their amenability to a consistent quatisation.

An important class of physical models in which symmetries have a readily identifiable geometric origin and their transcription to the quantum r\'egime is systematised by higher geometry and cohomology is formed by non-linear $\si$-models describing simple geometro-dynamics of compact $p$-dimensional distributions of topological charge and energy (such as, {\it e.g.}, charged and massive pointlike particles at $\,p=0$,\ loops and paths at $\,p=1$,\ membranes at $\,p=2$ {\it etc}.) in external fields permeating the ambient manifold $\,\cM\,$ in which the charge propagates and coupling to the charge currents thus induced, {\it i.e.}, a $(p+2)$-form field and a gravitational field. Geometrisation of the classes in the de Rham cohomology of $\,\cM\,$ determined by the $(p+2)$-form field, requisite for a rigorous definition of the Dirac--Feynman amplitude in these theories, {\it cp} Refs.\,\cite{Alvarez:1984es,Gawedzki:1987ak}, leads to the emergence of the so-called $p$-gerbes -- geometric structures that generalise, in a well-defined sense, principal $\bC^\x$-bundles encountered in the lagrangean description of the charged pointlike particle, {\it cp} Refs.\,\cite{Murray:1994db,Murray:1999ew,Stevenson:2000wj,Johnson:2003,Gajer:1996}. Being defined entirely in terms of smooth differential-geometric objects over the target space of the $\si$-model, the $p$-gerbes determine geometric (pre)quantisation of the $(p+1)$-dimensional field theory, {\it cp} \Rcite{Gawedzki:1987ak,Suszek:2011hg}, and in this manner enable us to distinguish, as quantum-mechanically consistent, those symmetries that lift (or \emph{transgress}) to isomorphisms in the (weak) $(p+1)$-category of $p$-gerbes over $\,\cM\,$ (in the case of global symmetries) resp.\ give rise to full-fledged equivariant structures over the nerve of the action groupoid engendered by the action of the symmetry group on the target space (in the case of gauged symmetries), {\it cp} Refs.\cite{Gawedzki:2008um,Gawedzki:2010rn,Gawedzki:2012fu,Suszek:2012ddg,Suszek:2013}. This universal higher-geometric scheme with a neat cohomological underpinning, readily extendible to more general dualities between \emph{different} theories (represented by a class of worldvolume defects) such as, {\it e.g.}, T-duality, gives us extensive control over (symmetries of) any given $\si$-model of interest and a robust method of charting the moduli space of theories of this type. The $\si$-models, finding numerous physical applications from the theory of condensed matter, through the description of long-distance phenomena in two-dimensional statistical mechanics at a critical point and the modelling of the effective field theory of certain collective excitations of discrete one-dimensional integrable systems (spin chains), to critical bosonic string theory, are prototypes of the field theories that we intend to study in the present paper.

Symmetry considerations acquire critical significance in the construction and study of field theories with the fibre of the covariant configuration bundle given by a single orbit of a symmetry group. Among these, we find the much studied Wess--Zumino--Witten (WZW) $\si$-model of \Rcite{Witten:1983ar} of charged-loop mechanics on a compact Lie group and the formally rather intricate Green--Schwarz(-type) super-$\si$-model of, {\it i.a.}, Refs.\,\cite{Green:1983wt,Green:1983sg,Bergshoeff:1985su,Bergshoeff:1987cm,Achucarro:1987nc,Bandos:1997ui,Metsaev:1998it,deWit:1998yu,Claus:1998fh,Arutyunov:2008if,Gomis:2008jt,Fre:2008qc,DAuria:2008vov,Suszek:2017xlw} for the super-$p$-brane in a homogeneous space $\,\txG/\txH\,$ of a Lie supergroup $\,\txG\,$ ({\it i.e.}, a group object in the category of supermanifolds, with the structure sheaf valued in the category of supercommutative $\bZ/2\bZ$-graded algebras) relative to a subgroup $\,\txH\subset|\txG|\,$ of its (Gra\ss mann-even) body $\,|\txG|$,\ the latter field theory being defined for mappings, from an (inner hom-)functorial mapping supermanifold $\,[\Om_p,\txG/\txH]\,$ ({\it cp} \Rcite{Freed:1999}), of the $(p+1)$-dimensional worldvolume $\,\Om_p\,$ into the supermanifold $\,\txG/\txH\,$ equipped with a transitive action of the supersymmetry Lie supergroup $\,\txG\,$ and tensorial data (a metric and a super-$(p+2)$-cocycle) descended from it. While the status of field theories of the latter type as models of natural phenomena seems at best dubitable at this moment, they do find a spectacular direct application in a predictive description of physical systems, closely akin to observable ones, involving strongly interacting elementary coloured particles, outside the perturbative r\'egime. A connection between the two universes: the perturbative superstring resp.\ super-$p$-brane theory and the non-perturbative (supersymmetric) Yang--Mills theory is established by means of the celebrated AdS/CFT correspondence whose in-depth elucidation, still missing to date, is one of the key motivations behind our interest in this particular class of supersymmetric field theories. Setting aside the subtle technical issue of a functional (or, indeed, functorial) definition of the super-$\si$-model\footnote{This will be discussed at great length in the main text.}, we immediately notice the fundamental difference between the two types of theories mentioned above, affecting their higher-geometric description: While the presence of a global symmetry in the $\si$-model described by the \emph{compact} target Lie group does not -- in the light of The Cartan--Eilenberg Theorem -- result in an actual choice of the cohomology in which to perform the aforementioned geometrisation of the 3-form field coupling to the loop current in the group manifold, the same structural property of the super-$\si$-model on (a homogeneous space of) the inherently \emph{noncompact} Lie supergroup confronts us with the \emph{non}trivial choice between the standard de Rham cohomology and its supersymmetric refinement. The dramatic discrepancy between the two is best exemplified by the super-Minkowski space with its trivial de Rham cohomology and nontrivial Cartan--Eilenberg cohomology, and the Green--Schwarz super-$(p+2)$-cocycle defining a \emph{non}-zero class in the latter. Based on a topological interpretation of the discrepancy and a reinterpretation of the super-$\si$-model (as a super-$\si$-model on a supermanifold with the topology of the Gra\ss mann-odd fibres faithfully encoding the discrepancy) implied by it, originally due to Rabin and Crane, {\it cp} Refs.\,\cite{Rabin:1984rm,Rabin:1985tv}, and further inspired by the canonical analysis of the field-theoretic realisation of supersymmetry in the super-$\si$-model in terms of a Poisson (super)algebra of Noether charges with a wrapping-charge anomaly sourced by monodromies of the embedding field along the Gra\ss mann-odd fibres (assumed compact), carried out by the Author in \Rcite{Suszek:2018bvx}, a geometrisation scheme for the Green--Schwarz super-$(p+2)$-cocycles was postulated in a series of papers \cite{Suszek:2017xlw,Suszek:2019cum,Suszek:2018bvx,Suszek:2018ugf} which essentially puts the standard construction of a $p$-gerbe from the Gra\ss mann-even category, originally due to Murray and Stevenson, {\it cp} Refs.\,\cite{Murray:1994db,Murray:1999ew}, and subsequently formally generalised by Gajer in \Rcite{Gajer:1996}, in the category of Lie supergroups, with surjective submersions replaced by Lie-supergroup extensions. The fundamental mechanism of the geometrisation scheme consists in (super)algebraisation of the Cartan--Eilenberg class of the Green--Schwarz super-$(p+2)$-cocycle on the supersymmetry Lie supergroup $\,\txG\,$ {\it via} the classical correspondence between the Cartan--Eilenberg cohomology $\,{\rm CaE}^\bullet(\txG)\,$ of $\,\txG\,$ and the Chevalley--Eilenberg cohomology $\,H^\bullet(\ggt,\bR)\,$ of its tangent Lie (super)algebra $\,\ggt\,$ with values in the trivial module $\,\bR$,\ augmented with the equally classical correspondence between classes in $\,H^2(\ggt,\bR)\,$ and (super)central Lie-(super)algebra extensions of $\,\ggt$.\ Thus, a sequence of Lie-superalgebra extensions, each corresponding to a partial (term-wise) cohomological trivialisation of the original Green--Schwarz super-$(p+2)$-cocycle and assumed integrable to a Lie-supergroup extension, should ultimately lead to the emergence of a Lie supergroup $\,\widehat\txG$,\ epimorphically mapped onto $\,\txG$,\ on which the pullback of the Green--Schwarz super-$(p+2)$-cocycle trivialises in the Cartan--Eilenberg cohomology $\,{\rm CaE}^\bullet(\widehat\txG)$.\ The surjective submersion $\,\widehat\txG\too\txG\,$ should then be taken as the basis of the standard construction of a $p$-gerbe, with the sequential-extension mechanism applied on each level of that construction. The resulting higher-geometric object -- the Cartan--Eilenberg super-$p$-gerbe -- still has to be descended to the physical target space $\,\txG/\txH$,\ for which the extensions indicated earlier ought to be equivariant with respect to (the adjoint action of) the tangent Lie algebra $\,\hgt\,$ of the isotropy group $\,\txH\,$ of the homogeneous space. The scheme was successfully applied in the super-Minkowskian setting (in \Rcite{Suszek:2017xlw}) and in the case of the super-2-cocycle for the Zhou super-$\si$-model of the super-0-brane in $\,{\rm s}({\rm AdS}_2\x\bS^2)\,$ (in \Rcite{Suszek:2018ugf}), resulting in a collection of concrete manifestly supersymmetric super-$p$-gerbes with various desirable propreties such as, in the super-Minkowskian case (for $\,p\in\{0,1\}$), equivariance with respect to the Lie supergroup of supertranslations realised in the adjoint, {\it cp} \Rcite{Suszek:2019cum}, in conformity with the intuition coming from the study of the WZW $\si$-model on the Lie group, {\it cp} Refs.\,\cite{Gawedzki:2010rn,Gawedzki:2012fu,Suszek:2012ddg}. The construction for the Zhou super-2-cocycle, on the other hand, was demonstrated to asymptote, in the limit of a homogeneous blow-up of the curved supertarget $\,{\rm s}({\rm AdS}_2\x\bS^2)$,\ dual to the standard \.In\"on\"u--Wigner contraction on the underlying Lie superalgebra, to the super-0-gerbe on the super-Minkowskian space. Contractibility of the super-$p$-gerbe over a curved Lie supergroup with a super-Minkowskian blow-up to its known\footnote{That is predicted by the correspondence on the level of the super-$\si$-models, {\it cp}, {\it e.g.}, \Rcite{Metsaev:1998it}.} Green--Schwarz counterpart over that blow-up was subsequently incorporated into the geometrisation scheme as a physically motivated organising principle. Just to re-emphasise, the crucial feature of the geometrisation scheme thus obtained is the canonical invariance of the super-$p$-gerbe under the action of the supersymmetry group of the super-$\si$-model, indispensable for the physical interpretation of the supersymmetry as a (pre)quantisable symmetry of the field theory.\medskip

Besides establishing `correspondence' between independent degrees of freedom of the embedding field of the super-$\si$-model of an arbitrary Gra\ss mann parity, which is the task of the global supersymmetry mentioned above, supersymmetry is also bound to take on the other r\^ole alluded to at the beginning of the present Introduction, {\it i.e.}, that of a mechanism of `reduction' of the excessive (Goldstone) degrees of freedom of the (Gra\ss mann-)odd type that arise in the process of localisation of the (Gra\ss mann-even) body of the embedding field in the (classical) vacuum of the theory. This is an elementary consequence of the universal feature of the supersymmetry algebra by which the anticommutator of (Gra\ss mann-odd) supercharges yields, in particular, the (Gra\ss mann-even) `momentum'. The `odd' gauge supersymmetry that does the job was discovered very early on by de Azc\'arraga and Lukierski in \Rcite{deAzcarraga:1982njd} (for the superparticle), and subsequently also by Siegel in Refs.\,\cite{Siegel:1983hh,Siegel:1983ke} as a `hidden' supersymmetry of the Green--Schwarz super-$\si$-model for the superparticle and the superstring with a super-Minkowskian supertarget. As it did \emph{not} preserve separately the metric or the topological terms of the super-$\si$-model in the standard Nambu--Goto formulation, but only a suitably relatively normalised combination thereof, its existence was later employed as a constraint in the construction of super-$\si$-models for higher-dimensional $p$-branes and for curved supertargets, {\it cp}, {\it e.g.}, Refs.\,\cite{Metsaev:1998it,Park:1998un,Zhou:1999sm}. For a long time, the `odd' gauge supersymmetry, dubbed $\k$-symmetry in the string-theory literature, has kept its status of a useful \emph{odd}ity, definable solely through its function, which is that of restoration of a supersymmetric `balance' in the localised vacuum of the super-$\si$-model through identification of (typically) a half of the Gra\ss mann-odd degrees of freedom as pure gauge, and otherwise exhibiting a variety of intertwined peculiarities: 
\bit
\item it appeared in the infinitesimal (un-integrated) form exclusively;
\item the anticommutator of two $\k$-symmetry transformations would yield, depending on the structure of the super-$\si$-model, either a bare worldvolume diffeomorphism, or one corrected by a local transformation from the sector of the Lie algebra $\,\hgt\,$ of the isotropy group of the target homogeneous space (the algebra $\,\hgt\,$ invariably acting as a model of an infinitesimal hidden gauge symmetry) \emph{not} preserving the vacuum, or both additionally augmented with an extra local transformation vanishing only in the vacuum of the theory;
\item the `$\k$-symmetry algebra' spanned on $\k$-symmetries, worldvolume diffeomorphisms and -- occasionally -- correcting hidden gauge transformations as above would close, if at all, only upon imposition of field equations of the theory and further constraints.
\eit
Its additional `peculiarity': the `wrong sign' of the $\k$-symmetry variation in comparison with the `right sign' of a global-supersymmetry transformation was soon interpreted as a \emph{sign} of its geometric origin, to wit, linearisation of a right translation on the Lie supergroup $\,\txG\,$ in which the homogeneous space $\,\txG/\txH\,$ is embedded (locally) with the help of a section of the surjective submersion $\,\txG\too\txG/\txH$,\ {\it cp} \Rcite{McArthur:1999dy,Gomis:2006wu}. The latter \emph{ad hoc} interpretation does not really seem to have enhanced our understanding of the actual geometric nature of $\k$-symmetry or explained the above peculiarities, and so as of this writing, it is still being described in the literature\footnote{It ought to be remarked that the so-called `superembedding formalism' has been developed by Sorokin, {\it cp} \Rcite{Sorokin:1989zi} and also \Rcite{Sorokin:1999jx} for a comprehensive review, in which a `canonical' Gra\ss mann-odd extension of the Gra\ss mann-even worldvolume permits to model $\k$-symmetry as an `odd' superdiffeomorphism of the extended (super-)worldvolume, and -- in this manner -- de-geometrise $\k$-symmetry entirely, as seen from the supertarget perspective. The purely internal supersymmetry is now transmitted back to the supertarget only upon embedding the (super-)worldvolume in it in a consistent manner. This reformulation goes, in a sense, in the direction precisely opposite to the one that we wish to pursue, motivated by the conviction of the necessity to gerbify (pre)quatisable symmetries, and the adherence to the original interpretation, due to Freed, of the super-$\si$-model as a theory defined for the mapping-supermanifold \emph{functor} rather than a set of (super)mappings with a rigidly fixed superdomain.} as `a ``hidden'' fermionic symmetry with no evident geometric interpretation' but with a well-defined field-theoretic r\^ole to play.\medskip

In the present paper, we intend to unequivocally establish the status of the above `odd' gauge supersymmetry entirely in terms of the geometry of the supertarget of the super-$\si$-model, without changing the latter's deep nature of a theory with a functorial mapping supermanifold as the domain, and subsequently verify the existence of a higher-geometric realisation of the gauge symmetry in the familiar form of an equivariant structure on the $p$-gerbe associated with the super-$\si$-model. In the light of the previous remarks on the behaviour, under a $\k$-symmetry transformation, of the two terms in the Dirac--Feynman amplitude of the super-$\si$-model in the Nambu--Goto formulation, this task requires a transcription of the field theory into an equivalent picture in which the information on the supertarget metric is carried by extra components of the embedding field and the metric term in the amplitude is replaced by the standard pullback, along the extended embedding field, of a topological object. This seemingly wild scenario is actually realised in a large class of super-$\si$-models of the type discussed for which purely topological duals exist with the information on the metric structure and its coupling to the super-$p$-brane worldvolume encoded in a choice of a $(p+1)$-dimensional Gra\ss mann-even subspace $\,\tgt_{\rm vac}^{(0)}\,$ in the complement $\,\tgt\,$ of the isotropy algebra $\,\hgt\,$ in $\,\ggt$,\ termed the vacuum subspace and $\ad$-stabilised by a subalgebra $\,\hgt_{\rm vac}\subset\hgt\,$ with the corresponding Lie subgroup $\,\txH_{\rm vac}\subset\txH$,\ a family of non-dynamical Goldstone modes associated with the spontaneous breakdown $\,\hgt\searrow\hgt_{\rm vac}\,$ of the hidden gauge symmetry and hence labelled by $\,\hgt/\hgt_{\rm vac}$,\ and a trivial $\hgt_{\rm vac}$-equivariant Cartan--Eilenberg super-$p$-gerbe determined uniquely by the geometrised volume form on the vacuum subspace, {\it cp} \Rcite{Suszek:2019cum}. Such a reformulation of the super-$\si$-model was first considered by Hughes and Polchinski in \Rcite{Hughes:1986dn} in the context of partial spontaneous breaking of global supersymmetry, later elaborated significantly by Gauntlett {\it et al.} in \Rcite{Gauntlett:1989qe}, and more recently used by McArthur ({\it cp} \Rcite{McArthur:1999dy,McArthur:2010zm}) and West {\it et al.} ({\it cp} Refs.\,\cite{West:2000hr,Gomis:2006xw,Gomis:2006wu}) in the construction of (super-)$\si$-models for (super-)$p$-branes in the broader context of nonlinear realisations of (super)symmetries that had been investigated from a variety of angles in Refs.\,\cite{Schwinger:1967tc,Weinberg:1968de,Coleman:1969sm,Callan:1969sn,Salam:1969rq,Salam:1970qk,Isham:1971dv,Volkov:1972jx,Volkov:1973ix,Ivanov:1978mx,Lindstrom:1979kq,Uematsu:1981rj,Ivanov:1982bpa,Samuel:1982uh,Ferrara:1983fi,Bagger:1983mv}. The general conditions to be satisfied by the triple $\,(\txG,\txH,\txH_{\rm vac})\,$ for the duality to obtain were identified by the Author in \Rxcite{Thms.\,5.1 \& 5.2}{Suszek:2019cum}, and the dual super-$\si$-model was given the name of the Hughes--Polchinski formulation of the Green--Schwarz super-$\si$-model. Its manifestly topological nature opens an avenue for geometrisation of the entire content of the field theory, that is its field equations, states and symmetries, and for a re-interpretation -- intuited from the study of well-known examples of topological gauge field theories (such as, {\it e.g.}, the three-dimensional Chern--Simons theory) -- of a distinguished class of its geometrised gauge supersymmetries in terms of the local (super)geometry of the critically embedded worldvolume, or the vacuum of the theory. The obvious expectation is that this class would contain the Hughes--Polchinski dual of the `odd' $\k$-symmetry of the Nambu--Goto formulation, and that in a distinguished r\^ole. Furthermore, with the higher-geometric object for the \emph{complete} dual Dirac--Feynman amplitude in hand, originally constructed in \Rxcite{Sec.\,6.2}{Suszek:2019cum} and dubbed the extended Hughes--Polchinski (HP) $p$-gerbe, the possibility arises to straightforwardly verify the existence of a lift of these distinguished gauge supersymmetries to an equivariant structure of sorts on the extended HP $p$-gerbe. A rigorous concretisation of the general ideas articulated above has been the prime objective of the work reported in the present paper.\medskip

The paper is organised as follows.
\bit
\item In Section \ref{sec:sCart}, we review the supergeometry of a homogeneous space of a Lie supergroup in the convenient language of super-Harish--Chandra pairs. In particular, we give a precise description of a decomposition of the supersymmetry algebra $\,\ggt\,$ required by the dual formulations of the super-$\si$-model of interest and subsequently describe in great detail the construction of a (complete) family of local trivialisations \eqref{eq:vacgauge} and \eqref{eq:novacgauge} of the principal superfibrations $\,\txG\too\txG/\txK\,$ with the Gra\ss mann-even structure group $\,\txK\in\{\txH,\txH_{\rm vac}\}\,$ associated with this decomposition and employed in the explicit statement of the duality. We also indicate those elements of the (super-)Cartan calculus on $\,\txG\,$ that descend to the homogeneous space, and so can be used in the construction of a field theory with $\,\txG/\txK\,$ as the fibre of the covariant configuration bundle.
\item In Section \ref{sec:physmod}, we define the field theory of interest in its both formulations alluded to above and state the duality between them in what constitutes an extension, given in Thm.\,\ref{thm:HpdualNGext}, of the former result of \Rxcite{Thms.\,5.1 \& 5.2}{Suszek:2019cum} to a larger class of super-$\si$-models. Upon giving the definition \eqref{eq:HPsec} of a supermanifold $\,\Si^{\rm HP}\,$ for the topological Hughes--Polchinski (HP) formulation (determined by a collection $\,\sgt\Bgt_{p,\la_p}^{\rm (HP)}\,$ of supergeometric data termed the HP superbackground) on which all our subsequent supergeometric and field-theoretic analysis takes place, and a convenient description of its tangent sheaf $\,\cT\Si^{\rm HP}\,$ in Prop.\,\ref{prop:bastanshHP}, we then provide a differential-geometric interpretation of the duality in terms of a sub-superdistribution $\,{\rm Corr}_{\rm HP/NG}(\sgt\Bgt_{p,\la_p}^{\rm (HP)})\subset\cT\Si^{\rm HP}\,$ introduced in Def.\,\ref{def:Corrsdistro}. The Section concludes with a detailed exposition, in Examples \ref{eq:s0gsMink}-\ref{eg:MT1}, of the relevant supergeometric structure in the existing super-$\si$-models of the type considered. 
\item In Section \ref{sec:vac}, the Euler--Lagrange equations of the Green--Schwarz super-$\si$-model for the super-$p$-brane in the HP formulation are derived under some natural assumptions, later shown to be satisfied in all the examples reviewed previously. Among the conditions, we find the physically motivated requirement of existence of the `reduction' mechanism in the vacuum by means of an `odd' gauge symmetry, mentioned above. The description of the classical vacuum of the field theory thus obtained, and summarised in Prop.\,\ref{prop:HPELeqs}, is subsequently reformulated in terms of a superdistribution $\,{\rm Vac}_{\rm HP/NG}(\sgt\Bgt_{p,\la_p}^{\rm (HP)})\subset{\rm Corr}_{\rm HP/NG}(\sgt\Bgt_{p,\la_p}^{\rm (HP)})$,\ introduced as the vacuum superdistribution in Def.\,\ref{def:vacHPsdistro}, to which the tangents of the critical embeddings are restricted by the field equations of the theory. Various symmetry and equivariance properties of this explicit \emph{geometrisation} of the field equations of the super-$\si$-model are considered, and the conditions of integrability of the vacuum superdistribution into what we are right to call, in Def.\,\ref{def:HPvacfol}, the vacuum foliation of the supertarget are established, in a purely superalgebraic language, in Prop.\,\ref{prop:vacsalg}. The abstract considerations are, once again, illustrated on the physical Examples \ref{eg:GSs0Vac}-\ref{eg:MTs1Vac}.  
\item In Section \ref{sec:susy}, a complete geometrisation of the supersymmetries -- both global and local -- of the super-$\si$-model in the HP formulation is worked out. A universal phenomenon of gauge-symmetry enhancement is discovered for the field configurations, distinguished by the HP/NG duality, with tangents in $\,{\rm Corr}_{\rm HP/NG}(\sgt\Bgt_{p,\la_p}^{\rm (HP)})$,\ leading to the definition of an enhanced gauge-symmetry superdistribution $\,\cG\cS(\sgt\Bgt_{p,\la_p}^{\rm (HP)})\subset{\rm Corr}_{\rm HP/NG}(\sgt\Bgt_{p,\la_p}^{\rm (HP)})\,$ that realises the gauge symmetries and its odd-generated sub-superdistribution aligned with the vacuum superdistribution -- the \textbf{$\k$-symmetry superdistribution} $\,\k(\sgt\Bgt_{p,\la_p}^{\rm (HP)})\subset{\rm Vac}_{\rm HP/NG}(\sgt\Bgt_{p,\la_p}^{\rm (HP)})$.\ The latter is \textbf{\emph{the sought-after geometrisation of $\k$-symmetry}}. Its explicit description formulated in Def.\,\ref{def:ksymmsdistro} (on the basis of a field-theoretic calculation), in conjunction with the definition of a distinguished global-supersymmetry subspace $\,S_\txG^{\rm HP}\subset\G(\cT\Si^{\rm HP})\,$ induced by the global symmetries and its restriction $\,S_\txG^{\rm HP, vac}\subset\G({\rm Vac}(\sgt\Bgt_{p,\la_p}^{\rm (HP)}))\,$ to the vacuum foliation described in Prop.\,\ref{prop:resglobsusysub} under the name of the residual global-supersymmetry subspace and modelled on a Lie superalgebra $\,\sgt_{\rm vac}$,\ are the concrete substance of the said geometrisation. The various components thereof are subsequently put in physically meaningful relations with one another and with the formerly estalished elements of geometrisation of the super-$\si$-model, whereby conditions for the global-supersymmetry invariance of the vacuum superdistribution and of its $\k$-symmetry sub-superdistribution are determined in Props.\,\ref{prop:descissusy} and \ref{prop:kglobsusy}, respectively. The `odd' gauge supersymmetry of the super-$\si$-model is definitively demystified in Prop.\,\ref{prop:kwdfinvac} which identifies the $\k$-symmetry superdistribution as a universal odd(-generated) component of the superdistribution of enhanced gauge symmetries of the super-$\si$-model which in the physically favoured circumstances of integrability (and then automatically global supersymmetry) of the vacuum superdistribution generates the latter -- through the so-called weak derived flag with the limit $\,\k^{-\infty}(\sgt\Bgt_{p,\la_p}^{\rm (HP)})\equiv{\rm Vac}_{\rm HP/NG}(\sgt\Bgt_{p,\la_p}^{\rm (HP)})\,$ modelled on a Lie su-superalgebra $\,\ggt\sgt_{\rm loc}(\sgt\Bgt_{p,\la_p}^{\rm (HP)})\subset\ggt\,$ of vacuum-generating gauge supersymmetries, or the \textbf{$\k$-symmetry superalgebra} -- and so envelops the vacuum foliation of the supertarget, which wins it its name: the \textbf{square root of the vacuum} featuring in the title of the present paper. The various findings of the Section of physical relevance are summarised in the fundamental Thm.\,\ref{thm:kdemyst}. The Section closes with physical Examples \ref{eg:GSs0k}-\ref{eg:MTs1k}. The identification stated above rests upon a number of assumptions concerning the superbackground, listed explicitely in this section and verified in the examples. An important scenario in which these assumptions are not satisfied -- the only such case among the concrete physical systems explored -- in scrutinised in Example \ref{eg:sqroots1bsMink}. 
\item In Section \ref{sec:hgeomphys}, we recall the essential aspects of the gerbe-theoretic approach to $\si$-models and recapitulate the geometrisation scheme of Refs.\,\cite{Suszek:2017xlw,Suszek:2019cum,Suszek:2018bvx,Suszek:2018ugf} for the super-$\si$-models under study, abstracting, along the way, several concrete definitions from the general considerations presented in the original papers. In the exposition, special emphasis is laid on the higher-geometric realisation of $\si$-model symmetries in both instantiations: families of $p$-gerbe (1-)isomorphisms labelled by the global-symmetry group for global symmetries and equivariant structures on the $p$-gerbe over nerves of the action groupoid of the symmetry group on the base of the $p$-gerbe for gauge symmetries.
\item In Section \ref{sec:susykappa}, we take the first step on the path towards a full-fledged gerbification of the vacuum-generating $\k$-symmetry, the latter being understood as the \emph{involutive} superdistribution $\,\k^{-\infty}(\sgt\Bgt_{p,\la_p}^{\rm (HP)})\,$ modelled on the \emph{Lie superalgebra} $\,\ggt\sgt_{\rm loc}(\sgt\Bgt_{p,\la_p}^{\rm (HP)})$,\ consistent with the residual global supersymmetry of the vacuum realised by the residual global-supersymmetry subspace $\,S_\txG^{\rm HP, vac}$.\ Thus, upon restricting the extended HP $p$-gerbe of the super-$\si$-model to the vacuum foliation $\,\Si^{\rm HP}_{\rm vac}\,$ of the HP section $\,\Si^{\rm HP}$,\ we equip the latter with a linearised $\k$-symmetry-equivariant structure, linearised-invariant with respect to the residual global supersymmetry of the vacuum. The definition of the latter object is derived through a cohomologically consistent linearisation (in the vicinity of the group unit of the local-symmetry group) of a standard equivariant structure on a $p$-gerbe in the Gra\ss mann-eve setting, and an analogous linearisation (in the vicinity of the group unit of the global-symmetry group) of a group-invariant structure on that equivariant structure. Our considerations lead to the statement of existence of a \emph{canonical} $\ggt\sgt_{\rm loc}(\sgt\Bgt_{p,\la_p}^{\rm (HP)})$-equivariant structure on the vacuum restriction of the extended HP $p$-gerbe in Prop.\,\ref{prop:canequivstr}, further proven \emph{canonically} $\sgt_{\rm vac}$-invariant in Prop.\,\ref{canextinvstr}. The results of immediate physical relevance are summarised in the fundamental Thm.\,\ref{thm:kappagerbified}.
\item In Section \ref{ref:CandO}, we give a concise summary of our findings and indicate directions of future research opened and motivated by them.
\item In the Appendix, we provide an overview of the differential-geometric and superalgebraic structure behind the super-$\si$-model for the Zhou super-0-brane in $\,{\rm s}({\rm AdS}_2\x\bS^2)\,$ that violates some of the previous assumptions made with regard to the super-$\si$-model data, and yet realises in an interestingly extended form the geometrisation scenario of delineated in Sections \ref{sec:vac}, \ref{sec:susy} and \ref{sec:susykappa}. The Appendix is to be regarded as an appetiser for a future investigation of a wider class of super-$\si$-models in the topological HP formulation in the spirit of the work reported herein.
\eit
\bigskip

\section{Supergeometry \` a la Cartan}\label{sec:sCart}

In the present paper, we aim to study distinguished lagrangean field theories that model simple geometric dynamics of charged extended objects of superstring theory, termed super-$p$-branes. The field theories of interest exhibit supersymmetry in a nonlinear realisation, and so it seems only natural to begin our discussion with a word on the supermanifolds of relevance to the construction of the corresponding configuration bundles in the adapted Cartan framework. For that purpose, we introduce a Lie supergroup $\,\txG\,$ that will play the r\^ole of the \textbf{supersymmetry group} of the physical model. A Lie supergroup can be understood abstractly (and naturally) as a group object in the category $\,\sMan \,$ of supermanifolds (cp \Rxcite{Sec.\,3}{Suszek:2017xlw}) with the body given by a Lie group $\,|\txG|$,\ and so, in particular, it comes equipped with a multiplication (supermanifold) morphism $\,\mu\ :\ \txG\x\txG\too\txG\,$ with the sheaf component $\,\mu^*\ :\ \cO_\txG\too\cO_{\txG\x\txG}\cong\cO_\txG\widehat\ox\cO_\txG\,$ (the tensor product is a suitable completion of the standard (super)tensor product of sheaves) and a unit morphism $\,e_\txG\equiv\widehat e\ :\ \bR^{0|0}\too\txG\,$ defining the group unit $\,e\in|\txG|$,\ one of the $|\txG|$-worth $\,\mor_\sMan (\bR^{0|0},\txG)\equiv\{\widehat g\ :\ \bR^{0|0}\too\txG\}_{g\in|\txG|}\,$ of topological points. There is an obvious notion of a Lie-supergroup morphism, and so there arises the \textbf{category of Lie supergroups} which we shall denote as $\,\sLieGrp$.

Among vector fields on the supermanifold $\,\txG$,\ {\it i.e.}, among global sections of the sheaf of superderivations $\,\tx{sDer}\,\cO_\txG\equiv\cT\txG\,$ of the structure sheaf, we have the left-invariant sections satisfying the condition 
\qq\nn
\bigl(\id_{\cO_\txG}\ox L\bigr)\circ\mu^*=\mu^*\circ L\,,
\qqq 
and their right-invariant counterparts for which
\qq\nn
\bigl(R\ox\id_{\cO_\txG}\bigr)\circ\mu^*=\mu^*\circ R\,.
\qqq 
The $\bR$-linear span of the former shall be denoted as $\,\G(\cT\txG)^\txG$.\ The Lie superbracket of vector fields closes on these sections and thus gives rise to the tangent Lie superalgebra $\,{\rm sLie}\,\txG\,$ of $\,\txG$,
\qq\nn
{\rm sLie}\,\txG=\bigl(\G(\cT\txG)^\txG,[\cdot,\cdot\}\bigr)\,,
\qqq
an object from the category $\,\sLieAlg\,$ of Lie superalgebras. We have a canonical isomorphism of supervector spaces 
\qq\nn
L_\cdot\ :\ \cT_e\txG\too\G(\cT\txG)^\txG\ :\ X\longmapsto\bigl(\id_{\cO_\txG}\ox X\bigr)\circ\mu^*\,.
\qqq
Similarly, every right-invariant vector field can be obtained from a unique vector tangent to $\,\txG\,$ at the topological unit as
\qq\nn
R_X=\bigl(X\ox\id_{\cO_\txG}\bigr)\circ\mu^*\,,\qquad X\in\ggt\,.
\qqq

Following Kostant (cp \Rcite{Kostant:1975}), we shall, equivalently, think of a Lie supergroup as an object described in 
\bedef
A \textbf{super-Harish--Chandra pair} is a pair $\,(|\txG|,\ggt)\,$ composed of
\bit
\item[-] a Lie group $\,|\txG|\,$ with the tangent Lie algebra $\,|\ggt|$;
\item[-] a Lie superalgebra $\,\ggt$,\ to be termed the \textbf{supersymmetry algebra}, with the Gra\ss mann-homogeneous components: $\,\ggt^{(0)}\,$ (even) and $\,\ggt^{(1)}\,$ (odd),
\qq\nn
\ggt=\ggt^{(0)}\oplus\ggt^{(1)}
\qqq
\eit
such that
\bit
\item[(sHCp1)] $\,\ggt^{(0)}\equiv|\ggt|$;
\item[(sHCp2)] there exists a realisation 
\qq\nn
\rho_\cdot\ :\ |\txG|\too{\rm Aut}_\sLieAlg\,\ggt\ :\ g\longmapsto\rho_g
\qqq 
of $\,|\txG|\,$ on $\,\ggt\,$ that extends its adjoint realisation on the tangent Lie algebra, 
\qq\nn
\forall_{g\in|\txG|}\ :\ \rho_g\rstr_{\ggt^{(0)}}\equiv\sfT_e\Ad_g\,.
\qqq
\eit

A \textbf{sHCp morphism} between super-Harish--Chandra pairs $\,(|\txG_A|,\ggt_A),\ A\in\{1,2\}$,\ with the respective units $\,e_A\,$ and realisations $\,\rho^A_\cdot\ :\ |\txG_A|\too{\rm Aut}_\sLieAlg\,\ggt_A$,\ is a pair 
\qq\nn
(\Phi,\phi)\ :\ \bigl(|\txG_1|,\ggt_1\bigr)\too\bigl(|\txG_2|,\ggt_2\bigr)
\qqq 
composed of 
\bit
\item[-] a Lie-group homomorphism $\,\Phi\ :\ |\txG_1|\too|\txG_2|$;
\item[-] a Lie-superalgebra homomorphism $\,\phi\ :\ \ggt_1\too\ggt_2$
\eit
such that 
\bit
\item[(sHCpm1)] $\,\phi\rstr_{\ggt_1^{(0)}}\equiv\sfT_{e_1}\Phi$;
\item[(sHCpm2)] $\,\forall_{g\in|\txG_1|}\ :\ \phi\circ\rho^1_g=\rho^2_{\Phi(g)}\circ\phi$.
\eit

Super-Harish--Chandra pairs together with the associated sHCp morphisms form the \textbf{category of super-Harish--Chandra pairs}, to be denoted as $\,\sHCp$.
\exdef 
\noindent We have the fundamental
\bethe[Kostant '77]
There exists an equivalence of categories
\qq\nn
\xcK\ :\ \sHCp\xrightarrow{\ \cong\ }\sLieGrp\,.
\qqq
\ethe
\brem
Given the ubiquity of Kostant's supergroups in field-theoretic models of systems with supersymmetry, and -- in particular -- with view to our subsequent considerations, it will be apposite to explicit at least some of the structures that arise in the constructive proof of the above theorem. In so doing, we adopt the conventions of \Rcite{Carmeli:2011}. 

Thus, in one direction, the equivalence $\,\xcK^{-1}\,$ assigns to a Lie supergroup $\,\txG\,$ the super-Harish--Chandra pair
\qq\nn
\bigl(|\txG|,{\rm sLie}\,\txG\bigr)
\qqq
composed of its body Lie group $\,|\txG|\,$ and its tangent Lie superalgebra together with the mapping $\,{\rm sAd}_\cdot\ :\ |\txG|\too{\rm Aut}_\sLieAlg({\rm sLie}\,\txG)\,$ given by
\qq\nn
{\rm sAd}_g\ :\ {\rm sLie}\,\txG\circlearrowleft\ :\ L\longmapsto c^*_{g^{-1}}\circ L\circ c^*_g\,,\qquad g\in|\txG|\,,
\qqq
where
\qq\nn
c^*_g=\bigl(\ev_g\ox\id_{\cO_\txG}\ox\ev_{g^{-1}}\bigr)\circ\bigl(\id_{\cO_\txG}\ox\mu^*\bigr)\circ\mu^*
\qqq
is expressed in terms of the \textbf{evaluation} $\,\ev_g\,$ of sections of $\,\cO_\txG\,$ at the topological point $\,g$,\ assigning to a section $\,f\,$ the unique real number $\,\ev_g(f)\equiv f(g)\in\bR\,$ such that the corrected section $\,f-f(g)\,$ is not invertible in any neighbourhood of $\,g$,\ {\it cp} \Rxcite{Lem.\,4.2.2}{Carmeli:2011}.

Going in the opposite direction is significantly more involved. Let $\,(|\txG|,\ggt)\,$ be a super-Harish--Chandra pair with the realisation $\,\rho_\cdot\ :\ |\txG|\too{\rm Aut}_\sLieAlg(\ggt)$,\ and denote by $\,\txU(\ggt)\,$ (resp.\ by $\,\txU(\ggt^{(0)})$) the \textbf{universal enveloping algebra} of the Lie superalgebra $\,\ggt\,$ (resp.\ of the Gra\ss mann-even component $\,\ggt^{(0)}\,$ of $\,\ggt$), as defined in \Rxcite{Chap.\,I\,\S\,2.1}{Scheunert:1979}. The latter is an example of an associative superalgebra, and we shall denote the category of all such superalgebras as $\,{\rm {\bf sAlg_{assoc}}}\,$ in what follows. Furthermore, the $\bZ/2\bZ$-graded vector space $\,\txU(\ggt)\,$ is equipped with a natural Hopf-superalgebra structure, with the product $\,\txm_{\txU(\ggt)}\equiv\cdot_{\txU(\ggt)}\ :\ \txU(\ggt)\ox\txU(\ggt)\too\txU(\ggt)\,$ descended from the (tensor) product on the tensor algebra of $\,\ggt$,\ and the coproduct $\,\D_{\txU(\ggt)}\ :\ \txU(\ggt)\too\txU(\ggt)\ox\txU(\ggt)\,$ and the antipode $\,S_{\txU(\ggt)}\ :\ \txU(\ggt)\circlearrowleft\,$ determined by the respective restrictions to the generators $\,X\in\ggt\,$ of $\,\txU(\ggt)\,$ (we are identifying $\,\ggt\,$ with its image in $\,\txU(\ggt)\,$ under the canonical injection $\,\ggt\emb\txU(\ggt)$) and the algebra unit,
\qq\nn
\D_{\txU(\ggt)}(\bd1):=\bd1\ox\bd1\,,\qquad\quad\D_{\txU(\ggt)}(X):=X\ox\bd1+\bd1\ox X\,,\qquad\qquad S_{\txU(\ggt)}(\bd1)=\bd1\,,\qquad\quad S_{\txU(\ggt)}(X)=-X\,,
\qqq
the former being a superalgebra homomorphism, and the latter -- an even super-antimultiplicative mapping, {\it i.e.}, one satisfying, for any homogeneous $\,u,v\in\txU(\ggt)$,\ the identity
\qq\nn
S_{\txU(\ggt)}\circ\txm_{\txU(\ggt)}(u,v)=(-1)^{|u|\cdot|v|}\,\txm_{\txU(\ggt)}\bigl(S_{\txU(\ggt)}(v),S_{\txU(\ggt)}(u)\bigr)\,,
\qqq
in which $\,|u|,|v|\in\{0,1\}\,$ are the Gra\ss mann parities. To begin with, we identify two distinguished objects in the object class of the category $\,\txU(\ggt^{(0)}){\rm {\bf -Mod}}\,$ of (left) $\txU(\ggt^{(0)})$-modules, to wit: the enveloping algebra $\,\txU(\ggt)\,$ of $\,\ggt\,$ with the action engendered by $\,\txm_{\txU(\ggt)}\,$ (and using the obvious embedding $\,\txU(\ggt^{(0)})\emb\txU(\ggt)$), and, for any open set $\,|\cU|\subset|\txG|$,\ the space $\,C^\infty(|\cU|,\bR)\,$ of smooth functions on $\,|\cU|\,$ with the action induced by the natural action of the left-invariant vector fields associated with elements of $\,\ggt^{(0)}$.\ Next, we form the \textbf{coinduced $\txU(\ggt)$-module}\footnote{These were called produced representations in \Rcite{Scheunert:1979}.}
\qq\nn 
\mor_{\txU(\ggt^{(0)}){\rm {\bf -Mod}}}\bigl(\txU(\ggt),C^\infty\bigl(|\cU|,\bR\bigr)\bigr)=:\cO_\txG\bigl(|\cU|\bigr)
\qqq
of \Rxcite{Sec.\,1}{Koszul:1982} with the (left) $\txU(\ggt)$-action 
\qq\nn
\txU\la_\cdot\ :\ \txU(\ggt)\x\cO_\txG\bigl(|\cU|\bigr)\too\cO_\txG\bigl(|\cU|\bigr)
\qqq
which, for a homogeneous $\,u\in\txU(\ggt)$,\ reads
\qq\label{eq:Ulau}
\txU\la_u\ :\ \cO_\txG\bigl(|\cU|\bigr)\circlearrowleft\ :\ \chi\longmapsto(-1)^{|u|}\,\chi\circ\txm_{\txU(\ggt)}(\cdot,u)\,,
\qqq
and use the coproduct on $\,\txU(\ggt)\,$ to induce on it the structure of a supercommutative superalgebra, 
\qq\nn
&\txm_{\cO_\txG(|\cU|)}\equiv\cdot_{\cO_\txG}\ :\ \cO_\txG\bigl(|\cU|\bigr)\ox\cO_\txG\bigl(|\cU|\bigr)\too\cO_\txG\bigl(|\cU|\bigr)\,,&\cr\cr
&\txm_{\cO_\txG(|\cU|)}(f_1\ox f_2)\equiv f_1\cdot_{\cO_\txG}f_2=\txm_{C^\infty(|\cU|,\bR)}\circ(f_1\ox f_2)\circ\D_{\txU(\ggt)}\,,&
\qqq
where
\qq\nn
&\txm_{C^\infty(|\cU|,\bR)}\equiv\cdot_{C^\infty}\ :\ C^\infty(|\cU|,\bR)\ox C^\infty(|\cU|,\bR)\too C^\infty(|\cU|,\bR)\,,&\cr\cr 
&\txm_{C^\infty(|\cU|,\bR)}(f_1\ox f_2)\equiv f_1\cdot_{C^\infty}f_2\ :\ |\cU|\ni g\longmapsto f_1(g)\cdot f_2(g)\in\bR&
\qqq
is the standard pointwise product of (smooth) functions. We then define the supermanifold
\qq\nn
\txG:=\bigl(|\txG|,\cO_\txG\bigr)
\qqq
with the structure sheaf 
\qq\nn
\cO_\txG\equiv\mor_{\txU(\ggt^{(0)}){\rm {\bf -Mod}}}\bigl(\txU(\ggt),C^\infty(\cdot,\bR)\bigr)\ :\ \xcT\bigl(|\txG|\bigr)\too\scommAlg\ :\ |\cU|\longmapsto\cO_\txG\bigl(|\cU|\bigr)
\qqq
(here, $\,\xcT(|\txG|)\,$ is the category of open sets in $\,|\txG|\,$ with inclusions as morphisms, and $\,\scommAlg\,$ is the category of supercommutative superalgebras). The canonical (parity-factorised) structure of the latter is determined by the isomorphism 
\qq\label{eq:locmodGsh}   
\mor_{\txU(\ggt^{(0)}){\rm {\bf -Mod}}}\bigl(\txU(\ggt),C^\infty\bigl(|\cU|,\bR\bigr)\bigr)\xrightarrow{\ \cong\ }C^\infty\bigl(|\cU|,\bR\bigr)\ox\bigwedge{}^\bullet\ggt^{(1)\,*}\ :\ f\longmapsto f\circ\g
\qqq 
of \Rxcite{Lem.\,2}{Koszul:1982}, written in terms of the supercoalgebra homomorphism 
\qq\label{eq:gamap}
\g\ :\ \bigwedge{}^\bullet\ggt^{(1)}\too\txU(\ggt)
\qqq
that $\bR$-linearly extends the assignment (here, $\,X_i\in\ggt^{(1)}\subset\txU(\ggt),\ i\in\ovl{1,k}$)
\qq\nn
\g\bigl(X_1\wedge X_2\wedge\cdots\wedge X_k\bigr):=\tfrac{1}{k!}\,\sum_{\si\in\Sgt_k}\,{\rm sgn}(\si)\,X_{\si(1)}\cdot_{\txU(\ggt)}X_{\si(2)}\cdot_{\txU(\ggt)}\cdots\cdot_{\txU(\ggt)}X_{\si(k)}\,.
\qqq
Finally, we promote $\,\txG\,$ to the rank of a Lie supergroup by defining the structure (supermanifold) morphisms $\,\mu\ :\ \txG\x\txG\too\txG,\ \Inv\ :\ \txG\circlearrowleft\,$ and $\,\vep\ :\ \bR^{0|0}\too\txG\,$ with the obvious (Lie-group) body components and with the sheaf components
\qq\nn
\mu^*\ :\ \cO_\txG\too\cO_\txG\widehat\ox\cO_\txG\,,\qquad\qquad\Inv^*\ :\ \cO_\txG\circlearrowleft\,,\qquad\qquad\vep^*\ :\ \cO_\txG\too\bR
\qqq
that evaluate on $\,\cO_\txG\ni f\,$ as
\qq\nn
\bigl(\mu^*(f)(u\ox v)\bigr)(g,h)&=&\bigl(f\circ\txm_{\txU(\ggt)}\bigl(\txU\rho_{h^{-1}}(u),v\bigr)\bigr)(g\cdot h)\,,\cr\cr
\bigl(\Inv^*(f)(u)\bigr)(g)&=&\bigl(f\circ\txU\rho_g\circ S_{\txU(\ggt)}(u)\bigr)\bigl(g^{-1}\bigr)\,,\cr\cr
\vep^*(f)&=&\bigl(f(\bd1)\bigr)(e)
\qqq
for $\,u,v\in\txU(\ggt),\ g,h\in|\txG|$,\ for $\,e\,$ the group unit of $\,|\txG|$,\ and for 
\qq\nn
\txU\rho_\cdot\ :\ |\txG|\too{\rm Aut}_{{\rm {\bf sAlg_{assoc}}}}\bigl(\txU(\ggt)\bigr)
\qqq 
the unique extension of $\,\rho_\cdot\,$ to the universal enveloping algebra of $\,\ggt$.
\erem

\brem
For the better part of our discussion, we shall need a geometric perspective on the formal concepts and constructions that we stumble upon in our considerations, on which we shall base our intuitions. Such a perspective is provided by a collection of fundamental category-theoretic results that introduce a familiar structure into the supergeometric universe. 

The first of these results is
\bethe[Yoneda's Lemma for $\,\sMan $]
There exists a fully faithful covariant functor, termed the {\bf Yoneda embedding}, from the category $\,\sMan\,$ into the category of presheaves\footnote{This is the category of contravariant functors from $\,\sMan\,$ to $\,\Set$,\ with natural transformations as morphisms.} $\,\Presh (\sMan )\,$ on the latter category,
\qq\nn
\Yon_\cdot\ :\ \sMan\emb\Presh \bigl(\sMan \bigr)\,,
\qqq
with the object component
\qq\nn
\Yon_\cM:=\mor_{\sMan }\bigl(\cdot,\cM\bigr)\,,\qquad\qquad\cM\in\obj\,\sMan 
\qqq
and the morphism component
\qq\nn
\Yon_\phi:=\mor_{\sMan }\bigl(\cdot,\phi\bigr)\equiv\phi\circ\,,\qquad\qquad\phi\in\mor_\sMan\bigl(\cM_1,\cM_2\bigr)\,.
\qqq
\ethe
\noindent The Lemma enables us to replace the study of a supermanifold $\,\cM\,$ with the study of the $\obj\,\sMan$-indexed family of sets $\,\{\Yon_\cM(\cS)\}_{\cS\in\obj\,\sMan}\,$ of the so-called {\bf $\cS$-points of $\,\cM$}. Note that the topological points in $\,\cM\,$ are among the $\cS$-points for any supermanifold $\,\cS$.\ Indeed, point $\,x\in|\cM|\,$ can be identified with a unique morphism $\,\widehat x\ :\ \bR^{0|0}\too\cM\,$ from the \emph{terminal} supermanifold $\,\bR^{0|0}\,$ into $\,\cM$,\ and the existence of $\,\widehat x\,$ implies the existence of the canonical morphism $\,\cS\dashrightarrow\bR^{0|0}\xrightarrow{\ \widehat x\ }\cM$.\ In the case of a Lie supergroup $\,\cM\equiv\txG\in\obj\,\sLieGrp$,\ every set $\,\Yon_\txG(\cS)\,$ of $\cS$-points of $\,\txG\,$ is actually a \emph{group}, and morphisms between Lie supergroups are mapped to group homomorphisms.

The next important result is
\bethe
There exists a fully faithful contravariant functor from the category $\,\sMan\,$ of supermanifolds into the category $\,\scommAlg$,
\qq\nn
\cA_\cdot\ :\ \sMan\emb\scommAlg\,,
\qqq
with the object component 
\qq\nn
\cA_\cM:=\cO_\cM\bigl(|\cM|\bigr)\,,\qquad\qquad\cM\in\obj\,\sMan
\qqq
and the morphism component
\qq\nn
\cA_{(|\phi|,\phi^*)}:=\phi_{|\cM_2|}^*\ :\ \cO_{\cM_2}\bigl(|\cM_2|\bigr)\too\cO_{\cM_1}\bigl(|\cM_1|\bigr)\,,\qquad\qquad\bigl(|\phi|,\phi^*\bigr)\in\mor_\sMan\bigl(\cM_1,\cM_2\bigr)\,.
\qqq
\ethe
\noindent The theorem reduces the analysis on supermanifolds to the analysis of the corresponding superalgebras of \emph{global} sections of their structure sheaves and superalgebra homomorphisms between them. Furthermore, it yields (in conjunction with the Hadamard Lemma) a natural description of the set $\,{\rm Yon}_{\bR^{p|q}}(\cS)\,$ of $\cS$-points in the model supermanifold $\,\bR^{p|q}$,\ and so -- when combined with the previous result -- paves the way to a `functional' formulation of the (local) differential calculus on supermanifolds. 

The way goes through yet another fundamental
\bethe[The Local Chart Theorem]
There is a bijection between -- on the one hand -- supermanifold morphisms $\,\phi\in\mor_\sMan(\cU_1,\cU_2)\,$ between superdomains $\,\cU_A\equiv(|\cU_A|,C^\infty(\cdot,\bR)\ox\bigwedge{}^\bullet\,\bR^{q_A}),\ |\cU_A|\subset\bR^{p_A},\ A\in\{1,2\}\,$ with the respective canonical coordinates $\,\{\theta_A^{\a_A},x_A^{a_A}\}^{(\a_A,a_A)\in\ovl{1,q_A}\x\ovl{1,p_A}}\,$ of Gra\ss mann parities $\,(|\theta_A^{\a_A}|,|x_A^{a_A}|)=(1,0)\,$ and -- on the other hand -- collections of global sections from $\,\cO_{\cU_1}(|\cU_1|)$:\ $\,q_2\,$ Gra\ss mann-odd ones $\,\{\widetilde\theta_2^{\a_2}\}^{\a_2\in\ovl{1,q_2}}\,$ and $\,p_2\,$ Gra\ss mann-even ones $\,\{\widetilde x^{a_2}_2\}^{a_2\in\ovl{1,p_2}}$,\ satisfying the condition 
\qq\nn
\forall_{m\in|\cU_1|}\ :\ \bigl(\widetilde x^1_2(m),\widetilde x^2_2(m),\ldots,\widetilde x^{p_2}_2(m)\bigr)\in|\cU_2|\,.
\qqq
\ethe
\noindent It leads to a straightforward extension of the standard local description of a manifold in terms of local coordinates and that of mappings between manifolds expressed in terms of such coordinates on the domain and codomain of the mapping, the extension in question consisting in the incorporation of Gra\ss mann-odd coordinates into the analysis on par with the standard Gra\ss mann-even ones. Thus, a morphism 
\qq\nn
\phi\equiv\bigl(|\phi|,\phi^*\bigr)\ :\ \cM_1\too\cM_2
\qqq
between supermanifolds $\,\cM_A,\ A\in\{1,2\}\,$ of the respective superdimensions $\,(p_A|q_A)$,\ with a restriction
\qq\nn
\phi\rstr_{\cU_1}\ :\ \cU_1\too\cU_2\,,\qquad\qquad\cU_A\equiv\bigl(|\cU_A|,\cO_{\cM_A}\rstr_{|\cU_A|}\bigr)\,,\quad A\in\{1,2\}
\qqq
to domains $\,\cU_A\,$ of the respective local coordinate charts 
\qq\nn
\k_A\ :\ \cU_A\xrightarrow{\ \cong\ }\bigl(|\cW_A|,C^\infty(\cdot,\bR)\ox\bigwedge{}^\bullet\bR^{\x q_A}\equiv\cO_{\cW_A}\bigr)\equiv\cW_A\,,\qquad\qquad|\cW_A|\subset\bR^{\x p_A}\,,
\qqq
with the corresponding canonical coordinates $\,\{\theta_A^{\a_A},x_A^{a_A}\}^{(\a_A,a_A)\in\ovl{1,q_A}\x\ovl{1,p_A}}$,\ acquires a coordinate presentation in the form of a collection of $\,p_2+q_2\,$ mappings 
\qq\nn
\bigl\{\phi_{21}^*\bigl(\theta_2^{\a_2}\bigr),\phi_{21}^*\bigl(x_2^{a_2}\bigr)\bigr\}^{(\a_2,a_2)\in\ovl{1,q_2}\x\ovl{1,p_2}}
\qqq
of the respective parities $\,|\phi_{21}^*(\theta_2^{\a_2})|=1\,$ and $\,|\phi_{21}^*(x_2^{a_2})|=0$,\ determined by the composite morphism
\qq\nn
\phi_{21}:=\k_2\circ\phi\circ\k_1^{-1}\,.
\qqq
As sections of $\,\cO_{\cW_1}$,\ the $\,\phi_{21}^*(\theta_2^{\a_2})\,$ and $\,\phi_{21}^*(x_2^{a_2})\,$ admit coordinate presentations
\qq\nn
\phi_{21}^*\bigl(\theta_2^{\a_2}\bigr)=\sum_{k=0}^{q_1}\,\theta_1^{\a_1^1}\,\theta_1^{\a_1^2}\,\cdots\,\theta_1^{\a_1^k}\,\phi_{21\,\a_1^1\a_1^2\ldots\a_1^k}^{(1)}\bigl(x_1^{a_1}\bigr)\,,\qquad\qquad\phi_{21}^*\bigl(x_2^{a_2}\bigr)=\sum_{l=0}^{q_1}\,\theta_1^{\a_1^1}\,\theta_1^{\a_1^2}\,\cdots\,\theta_1^{\a_1^l}\,\phi_{21\,\a_1^1\a_1^2\ldots\a_1^l}^{(0)}\bigl(x_1^{b_1}\bigr)\,,
\qqq
with $\,\phi_{21\,\a_1^1\a_1^2\ldots\a_1^l}^{(1)}=0=\phi_{21\,\a_1^1\a_1^2\ldots\a_1^{l+1}}^{(0)}\,$ for $\,l\in 2\bN$.\ This imitates a `mapping' 
\qq\nn
\bigl(\theta_1,x_1\bigr)\longmapsto\bigl(\theta_2\bigl(\theta_1,x_1\bigr),x_2\bigl(\theta_1,x_1\bigr)\bigr)\equiv\bigl(\phi_{21}^*\bigl(\theta_2\bigr),\phi_{21}^*\bigl(x_2\bigr)\bigr)
\qqq 
of `points' in the $\,\cM_A$.\ The scheme induces a (local) mapping between $\cS$-points
\qq\nn
\mor_\sMan(\cS,\cU_1)\ni\psi\too\phi\circ\psi\in\mor_\sMan(\cS,\cU_2)\,,
\qqq
with a coordinate description
\qq\nn
\mor_\sMan(\cS,\cW_1)\ni\psi_1\equiv\k_1\circ\psi\too\phi_{21}\circ\psi_1\in\mor_\sMan(\cS,\cW_2)
\qqq
fixed by the formul\ae
\qq\nn
\bigl(\phi_{21}\circ\psi_1\bigr)^*\bigl(\theta_2^{\a_2}\bigr)&=&\sum_{k=0}^{q_1}\,\psi_1^*\bigl(\theta_1^{\a_1^1}\bigr)\,\psi_1^*\bigl(\theta_1^{\a_1^2}\bigr)\,\cdots\,\psi_1^*\bigl(\theta_1^{\a_1^k}\bigr)\,\phi_{21\,\a_1^1\a_1^2\ldots\a_1^k}^{(1)}\bigl(\psi_1^*\bigl(x_1^{a_1}\bigr)\bigr)\,,\cr\cr
\bigl(\phi_{21}\circ\psi_1\bigr)\bigl(x_2^{a_2}\bigr)&=&\sum_{l=0}^{q_1}\,\psi_1^*\bigl(\theta_1^{\a_1^1}\bigr)\,\psi_1^*\bigl(\theta_1^{\a_1^2}\bigr)\,\cdots\,\psi_1^*\bigl(\theta_1^{\a_1^l}\bigr)\,\phi_{21\,\a_1^1\a_1^2\ldots\a_1^l}^{(0)}\bigl(\psi_1^*\bigl(x_1^{b_1}\bigr)\bigr)
\qqq
that correspond to a `mapping'
\qq\nn
\bigl(\psi_1^*\bigl(\theta_1\bigr),\psi_1^*\bigl(x_1\bigr)\bigr)\longmapsto\bigl(\bigl(\phi_{21}\circ\psi_1\bigr)^*\bigl(\theta_2\bigr),\bigl(\phi_{21}\circ\psi_1\bigr)^*\bigl(x_2\bigr)\bigr)\,.
\qqq
These considerations define the point of departure for a `standard' differential calculus augmented with the sign conventions of \Rxcite{App.\,A}{Suszek:2017xlw}.

The supermanifolds appearing in the field-theoretic setting of immediate interest come each with a distinguished atlas of local coordinate charts modelled on (a subspace of) the tangent Lie superalgebra $\,\ggt=\ggt^{(0)}\oplus\ggt^{(1)}\,$ of the supersymmetry supergroup, and so the above reasoning reduces our task to the differential calculus of (typically quite explicit) `functions' of $\,\dim\,\ggt^{(0)}\,$ Gra\ss mann-even and $\,\dim\,\ggt^{(1)}\,$ Gra\ss mann-odd coordinates, with the relevant vector fields and differential forms locally spanned on the corresponding coordinate derivations resp.\ differentials. Details follow below.
\erem

Having introduced the supersymmetry group, we may, next, consider a Lie subgroup $\,\txH\subset|\txG|$,\ to be interpreted as the (hidden) gauge group of the physical theory and termed the \textbf{isotropy group}, with the tangent Lie algebra $\,\hgt\subset\ggt^{(0)}$,\ and its Lie subgroup $\,\txH_{\rm vac}\subset\txH\,$ (with the tangent Lie algebra $\,\hgt_{\rm vac}\subset\hgt$),\ to be termed the \textbf{vacuum isotropy group}, the latter modelling the gauge symmetries preserved by the classical solution of the field theory, to be called the \textbf{vacuum} in what follows. With these ingredients in hand, we may finally indicate the supermanifolds of immediate interest to us -- these are the homogeneous spaces 
\qq\nn
\cM_\txH:=\txG/\txH\qquad\qquad\textrm{and}\qquad\qquad \cM_{\txH_{\rm vac}}:=\txG/\txH_{\rm vac}\,.
\qqq
As in the purely Gra\ss mann-even setting, they can be regarded as bases of the respective principal (super)fibrations
\qq\label{eq:homasprinc}
\alxydim{@C=1.5cm@R=1.5cm}{ \txK \ar[r] & \txG \ar[d]^{\pi_{\txG/\txK}} \\ & \cM_\txK}\,,\qquad\txK\in\{\txH,\txH_{\rm vac}\}
\qqq
with the structure group $\,\txK\,$ (with the tangent Lie algebra $\,\kgt\in\{\hgt,\hgt_{\rm vac}\}$), {\it cp} Refs.\,\cite{Kostant:1975,Koszul:1982} (but see also, {\it e.g.}, \Rcite{Carmeli:2011} for a modern perspective). This paves the way to the (redundant) realisation of the homogeneous spaces as collections of local (super)sections 
\qq\nn
\si^\txK_i\ :\ \cU^\txK_i\too\txG\,,\qquad\qquad\pi_{\txG/\txK}\circ\si_i^\txK=\id_{\cU_i^\txK}\,,\qquad\qquad i\in I_\txK
\qqq
of the surjective submersion $\,\pi_{\txG/\txK}\,$ associated with a covering $\,\{\cU_i^\txK\}_{i\in I_\txK}\equiv\cU^\txK\,$ of the base $\,\cM_\txK\,$ by open superdomains. The latter are supermanifolds of the form $\,\cU_i^\txK=(|\cU_i^\txK|,\cO_{\cM_\txK}\rstr_{|\cU_i^\txK|})\,$ with $\,\{|\cU_i^\txK|\}_{i\in I_\txK}\equiv|\cU^\txK|\,$ a trivialising (open) cover of the body principal $\txK$-bundle  
\qq\label{eq:bodyhomasprinc}
\alxydim{@C=1.5cm@R=1.5cm}{ \txK \ar[r] & |\txG| \ar[d]^{\pi_{|\txG|/\txK}\equiv|\pi_{\txG/\txK}|} \\ & |\cM_\txK|\equiv|\txG|/\txK}\,,\qquad\txK\in\{\txH,\txH_{\rm vac}\}\,,
\qqq
and with 
\qq\nn
\cO_{\cM_\txK}\ &:&\ \cT(|\cM_\txK|)\too{\rm {\bf Alg}}_{\rm {\bf scomm}}\cr\cr 
&:&\ |\cU|\longmapsto\bigl\{\ f\in\cO_\txG\bigl(\pi_{|\txG|/\txK}^{-1}\bigl(|\cU|\bigr)\bigr) \quad\vert\quad \forall_{J\in\kgt}\ :\ L_J(f)=0\quad\land\quad\forall_{k\in\txK}\ :\ r_k^*(f)=f\ \bigr\}\equiv\cO_{\cM_\txK}\bigl(|\cU|\bigr)
\qqq
the structure sheaf of the homogeneous space $\,\cM_\txK$,\ introduced in \Rxcite{Sec.\,III.A}{Fioresi:2007zz} ({\it cp} also \Rxcite{Sec.\,9.3}{Carmeli:2011}) and determined in terms of the left-invariant vector fields $\,L_J\,$ on $\,\txG\,$ associated with vectors of $\,\cT_e\txG\equiv\ggt\supset\kgt\ni J\,$ as per $\,L_J\equiv(\id_{\cO_\txG}\ox J)\circ\mu^*\,$ and of the sheaf component of the natural right action 
\qq\nn
r_g\equiv\mu\circ\bigl(\id_\txG\x\widehat g\bigr)\ :\ \txG\x\bR^{0|0}\cong\txG\too\txG\,,\qquad g\in|\txG|
\qqq
of $\,|\txG|\supset\txK\,$ on $\,\txG\,$ that descends to the usual right regular action of the body group on itself, 
\qq\nn
|r_\cdot|\equiv|r|_\cdot\ :\ |\txG|\x|\txG|\too|\txG|\ :\ (g,f)\longmapsto g\cdot f\equiv|\mu|(g,f)\equiv|r_f|(g)\equiv|r|_f(g)\,,
\qqq
and admits an explicit realisation in Kostant's model (locally) given by
\qq\nn
r_g^*\ :\ \cO_\txG\bigl(|\cU|\bigr)\circlearrowleft\ :\ f\longmapsto|r|_g^*\circ f\circ\txU\rho_{g^{-1}}\,,
\qqq
The first indication of the existence of such sections appeared already in \Rxcite{Prop.\,3.9.2}{Kostant:1975}. A more tractable construction, which we are about to recapitulate and elaborate below, was given in \Rxcite{Sec.\,III.A}{Fioresi:2007zz}. It is along these sections, in the $\cS$-point picture advocated before, that we pull back suitable $\txK$-basic covariant tensor fields from the supergroup supermanifold $\,\txG\,$ to $\,\cM_\txK\,$ and ultimately employ them, in a manner reviewed at length in \Rxcite{Sec.\,5}{Suszek:2019cum}, in the construction of a lagrangean model of the dynamics of interest. While the models that we have in mind do not depend on the choice of the sections, establishing a correspondence between them that will be instrumental in our considerations calls for a judicious choice thereof in which that correspondence is particularly easy to write out. Towards this end, we take a closer look at and make -- with hindsight -- certain further assumptions with regard to the decomposition of the supersymmetry algebra $\,\ggt\,$ in the presence of its distinguished Lie subalgebras $\,\hgt\,$ and $\,\hgt_{\rm vac}\subset\hgt$.\ Thus, we write 
\qq\nn
\ggt=\tgt\oplus\hgt\,,
\qqq
where the supervector space 
\qq\nn
\tgt=\tgt^{(0)}\oplus\tgt^{(1)}\,,
\qqq 
a direct-sum complement of the \textbf{isotropy algebra} $\,\hgt\,$ within $\,\ggt$,\ is assumed to be an $\hgt$-module,
\qq\nn
[\hgt,\tgt]\subset\tgt\,,
\qqq 
which qualifies the decomposition as \textbf{reductive}. The isotropy algebra decomposes further as
\qq\label{eq:isovacsplit}
\hgt=\dgt\oplus\hgt_{\rm vac}
\qqq
into the \textbf{vacuum isotropy algebra} $\,\hgt_{\rm vac}\,$ and its direct-sum complement $\,\dgt$.\ The former is determined as the isotropy subalgebra within the Lie algebra $\,\hgt\,$ of the Gra\ss mann-even component of a (physically) distinguished \textbf{vacuum subspace}\label{ref:vacpresc}
\qq\nn
\tgt_{\rm vac}=\tgt_{\rm vac}^{(0)}\oplus\tgt_{\rm vac}^{(1)}\subset\tgt
\qqq
in the $\hgt$-module $\,\tgt$,
\qq\nn
\bigl[\hgt_{\rm vac},\tgt^{(0)}_{\rm vac}\bigr]\subset\tgt^{(0)}_{\rm vac}\,.
\qqq
We write the corresponding direct-sum complements within $\,\tgt\,$ as
\qq\nn
\tgt=\egt\oplus\tgt_{\rm vac}\,,\qquad\qquad\egt=\egt^{(0)}\oplus\egt^{(1)}\,.
\qqq
The decompositions 
\qq\nn
\tgt^{(0)}=\egt^{(0)}\oplus\tgt_{\rm vac}^{(0)}\,,\qquad\qquad\tgt^{(1)}=\egt^{(1)}\oplus\tgt_{\rm vac}^{(1)}
\qqq
correspond to a pair of projectors $\,\sfP^{(A)}=\sfP^{(A)}\circ\sfP^{(A)}\in\End\,\tgt^{(A)},\ A\in\{0,1\}\,$ such that 
\qq\nn
\tgt_{\rm vac}^{(0)}\equiv{\rm Im}\,\sfP^{(0)}\,,\qquad\qquad\tgt_{\rm vac}^{(1)}\equiv{\rm Im}\,\sfP^{(1)}\,.
\qqq
In what follows, we assume both projections to be nontrivial, {\it i.e.},
\qq\nn
\tgt_{\rm vac}^{(0)}\subsetneq\tgt^{(0)}\,,\qquad\qquad\tgt_{\rm vac}^{(1)}\subsetneq\tgt^{(1)}\,.
\qqq
Assuming $\,\dgt\,$ to be an $\hgt_{\rm vac}$-module, 
\qq\nn
[\hgt_{\rm vac},\dgt]\subset\dgt\,,
\qqq
as well as independent preservation of $\,\egt^{(0)}\,$ by $\,\hgt_{\rm vac}$,
\qq\nn
\bigl[\hgt_{\rm vac},\egt^{(0)}\bigr]\subset\egt^{(0)}\,,
\qqq
we obtain another reductive decomposition of the supersymmetry algebra, to wit,
\qq\nn
\ggt=\fgt\oplus\hgt_{\rm vac}\,,\qquad\fgt=\tgt\oplus\dgt\,.
\qqq

\noindent With view to subsequent field-theoretic applications, and in particular to the analysis of a correspondence between various formulations of the field theory of interest, we augment the above and impose\medskip

\noindent\textbf{The Even Effective-Mixing Constraints:} We assume the vacuum isotropy algebra $\,\hgt_{\rm vac}\,$ to preserve the decomposition of  $\,\tgt^{(0)}\,$ into the Gra\ss mann-even vacuum subspace $\,\tgt^{(0)}_{\rm vac}\,$ and its direct-sum complement $\,\egt^{(0)}$, 
\qq\label{eq:EMC0}
[\hgt_{\rm vac},\tgt^{(0)}_{\rm vac}]\stackrel{!}{\subset}\tgt^{(0)}_{\rm vac}\,,\qquad\qquad[\hgt_{\rm vac},\egt^{(0)}]\stackrel{!}{\subset}\egt^{(0)}\,,
\qqq
and its direct-sum complement $\,\dgt\,$ in the isotropy algebra $\,\hgt\,$ to $\ad$-rotate the two subspaces into one another,
\qq\label{eq:EMC1}
[\dgt,\tgt^{(0)}_{\rm vac}]\stackrel{!}{\subset}\egt^{(0)}\,,\qquad\qquad[\dgt,\egt^{(0)}]\stackrel{!}{\subset}\tgt_{\rm vac}^{(0)}\,.
\qqq
\medskip

\brem
As a consistency condition, implied by the Jacobi identity for triples from $\,\dgt\x\dgt\x\tgt_{\rm vac}^{(0)}\,$ and $\,\dgt\x\dgt\x\egt^{(0)}$,\ we derive from the above the additional constraints:
\qq\label{eq:EMC3}
[\dgt,\dgt]\subset\hgt_{\rm vac}\,.
\qqq
\erem

\noindent For the sake of later bookkeeping, we set 
\qq\nn
(D,\d,\unl\d,d,p,q):=(\dim\,\ggt-1,\dim\,\tgt-1,\dim\,\fgt-1,\dim\,\tgt^{(0)}-1,\dim\,\tgt^{(0)}_{\rm vac}-1,\dim\,\tgt^{(1)}_{\rm vac})
\qqq
and denote the respective homogeneous basis vectors (generators) of the various subalgebras and subspaces as 
\qq
&\ggt=\bigoplus_{A=0}^D\,\corr{t_A}\,,&\cr\cr
&\hgt=\bigoplus_{S=1}^{D-\d}\,\corr{J_S}\,,\qquad\,,\qquad\hgt_{\rm vac}=\bigoplus_{\unl S=1}^{D-\unl\d}\,\corr{J_{\unl S}}\qquad\,,\qquad\dgt=\bigoplus_{\widehat S=D-\unl\d+1}^{D-\d}\,\corr{J_{\widehat S}}\,,&\cr\cr
&\fgt=\bigoplus_{\mu=0}^{\unl\d}\,\corr{t_\mu}\qquad\,,\qquad\tgt=\bigoplus_{\unl A=0}^\d\,\corr{t_{\unl A}}\qquad\,,\qquad\tgt_{\rm vac}=\bigoplus_{\unl{\unl A}=0}^{p+q}\,\langle t_{\unl{\unl A}}\rangle\,,&\cr\cr
&\tgt^{(0)}=\bigoplus_{a=0}^d\,\corr{P_a}\,,\qquad\qquad\tgt^{(1)}=\bigoplus_{\a=1}^{\d-d}\,\corr{Q_\a}\,,&\label{eq:Liesalgbas}\\ \cr
&\tgt_{\rm vac}^{(0)}=\bigoplus_{\unl a=0}^p\,\corr{P_{\unl a}}\,,\qquad\qquad\tgt_{\rm vac}^{(1)}=\bigoplus_{\unl\a=1}^q\,\corr{Q_{\unl\a}}\,,&\cr\cr
&\egt^{(0)}=\bigoplus_{\widehat a=p+1}^d\,\corr{P_{\widehat a}}\,,\qquad\qquad\egt^{(1)}=\bigoplus_{\widehat\a=q+1}^{\d-d}\,\corr{Q_{\widehat\a}}\,.&\nonumber
\qqq
We also introduce the structure constants
\qq\nn
[t_A,t_B\}=f_{AB}^{\ \ \ C}\,t_C
\qqq
of $\,\ggt\,$ in the above basis, with the obvious symmetries:
\qq\nn
f_{BA}^{\ \ \ C}=(-1)^{|A|\cdot|B|+1}\,f_{AB}^{\ \ \ C}\,,
\qqq
written in terms of the Gra\ss mann parities $\,|A|\equiv|t_A|\,$ and $\,|B|\equiv|t_B|\,$ of the homogeneous generators $\,t_A$.\ Finally, we write, for any $\,g\in|\txG|$,
\qq\nn
\rho_g(Q_\a)=S(g)_{\ \a}^\b\,Q_\b\,.
\qqq

We may, now, use the above algebra to give a very natural and convenient definition of local sections of the two principal bundles \eqref{eq:homasprinc}. Thus, first of all, we pick up an open neighbourhood $\,|\cU_e|\subset|\txG|\,$ of the group unit $\,e\in|\txG|\,$ sufficiently small to support local coordinates 
\qq\nn
\{\chi^A\}^{A\in\ovl{0,D}}\equiv\{\chi^a\equiv x^a\}^{a\in\ovl{0,d}}\cup\{\chi^\a\equiv\theta^\a\}^{\a\in\ovl{1,\d-d}}\cup\{\chi^{\widehat S}\equiv\phi^{\widehat S}\}^{\widehat S\in\ovl{D-\unl\d+1,D-\d}}\cup\{\chi^{\unl T}\equiv\psi^{\unl T}\}^{\unl T\in\ovl{1,D-\unl\d}}
\qqq 
in which the involutive (super)distribution $\,\xcD_{\hgt_{\rm vac}}\,$ generated by the left-invariant vector fields $\,L_{J_{\unl T}}\equiv L_{\unl T},\ \unl T\in\ovl{1,D-\unl\d}\,$ is spanned by the coordinate derivations $\,\{\frac{\p\ }{\p\psi^{\unl T}}\}_{\unl T\in\ovl{1,D-\unl\d}}\,$ and the Gra\ss mann-even coordinates $\,\{x^a,\phi^{\widehat S}\}^{(a,\widehat S)\in\ovl{0,d}\x\ovl{D-\unl\d+1,D-\d}}\,$ chart $\,|\cU_e|\,$ faithfully, that is each (vector) value of the coordinates from a parameter domain $\,|\cW_e|\subset\bR^{\x d+1}\x\bR^{\x\unl\d-\d}\,$ corresponds to a different $\txH_{\rm vac}$-coset. The existence of such a neighbourhood follows directly from the supergeometric variant of The Local Frobenius Theorem \cite[Thm.\,A.9]{Fioresi:2007zz} ({\it cp} also \Rxcite{Thms.\,6.1.12}{Carmeli:2011}). Consider the sub-supermanifold $\,\cV_e\,$ of $\,\txG\,$ defined as the common zero locus of the coordinates $\,\psi^{\unl T}$,\ that is the supermanifold with the body given by the corresponding submanifold $\,|\cV_e|\subset|\cU_e|\,$ of the body Lie group $\,|\txG|\,$ and the structure sheaf induced from $\,\cO_\txG\rstr_{|\cU_e|}\,$ by setting all the $\,\psi^{\unl T}\,$ to zero. The sub-supermanifold is modelled on the superdomain
\qq\label{eq:modvice}
\cW_e:=\bigl(|\cW_e|,C^\infty(\cdot,\bR)\ox\bigwedge{}^\bullet\,\bR^{\x\d-d}\equiv\cO_{\cW_e}\bigr)\,,
\qqq
that is there exists a superdiffeomorphism 
\qq\nn
\varpi_e\ :\ \cW_e\xrightarrow{\ \cong\ }\cV_e\,.
\qqq
Its body component $\,|\varpi_e|\ :\ |\cW_e|\xrightarrow{\ \cong\ }|\cV_e|\,$ may be chosen in the form of a smooth section of the body projection $\,|\pi_{\txG/\txH_{\rm vac}}|$,
\qq\nn
|\pi_{\txG/\txH_{\rm vac}}|\circ|\varpi_e|=\id_{|\cW_e|}\,.
\qqq
It is then particularly straightforward to prove, as was done in \Rxcite{Sec.\,III.A}{Fioresi:2007zz}, the existence of a superdiffeomorphism 
\qq\nn
\xi_e\equiv\bigl(\id_{|\cW_e|},\xi_e^*\bigr)\ :\ \cU_0^{\txH_{\rm vac}}\equiv\bigl(|\cW_e|,\cO_{\txG/\txH_{\rm vac}}\rstr_{|\cW_e|}\bigr)\xrightarrow{\ \cong\ }\cW_e\,,
\qqq
which identifies $\,\cW_e\,$ as a model of the neighbourhood $\,\cU_0^{\txH_{\rm vac}}\,$ of the unital coset $\,\txH_{\rm vac}\equiv e\cdot\txH_{\rm vac}\,$ in $\,\txG/\txH_{\rm vac}\,$ (a right $\txH_{\rm vac}$-invariant section of $\,\cO_\txG\,$ over $\,|\cW_e|\,$ is determined uniquely by its evaluation at the section $\,|\cV_e|$). Central to the proof is the superdiffeomorphic nature of the supermanifold morphism 
\qq\nn
\g_e\equiv\mu\circ\bigl(\jmath_{\cV_e}\x\jmath_{\txH_{\rm vac}}\bigr)\ :\ \cV_e\x\txH_{\rm vac}\xrightarrow{\ \cong\ }\bigl(|\cV_e|\cdot\txH_{\rm vac},\cO_\txG\rstr_{|\cV_e|\cdot\txH_{\rm vac}}\bigr)\equiv\pi_{\txG/\txH_{\rm vac}}^{-1}\bigl(\cU_0^{\txH_{\rm vac}}\bigr)\subset\txG
\qqq
written in terms of the canonical superembeddings
\qq\nn
\jmath_X\ :\ X\emb\txG\,,\qquad\qquad X\in\{\txH_{\rm vac},\cV_e\}\,.
\qqq
The latter defines (the inverse of) a local trivialisation of the principal $\txH_{\rm vac}$-bundle \eqref{eq:homasprinc} over $\,\cU_0^{\txH_{\rm vac}}\,$ as per
\qq\nn
\t_0^{-1}\equiv\g_e\circ\bigl(\varpi_e\circ\xi_e\x\id_{\txH_{\rm vac}}\bigr)\ :\ \cU_0^{\txH_{\rm vac}}\x\txH_{\rm vac}\xrightarrow{\ \cong\ }\pi_{\txG/\txH_{\rm vac}}^{-1}\bigl(\cU_0^{\txH_{\rm vac}}\bigr)\subset\txG\,,
\qqq
and so also a local section of the principal $\txH_{\rm vac}$-bundle,
\qq\nn
\si_0^{\txH_{\rm vac}}\equiv\t_0^{-1}\circ\bigl(\id_{\cU_0^{\txH_{\rm vac}}}\x\widehat e\bigr)\ :\ \cU_0^{\txH_{\rm vac}}\x\bR^{0|0}\cong\cU_0^{\txH_{\rm vac}}\too\txG\,,
\qqq
{\it cp} \Rxcite{Lem.\,3.2}{Fioresi:2007zz}. It is to be noted that the entire non-canonical information on the latter is encoded in the superparametrisation $\,\varpi_e$.\ Below, we present one such superparametrisation that is particulary suited to subsequent physical considerations. It was introduced operationally in the functor-of-point picture in Refs.\,\cite{Suszek:2017xlw,Suszek:2019cum} and termed the \textbf{exponential superparametrisation}\footnote{Note its structural relation with the normal-coordinate description of the Lie supergroup (with a connected body) due to Berezin and Ka\v c given in \Rcite{Berezin:1970}.} {\it ibidem}. It was subsequently employed in concrete calculations in Refs.\,\cite{Suszek:2018bvx,Suszek:2018ugf}.

The point of departure of the explicit construction of a local section of the principal $\txH_{\rm vac}$-bundle \eqref{eq:homasprinc} over $\,\cU_0^{\txH_{\rm vac}}\,$ is a local section of the body fibration \eqref{eq:bodyhomasprinc} of the standard form
\qq\nn
|\si^{\txH_{\rm vac}}_0|\ :\ |\cW_e|\too|\txG|\ :\ \bigl(x^a,\phi^{\widehat S}\bigr)\longmapsto\ee^{x^aP_a}\cdot\ee^{\phi^{\widehat S}J_{\widehat S}}\,,
\qqq
with the exponential map defined, as usual, in terms of the (unital-time) flows
\qq\nn
\ee^X(g)\equiv\Phi_{|L|_X}(t=1;g)
\qqq
of the point $\,g\in|\txG|\,$ along the integral lines of the left-invariant vector fields $\,|L|_X\in\G(\sfT|\txG|)\,$ associated with the respective elements $\,X\in\ggt^{(0)}$,\ so that 
\qq\nn
|\si^{\txH_{\rm vac}}_0|\bigl(x^a,\psi^{\widehat S}\bigr)\equiv\Phi_{|L|_{\phi^{\widehat S}J_{\widehat S}}}\bigl(1;\Phi_{|L|_{x^aP_a}}\bigl(1;e\bigr)\bigr)\,.
\qqq
We may, next, extend the above body section to a full-blown mapping between the corresponding supermanifolds upon -- by a mild abuse of the notation -- replacing the points $\,(x^a,\phi^{\widehat S})\in|\cW_e|\,$ with coordinate functions on $\,|\cW_e|\,$ denoted by the same symbols,
\qq\nn
\ee^{x^a\ox P_a}\cdot\ee^{\phi^{\widehat S}\ox J_{\widehat S}}\equiv\bigl(|\si^{\txH_{\rm vac}}_0|(\cdot)\equiv\ee^{x^a(\cdot)P_a}\cdot\ee^{\phi^{\widehat S}(\cdot)J_{\widehat S}},|\si^{\txH_{\rm vac}}_0|^*\bigr)\ :\ \bigl(|\cW_e|,C^\infty(\cdot,\bR)\bigr)\too\bigl(|\txG|,C^\infty(\cdot,\bR)\bigr)\,,
\qqq
where -- as the notation suggests -- for any $\,f\in C^\infty(|\cU|;\bR)\,$ on $\,|\cU|\subset|\si^{\rm vac}_0|(|\cW_e|)$,
\qq\nn
|\si^{\txH_{\rm vac}}_0|^*f\equiv f\circ|\si^{\txH_{\rm vac}}_0|\,.
\qqq
In order to make direct contact with Kostant's construction, it suffices to invoke isomorphism \eqref{eq:locmodGsh} for $\,\ggt^{(1)}\equiv 0$,\ whereby we obtain the identity
\qq\nn
\bigl(|\txG|,C^\infty(\cdot,\bR)\bigr)\cong\bigl(|\txG|,\mor_{\txU(\ggt^{(0)}){\rm {\bf -Mod}}}\bigl(\txU(\ggt^{(0)}),C^\infty\bigl(\cdot,\bR\bigr)\bigr)\bigr)\,.
\qqq 
At this stage, it is fairly straightforward to write out a super-extension of the body morphism $\,\ee^{x^a\ox P_a}\cdot\ee^{\phi^{\widehat S}\ox J_{\widehat S}}$.\ Indeed, as long as we work with the local models \eqref{eq:locmodGsh} and \eqref{eq:modvice}, all we need is an isomorphism $\,\bigwedge{}^\bullet\ggt^{(1)\,*}\cong\bigwedge{}^\bullet\,\bR^{\x\d-d}\,$ for $\,\ggt^{(1)}\equiv\bigoplus_{\a=1}^{\d-d}\,\corr{Q_\a}\,$ that yields a superalgebra homomorphism when tensored with the body morphism $\,|\si^{\txH_{\rm vac}}_0|^*$.\ This we choose, in keeping with the original considerations in Refs.\,\cite{Suszek:2017xlw,Suszek:2019cum,Suszek:2018bvx,Suszek:2018ugf} and the physics literature\footnote{The status of the exponential superparametrisation is very rarely discussed rigorously in the physics (mainly superstring-theoretic) literature, which is where the specific sections considered herein are employed in the construction of the relevant action functionals, and then it is usually related to Berezin's concept of an exponential mapping.}, in the distinguished form
\qq\nn
\unl\ee^{\theta^\a\ox Q_\a}\ &:&\ \bigwedge{}^\bullet\ggt^{(1)\,*}\xrightarrow{\ \cong\ }\bigwedge{}^\bullet\,\bR^{\x\d-d}\cr\cr 
&:&\ \la\,\bd1+\sum_{k=1}^{\d-d}\,\la_{\b_1\b_2\ldots\b_k}\,q^{\b_1}\wedge q^{\b_2}\wedge\cdots\wedge q^{\b_k}\longmapsto\la\,\bd1+\sum_{k=1}^{\d-d}\,(-1)^{\frac{k(k+1)}{2}}\,\la_{\b_1\b_2\ldots\b_k}\,\theta^{\b_1}\,\theta^{\b_2}\cdots\theta^{\b_k}\,,
\qqq
expressed in terms of the dual basis $\,\{q^\a\}_{\a\in\ovl{1,\d-d}}\,$ of $\,\ggt^{(1)\,*}$,\
\qq\label{eq:dualQ}
q^\a(Q_\b)=\d^\a_{\ \ \b}\,,
\qqq 
and with the sums taken over sequences of spinor indices ordered as $\,1\leq\b_1<\b_2<\ldots<\b_l\leq\d-d$.\ In order to explain the somewhat non-obvious signs appearing in the above formula and justify the suggestive notation $\,\unl\ee^{\theta^\a\ox Q_\a}$,\ we need to pull back the above isomorphism to Kostant's structure sheaf. Below, we do that for the composite mapping $\,|\si^{\txH_{\rm vac}}_0|^*\ox\unl\ee^{\theta^\a\ox Q_\a}$.\ To this end, we employ \eqref{eq:locmodGsh}, with the explicit definition \eqref{eq:gamap} of $\,\g$,\ alongside the {\bf evaluation map} 
\qq\nn
\ev_g\ :\ \cO_\txG(|\cU|)\too\bR\ :\ f\longmapsto f(\bd1)(g)\,.
\qqq 
After some trivial manipulations, we obtain a mapping
\qq\nn
\ee^{\theta^\a\ox Q_\a}\cdot\ee^{x^a\ox P_a}\cdot\ee^{\phi^{\widehat S}\ox J_{\widehat S}}\equiv\varpi_e=\bigl(|\si^{\txH_{\rm vac}}_0|,\varpi_e^*\bigr)\ :\ \cW_e\too\txG
\qqq
with the sheaf component
\qq\nn
\varpi_e^*:=\bigl(\id_{\cO_{\cW_e}}\ox\ev_e\bigr)\circ\bigl(\sum_{k=0}^{\d-d}\,\tfrac{1}{k!}\,\bigl(\theta^{\a_1}\ox Q_{\a_1}\bigr)\circ\bigl(\theta^{\a_2}\ox Q_{\a_2}\bigr)\circ\cdots\circ(\theta^{\a_k}\ox Q_{\a_k}\bigr)\bigr)\circ r_{\ee^{x^a(\cdot)\ox P_a}}^*\circ r_{\ee^{\psi^{\widehat S}(\cdot)\ox J_{\widehat S}}}^*\,.
\qqq
Here, it is understood that the above evaluates on an arbitrary $\,f\in\cO_\txG(|\cU|),\ |\cU|\subset|\si^{\txH_{\rm vac}}_0|(|\cW_e|)$,\ with the decomposition (in which the sum is taken over sequences of spinor indices ordered as $\,1\leq\b_1<\b_2<\ldots<\b_l\leq\d-d\,$ for which the corresponding $l$-forms $\,q^{\b_1}\wedge q^{\b_2}\wedge\cdots\wedge q^{\b_l}\,$ are linearly independent)
\qq\nn
f\circ\g=\sum_{l=0}^{\d-d}\,f_{\b_1\b_2\ldots\b_l}\ox q^{\b_1}\wedge q^{\b_2}\wedge\cdots\wedge q^{\b_l}
\qqq
as
\qq\nn
&&\varpi_e^*(f)\cr\cr
&=&\sum_{k=0}^{\d-d}\,\tfrac{(-1)^{\frac{k(k-1)}{2}}}{k!}\,\theta^{\a_1}\,\theta^{\a_2}\cdots\theta^{\a_k}(\cdot)\,\txU\la_{Q_{\a_1}\cdot_{\txU(\ggt)}Q_{\a_2}\cdot_{\txU(\ggt)}\cdots\cdot_{\txU(\ggt)}Q_{\a_k}}\bigl(f\circ\txU\rho_{\ee^{-x^a(\cdot)\,P_a}\cdot\ee^{\phi^{\widehat S}(\cdot)\,J_{\widehat S}}}\bigr)(\bd1)\bigl(\ee^{x^a(\cdot)\,P_a}\cdot\ee^{\phi^{\widehat S}(\cdot)\,J_{\widehat S}}\bigr)\cr\cr
&=&\sum_{k=0}^{\d-d}\,\tfrac{(-1)^{\frac{k(k+1)}{2}}}{k!}\,\theta^{\a_1}\,\theta^{\a_2}\cdots\theta^{\a_k}(\cdot)\,f\bigl(\rho_{\ee^{-x^a(\cdot)\,P_a}\cdot\ee^{\phi^{\widehat S}(\cdot)\,J_{\widehat S}}}\bigl(Q_{\a_1}\bigr)\cdot_{\txU(\ggt)}\rho_{\ee^{-x^a(\cdot)\,P_a}\cdot\ee^{\phi^{\widehat S}(\cdot)\,J_{\widehat S}}}\bigl(Q_{\a_2}\bigr)\cr\cr
&&\hspace{5cm}\cdot_{\txU(\ggt)}\cdots\cdot_{\txU(\ggt)}\rho_{\ee^{-x^a(\cdot)\,P_a}\cdot\ee^{\phi^{\widehat S}(\cdot)\,J_{\widehat S}}}\bigl(Q_{\a_k}\bigr)\bigr)\bigl(\ee^{x^a(\cdot)\,P_a}\cdot\ee^{\phi^{\widehat S}(\cdot)\,J_{\widehat S}}\bigr)\cr\cr
&=&\sum_{k=0}^{\d-d}\,(-1)^{\frac{k(k+1)}{2}}\,\theta^{\b_1<}\,\theta^{\b_2<}\cdots{}^{<}\theta^{\b_k}(\cdot)\,S\bigl(\ee^{x^a(\cdot)\,P_a}\cdot\ee^{\phi^{\widehat S}(\cdot)\,J_{\widehat S}}\bigr)_{\ \b_1}^{\g_1}\,S\bigl(\ee^{x^a(\cdot)\,P_a}\cdot\ee^{\phi^{\widehat S}(\cdot)\,J_{\widehat S}}\bigr)_{\ \b_2}^{\g_2}\cr\cr
&&\hspace{3.5cm}\cdots\,\,S\bigl(\ee^{x^a(\cdot)\,P_a}\cdot\ee^{\phi^{\widehat S}(\cdot)\,J_{\widehat S}}\bigr)_{\ \b_k}^{\g_k}\,f\circ\g\bigl(Q_{\g_1}\wedge Q_{\g_2}\wedge\cdots\wedge Q_{\g_k}\bigr)\bigl(\ee^{x^a(\cdot)\,P_a}\cdot\ee^{\phi^{\widehat S}(\cdot)\,J_{\widehat S}}\bigr)\cr\cr
&=&\sum_{k=0}^{\d-d}\,(-1)^{\frac{k(k+1)}{2}}\,\theta^{\b_1}\,\theta^{\b_2}\cdots\theta^{\b_k}(\cdot)\,S\bigl(\ee^{x^a(\cdot)\,P_a}\cdot\ee^{\phi^{\widehat S}(\cdot)\,J_{\widehat S}}\bigr)_{\ \b_1}^{\g_1}\,S\bigl(\ee^{x^a(\cdot)\,P_a}\cdot\ee^{\phi^{\widehat S}(\cdot)\,J_{\widehat S}}\bigr)_{\ \b_2}^{\g_2}\cr\cr
&&\hspace{3.5cm}\cdots\,\,S\bigl(\ee^{x^a(\cdot)\,P_a}\cdot\ee^{\phi^{\widehat S}(\cdot)\,J_{\widehat S}}\bigr)_{\ \b_k}^{\g_k}\,f_{\g_1\g_2\ldots\g_k}\bigl(\ee^{x^a(\cdot)\,P_a}\cdot\ee^{\phi^{\widehat S}(\cdot)\,J_{\widehat S}}\bigr)\,,
\qqq
which, incidentally, explains the signs introduced earlier. Given the last formula, it is easy to convince oneself that $\,\varpi_e^*\,$ is (or, more precisely, gives rise to) a superalgebra homomorphism. Indeed, we have, for any $\,f_1,f_2\in\cO_\txG(|\cU|)$,
\qq\nn
\varpi_e^*(f_1\cdot_{\cO_\txG}f_2)&\equiv&\sum_{k=0}^{\d-d}\,\tfrac{(-1)^{\frac{k(k+1)}{2}}}{k!}\,\theta^{\a_1}\,\theta^{\a_2}\cdots\theta^{\a_k}(\cdot)\,\txm_{C^\infty(|\cU|,\bR)}\bigl((f_1\ox f_2)\cr\cr
&&\bigl(\prod_{l=1}^k\bigl(\rho_{\ee^{-x^a(\cdot)\,P_a}\cdot\ee^{\phi^{\widehat S}(\cdot)\,J_{\widehat S}}}\bigl(Q_{\a_l}\bigr)\ox\bd1+\bd1\ox\rho_{\ee^{-x^a(\cdot)\,P_a}\cdot\ee^{\phi^{\widehat S}(\cdot)\,J_{\widehat S}}}\bigl(Q_{\a_l}\bigr)\bigr)\bigr)\bigr)\bigl(\ee^{x^a(\cdot)\,P_a}\cdot\ee^{\phi^{\widehat S}(\cdot)\,J_{\widehat S}}\bigr)\,,
\qqq
which -- in consequence of the {\it anti}commutativity of the Gra\ss mann-odd sections $\,\theta^\a\,$ and due to the implication $\,f_2(u)\neq 0\ \Longrightarrow\ |f_2|=|u|\,$ -- can be rewritten in the form
\qq\nn
\varpi_e^*(f_1\cdot_{\cO_\txG}f_2)&=&\sum_{k=0}^{\d-d}\,\tfrac{(-1)^{\frac{k(k+1)}{2}}}{k!}\,\theta^{\a_1}\,\theta^{\a_2}\cdots\theta^{\a_k}(\cdot)\,\txm_{C^\infty(|\cU|,\bR)}\bigl((f_1\ox f_2)\bigl(\sum_{l=0}^k\,\binom{k}{l}\,\bigl(\rho_{\ee^{-x^a(\cdot)\,P_a}\cdot\ee^{\phi^{\widehat S}(\cdot)\,J_{\widehat S}}}\bigl(Q_{\a_1}\bigr)\cr\cr
&&\cdot_{\txU(\ggt)}\rho_{\ee^{-x^a(\cdot)\,P_a}\cdot\ee^{\phi^{\widehat S}(\cdot)\,J_{\widehat S}}}\bigl(Q_{\a_2}\bigr)\cdot_{\txU(\ggt)}\cdots\cdots \rho_{\ee^{-x^a(\cdot)\,P_a}\cdot\ee^{\phi^{\widehat S}(\cdot)\,J_{\widehat S}}}\bigl(Q_{\a_l}\bigr)\cr\cr
&&\ox\rho_{\ee^{-x^a(\cdot)\,P_a}\cdot\ee^{\phi^{\widehat S}(\cdot)\,J_{\widehat S}}}\bigl(Q_{\a_{l+1}}\bigr)\cdot_{\txU(\ggt)}\rho_{\ee^{-x^a(\cdot)\,P_a}\cdot\ee^{\phi^{\widehat S}(\cdot)\,J_{\widehat S}}}\bigl(Q_{\a_{l+2}}\bigr)\cdot_{\txU(\ggt)}\cdots\cr\cr
&&\cdot_{\txU(\ggt)}\rho_{\ee^{-x^a(\cdot)\,P_a}\cdot\ee^{\phi^{\widehat S}(\cdot)\,J_{\widehat S}}}\bigl(Q_{\a_k}\bigr)\bigr)\bigr)\bigl(\ee^{x^a(\cdot)\,P_a}\cdot\ee^{\phi^{\widehat S}(\cdot)\,J_{\widehat S}}\bigr)\cr\cr
&=&\sum_{k=0}^{\d-d}\,\sum_{l=0}^k\,\tfrac{(-1)^{\frac{k(k+1)}{2}+l\,|f_2|}}{l!\,(k-l)!}\,\theta^{\a_1}\,\theta^{\a_2}\cdots\theta^{\a_l}\,\theta^{\a_{l+1}}\,\theta^{\a_{l+2}}\cdots\theta^{\a_k}(\cdot)\,\bigl(f_1\bigl(\rho_{\ee^{-x^a(\cdot)\,P_a}\cdot\ee^{\phi^{\widehat S}(\cdot)\,J_{\widehat S}}}\bigl(Q_{\a_1}\bigr)\cr\cr
&&\cdot_{\txU(\ggt)}\rho_{\ee^{-x^a(\cdot)\,P_a}\cdot\ee^{\phi^{\widehat S}(\cdot)\,J_{\widehat S}}}\bigl(Q_{\a_2}\bigr)\cdot_{\txU(\ggt)}\cdots\cdots\rho_{\ee^{-x^a(\cdot)\,P_a}\cdot\ee^{\phi^{\widehat S}(\cdot)\,J_{\widehat S}}}\bigl(Q_{\a_l}\bigr)\bigr)\cr\cr
&&\cdot_{C^\infty}f_2\bigl(\rho_{\ee^{-x^a(\cdot)\,P_a}\cdot\ee^{\phi^{\widehat S}(\cdot)\,J_{\widehat S}}}\bigl(Q_{\a_{l+1}}\bigr)\cdot_{\txU(\ggt)}\rho_{\ee^{-x^a(\cdot)\,P_a}\cdot\ee^{\phi^{\widehat S}(\cdot)\,J_{\widehat S}}}\bigl(Q_{\a_{l+2}}\bigr)\cdot_{\txU(\ggt)}\cdots\cr\cr
&&\cdot_{\txU(\ggt)}\rho_{\ee^{-x^a(\cdot)\,P_a}\cdot\ee^{\phi^{\widehat S}(\cdot)\,J_{\widehat S}}}\bigl(Q_{\a_k}\bigr)\bigr)\bigr)\bigl(\ee^{x^a(\cdot)\,P_a}\cdot\ee^{\phi^{\widehat S}(\cdot)\,J_{\widehat S}}\bigr)\cr\cr
&\equiv&\sum_{k=0}^{\d-d}\,\sum_{l=0}^k\,\tfrac{(-1)^{\frac{k(k+1)}{2}+l(k-l)}}{l!\,(k-l)!}\,\theta^{\a_1}\,\theta^{\a_2}\cdots\theta^{\a_l}(\cdot)\,f_1\bigl(\rho_{\ee^{-x^a(\cdot)\,P_a}\cdot\ee^{\phi^{\widehat S}(\cdot)\,J_{\widehat S}}}\bigl(Q_{\a_1}\bigr)\cr\cr
&&\cdot_{\txU(\ggt)}\rho_{\ee^{-x^a(\cdot)\,P_a}\cdot\ee^{\phi^{\widehat S}(\cdot)\,J_{\widehat S}}}\bigl(Q_{\a_2}\bigr)\cdot_{\txU(\ggt)}\cdots\cdots\rho_{\ee^{-x^a(\cdot)\,P_a}\cdot\ee^{\phi^{\widehat S}(\cdot)\,J_{\widehat S}}}\bigl(Q_{\a_l}\bigr)\bigr)\bigl(\ee^{x^a(\cdot)\,P_a}\cdot\ee^{\phi^{\widehat S}(\cdot)\,J_{\widehat S}}\bigr)\cr\cr
&&\cdot\theta^{\a_{l+1}}\,\theta^{\a_{l+2}}\cdots\theta^{\a_k}\,f_2\bigl(\rho_{\ee^{-x^a(\cdot)\,P_a}\cdot\ee^{\phi^{\widehat S}(\cdot)\,J_{\widehat S}}}\bigl(Q_{\a_{l+1}}\bigr)\cdot_{\txU(\ggt)}\rho_{\ee^{-x^a(\cdot)\,P_a}\cdot\ee^{\phi^{\widehat S}(\cdot)\,J_{\widehat S}}}\bigl(Q_{\a_{l+2}}\bigr)\cdot_{\txU(\ggt)}\cdots\cr\cr
&&\cdot_{\txU(\ggt)}\rho_{\ee^{-x^a(\cdot)\,P_a}\cdot\ee^{\phi^{\widehat S}(\cdot)\,J_{\widehat S}}}\bigl(Q_{\a_k}\bigr)\bigr)\bigl(\ee^{x^a(\cdot)\,P_a}\cdot\ee^{\phi^{\widehat S}(\cdot)\,J_{\widehat S}}\bigr)\,.
\qqq
However,
\qq\nn
\frac{k(k+1)}{2}+l(k-l)=\frac{l(l+1)}{2}+\frac{(k-l)(k-l+1)}{2}+2l(k-l)\,,
\qqq
and so upon invoking the trivial implication $\,k>\d-d\Longrightarrow\prod_{i=1}^k\,\theta^{\a_i}\equiv 0$,\ we finally obtain 
\qq\nn
\varpi_e^*(f_1\cdot_{\cO_\txG}f_2)&=&\sum_{k=0}^{\d-d}\,\sum_{l=0}^k\,\tfrac{(-1)^{\frac{l(l+1)}{2}}}{l!}\,\theta^{\a_1}\,\theta^{\a_2}\cdots\theta^{\a_l}(\cdot)\,f_1\bigl(\rho_{\ee^{-x^a(\cdot)\,P_a}\cdot\ee^{\phi^{\widehat S}(\cdot)\,J_{\widehat S}}}\bigl(Q_{\a_1}\bigr)\cr\cr
&&\cdot_{\txU(\ggt)}\rho_{\ee^{-x^a(\cdot)\,P_a}\cdot\ee^{\phi^{\widehat S}(\cdot)\,J_{\widehat S}}}\bigl(Q_{\a_2}\bigr)\cdot_{\txU(\ggt)}\cdots\cdots\rho_{\ee^{-x^a(\cdot)\,P_a}\cdot\ee^{\phi^{\widehat S}(\cdot)\,J_{\widehat S}}}\bigl(Q_{\a_l}\bigr)\bigr)\bigl(\ee^{x^a(\cdot)\,P_a}\cdot\ee^{\phi^{\widehat S}(\cdot)\,J_{\widehat S}}\bigr)\cr\cr
&&\cdot\tfrac{(-1)^{\frac{(k-l)(k-l+1)}{2}}}{(k-l)!}\,\theta^{\a_{l+1}}\,\theta^{\a_{l+2}}\cdots\theta^{\a_k}\,f_2\bigl(\rho_{\ee^{-x^a(\cdot)\,P_a}\cdot\ee^{\phi^{\widehat S}(\cdot)\,J_{\widehat S}}}\bigl(Q_{\a_{l+1}}\bigr)\cdot_{\txU(\ggt)}\cr\cr
&&\rho_{\ee^{-x^a(\cdot)\,P_a}\cdot\ee^{\phi^{\widehat S}(\cdot)\,J_{\widehat S}}}\bigl(Q_{\a_{l+2}}\bigr)\cdot_{\txU(\ggt)}\cdots\cdot_{\txU(\ggt)}\rho_{\ee^{-x^a(\cdot)\,P_a}\cdot\ee^{\phi^{\widehat S}(\cdot)\,J_{\widehat S}}}\bigl(Q_{\a_k}\bigr)\bigr)\bigl(\ee^{x^a(\cdot)\,P_a}\cdot\ee^{\phi^{\widehat S}(\cdot)\,J_{\widehat S}}\bigr)\cr\cr
&\equiv&\varpi_e^*(f_1)\cdot\varpi_e^*(f_2)\,,
\qqq
as claimed.

In fact, a moment's thought\footnote{It suffices to consider Taylor's expansion of the $\,f_{\b_1\b_2\ldots\b_k}(\ee^{x^a(\cdot)\,P_a}\cdot\ee^{\phi^{\widehat S}(\cdot)\,J_{\widehat S}})\,$ around $\,(x,\phi)=(0,0)\,$ and invoke \Reqref{eq:Ulau}.} reveals that the Gra\ss mann-odd component $\,\ee^{\theta^\a\ox Q_\a}\,$ of $\,\varpi_e\,$ is a natural super-completion of the standard exponential parametrisation of a neighbourhood of the Lie-group unit, taking into account the polynomial nature of the Gra\ss mannian component $\,\bigwedge{}^\bullet\,\ggt^{(1)\,*}\,$ of the structure sheaf, and that the notation $\,\ee^{\theta^\a\ox Q_\a}\cdot\ee^{x^a\ox P_a}\cdot\ee^{\phi^{\widehat S}\ox J_{\widehat S}}\,$ used above acquires the same status as its classical counterpart $\,\ee^{x^a\ox P_a}\cdot\ee^{\phi^{\widehat S}\ox J_{\widehat S}}\,$ in the $\cS$-point picture to which we now pass. In particular, we have the expected
\berop
In the above notation, the supermanifold morphism
\qq\nn
\ee^{x^a\ox P_a}\cdot\ee^{\phi^{\widehat S}\ox J_{\widehat S}}\cdot\ee^{\theta^\a\ox Q_\a}\equiv\widetilde\varpi_e=\bigl(|\si^{\txH_{\rm vac}}_0|,\widetilde\varpi_e^*\bigr)\ :\ \cW_e\too\txG
\qqq
with the sheaf component
\qq\nn
\widetilde\varpi_e^*:=\bigl(\id_{\cO_{\cW_e}}\ox\ev_e\bigr)\circ r_{\ee^{x^a(\cdot)\ox P_a}}^*\circ r_{\ee^{\psi^{\widehat S}(\cdot)\ox J_{\widehat S}}}^*\circ\bigl(\sum_{k=0}^{\d-d}\,\tfrac{1}{k!}\,\bigl(\theta^{\a_1}\ox Q_{\a_1}\bigr)\circ\bigl(\theta^{\a_2}\ox Q_{\a_2}\bigr)\circ\cdots\circ(\theta^{\a_k}\ox Q_{\a_k}\bigr)\bigr)
\qqq
satisfies the identity
\qq\nn
\ee^{x^a\ox P_a}\cdot\ee^{\phi^{\widehat S}\ox J_{\widehat S}}\cdot\ee^{\theta^\a\ox Q_\a}=\ee^{\theta^\a\ox\rho_{\ee^{x^a\ox P_a}\cdot\ee^{\phi^{\widehat S}\ox J_{\widehat S}}} (Q_\a)}\cdot\ee^{x^a\ox P_a}\cdot\ee^{\phi^{\widehat S}\ox J_{\widehat S}}\,.
\qqq
\eerop
\beroof 
Straightforward.
\eroof
Accordingly, we shall -- by a mild abuse of the notation in which the superdiffeomorphism $\,\xi_e\,$ is dropped and $\,\cW_e\,$ is taken as an element of the trivialising cover of the principal $\txH_{\rm vac}$-bundle \eqref{eq:homasprinc} near $\,\txH_{\rm vac}\,$ -- write 
\qq\label{expodd}
\si_0^{\txH_{\rm vac}}(\theta,x,\phi)\equiv\ee^{\theta^\a\ox Q_\a}\cdot\ee^{x^a\ox P_a}\cdot\ee^{\phi^{\widehat S}\ox J_{\widehat S}}
\qqq
and interpret this formula as an explicit coordinate description of a trivialising section of the principal $\txH_{\rm vac}$-bundle \eqref{eq:homasprinc} over $\,\cU_0^{\rm vac}$.

Next, we fix a collection $\,\{g_i\}_{i\in I_{\txH_{\rm vac}}}$,\ including $\,g_0\equiv e\,$ ({\it i.e.}, $\,I_{\txH_{\rm vac}}\ni 0$), of (pairwise distinct) `reference' points $\,g_i\in|\txG|\,$ with the property
\qq\nn
\bigcup_{i\in I_{\txH_{\rm vac}}}\,[|l|]_{g_i}\bigl(|\cU_0^{\txH_{\rm vac}}|\bigr)=|\cM_{\txH_{\rm vac}}|\,,
\qqq
written in terms of the natural left action $\,[|l|]_\cdot\,$ of the body group $\,|\txG|\,$ on the body homogeneous space $\,|\cM_{\txH_{\rm vac}}|\,$ induced from the left regular action 
\qq\nn
|l|_\cdot\ :\ |\txG|\x|\txG|\too|\txG|\ :\ (f,g)\longmapsto f\cdot g\equiv|\mu|(f,g)\equiv|l|_f(g)
\qqq
of that group on itself as 
\qq\nn
[|l|]_\cdot\ :\ |\txG|\x|\cM_{\txH_{\rm vac}}|\too|\cM_{\txH_{\rm vac}}|\ :\ (f,g\txH_{\rm vac})\longmapsto|l|_f(g)\txH_{\rm vac}\,.
\qqq
The supermanifolds 
\qq\nn
\cU_i^{\txH_{\rm vac}}:=\bigl(|\cU_i^{\txH_{\rm vac}}|\equiv[|l|]_{g_i}\bigl(|\cU_0^{\txH_{\rm vac}}|\bigr),\cO_{\cM_{\txH_{\rm vac}}}\rstr_{|\cU_i^{\txH_{\rm vac}}|}\bigr)\,,\qquad i\in I_{\txH_{\rm vac}}
\qqq
compose the desired trivialising cover for the fibration $\,\txG\too\txG/\txH_{\rm vac}$.\ Indeed, we may now propagate the previously constructed local section 
\qq\nn
\si_0^{\rm vac}\equiv\si_0^{\txH_{\rm vac}}\ :\ \cU_0^{\txH_{\rm vac}}\too\txG\,,
\qqq
with the fundamental property
\qq\nn
\pi_{\txG/\txH_{\rm vac}}\circ\si_0^{\rm vac}=\id_{\cU^{\txH_{\rm vac}}_0}\,,
\qqq
to all the remaining superdomains of the cover by means of the superdiffeomorphisms 
\qq\nn
l_{g_i}:=\mu\circ\bigl(\widehat{g_i}\x\id_\txG\bigr)\ :\ \bR^{0|0}\x\txG\cong\txG\too\txG\,,\qquad i\in I_{\txH_{\rm vac}}\,,
\qqq
lifting the formerly introduced body action $\,|l|_\cdot\,$ of $\,|\txG|\,$ from $\,|\txG|\,$ to $\,\txG$,\ together with its quotient counterpart
\qq\nn
[l]_{g_i}:=[\ell]^{\txH_{\rm vac}}_\cdot\circ\bigl(\widehat{g_i}\x\id_{\cM_{\txH_{\rm vac}}}\bigr)\ :\ \bR^{0|0}\x \cM_{\txH_{\rm vac}}\cong \cM_{\txH_{\rm vac}}\too \cM_{\txH_{\rm vac}}\,,
\qqq
the latter being defined in terms of the unique action of $\,\txG\,$ on $\,\cM_{\txH_{\rm vac}}$,
\qq\nn
[\ell]^{\txH_{\rm vac}}_\cdot\ :\ \txG\x \cM_{\txH_{\rm vac}}\too \cM_{\txH_{\rm vac}}\,,
\qqq
that closes, for $\,\txK=\txH_{\rm vac}$,\ the commutative diagram (in $\,\sMan $)
\qq\label{eq:leftrans}
\alxydim{@C=1.5cm@R=1.5cm}{ \txG\x\txG \ar[r]^{\ell_\cdot\equiv\mu} \ar[d]_{\id_\txG\x\pi_{\txG/\txK}} & \txG \ar[d]^{\pi_{\txG/\txK}} \\ \txG\x \cM_\txK \ar[r]_{[\ell]^\txK_\cdot} & \cM_\txK}\,.
\qqq
Its existence was stated in \Rxcite{Prop.\,3.10.1}{Kostant:1975} ({\it cp} also \Rxcite{Prop.\,3.4 \& 3.5}{Fioresi:2007zz} and \Rxcite{Prop.\,9.3.5}{Carmeli:2011}). With these in hand, we define
\qq\nn
\si_i^{\txH_{\rm vac}}\equiv\si_i^{\rm vac}:=l_{g_i}\circ\si_0^{\rm vac}\circ\bigl([l]_{g_i}^{\txH_{\rm vac}}\bigr)^{-1}\ :\ \cU_i^{\txH_{\rm vac}}\too\txG\qqq 
and readily check
\qq\nn
\pi_{\txG/\txH_{\rm vac}}\circ\si_i^{\rm vac}\equiv\pi_{\txG/\txH_{\rm vac}}\circ l_{g_i}\circ\si_0^{\rm vac}\circ\bigl([l]_{g_i}^{\txH_{\rm vac}}\bigr)^{-1}=[l]^{\txH_{\rm vac}}_{g_i}\circ\pi_{\txG/\txH_{\rm vac}}\circ\si_0^{\rm vac}\circ\bigl([l]_{g_i}^{\txH_{\rm vac}}\bigr)^{-1}=\id_{\cU_i^{\txH_{\rm vac}}}\,.
\qqq
In the $\cS$-point picture, and in the local coordinates 
\qq\nn
\bigl(\unl\chi{}_i^\mu\bigr)\equiv\bigl(\theta_i^\a,x_i^a,\phi_i^{\widehat S}\bigr)
\qqq
introduced earlier (assigned an extra index $\,i\in I_{\txH_{\rm vac}}\,$ to formally distinguish the various local incarnations of the same coordinate), we may, therefore, write\label{page:sivacnovac}
\qq\label{eq:vacgauge}
\si_i^{\rm vac}\bigl(\unl\chi{}_i\bigr)=g_i\cdot\ee^{\theta_i^\a\ox Q_\a}\cdot\ee^{x^a\ox P_a}\cdot\ee^{\phi_i^{\widehat S}\ox J_{\widehat S}}\,,\qquad i\in I_{\txH_{\rm vac}}\,.
\qqq

It is crucial to note that upon freezing the last subset of coordinates at $\,\phi_i^{\widehat S}=0,\ \widehat S\in\ovl{D-\unl\d+1,D-\d}$,\ we induce local sections of the other principal bundle
\qq\nn
\si^\txH_i\equiv\si_i^{\xcancel{\rm vac}}:=\si_i^{\rm vac}\rstr_{\phi_i=0}\ :\ \pi_{\txG/\txH}\bigl(\si_i^{\rm vac}\bigl(\cU^{\txH_{\rm vac}}_i\bigr)\bigr)\equiv\cU_i^\txH\too\txG\,,
\qqq
with the $\cS$-point presentation
\qq\label{eq:novacgauge}
\si_i^{\xcancel{\rm vac}}\bigl(\theta_i,x_i\bigr):=\si_i^{\rm vac}\bigl(\theta_i,x_i,0\bigr)\equiv g_i\cdot\ee^{\theta_i^\a\ox Q_\a}\cdot\ee^{x^a\ox P_a}\,,\qquad i\in I_{\txH_{\rm vac}}
\qqq
in terms of the local coordinates
\qq\nn
\bigl(\chi{}_i^{\unl A}\bigr)\equiv\bigl(\theta_i^\a,x_i^a\bigr)
\qqq
This particular induction mechanism shall be encountered in the field-theoretic analysis to follow.

\brem
Our findings can be rephrased as follows: The canonical surjective submersion \eqref{eq:homasprinc} defines a family of principal $\txK$-bundles indexed by $\,\obj\,\sMan$,\ each coming with a preferred trivialising cover and the associated local trivialisations described in terms of the ($|\txG|$-shifted) exponential superparametrisations discussed above.
\erem

We are now ready to study, in the convenient geometric $\cS$-point picture, the differential calculus on the homogeneous spaces $\,\cM_\txK,\ \txK\in\{\txH,\txH_{\rm vac}\}$.\ Various physically relevant elements thereof can be descended from the super-variant of the standard Cartan calculus on the mother supersymmetry group $\,\txG\,$ along the local sections $\,\si_i^{\txH_{\rm vac}}\,$ (resp.\ $\,\si_i^\txH$), whenever the pair $\,(\txG,\txK)\,$ is reductive, as assumed above. In order to facilitate the discussion of the descent, we denote 
\qq\nn
&\ggt=\lgt\oplus\kgt\,,\qquad\qquad(\lgt,\kgt)\in\{(\tgt,\hgt),(\fgt,\hgt_{\rm vac})\}\,,&\cr\cr
&\lgt=\bigoplus_{\z=0}^{\dim\,\lgt-1}\,\corr{T_\z}\,,\qquad\qquad\kgt=\bigoplus_{Z=1}^{\dim\,\kgt}\,\corr{J_Z}&
\qqq
and decompose the $\ggt$-valued left-invariant Maurer--Cartan 1-form on $\,\txG\,$ as
\qq\nn
\theta_{\rm L}=\theta_{\rm L}^\z\ox T_\z+\theta_{\rm L}^Z\ox J_Z\,.
\qqq
whereupon we identify the $\kgt$-valued 1-form 
\qq\nn
\Theta_\txK:=\theta_{\rm L}^Z\ox J_Z
\qqq
as a principal connection 1-form on the total space $\,\txG\,$ of the principal $\txK$-bundle \eqref{eq:homasprinc}, and the remaining components as $\rho_\cdot$-tensors with respect to the defining fibrewise right action of the structure group $\,\txK\,$ on $\,\txG$,
\qq\label{eq:ractdef}
r^\txK_\cdot\ :\ \txG\x\txK\too\txG\ :\ (g,k)\longmapsto r^\txK_k(g)\equiv g\cdot k\,,
\qqq
that is we have
\qq\nn
r_\cdot^{\txK\,*}\theta_{\rm L}^\z(g,k)=\bigl(\rho_{k^{-1}}\bigr)^\z_{\ \ \z'}\,\theta_{\rm L}^{\z'}(g)
\qqq
for 
\qq\nn
\rho_k(T_{\z'})=:(\rho_k)^\z_{\ \ \z'}\,T_\z\,.
\qqq
These observations give us a definition of a horizontal lift of sections of the tangent sheaf (or vector fields) over $\,\cU^\txK$,
\qq\label{eq:Horhom}\qquad\qquad
{\rm Hor}_{\si_i^\txK(\cdot)}\ :\ \cT_\cdot\cU^\txK_i\too\cT_{\si_i^\txK(\cdot)}\txG\ :\ \cV(\cdot)\longmapsto\bigl(\id_{\cT\txG}-\widehat\Theta_\txK\bigr)\bigl(\si_i^\txK(\cdot)\bigr)\circ\sfT_\cdot\si_i^\txK(\cV)\,,
\qqq
expressed in terms of the identity endomorphism $\,\id_{\cT\txG}\equiv\theta_{\rm L}^A\ox L_A,\ L_A\equiv L_{t_A}\,$ and of the vertical-projector field
\qq\nn
\widehat\Theta_\txK:=\theta_{\rm L}^Z\ox L_Z\,,
\qqq
alongside a natural class of $\txK$-basic contravariant $n$-tensor fields on $\,\txG\,$ that descend to the homogeneous space $\,\cM_\txK$,\ to wit, those given by ($\bR$-)linear combinations 
\qq\nn
\underset{\tx{\ciut{(n)}}}{\om}=\om_{\z_1\z_2\ldots\z_n}\,\theta_{\rm L}^{\z_1}\ox\theta_{\rm L}^{\z_2}\ox\cdots\ox\theta_{\rm L}^{\z_n}\,,\qquad\z_1,\z_2,\ldots,\z_n\in\ovl{0,\dim\,\lgt-1}\,,
\qqq
of the $\txK$-horizontal left-invariant (component) 1-forms $\,\theta_{\rm L}^\z,\ \z\in\ovl{0,\dim\,\lgt-1}$,\ dual to the left-invariant vector fields $\,L_\z\equiv L_{T_\z}\,$ generated by vectors from $\,\lgt$,
\qq\nn
L_{T_\z}\con\theta_{\rm L}^{\z'}=\d_\z^{\ \z'}\,,
\qqq
with (constant) $\txK$-invariant tensors as coefficients, {\it i.e.}, with -- for any $\,k\in\txK\,$ --
\qq\nn
\om_{\z_1\z_2\ldots\z_n}\,(\rho_k)^{\z_1}_{\ \z_1'}\,(\rho_k)^{\z_2}_{\ \z_2'}\,\cdots\,(\rho_k)^{\z_n}_{\ \z_n'}=\om_{\z_1'\z_2'\ldots\z_n'}\,.
\qqq
Locally, the descent is effected by the sections $\,\si_i^{\txH_{\rm vac}},\ i\in I^{\txH_{\rm vac}}$,\ providing us -- in the $\cS$-point picture -- with the standard local Vielbeine,
\qq\label{eq:Vielbein}
\si_i^{\txK\,*}\theta_{\rm L}^A(\xi_i)=:\sfd\xi_i^\z\,{}^\txK\hspace{-2pt}E^{\ A}_\z(\xi_i)\,,\qquad A\in\ovl{0,D}\,,
\qqq
on $\,\cU_i^\txK\,$ coordinatised by $\,\xi_i=\unl\chi{}_i\,$ (for $\,\txK=\txH_{\rm vac}$) resp.\ $\,\xi_i=\chi_i\,$ (for $\,\txK=\txH$). In what follows, we denote the Vielbeine $\,{}^{\txH_{\rm vac}}\hspace{-2pt}E^{\ A}_\mu\,$ relevant to one of the formulations of the field theory of interest as
\qq\label{eq:vacVielbein}
{}^{\txH_{\rm vac}}\hspace{-2pt}E^{\ A}_\mu\equiv E^{\ A}_\mu
\qqq
to unclutter the notation. The ensuing $n$-tensor fields on the homogeneous space $\,\cM_\txK\,$ do \emph{not} depend on the choice of the local section along which we pull them back to it, and hence glue smoothly over non-empty intersections $\,\cU^\txK_{ij}\,$ to \emph{globally} smooth tensor fields on the homogeneous space. As such, they become natural building blocks of a field theory with the typical fibre of the configuration bundle given by $\,\txG/\txK$. 
\medskip

We close this differential-(super)geometric intermezzo by commenting on the structure of the tangent sheaf $\,\cT\cM_\txK\,$ of $\,\cM_\txK$.\ Among its global sections, we find the distinguished vector fields 
\qq\nn
\cK^{[\ell]^\txK_\cdot}_X:=\bigl(X\ox\id_{\cO_{\cM_\txK}}\bigr)\circ\bigl([\ell]^\txK_\cdot\bigr)^*\,,\qquad X\in\cT_e\txG\,.
\qqq
This is a special instance of the general situation in which an action 
\qq\nn
\la_\cdot\equiv\bigl(|\la|_\cdot,\la_\cdot^*\bigr)\ :\ \txG\x\cM\too\cM
\qqq
of a Lie supergroup $\,\txG\,$ on a supermanifold $\,\cM\,$ gives rise to the \textbf{fundamental vector fields}
\qq\label{eq:fundvecfields}
\cK^{\la_\cdot}_X:=\bigl(X\ox\id_{\cO_\cM}\bigr)\circ\la_\cdot^*\,,\qquad X\in\cT_e\txG\,,
\qqq
with the property
\berop\cite[Thm.\,8.2.3]{Carmeli:2011}\label{prop:fundvecfields}
In the hitherto notation, the mapping
\qq\nn
\cK^{\la_\cdot}_\cdot\ :\ \ggt\too\G(\cT\cM)\ :\ L_X\longmapsto\cK^{\la_\cdot}_X
\qqq
is an antimorphism of Lie superalgebras, satisfying the identity
\qq\nn
\la_\cdot^*\circ\cK^{\la_\cdot}_X=\bigl(R_X\ox\id_{\cO_\cM}\bigr)\circ\la_\cdot^*\,.
\qqq
\eerop
\noindent An obvious example of a supermanifold with an action of a Lie supergroup is $\,\txG\,$ itself -- we have the left regular action 
\qq\nn
\ell_\cdot\equiv\mu\ :\ \txG\x\txG\too\txG
\qqq
and its right regular counterpart
\qq\nn
\wp_\cdot\equiv\mu\circ\t\ :\ \txG\x\txG\too\txG\,,
\qqq
where $\,\t\ :\ \txG\x\txG\circlearrowleft\,$ is the standard transposition. In these cases, we have the intuitive results
\qq\label{eq:LRasfund}
\cK^{\ell_\cdot}_X\equiv R_X\,,\qquad\qquad\cK_X^{\wp_\cdot}\equiv L_X\,.
\qqq
The important peculiarity of the homogeneous space $\,\cM_\txK\,$ is that the fundamental vector fields $\,\cK^{[\ell]^\txK_\cdot}_X,\ X\in\cT_e\txG\,$ actually span the tangent sheaf.

\section{The two faces of the physical model}\label{sec:physmod}

Our hitherto considerations provide us with all the conceptual and computational tools requisite for the formulation and canonical analysis of two classes of supersymmetric field theories of immediate interest. Both are lagrangean (field) theories of $\cS$-points $\,\xi\in[\Om_p,\cM_\txK](\cS)\,$ of the \textbf{mapping supermanifold} $\,[\Om_p,\cM_\txK]\,$ defined, for $\,\Om_p\,$ an arbitrary $p$-dimensional \emph{closed} ($\p\Om_p=\emptyset$) oriented manifold (the \textbf{worldvolume}) and $\,\cM_\txK\,$ as introduced previously (and termed the \textbf{supertarget} in this context), as the inner-$\mor$ functor
\qq\nn
[\Om_p,\cM_\txK]\equiv\mor_\sMan\bigl(\cdot\x\Om_p,\cM_\txK\bigr)\ :\ \sMan\too\Set\,,
\qqq
to be evaluated on odd hyperplanes $\,\cS\equiv\bR^{0|N}\,$ of an arbitrary superdimension $\,(0|N),\ N\in\bN^\x$,\ {\it cp} \Rcite{Freed:1999}. In other words, we may think of the theories in question as countable families (indexed by $\,\bN^\x$) of theories of generalised superembeddings of the (Gra\ss mann-)odd-extended worldvolume $\,\bR^{0|N}\x\Om_p\,$ in the supertarget $\,\cM_\txK$,\ and we shall write $\,\xi\in[\Om_p,\cM_\txK]\,$ with this understanding. Their definition calls for a pair of $\txH$-basic (and so also $\txH_{\rm vac}$-basic) contravariant tensors on $\,\txG$,\ to wit, a degenerate symmetric rank-2 tensor 
\qq\nn
\txg=\txg_{ab}\,\theta_{\rm L}^a\ox\theta_{\rm L}^b\,,\qquad\qquad\txg_{ba}=\txg_{ab}\,,
\qqq
with -- for any $\,k\in\txG\,$ --
\qq\nn
\txg_{ab}\,(\rho_k)^a_{\ c}\,(\rho_k)^b_{\ d}=\txg_{cd}\,,
\qqq
and a de Rham-exact\footnote{Actually, we only need $\,\underset{\tx{\ciut{(p+2)}}}{\chi}\,$ to be de Rham-closed, and hence \emph{locally} exact. In the best studied examples of physical relevance, though, the Green--Schwarz super-$(p+2)$-cocycle is exact and admits a global primitive, non-supersymmetric in general. We choose to incorporate this empirical fact in the very definition of the superbackground, so that the usual de Rham-cohomological issues do not obscure the Cartan--Eilenberg-cohomological ones that arise as we try to understand the supersymmetry of the super-$\si$-model from the higher-geometric perspective.} Gra\ss mann-even (super-)$(p+2)$-form
\qq\nn
\underset{\tx{\ciut{(p+2)}}}{\chi}=\tfrac{1}{(p+2)!}\,\unl\chi{}_{\unl A_1\unl A_2\ldots\unl A_{p+2}}\,\theta_{\rm L}^{\unl A_1}\wedge\theta_{\rm L}^{\unl A_2}\wedge\cdots\wedge\theta_{\rm L}^{\unl A_{p+2}}\in\Om^{p+2}(\txG)\,,
\qqq
termed the \textbf{Green--Schwarz super-$(p+2)$-cocycle}, with -- for any $\,k\in\txG\,$ --
\qq\nn
\chi_{\unl A_1\unl A_2\ldots\unl A_{p+2}}\,(\rho_k)^{\unl A_1}_{\ \unl B_1}\,(\rho_k)^{\unl A_2}_{\ \unl B_2}\,\cdots\,(\rho_k)^{\unl A_{p+2}}_{\ \unl B_{p+2}}=\chi_{\unl B_1\unl B_2\ldots\unl B_{p+2}}
\qqq
and with an $\txH$-basic global primitive
\qq\nn
\underset{\tx{\ciut{(p+1)}}}{\b}=\b_{\unl A_1\unl A_2\ldots\unl A_{p+1}}\,\theta_{\rm L}^{\unl A_1}\wedge\theta_{\rm L}^{\unl A_2}\wedge\cdots\wedge\theta_{\rm L}^{\unl A_{p+1}}\in\Om^{p+1}(\txG)\,,\qquad\qquad\sfd\underset{\tx{\ciut{(p+1)}}}{\b}=\underset{\tx{\ciut{(p+2)}}}{\chi}
\qqq
such that -- for any $\,k\in\txG\,$ --
\qq\nn
\b_{\unl A_1\unl A_2\ldots\unl A_{p+1}}\,(\rho_k)^{\unl A_1}_{\ \unl B_1}\,(\rho_k)^{\unl A_2}_{\ \unl B_2}\,\cdots\,(\rho_k)^{\unl A_{p+1}}_{\ \unl B_{p+1}}=\b_{\unl B_1\unl B_2\ldots\unl B_{p+1}}\,.
\qqq
These define uniquely the corresponding tensors: $\,\unl\txg,\underset{\tx{\ciut{(p+2)}}}{\txH}\,$ and $\,\underset{\tx{\ciut{(p+1)}}}{\txB}\,$ on $\,\cM_\txH\,$ satisfying the identities 
\qq\label{eq:Hachi}\qquad\qquad
\pi_{\txG/\txH}^*\unl\txg=\txg\,,\qquad\qquad\pi_{\txG/\txH}^*\underset{\tx{\ciut{(p+2)}}}{\txH}=\underset{\tx{\ciut{(p+2)}}}{\chi}\,,\qquad\qquad\pi_{\txG/\txH}^*\underset{\tx{\ciut{(p+1)}}}{\txB}=\underset{\tx{\ciut{(p+1)}}}{\b}\,.
\qqq
By the end of the day, then, we arrive at what has been and shall be referred to as the \textbf{Green--Schwarz super-$p$-brane superbackground},
\qq\label{eq:GSspbgrnd}
\sgt\Bgt^{\rm (GS)}_p=\bigl(\cM_\txH,\txg,\underset{\tx{\ciut{(p+2)}}}{\chi}\bigr)\,.
\qqq
Besides the above, the definition of the theories that we have in mind employs a canonical object associated with the pair $\,(\txG,\tgt_{\rm vac}^{(0)})\,$ given by the (rescaled) volume form on the Gra\ss mann-even component $\,\tgt_{\rm vac}^{(0)}\,$ of the vacuum subspace $\,\tgt_{\rm vac}\subset\tgt\subset\ggt$,
\qq\label{eq:HPcurv}
\underset{\tx{\ciut{(p+1)}}}{\b}\hspace{-7pt}{}^{\rm (HP)}=\tfrac{1}{(p+1)!}\,\ep_{\unl a{}_0\unl a{}_1\ldots\unl a{}_p}\,\theta^{\unl a{}_0}_{\rm L}\wedge\theta^{\unl a{}_1}_{\rm L}\wedge\cdots\wedge\theta^{\unl a{}_p}_{\rm L}\,,
\qqq
written in terms of the standard totally antisymmetric symbol
\qq\nn
\ep_{\unl a{}_0\unl a{}_1\ldots\unl a{}_p}=\left\{\barr{cl} \sign\left(\barr{cccc}0 & 1 &\ldots& p \\ \unl a{}_0 & \unl a{}_1 & \ldots & \unl a{}_p\earr\right) & \tx{ if } \{\unl a{}_0,\unl a{}_1,\ldots,\unl a{}_p\}=\ovl{0,p} \\ \\
0 & \tx{ otherwise}\earr\right.
\qqq
and called the \textbf{Hughes--Polchinski super-$(p+1)$-form}. If, as we shall assume henceforth, the restriction of the adjoint action of $\,\txH_{\rm vac}\,$ on $\,\ggt^{(0)}\,$ to the Gra\ss mann-even component $\,\tgt^{(0)}_{\rm vac}\,$ of the vacuum subspace is \textbf{unimodular}, {\it i.e.}, we have
\qq\label{eq:unimodvac}
\forall_{h\in\txH_{\rm vac}}\ :\ \det\,\bigl(\sfT_e\Ad_h\rstr_{\tgt_{\rm vac}^{(0)}}\bigr)=1\,,
\qqq
the HP super-$(p+1)$-form descends to the homogeneous space $\,\cM_{\txH_{\rm vac}}$,\ that is there exists a unique super-$(p+1)$-form $\,\underset{\tx{\ciut{(p+1)}}}{\txB}\hspace{-7pt}{}^{\rm (HP)}\,$ on $\,\cM_{\txH_{\rm vac}}\,$ such that 
\qq\nn
\pi_{\txG/\txH_{\rm vac}}^*\underset{\tx{\ciut{(p+1)}}}{\txB}\hspace{-7pt}{}^{\rm (HP)}=\underset{\tx{\ciut{(p+1)}}}{\b}\hspace{-7pt}{}^{\rm (HP)}\,.
\qqq
The latter, in conjunction with the descendant $\,\underset{\tx{\ciut{(p+2)}}}{\txH^{\rm vac}}\,$ such that 
\qq\nn
\pi_{\txG/\txH_{\rm vac}}^*\underset{\tx{\ciut{(p+2)}}}{\txH}\hspace{-7pt}{}^{\rm vac}=\underset{\tx{\ciut{(p+2)}}}{\chi}
\qqq
and
\qq\nn
\underset{\tx{\ciut{(p+2)}}}{\txH}\hspace{-7pt}{}^{\rm vac}=\sfd\underset{\tx{\ciut{(p+1)}}}{\txB}\hspace{-7pt}{}^{\rm vac}
\qqq
for $\,\underset{\tx{\ciut{(p+1)}}}{\txB}\hspace{-7pt}{}^{\rm vac}\,$ such that
\qq\nn
\pi_{\txG/\txH_{\rm vac}}^*\underset{\tx{\ciut{(p+1)}}}{\txB}\hspace{-7pt}{}^{\rm vac}=\underset{\tx{\ciut{(p+1)}}}{\b}\,,
\qqq
determines the \textbf{Hughes--Polchinski super-$p$-brane background}
\qq\label{eq:HPspbgrnd}
\sgt\Bgt^{{\rm (HP)}}_{p,\la_p}=\bigl(\cM_\txH,\underset{\tx{\ciut{(p+2)}}}{\chi}+\la_p\,\sfd\underset{\tx{\ciut{(p+1)}}}{\b}\hspace{-7pt}{}^{\rm (HP)}\equiv\underset{\tx{\ciut{(p+2)}}}{\widehat\chi}\bigr)\,.
\qqq

Under circumstances to be made precise in what follows, the field theories come in (essentially) dual pairs consisting of a theory with $\,\txK=\txH\,$ and another one of the latter type for the corresponding $\,\txK=\txH_{\rm vac}$.\ Accordingly, we define pairs of super-$\si$-models related by a duality and, in the subsequent sections, exploit that duality towards an elucidation of the (super)symmetry content of the theories. We begin with
\bedef\label{def:GSinNG}
The \textbf{Green--Schwarz super-$\si$-model in the Nambu--Goto formulation for the super-$p$-brane} in the Green--Schwarz superbackground $\,\sgt\Bgt^{{\rm (NG)}}_p\,$ as above, with -- in particular -- $\,\ggt=\tgt\oplus\hgt\,$ reductive, is the lagrangean theory of mappings $\,\xi\in[\Om_p,\cM_\txH]\,$ determined by the principle of least action applied, in the $\cS$-point picture, to the Dirac--Feynman amplitude
\qq\nn
\cA_{\rm DF}^{{\rm (NG)},p}[\xi]:=\ee^{\sfi\,S_{{\rm GS,p}}^{{\rm (NG)}}[\xi]}
\qqq
determined by the action functional 
\qq\nn
S_{{\rm GS},p}^{{\rm (NG)}}[\xi]:=\int_{\Om_p}\,\sqrt{\det_{(p)}\,\bigl(\xi^*\unl\txg\bigr)}+\int_{\Om_p}\,\xi^*\underset{\tx{\ciut{(p+1)}}}{\txB}\,.
\qqq

\noindent Equivalently, upon picking up an arbitrary open (superdomain) cover $\,\cU^\txH\equiv\{\cU^\txH_i\}_{i\in I^\txH}\,$ of $\,\cM_\txH\,$ that trivialises the principal $\txH$-bundle \eqref{eq:homasprinc} (for $\,\txK=\txH$) together with the corresponding local sections $\,\si_i^\txH\ :\ \cU_i^\txH\too\txG\,$ of $\,\pi_{\txG/\txH}$,\ and an arbitrary tessellation $\,\triangle_{\Om_p}\,$ of $\,\Om_p$,\ with the $k$-cell sets $\,\Tgt_k,\ k\in\ovl{0,p+1}$,\ subordinate to $\,\cU^\txH\,$ along $\,\xi$,\ as captured by the existence of a map $\,\imath_\cdot\ :\ \triangle_{\Om_p}\too I^\txH\,$ satisfying the condition
\qq\nn
\forall_{\t\in\Tgt_{p+1}}\ :\ |\xi|(\t)\subset\bigl|\cU^\txH_{\imath_\t}\bigr|\,,
\qqq
the action functional of the theory can be expressed in terms of tensors $\,\txg\,$ and $\,\underset{\tx{\ciut{(p+1)}}}{\b}\,$ as ($\xi_\t\equiv\xi\rstr_\t$)
\qq\nn
S_{{\rm GS},p}^{{\rm (NG)}}[\xi]\equiv\sum_{\t\in\Tgt_{p+1}}\,\bigl(\int_\t\,\sqrt{\det_{(p)}\,\bigl(\bigl(\si_{\imath_\t}^\txH\circ\xi_\t\bigr)^*\txg\bigr)}+\int_\t\,\bigl(\si_{\imath_\t}^\txH\circ\xi_\t\bigr)^*\underset{\tx{\ciut{(p+1)}}}{\b}\bigr)\,.
\qqq
\exdef
\noindent The previous definition is accompanied by
\bedef\label{def:GSinHP}
The \textbf{Green--Schwarz super-$\si$-model in the Hughes--Polchinski formulation for the super-$p$-brane} in the Hughes--Polchinski superbackground $\,\sgt\Bgt^{{\rm (HP)}}_p\,$ as above at $\,\la_p\in\bR^\x$,\ with -- in particular -- $\,\ggt=\fgt\oplus\hgt_{\rm vac}\,$ reductive and for a unimodular adjoint action of $\,\txH_{\rm vac}\,$ on $\,\tgt^{(0)}_{\rm vac}$,\ is the lagrangean theory of mappings $\,\widehat\xi\in[\Om_p,\cM_{\txH_{\rm vac}}]\,$ determined by the principle of least action applied, in the $\cS$-point picture, to the Dirac--Feynman amplitude
\qq\nn
\cA_{\rm DF}^{{\rm (HP)},p,\la_p}\bigl[\widehat\xi\bigr]:=\ee^{\sfi\,S_{{\rm GS},p}^{{\rm (HP)},\la_p}[\widehat\xi]}
\qqq
determined by the action functional
\qq\nn
S_{{\rm GS},p}^{{\rm (HP)},\la_p}\bigl[\widehat\xi\bigr]:=\int_{\Om_p}\,\widehat\xi^*\bigl(\la_p\,\underset{\tx{\ciut{(p+1)}}}{\txB}\hspace{-7pt}{}^{\rm (HP)}+\underset{\tx{\ciut{(p+1)}}}{\txB}\hspace{-7pt}{}^{\rm vac}\bigr)\,.
\qqq

\noindent Equivalently, upon picking up an arbitrary open (superdomain) cover $\,\cU^{\txH_{\rm vac}}\equiv\{\cU^{\txH_{\rm vac}}_i\}_{i\in I^{\txH_{\rm vac}}}\,$ of $\,\cM_{\txH_{\rm vac}}\,$ that trivialises the principal $\txH_{\rm vac}$-bundle \eqref{eq:homasprinc} (for $\,\txK=\txH_{\rm vac}$) together with the corresponding local sections $\,\si_i^{\txH_{\rm vac}}\ :\ \cU_i^{\txH_{\rm vac}}\too\txG\,$ of $\,\pi_{\txG/\txH_{\rm vac}}$,\ and an arbitrary tessellation $\,\triangle_{\Om_p}\,$ of $\,\Om_p$,\ with the $k$-cell sets $\,\Tgt_k,\ k\in\ovl{0,p+1}$,\ subordinate to $\,\cU^{\txH_{\rm vac}}\,$ along $\,\widehat\xi$,\ as captured by the existence of a map $\,\imath_\cdot\ :\ \triangle_{\Om_p}\too I^{\txH_{\rm vac}}\,$ satisfying the condition
\qq\nn
\forall_{\t\in\Tgt_{p+1}}\ :\ |\widehat\xi|(\t)\subset\bigl|\cU^{\txH_{\rm vac}}_{\imath_\t}\bigr|\,,
\qqq
the action functional of the theory can be expressed in terms of tensors $\,\underset{\tx{\ciut{(p+1)}}}{\b}\hspace{-7pt}{}^{\rm (HP)}\,$ and $\,\underset{\tx{\ciut{(p+1)}}}{\b}\,$ as ($\widehat\xi_\t\equiv\widehat\xi\rstr_\t$)
\qq\nn
S_{{\rm GS},p}^{{\rm (HP)}}\bigl[\widehat\xi\bigr]\equiv\sum_{\t\in\Tgt_{p+1}}\,\int_\t\,\bigl(\si_{\imath_\t}^{\txH_{\rm vac}}\circ\widehat\xi_\t\bigr)^*\bigl(\la_p\,\underset{\tx{\ciut{(p+1)}}}{\b}\hspace{-7pt}{}^{\rm (HP)}+\underset{\tx{\ciut{(p+1)}}}{\b}\bigr)\,.
\qqq
\exdef

\noindent\textbf{The Dimensional Constraint:} We may further constrain the admissible choices of superbackgrounds by restoring linear dimensions of the various coordinates and -- as a consequence -- also of the corresponding basis super-1-forms entering the definition of the GS super-$(p+2)$-cocycle. Thus, upon setting 
\qq\nn
\bigl[x^a\bigr]\equiv\bigl[\theta^a_{\rm L}\bigr]=1 \txm=2\bigl[\theta_{\rm L}^\a\bigr]\equiv2\bigl[\theta^\a\bigr]\,,
\qqq
we readily conclude that the only super-$\si$-models (of the type considered) with \emph{both} terms in the action functional of dimensionality $\,\txm^{p+1}\,$ (or, equivalently, with a dimension{\it less} relative normalisation coefficient) are those whose GS super-$(p+2)$-cocycles are wedge products of $2s$ spinorial components $\,\theta_{\rm L}^\a\,$ and $\,p+2-2s\,$ vectorial components $\,\theta_{\rm L}^a\,$ of the Maurer--Cartan super-1-form $\,\theta_{\rm L}\,$ (recall that the super-$(p+2)$-cocycles are $\hgt$-horizontal by assumption), with $s$ constrained by the equality
\qq\nn
\tfrac{1}{2}\,2s+p+2-2s\must p+1\,,
\qqq
whence the requirement
\qq\label{eq:DimConstr}
s\must 1\,,
\qqq
which we impose henceforth (resp.\ verify in the most studied examples). Consequently, we always write
\qq\nn
\underset{\tx{\ciut{(p+2)}}}{\chi}=\tfrac{1}{2p!}\,\chi_{\a\b a_1 a_2\ldots a_p}\,\Si_{\rm L}^\a\wedge\Si_{\rm L}^\b\wedge\theta_{\rm L}^{a_1}\wedge\theta_{\rm L}^{a_2}\wedge\cdots\wedge\theta_{\rm L}^{a_p}
\qqq
for some $\txH$-invariant tensors $\,\chi_{\a\b a_1 a_2\ldots a_p}\equiv\chi_{(\a\b)[a_1 a_2\ldots a_p]},\ \a,\b\in\ovl{1,\d-d},\ a_1,a_2,\ldots,a_p\in\ovl{0,d}$.\medskip

It ought to be noted that among the known consistent superbackgrounds there are also those that explicitly violate the above constraint, to wit, the Zhou super-0-brane of \Rcite{Zhou:1999sm} in $\,{\rm s}({\rm AdS}_2\x\bS^2)$,\ the Metsaev--Tseytlin D3-brane of \Rcite{Metsaev:1998hf} in $\,{\rm s}({\rm AdS}_5\x\bS^5)$,\ the M2-brane of \Rcite{deWit:1998yu} in $\,{\rm s}({\rm AdS}_4\x\bS^7)\,$ or in $\,{\rm s}({\rm AdS}_7\x\bS^4)$,\ and the M5-brane of \Rcite{Claus:1998fh} in the same supertargets. In the first of these `counterexamples', the GS super-2-cocycle is a linear combination of a bi-spinorial term and a bi-vectorial term, conspiring to render the combination de Rham-closed. The construction of a supersymmetric topological (Wess--Zumino) term of the super-$\si$-model for the D3-brane, on the other hand, calls for the incorporation of the curvature of a principal $\bC^\x$-bundle over the D3-brane worldvolume, the latter super-2-form admitting only an implicit (homotopy-formula) presentation in terms of the components of the LI Maurer--Cartan super-1-form on $\,{\rm SU}(2,2|4)$.\ Finally, in the eleven-dimensional setting, the super-$\si$-model uses the M-theory 4-form with a non-trivial geometrisation reflected by a subtle Dirac's charge-quantisation condition\footnote{The de Rham class of the 4-form becomes integral only after a twist by one quarter of the first Pontryagin class of the supertarget.} imposed upon it, {\it cp} Refs.\,\cite{Witten:1996md,Witten:1996hc}. We hope to return to these more involved structures in the future. In order to give the Reader a foretaste of what we may stumble upon along the way, we scrutinise the much tractable four-dimensional `counterexample' from the above list in the Appendix.
\medskip

\noindent The two field-theoretic constructs introduced above were first set in correspondence by 
\bethe[\Rxcite{Thms.\,5.1 \& 5.2}{Suszek:2019cum}]\label{thm:corrNGHP} If the superbackgrounds $\,\sgt\Bgt_p^{\rm (NG)}\,$ and $\,\sgt\Bgt_{p,\la_p}^{\rm (HP)}\,$ are as described above (that is -- in particular -- both decompositions $\,\ggt=\fgt\oplus\kgt\,$ are reductive and the adjoint action of $\,\txH_{\rm vac}\,$ on $\,\tgt^{(0)}_{\rm vac}\,$ is unimodular), there exists a $\sfT_e\Ad_\txH$-invariant scalar product $\,\txg\,$ on $\,\tgt^{(0)}\,$ such that\footnote{In what folows, we refer to this situation as the \textbf{$\sfT_e\Ad_\txH$-invariance of the vacuum splitting}.}
\qq\nn
\tgt^{(0)}_{\rm vac}\perp_\txg\egt^{(0)}\,,
\qqq
and the \textbf{Maximal Mixing Constraint}
\qq\label{eq:MMC}
\corr{\,P_{\widehat a}\,\vert\,\exists_{(\unl b,\widehat S)\in\ovl{0,p}\x\ovl{D-\unl\d+1,D-\d}}\ :\ f_{\widehat S\widehat a}^{\ \ \ \ \unl b}\neq 0\,}=\egt^{(0)}
\qqq
is obeyed, then there exists a unique value of the parameter $\,\la_p\,$ for which the GS super-$\si$-model in the NG formulation of Def.\,\ref{def:GSinNG} with $\,\unl\txg=\txg\,$ is (classically) equivalent  to the GS super-$\si$-model in the HP formulation of Def.\,\ref{def:GSinHP} partially reduced through imposition of the \textbf{Inverse Higgs Constraints}
\qq\nn
\bigl(\si_{\imath_\t}^{\txH_{\rm vac}}\circ\widehat\xi\bigr)^*\theta_{\rm L}^{\widehat a}\must 0\,,\qquad\qquad\widehat a\in\ovl{p+1,d}\,.
\qqq
More specifically, the DF amplitude $\,\cA_{\rm DF}^{{\rm (HP),p},\la_p}\,$ written in the \emph{gauge} $\,\si_i^{\txH_{\rm vac}}\equiv\si_i^{\rm vac},\ i\in I_{\txH_{\rm vac}}\,$ of \Reqref{eq:vacgauge} then reduces to the DF amplitude $\,\cA_{\rm DF}^{{\rm (NG),p}}\,$ written in the \emph{gauge} $\,\si_i^\txH\equiv\si_i^{\xcancel{\rm vac}},\ i\in I_{\txH_{\rm vac}}\,$ of \Reqref{eq:novacgauge}.
\ethe
\noindent Taking into account the manifest violation of the Maximal Mixing Constraint in many physically interesting and important situations, we readily generalise the result of \Rcite{Suszek:2019cum} in the form of
\bethe\label{thm:HpdualNGext}
Let the superbackgrounds $\,\sgt\Bgt_p^{\rm (NG)}\,$ and $\,\sgt\Bgt_{p,\la_p}^{\rm (HP)}\,$ be as described above (that is -- in particular -- assume both decompositions $\,\ggt=\fgt\oplus\kgt\,$ to be reductive and the adjoint action of $\,\txH_{\rm vac}\,$ on $\,\tgt^{(0)}_{\rm vac}\,$ to be unimodular), write 
\qq\nn
\dgt^{-1}\tgt^{(0)}_{\rm vac}:=\corr{\,P_{\widehat a}\,\vert\,\exists_{(\unl b,\widehat S)\in\ovl{0,p}\x\ovl{D-\unl\d+1,D-\d}}\ :\ f_{\widehat S\widehat a}^{\ \ \ \ \unl b}\neq 0\,}\equiv\bigoplus_{\widehat{\unl a}=p+1}^{d-L}\,\corr{P_{\widehat{\unl a}}}
\qqq
for some $\,0\leq L\leq d-p\,$ (with $\,L=d-p\,$ corresponding to the degenerate $\,\dgt^{-1}\tgt^{(0)}_{\rm vac}=\brd0$) and subsequently decompose
\qq\nn
\egt^{(0)}\equiv\dgt^{-1}\tgt^{(0)}_{\rm vac}\oplus\lgt^{(0)}\,,
\qqq
assuming that both
\qq\nn
\dgt^{-1}\tgt^{(0)}\equiv\tgt^{(0)}_{\rm vac}\oplus\dgt^{-1}\tgt^{(0)}_{\rm vac}
\qqq 
and $\,\lgt^{(0)}\,$ are $\sfT_e\Ad_\txH$-invariant (and so also ${\rm ad}_\hgt$-invariant), and -- finally --  let $\,\txg_{\rm vac}\,$ be a $\sfT_e\Ad_\txH$-invariant scalar product on $\,\dgt^{-1}\tgt^{(0)}\,$ such that
\qq\nn
\tgt^{(0)}_{\rm vac}\perp_{\txg_{\rm vac}}\dgt^{-1}\tgt^{(0)}_{\rm vac}\,.
\qqq
Then, there exists a unique value of the parameter $\,\la_p\,$ for which the GS super-$\si$-model in the HP formulation of Def.\,\ref{def:GSinHP} restricted to field configurations obeying the \textbf{Body-Localisation Constraints}
\qq\label{eq:BLC-EL}
\bigl(\si_{\imath_\t}^{\txH_{\rm vac}}\circ\widehat\xi\bigr)^*\theta_{\rm L}^{\widehat a}\must 0\,,\qquad\widehat a\in\ovl{d-L+1,d}
\qqq  
and further reduced (partially) through imposition of the \textbf{Inverse Higgs Constraints}
\qq\label{eq:IHC-EL}
\bigl(\si_{\imath_\t}^{\txH_{\rm vac}}\circ\widehat\xi\bigr)^*\theta_{\rm L}^{\widehat{\unl a}}\must 0\,,\qquad\widehat{\unl a}\in\ovl{p+1,d-L}
\qqq
is (classically) equivalent to the GS super-$\si$-model in the NG formulation of Def.\,\ref{def:GSinNG} with $\,\txg\,$ such that $\,\txg\rstr_{\dgt^{-1}\tgt^{(0)}}\equiv\txg_{\rm vac}$,\ restricted to field configurations subject to the (same) Body-Localisation Constraints
\qq\nn
\bigl(\si_{\imath_\t}^\txH\circ\xi\bigr)^*\theta_{\rm L}^{\widehat a}\must 0\,,\qquad\widehat a\in\ovl{d-L+1,d}\,,
\qqq
where it is to be understood that the DF amplitude $\,\cA_{\rm DF}^{{\rm (HP),p},\la_p}\,$ written in the \emph{gauge} $\,\si_i^{\txH_{\rm vac}}\equiv\si_i^{\rm vac},\ i\in I_{\txH_{\rm vac}}\,$ of \Reqref{eq:vacgauge} reproduces the DF amplitude $\,\cA_{\rm DF}^{{\rm (NG),p}}\,$ written in the \emph{gauge} $\,\si_i^\txH\equiv\si_i^{\xcancel{\rm vac}},\ i\in I_{\txH_{\rm vac}}\,$ of \Reqref{eq:novacgauge} under the correspondence.
\ethe
\beroof
Obvious.
\eroof
\brem
The generalisation formulated above is intended to cover a situation in which the body of the supertarget is a cartesian product
\qq\nn
|\txG/\txH|=\bigl(|\txG_1|/\txH_1\bigr)\x\bigl(|\txG_2|/\txH_2\bigr)
\qqq
of two \emph{reductive} homogeneous spaces of the Lie groups $\,|\txG_A|,\ A\in\{1,2\}\,$ (with the respective Lie algebras $\,|\ggt_A|$) relative to the respective cartesian factors $\,\txH_A\,$ of the isotropy group
\qq\nn
\txH=\txH_1\x\txH_2\,,
\qqq
and the vacuum of the super-$\si$-model is constrained to be embedded entirely in one of the factors, say $\,|\txG_1|/\txH_1$,\ in such a manner that in the obvious notation (referring directly to the previously introduced one)
\qq\nn
|\ggt_A|=\tgt^{(0)}_A\oplus\hgt_A\,,
\qqq
with
\qq\nn
\tgt^{(0)}_1\oplus\tgt^{(0)}_2\equiv\tgt^{(0)}\,,\qquad\qquad\hgt_1\oplus\hgt_2\equiv\hgt\,,
\qqq
we have
\qq\nn
\hgt_{\rm vac}=\hgt_{1\,{\rm vac}}\oplus\hgt_2\,,\qquad\qquad\hgt_{1\,{\rm vac}}\subset\hgt_{1\,{\rm vac}}\oplus\dgt\equiv\hgt_1
\qqq
and
\qq\nn
\tgt_{\rm vac}^{(0)}\subset\tgt_{\rm vac}^{(0)}\oplus\egt_1^{(0)}\equiv\tgt_1^{(0)}\,,\qquad\qquad\egt^{(0)}\equiv\egt_1^{(0)}\oplus\tgt_2^{(0)}\,,
\qqq
with
\qq\nn
\dgt^{-1}\tgt_{\rm vac}^{(0)}\subset\egt^{(0)}_1\subsetneq\egt^{(0)}\,.
\qqq
Under such circumstances, switching on and subsequently integrating out Goldstone modes along $\,\dgt\,$ will not induce the terms of the metric integrand in the NG super-$\si$-model along $\,\tgt_2^{(0)}$,\ and so the latter have to be frozen out by hand in the metric term of the NG super-$\si$-model through imposition of the Body-Localisation Constraints. This certainly is an invasive manipulation on the GS field theory under consideration. However, we should bear in mind that the correspondence is employed solely towards geometrisation of the description of the vacuum of that field theory ensuing from the variational analysis of its DF amplitude, and that of its local supersymmetry. When considered from this vantage point, the manipulation acquires the interpretation of a mere (partial) localisation of (the body of) the vacuum.
\erem

The correspondence established in the last theorem admits a reformulation that opens up an avenue for a geometrisation of field-theoretic statements made in the purely topological setting of the HP formulation that we shall find particularly convenient and robust in our subsequent considerations. It begins with the definition of the family of sub-supermanifolds
\qq\label{eq:secHP}
\cV_i:=l_{g_i}\bigl(\cV_e\bigr)\equiv\si_i^{\rm vac}\bigl(\cU_i^{\txH_{\rm vac}}\bigr)\subset\txG\,,\qquad i\in I_{\txH_{\rm vac}}\,,
\qqq
with
\qq\nn
\cV_0\equiv\cV_e\,,
\qqq
that faithfully present the supertarget $\,\cM_{\txH_{\rm vac}}\,$ patchwise. Their disjoint union 
\qq\label{eq:HPsec}
\Si^{\rm HP}:=\bigsqcup_{i\in I_{\txH_{\rm vac}}}\,\cV_i\,,
\qqq
which we shall call the \textbf{Hughes--Polchinski section} in what follows, is the arena on which all supergeometric phenomena of interest to us take place, and so it is certainly worth a closer inspection. We have
\berop\label{prop:bastanshHP}
In the hitherto notation, the tangent sheaf 
\qq\nn
\cT\Si^{\rm HP}\equiv\bigsqcup_{i\in I_{\txH_{\rm vac}}}\,\cT\cV_i\,,\qquad\qquad\cT\cV_i=\sfT\si_i^{\rm vac}\bigl(\cT\cM_{\txH_{}\rm vac}\rstr_{\cU_i^{\txH_{\rm vac}}}\bigr)
\qqq
of the Hughes--Polchinski section $\,\Si^{\rm HP}\,$ is spanned, over its component $\,\cV_i$,\ on vector fields
\qq\label{eq:TmuasL}
\cT_{\mu\,i}=L_\mu\rstr_{\cV_i}+T_{\mu\,i}^{\ \ \ \unl S}\,L_{\unl S}\,,\qquad\mu\in\ovl{0,\unl\d}\,,
\qqq
with sections $\,T_{\mu\,i}^{\ \ \ \unl S}\in\cO_\txG(\cV_i)\,$ uniquely fixed by the tangency condition 
\qq\nn
\cT_{\mu\,i}\stackrel{!}{\in}\cT\cV_i
\qqq
in the form
\qq\nn
T_{\mu\,i}^{\ \ \ \unl S}\bigl(\si_i^{\rm vac}(\unl\chi{}_i)\bigr)=\bigl(\unl E(\unl\chi{}_i)^{-1}\bigr)_\mu^{\ \nu}\,E^{\ \unl S}_\nu(\unl\chi{}_i)\,,
\qqq
where $\,E_\mu^{\ A}\,$ is the Vielbein field introduced in Eqs.\,\eqref{eq:Vielbein} and \eqref{eq:vacVielbein} and $\,\unl E(\unl\chi{}_i)^{-1}\,$ is the inverse of the quadratic matrix 
\qq\nn
\unl E(\unl\chi{}_i)\equiv\bigl(E_\mu^{\ \nu}(\unl\chi{}_i)\bigr)_{\mu\in\ovl{0,\unl\d}}^{\nu\in\ovl{0,\unl\d}}\,.
\qqq
\eerop
\beroof
We adapt the reasoning given in \Rxcite{Sec.\,2}{Suszek:2018bvx}. Present $\,\cT_{\mu\,i}\,$ as the pushforward of a vector field tangent to the base,
\qq\nn
\cT_{\mu\,i}\bigl(\si_i^{\rm vac}(\unl\chi{}_i)\bigr)\equiv\sfT_{\unl\chi{}_i}\si_i^{\rm vac}\bigl(\unl\cT{}_{\mu\,i}(\unl\chi{}_i)\bigr)\,,
\qqq
that we write in a local coordinate system as
\qq\nn
\unl\cT{}_{\mu\,i}(\unl\chi{}_i)=\D^\nu_{\mu\,i}(\unl\chi{}_i)\,\tfrac{\vec\p\ }{\p\unl\chi{}_i^\nu}\,,
\qqq
and 
\qq\nn
\D^\nu_{\mu\,i}\,E^{\ A}_\nu(\unl\chi{}_i)\equiv\unl\cT{}_{\mu\,i}\con\si_i^{{\rm vac}\,*}\theta_{\rm L}^A(\unl\chi{}_i)=\cT_{\mu\,i}\con\theta_{\rm L}^A\bigl(\si_i^{\rm vac}(\unl\chi{}_i)\bigr)\equiv\d_\mu^{\ A}+T_{\mu\,i}^{\ \ \ \unl S}\bigl(\si_i^{\rm vac}(\unl\chi{}_i)\bigr)\,\d_{\unl S}^{\ A}\,,
\qqq
valid for any $\,A\in\ovl{0,D}$.\ Setting $\,A\equiv\la\in\ovl{0,\unl\d}$,\ we obtain 
\qq\label{eq:hatlasrho}
\D^\nu_{\mu\,i}\,E^{\ \la}_\nu(\unl\chi{}_i)=\d_{\mu}^{\ \la}\,.
\qqq 
At this stage, the only thing that has to be proven is the invertibility of $\,\unl E$.\ To this end, we compute, by a variation on the above,
\qq\nn
\sfT_{\unl\chi{}_i}\si_i^{\rm vac}\bigl(\tfrac{\vec\p\ }{\p\unl\chi{}_i^\mu}\bigr)=E^{\ \nu}_\mu(\unl\chi{}_i)\,L_\nu\bigl(\si_i^{\rm vac}(\unl\chi{}_i)\bigr)+E^{\ \unl S}_\mu(\unl\chi{}_i)\,L_{\unl S}\bigl(\si_i^{\rm vac}(\unl\chi{}_i)\bigr)\,.
\qqq
Now, the horizontal lift of the coordinate vector fields determined by the principal $\txH_{\rm vac}$-connection as in \Reqref{eq:Horhom}, 
\qq\nn
{\rm Hor}_{\si_i^{\rm vac}(\unl\chi{}_i)}\bigl(\tfrac{\vec\p\ }{\p\unl\chi{}_i^\mu}\bigr)=E^{\ \nu}_\mu(\unl\chi{}_i)\,L_\nu\bigl(\si_i^{\rm vac}(\unl\chi{}_i)\bigr)\,,
\qqq
yields a basis of the horizontal subspace $\,\ceH_{\si_i^{\rm vac}(\unl\chi{}_i)}\txG\,$ in the horizontal subsheaf $\,\ceH\txG\,$ of the tangent sheaf $\,\cT\txG$,\ of dimension
\qq\nn
\dim\,\ceH_{\si_i^{\rm vac}(\unl\chi{}_i)}\txG=\dim\,\fgt\,,
\qqq
whence the anticipated property of the reduced Vielbein $\,\unl E$.
\eroof

\noindent It is natural, from our field-theoretic point of view, to distinguish those among global sections of the tangent sheaf of the HP section that descend to the homogeneous space $\,\cM_{\txH_{\rm vac}}$.\ This we do in
\bedef
Adopt the hitherto notation and let $\,\{h_{ij}\}_{i,j\in I_{\txH_{\rm vac}}}\,$ be the transition mappings of the principal $\txH_{\rm vac}$-bundle \eqref{eq:homasprinc} (with $\,\txK=\txH_{\rm vac}$) for the family $\,\{\si_i^{\rm vac}\}_{i\in I_{\txH_{\rm vac}}}\,$ of the local sections introduced previously, so that we have 
\qq\nn
\si_j^{\rm vac}(\unl\chi{}_j)=\si_i^{\rm vac}(\unl\chi{}_i)\cdot h_{ij}\bigl(\unl\chi{}_i\bigr)
\qqq
(at every $\cS$-point) in the intersection of $\,\cU_i^{\txH_{\rm vac}}\,$ and $\,\cU_j^{\txH_{\rm vac}}$.\ Consider an arbitrary vector field $\,\cW\in\G(\cT\Si^{\rm HP})\,$ with restrictions
\qq\nn
\cW\rstr_{\cV_i}=W^\mu\,\cT_{\mu\,i}
\qqq
expressed in terms of sections $\,W^\mu\in\cO_\txG(\cV_i)$.\ We call $\,\cW\,$ an \textbf{$\txH_{\rm vac}$-descendable vector field on} $\,\Si^{\rm HP}\,$ if the identity
\qq\nn
\bigl(\rho_{h_{ij}(\unl\chi{}_i)^{-1}}\bigr)^\nu_{\ \mu}\,W^\mu\bigl(\unl\chi{}_i\bigr)=W^\nu\bigl(\unl\chi{}_j\bigr)
\qqq
holds true over the intersection of $\,\cU_i^{\txH_{\rm vac}}\,$ and $\,\cU_j^{\txH_{\rm vac}}\,$ for any $\,i,j\in I_{\txH_{\rm vac}}$.
\exdef
\noindent A related concept of prime relevance to our later analysis is introduced in the next
\bedef\label{def:Hvacdescsdistro}
In the hitherto notation, let 
\qq\nn
\cD\subset\cT\Si^{\rm HP}
\qqq
be an arbitrary superdistribution (in the sense of \Rxcite{Def.\,6.1.1}{Carmeli:2011}). We call $\,\cD\,$ an \textbf{$\txH_{\rm vac}$-descendable superdistribution over} $\,\Si^{\rm HP}\,$ if the relation 
\qq\nn
\sfT_{\unl\chi{}_i}r^{\txH_{\rm vac}}_{h_{ij}(\unl\chi{}_i)}\bigl(\cD_{\si_i^{\rm vac}(\unl\chi{}_i)}\bigr)\subseteq\cD_{\si_j^{\rm vac}(\unl\chi{}_j)}\,,
\qqq 
written in terms of the defining action $\,r^{\txH_{\rm vac}}_\cdot\,$ of \Reqref{eq:ractdef}, obtains over the intersection of $\,\cU_i^{\txH_{\rm vac}}\,$ and $\,\cU_j^{\txH_{\rm vac}}\,$ for any $\,i,j\in I_{\txH_{\rm vac}}$,\ or -- equivalently -- if the following implications hold true
\qq\nn
W^\mu\,\cT_{\mu\,i}\bigl(\si_i^{\rm vac}\bigl(\unl\chi{}_i\bigr)\bigr)\in\cD_{\unl\chi{}_i}\qquad\Longrightarrow\qquad \bigl(\rho_{h_{ij}(\unl\chi{}_i)^{-1}}\bigr)^\nu_{\ \mu}\,W^\mu\,\cT_{\nu\,j}\bigl(\si_j^{\rm vac}\bigl(\unl\chi{}_j\bigr)\bigr)\in\cD_{\unl\chi{}_j}
\qqq 
over the intersections.
\exdef

\brem\label{rem:hvacsymsdistro}
The notion of an $\txH_{\rm vac}$-descendable superdistribution is closely related to the familiar notion of an \textbf{$\hgt_{\rm vac}$-invariant superdistribution}, {\it i.e.}, of a distribution $\,\cD\,$ with the property
\qq\nn
[\hgt_{\rm vac}\rstr_{\Si^{\rm HP}},\cD]\subseteq\cD\,,
\qqq
with $\,\hgt_{\rm vac}\rstr_{\Si^{\rm HP}}\,$ standing for the restriction of the vertical distribution of the principal $\txH_{\rm vac}$-bundle \eqref{eq:homasprinc} (with $\,\txK=\txH_{\rm vac}$) to the HP section.
\erem

We may, at last, return to our discussion of the super-$\si$-model and rephrase the thesis of Thm.\,\ref{thm:HpdualNGext} upon giving one last
\bedef\label{def:Corrsdistro}
Adopt the hitherto notation. The \textbf{correspondence superdistribution of} $\,\sgt\Bgt^{{\rm (HP)}}_{p,\la_p}\,$ is the superdistribution within the tangent sheaf $\,\cT\Si^{\rm HP}\,$ of the Hughes--Polchinski section of Prop.\,\ref{prop:bastanshHP} defined as
\qq\label{eq:corresdistro}
{\rm Corr}_{\rm HP/NG}\bigl(\sgt\Bgt_{p,\la_p}^{\rm (HP)}\bigr):={\rm Ker}\,\bigl(\sfP^\ggt_{\ \egt^{(0)}}\circ\theta_{\rm L}\rstr_{\cT\Si^{\rm HP}}\bigr)\subset\cT\Si^{\rm HP}
\qqq
in terms of the projector 
\qq\nn
\sfP^\ggt_{\ \egt^{(0)}}\ :\ \ggt\circlearrowleft
\qqq 
onto $\,\egt^{(0)}$,\ with the kernel
\qq\nn
{\rm Ker}\,\sfP^\ggt_{\ \egt^{(0)}}=\tgt^{(1)}\oplus\tgt^{(0)}_{\rm vac}\oplus\hgt
\qqq
whose proper subspace
\qq\nn
\tgt^{(1)}\oplus\tgt^{(0)}_{\rm vac}\oplus\dgt\equiv\gt{corr}_{\rm HP/NG}\bigl(\sgt\Bgt_{p,\la_p}^{\rm (HP)}\bigr)
\qqq
models $\,{\rm Corr}_{\rm HP/NG}(\sgt\Bgt_{p,\la_p}^{\rm (HP)})\,$ locally. 
\exdef 

\noindent The correspondence between the two formulations of the GS super-$\si$-model stated in Thm.\,\ref{thm:HpdualNGext} pertains to those mappings $\,\widehat\xi\in[\Om_p,\cM_{\txH_{\rm vac}}]\,$ for which the images of the tangents of the superpositions $\,\si_i^{\rm vac}\circ\widehat\xi\,$ are contained in the correspondence superdistribution -- we shall refer to the entirety of such mappings (in the context of the opening paragraph of the present section) as the \textbf{HP/NG correspondence sector}. In order for the correspondence to be physically meaningful, it is necessary that the localisation constraints: \eqref{eq:BLC-EL} and \eqref{eq:IHC-EL} which we impose \emph{locally} over $\,\Om_p\,$ can be continued across the trivialising patches $\,\cU_i^{\txH_{\rm vac}}\,$ and do not depend on the choice of the local gauge ({\it i.e.}, on the choice of the $\,\si_i^{\rm vac}$). Thus, altogether, we demand that the correspondence superdistribution descend to $\,\cM_{\txH_{\rm vac}}$.\ Inspection of the structure of the modelling supervector space $\,\gt{corr}_{\rm HP/NG}(\sgt\Bgt_{p,\la_p}^{\rm (HP)})\,$ readily shows that our demand should be formulated as\medskip

\noindent\textbf{The Descendability Constraint:} 
We require the splitting \eqref{eq:isovacsplit} of the isotropy algebra to be $\ad_{\hgt_{\rm vac}}$-invariant in the sense expressed by the relation
\qq\label{eq:DesConstr}
[\hgt_{\rm vac},\dgt]\subset\dgt\,.
\qqq
\medskip

The last few definitions, related by the correspondence theorem (and subject to the Dimensional Constraint), demarcate the environment in which all our subsequent physical considerations are placed, and the rest of the present section provides the relevant supergeometric substrate. In order to put some flesh on the abstract logical skeleton laid thus, let us, prior to launching a canonical (super)symmetry analysis of the field theories of interest and identifying the higher geometry behind them, take a look at a bunch of physically relevant examples that illustrate the abstract ideas and constructions.\medskip

\beg\textbf{The Green--Schwarz super-0-brane in $\,{\rm sMink}(9,1\,|\,32)$.}\label{eq:s0gsMink}
\bit
\item[$\bullet$] The mother super-Harish--Chandra pair: 
\qq\nn
{\rm sISO}(9,1\,|\,32)\equiv\bigl({\rm ISO}(9,1)\equiv\bR^{9,1}\rx{\rm SO}(9,1),\gt{siso}(9,1\,|\,32)\bigr)\,,
\qqq 
consisting of the Poincar\'e group $\,{\rm ISO}(9,1)\,$ of the Minkowski space $\,\bR^{9,1}\,$ and the super-Poincar\'e algebra
\qq\nn
\gt{siso}(9,1|32)=\bigoplus_{\a=1}^{32}\,\corr{Q_\a}\oplus\bigoplus_{a=0}^9\,\corr{P_a}\oplus\bigoplus_{a,b=0}^9\,\corr{J_{ab}=-J_{ba}}
\qqq
with the structure equations
\qq\nn
&\{Q_\a,Q_\b\}=\bigl(C\,\G^a\bigr)_{\a\b}\,P_a\,,\qquad\qquad[P_a,P_b]=0\,,&\cr\cr\cr
&[J_{ab},J_{cd}]=\eta_{ad}\,J_{bc}-\eta_{ac}\,J_{bd}+\eta_{bc}\,J_{ad}-\eta_{bd}\,J_{ac}\,,\qquad\qquad[J_{ab},P_c]=\eta_{bc}\,P_a-\eta_{ac}\,P_b\,,&\cr\cr\cr
&[P_a,Q_\a]=0\,,\qquad\qquad[J_{ab},Q_\a]=\tfrac{1}{2}\,(\G_{ab})^\b_{\ \a}\,Q_\b\,,&
\qqq
expressed in terms of the generators 
\qq\nn
\{\G^a\}^{a\in\ovl{0,9}}
\qqq 
of the Clifford algebra $\,{\rm Cliff}(\bR^{9,1})\,$ of the Minkowski space $\,\bR^{9,1}\,$ with the metric $\,\eta=\diag(-1,+1,$ $+1,+1,+1,+1,+1,+1,+1,+1)\,$ (used to lower and raise vector indices throughout), their commutators 
\qq\nn
\G_{ab}=\tfrac{1}{2}\,[\G_a,\G_b]\,,
\qqq 
the chirality operator
\qq\nn
\G_{11}=\G^0\,\G^1\,\cdots\,\G^9
\qqq
and a charge-conjugation matrix $\,C\,$ (used to lower and raise spinor indices throughout) in a Majorana-spinor representation in which 
\qq\nn
C^{\rm T}=-C
\qqq
and
\qq\nn
\ovl\G{}^a\equiv C\,\G^a=\bigl(C\,\G^a\bigr)^{\rm T}\,,
\qqq 
so that -- in particular -- 
\qq\nn
C\,\G^a\,C^{-1}=-\G^{a\,{\rm T}}
\qqq
and 
\qq\nn
\ovl\G_{11}\equiv C\,\G_{11}=-\G_{11}^{\rm T}\,C\,;
\qqq
\item[$\bullet$] The $\sfT_e\Ad_{{\rm SO}(9,1)}$-invariance of the vacuum splitting -- obvious;
\item[$\bullet$] The homogeneous spaces: the NG one
\qq\nn
&{\rm sMink}(9,1\,|\,32)\equiv{\rm sISO}(9,1\,|\,32)/{\rm SO}(9,1)\equiv\bR(9,1\,|\,32)\,,&\cr\cr
&\txH\equiv{\rm SO}(9,1)\,,\qquad\qquad\hgt\equiv\gt{so}(9,1)=\bigoplus_{a,b=0}^9\,\corr{J_{ab}=-J_{ba}}\,,&
\qqq 
with the body
\qq\nn
|{\rm sMink}(9,1\,|\,32)|=\bR^{9,1}\equiv{\rm Mink}(9,1)\,,
\qqq
which happens to be a Lie supergroup, with global coordinates $\,\{\theta^\a,x^a\}^{(\a,a)\in\ovl{1,32}\x\ovl{0,9}}\,$ in which 
\qq\nn
\mu^*\ &:&\ \bigl(\theta^\a,x^a\bigr)\longmapsto\bigl(\theta^\a\ox\bd1+\bd1\ox\theta^\a,x^a\ox\bd1+\bd1\ox x^a-\tfrac{1}{2}\,\theta^\a\ox\ovl\G{}^a_{\a\b}\,\theta^\b\bigr)\,,\cr\cr
\Inv^*\ &:&\ \bigl(\theta^\a,x^a\bigr)\longmapsto\bigl(-\theta^\a,-x^a\bigr)\,,
\qqq
or, equivalently (in the $\cS$-point picture), with the group operations
\qq\nn
\bigl(\theta_1^\a,x_1^a\bigr)\cdot\bigl(\theta_2^\b,x_2^b\bigr)=\bigl(\theta_1^\a+\theta_2^\a,x_1^a+x_2^a-\tfrac{1}{2}\,\theta_1^\a\,\ovl\G{}^a_{\a\b}\,\theta_2^\b\bigr)\,,\qquad\qquad\bigl(\theta^\a,x^a\bigr)^{-1}=\bigl(-\theta^\a,-x^a\bigr)\,,
\qqq
and the HP one
\qq\nn
&{\rm sISO}(9,1\,|\,32)/{\rm SO}(9)\,,&\cr\cr
&\txH_{\rm vac}\equiv{\rm SO}(9)\,,\qquad\qquad\hgt_{\rm vac}\equiv\bigoplus_{\widehat a,\widehat b=1}^9\,\corr{J_{\widehat a\widehat b}=-J_{\widehat b\widehat a}}\,,&\cr\cr
&\dgt=\bigoplus_{\widehat a=1}^9\,\corr{J_{0\widehat a}}\,,&\cr\cr
&\tgt_{\rm vac}^{(0)}=\corr{P_0}\,,\qquad\qquad\dgt^{-1}\tgt_{\rm vac}^{(0)}=\bigoplus_{\widehat a=1}^9\,\corr{P_{\widehat a}}\equiv\egt^{(0)}\,;&
\qqq
\item[$\bullet$] The exponential superparametrisation(s) ($\widehat b\in\ovl{1,9}$):
\qq\nn
\si_0^{\rm vac}\bigl(\theta^\a,x^a,\phi^{0\widehat b}\bigr)=\ee^{\theta^\a\ox Q_\a}\cdot\ee^{x^a\ox P_a}\cdot\ee^{\phi^{0\widehat b}\ox J_{0\widehat b}}\,;
\qqq
\item[$\bullet$] The superbackgrounds: the NG one
\qq\nn
\sgt\Bgt_0^{{\rm (NG)}}&=&\bigl({\rm sMink}(9,1\,|\,32),\eta_{ab}\,\theta^a_{\rm L}\ox\theta_{\rm L}^b,\Si_{\rm L}\wedge\ovl\G_{11}\,\Si_{\rm L}\equiv\underset{\tx{\ciut{(2)}}}{\chi}\hspace{-1pt}{}^{\rm GS}\bigr)\,,\cr\cr
&&\theta_{\rm L}=\Si_{\rm L}^\a\ox Q_\a+\theta_{\rm L}^a\ox P_a+\theta^{ab}_{\rm L}\ox J_{a<b}
\qqq
with a \emph{non}supersymmetric global curving
\qq\nn
\underset{\tx{\ciut{(1)}}}{\b}\hspace{-1pt}{}^{\rm GS}(\theta,x,\phi)=\theta\,\ovl\G_{11}\,\sfd\theta
\qqq
(and no supersymmetric one),  and the HP one
\qq\nn
\sgt\Bgt_{0,\la_0}^{{\rm (HP)}}&=&\bigl({\rm sISO}(9,1\,|\,32)/{\rm SO}(9),\underset{\tx{\ciut{(2)}}}{\chi}\hspace{-1pt}{}^{\rm GS}+\tfrac{\la_0}{2}\,\bigl(\Si_{\rm L}\wedge\ovl\G{}^0\,\Si_{\rm L}-2\d_{\widehat a\widehat b}\,\theta_{\rm L}^{0\widehat a}\wedge\theta_{\rm L}^{\widehat b}\bigr)\equiv\underset{\tx{\ciut{(2)}}}{\widehat\chi}\hspace{-1pt}{}^{\rm GS}\bigr)\,,\cr\cr
&&\corr{P_0}\perp_\eta\bigoplus_{\widehat a=1}^9\,\corr{P_{\widehat a}}\,;
\qqq
\item[$\bullet$] The Body-Localisation Constraints: none.
\eit
\eeg

\beg\textbf{The Green--Schwarz super-$p$-brane with $\,p\in\ovl{1,9}\,$ in $\,{\rm sMink}(d,1\,|\,D_{d,1})$.}\label{eq:spgsMink}
\bit
\item[$\bullet$] The mother super-Harish--Chandra pair: 
\qq\nn
{\rm sISO}(d,1\,|\,D_{d,1})\equiv\bigl({\rm ISO}(d,1)\equiv\bR^{d,1}\rx{\rm SO}(d,1),\gt{siso}(d,1\,|\,D_{d,1})\bigr)\,,
\qqq 
consisting of the Poincar\'e group $\,{\rm ISO}(d,1)\,$ of the Minkowski space $\,\bR^{d,1}\,$ and the super-Poincar\'e algebra
\qq\nn
\gt{siso}(d,1\,|\,D_{d,1})=\bigoplus_{\a=1}^{D_{d,1}}\,\corr{Q_\a}\oplus\bigoplus_{a=0}^d\,\corr{P_a}\oplus\bigoplus_{a,b=0}^d\,\corr{J_{ab}=-J_{ba}}
\qqq
with the structure equations
\qq\nn
&\{Q_\a,Q_\b\}=\bigl(C\,\G^a\bigr)_{\a\b}\,P_a\,,\qquad\qquad[P_a,P_b]=0\,,&\cr\cr\cr
&[J_{ab},J_{cd}]=\eta_{ad}\,J_{bc}-\eta_{ac}\,J_{bd}+\eta_{bc}\,J_{ad}-\eta_{bd}\,J_{ac}\,,\qquad\qquad[J_{ab},P_c]=\eta_{bc}\,P_a-\eta_{ac}\,P_b\,,&\cr\cr\cr
&[P_a,Q_\a]=0\,,\qquad\qquad[J_{ab},Q_\a]=\tfrac{1}{2}\,(\G_{ab})^\b_{\ \a}\,Q_\b\,,&
\qqq
expressed in terms of the generators $\,\{\G^a\}^{a\in\ovl{0,d}}\,$ of the Clifford algebra $\,{\rm Cliff}(\bR^{d,1})\,$ of the Minkowski space $\,\bR^{d,1}\,$ with the metric $\,\eta=\diag(-1,\underbrace{+1,+1,\ldots,+1}_{d\ \tx{times}})$,\ their commutators 
\qq\nn
\G_{ab}=\tfrac{1}{2}\,[\G_a,\G_b]
\qqq 
and a charge-conjugation matrix $\,C$ in a Majorana-spinor representation of dimension $\,D_{d,1}\,$ in which\footnote{We adopt the conventions of Refs.\,\cite{West:1998ey,Chryssomalakos:2000xd}.} 
\qq\nn
C\,\G^a\,C^{-1}=-\G^a{}^{\rm T}\,,\qquad\qquad C^{\rm T}=-\ep_d\,C\,,\quad\ep_d=\left\{ \barr{cl} -\sqrt{2}\,\cos\left(\tfrac{(d+2)\,\pi}{4}\right) & \tx{ if } d\in\{1,3,5,7,9\} \cr\cr -\sqrt{2}\,\cos\left(\tfrac{(d+1)\,\pi}{4}\right) & \tx{ if } d\in\{2,6,10\} \earr\right.\,,
\qqq
with $\,(d,p)\,$ chosen in such a manner that
\qq\nn
\left(C\,\G^{a_1 a_2\ldots a_p}\right)^{\rm T}=C\,\G^{a_1 a_2\ldots a_p}\,,
\qqq
that is with
\qq\nn
\ep_d\must(-1)^{\frac{(p-1)\cdot(p-2)}{2}}\,,
\qqq
and in which the following Fierz identity obtains:
\qq\nn
\eta_{ab}\,\ovl\G{}^a_{(\a\b}\,\ovl\G{}^{ba_1a_2\ldots a_{p-1}}_{\g\d)}=0\,;
\qqq
\item[$\bullet$] The $\sfT_e\Ad_{{\rm SO}(d,1)}$-invariance of the vacuum splitting -- obvious;
\item[$\bullet$] The homogeneous spaces: the NG one
\qq\nn
&{\rm sMink}(d,1\,|\,D_{d,1})\equiv{\rm sISO}(d,1\,|\,D_{d,1})/{\rm SO}(d,1)\equiv\bR(d,1\,|\,D_{d,1})\,,&\cr\cr
&\txH\equiv{\rm SO}(d,1)\,,\qquad\qquad\hgt\equiv\gt{so}(d,1)=\bigoplus_{a,b=0}^d\,\corr{J_{ab}=-J_{ba}}\,,&
\qqq 
with the body
\qq\nn
|{\rm sMink}(d,1\,|\,D_{d,1})|=\bR^{d,1}\equiv{\rm Mink}(d,1)\,,
\qqq
which is, again, a Lie supergroup with the structure as in the previous example, and the HP one
\qq\nn
&{\rm sISO}(d,1\,|\,D_{d,1})/\bigl({\rm SO}(p,1)\x{\rm SO}(d-p)\bigr)\,,&\cr\cr
&\txH_{\rm vac}\equiv{\rm SO}(p,1)\x{\rm SO}(d-p)\,,\qquad\qquad\hgt_{\rm vac}\equiv\bigoplus_{\unl a,\unl b=0}^p\,\corr{J_{\unl a\unl b}}\oplus\bigoplus_{\widehat a,\widehat b=p+1}^d\,\corr{J_{\widehat a\widehat b}}\,,&\cr\cr
&\dgt=\bigoplus_{\unl a=0}^p\,\bigoplus_{\widehat b=p+1}^d\,\corr{J_{\unl a\widehat b}}\,,&\cr\cr
&\tgt_{\rm vac}^{(0)}=\bigoplus_{\unl a=0}^p\,\corr{P_{\unl a}}\,,\qquad\qquad\dgt^{-1}\tgt_{\rm vac}^{(0)}=\bigoplus_{\widehat a=p+1}^d\,\corr{P_{\widehat a}}\equiv\egt^{(0)}\,;&
\qqq
\item[$\bullet$] The exponential superparametrisation(s) ($(\unl b,\widehat c)\in\{0,1\}\x\ovl{2,d}$):
\qq\nn
\si_0^{\rm vac}\bigl(\theta^\a,x^a,\phi^{\unl b\widehat c}\bigr)=\ee^{\theta^\a\ox Q_\a}\cdot\ee^{x^a\ox P_a}\cdot\ee^{\phi^{\unl b\widehat c}\ox J_{\unl b\widehat c}}\,;
\qqq
\item[$\bullet$] The superbackgrounds: the NG one
\qq\nn
\sgt\Bgt_p^{{\rm (NG)}}=\bigl({\rm sMink}(d,1\,|\,D_{d,1}),\eta_{ab}\,\theta^a_{\rm L}\ox\theta_{\rm L}^b,\Si_{\rm L}\wedge\ovl\G_{a_1a_2\ldots a_p}\,\Si_{\rm L}\wedge\theta_{\rm L}^{a_1}\wedge\theta_{\rm L}^{a_1}\wedge\cdots\wedge\theta_{\rm L}^{a_p}\equiv\underset{\tx{\ciut{(p+2)}}}{\chi}\hspace{-7pt}{}^{\rm GS}\bigr)
\qqq
with a \emph{non}supersymmetric global curving 
\qq\nn
&\underset{\tx{\ciut{(p+1)}}}{\b}\hspace{-7pt}{}^{\rm GS}(\theta,x,\phi)=\tfrac{1}{p+1}\,\sum_{k=0}^p\,\theta\,\ovl\G_{a_1 a_2\ldots a_k a_{k+1} a_{k+2}\ldots a_p}\,\sfd\theta\wedge\sfd x^{a_1 a_2\ldots a_k}\wedge e^{a_{k+1} a_{k+2}\ldots a_p}(\theta,x)\,,&\cr\cr
&\sfd x^{a_1 a_2\ldots a_k}\equiv\sfd x^{a_1}\wedge\sfd x^{a_2}\wedge\cdots\wedge\sfd x^{a_k}\,,&\cr\cr 
&e^{a_{k+1} a_{k+2}\ldots a_p}\equiv e^{a_{k+1}}\wedge e^{a_{k+2}}\wedge\cdots\wedge e^{a_p}\,,\qquad\qquad e^a(\theta,x)=\sfd x^a+\tfrac{1}{2}\,\theta\,\ovl\G{}^a\,\sfd\theta&
\qqq
(and no supersymmetric one), and the HP one
\qq\nn
\sgt\Bgt_{p,\la_p}^{{\rm (HP)}}&=&\bigl({\rm sISO}(d,1\,|\,D_{d,1})/\bigl({\rm SO}(p,1)\x{\rm SO}(d-p)\bigr),\cr\cr
&&\underset{\tx{\ciut{(p+2)}}}{\chi}\hspace{-7pt}{}^{\rm GS}+\tfrac{\la_p}{2p!}\,\ep_{\unl a{}_0\unl a{}_1\ldots\unl a{}_p}\,\bigl(\Si_{\rm L}\wedge\ovl\G{}^{\unl a{}_0}\,\Si_{\rm L}-2\d_{\widehat c\widehat d}\,\theta_{\rm L}^{\unl a{}_0\widehat c}\wedge\theta_{\rm L}^{\widehat d}\bigr)\wedge\theta_{\rm L}^{\unl a{}_1}\wedge\theta_{\rm L}^{\unl a{}_2}\wedge\cdots\wedge\theta_{\rm L}^{\unl a{}_p}\equiv\underset{\tx{\ciut{(p+2)}}}{\widehat\chi}\hspace{-6pt}{}^{\rm GS}\bigr)\,,&\cr\cr
&&\bigoplus_{\unl a=0}^p\,\corr{P_{\unl a}}\perp_\eta\bigoplus_{\widehat a=p+1}^d\,\corr{P_{\widehat a}}\,;
\qqq
\item[$\bullet$] The Body-Localisation Constraints: none.
\eit
\eeg

\brem\label{rem:4krule}
Implicit in the structure of the Lie superalgebra $\,\gt{siso}(d,1\,|\,D_{d,1})\,$ is the assumption 
\qq\nn
\bigl(C\,\G^a\bigr)^{\rm T}=C\,\G^a
\qqq 
which further constrains $\,\ep_d\,$ to be equal to 1. This assumption can be maintained, in conjunction with the constraints already imposed, only for $\,p\in 4\bN+1\,$ or $\,p\in 4\bN+2$.\ The remaining possibilities for which we might contemplate more general symmetry conditions for the $\,\G^a\,$ and a supersymmetry algebra with the anticommutator of the supercharges spanned exclusively on topological charges turn out to be ruled out by a simple algebraic argument given in \Rxcite{Sec.\,1.5}{West:1998ey}. The argument not only restricts the admissible values of $\,p\,$ as $\,p\equiv 1\mod 4\,$ or $\,p\equiv 2\mod 4$,\ but it also constrains the spatial dimension of the body of the supertarget as $\,d-p\equiv 0\mod 4\,$ or $\,d-p\equiv 1\mod 4$.
\erem

\beg\textbf{The Zhou super-1-brane in $\,{\rm s}({\rm AdS}_2\x\bS^2)$.}\label{eg:Zhou1}
\bit
\item[$\bullet$] The mother super-Harish--Chandra pair:
\qq\nn
{\rm SU}(1,1\,|\,2)_2\equiv\bigl({\rm SO}(1,2)\x{\rm SO}(3),\gt{su}(1,1\,|\,2)_2\bigr)\,,
\qqq
consisting of the product Lie group $\,{\rm SO}(1,2)\x{\rm SO}(3)\,$ and the Lie superalgebra
\qq\nn
\gt{su}(1,1\,\vert\,2)_2&=&\bigoplus_{(\a',\a'', I)\in\{1,2\}^{\x 3}}\,\corr{Q_{\a'\a'' I}}\oplus\bigoplus_{a'\in\{0,1\}}\,\corr{P_{a'}}\oplus\bigoplus_{a''\in\{2,3\}}\,\corr{P_{a''}}\cr\cr
&&\oplus\corr{J_{01}=-J_{10}}\oplus\corr{J_{23}=-J_{32}}
\qqq
with the structure relations
\qq
\{Q_{\a'\a'' I},Q_{\b'\b'' J}\}=2\bigl(\bigl(\unl C\,\unl\g^a\ox\bd1_2\bigr)_{\a'\a''I\b'\b''J}\,P_a-\sfi\,\bigl(\unl C\ox\si_2\bigr)_{\a'\a''I\b'\b''J}\,J_{01}-\sfi\,\bigl(\unl C\,\unl\g{}_5\ox\si_2\bigr)_{\a'\a''I\b'\b''J}\,J_{23}\bigr)\,,\cr\cr\cr
[P_0,P_1]=J_{01}\,,\qquad\qquad[P_2,P_3]=-J_{23}\,,\qquad\qquad[P_{a'},P_{a''}]=0\,,\cr\cr\cr
[J_{01},J_{23}]=0\,,\label{eq:AdSS2}\\ \cr\cr
[J_{01},P_{a'}]=\eta_{a'1}\,P_0-\eta_{a'0}\,P_1\,,\qquad\qquad[J_{23},P_{a''}]=\d_{a''3}\,P_2-\d_{a''2}\,P_3\,,\qquad\qquad[J_{23},P_{a'}]=0=[J_{01},P_{a''}]\,,\cr\cr\cr
[P_a,Q_{\a'\a'' I}]=\tfrac{\sfi}{2}\,\bigl(\widetilde\g_3'\,\unl\g{}_a\ox\si_2\bigr)^{\b'\b''J}_{\ \ \a'\a''I}\,Q_{\b'\b'' J}\,,\qquad\qquad[J_{ab},Q_{\a'\a'' I}]=\tfrac{1}{2}\,\bigl(\unl\g{}_{ab}\ox\bd1_2\bigr)^{\b'\b''J}_{\ \ \a'\a''I}\,Q_{\b'\b''J}\,,\nn
\qqq
expressed in terms of the Pauli matrix
\qq\nn
\si_2=\left(\barr{cc} 0 & -\sfi \\ \sfi & 0 \earr\right)\,,
\qqq
the generators 
\qq\nn
\bigl\{\unl\g^{a'}\equiv\G^{a'}\ox\bd1_2,\unl\g^{b''}\equiv\G_0\,\G_1\ox\G^{b''}\bigr\}^{(a',b'')\in\{0,1\}\x\{2,3\}}
\qqq
of the Clifford algebra $\,{\rm Cliff}(\bR^{3,1})\,$ of the Minkowski space $\,\bR^{3,1}\,$ with the metric $\,\eta=\diag(-1,+1,$ $+1,+1)\,$ and their commutators
\qq\nn
\unl\g^{ab}=\tfrac{1}{2}\,\bigl[\unl\g^a,\unl\g^b\bigr]\,,\qquad a,b\in\ovl{0,3}
\qqq
built of the generators 
\qq\nn
\{\G^{a'}\}^{a'\in\{0,1\}}
\qqq
of the Clifford algebra $\,{\rm Cliff}(\bR^{1,1})\,$ (in the 2-dimensional spinor representation) and the generators 
\qq\nn
\{\G^{a''}\}^{a''\in\{2,3\}}
\qqq
of the Clifford algebra $\,{\rm Cliff}(\bR^{2,0})\,$ (also in the 2-dimensional spinor representation), and in terms of (the tensor components of) the chirality operator 
\qq\nn
\unl\g{}_5\equiv\unl\g{}_0\,\unl\g{}_1\,\unl\g{}_2\,\unl\g{}_3=\widetilde\g_3'\,\widetilde\g_3''\,,\qquad\qquad\widetilde\g_3'=\G_0\,\G_1\ox\bd1_2\,,\qquad\qquad\widetilde\g_3''=\bd1_2\ox\G_2\,\G_3
\qqq
of $\,{\rm Cliff}(\bR^{3,1})$,\ as well as of the charge conjugation matrix 
\qq\nn
\unl C=C'\ox C''=-\unl C^{\rm T}
\qqq
given as the product of the charge conjugation matrices 
\qq\nn
C'=-C'{}^{\rm T}
\qqq
of $\,{\rm Cliff}(\bR^{1,1})\,$ and 
\qq\nn
C''=C''{}^{\rm T}
\qqq
of $\,{\rm Cliff}(\bR^{2,0})$,\ and such that 
\qq\nn
C'\,\G^{a'}=\bigl(C'\,\G^{a'}\bigr)^{\rm T}\,,\qquad\qquad C''\,\G^{d''}=\bigl(C''\,\G^{d''}\bigr)^{\rm T}\,,\qquad\qquad C'\,\G^{b'c'}=\bigl(C'\,\G^{b'c'}\bigr)^{\rm T}\,,
\qqq
whereas 
\qq\nn
C''\,\G^{a''b''}=-\bigl(C''\,\G^{a''b''}\bigr)^{\rm T}\,,
\qqq
so that -- in particular -- 
\qq\nn
\unl C\,\unl\g^a\,\unl C^{-1}=-\unl\g^{a\,{\rm T}}\,;
\qqq
\item[$\bullet$] The $\sfT_e\Ad_{{\rm SO}(1,1)\x{\rm SO}(2)}$-invariance of the vacuum splitting -- obvious;
\item[$\bullet$] The homogeneous spaces: the NG one
\qq\nn
&{\rm s}\bigl({\rm AdS}_2\x\bS^2\bigr)={\rm SU}(1,1\,|\,2)_2/\bigl({\rm SO}(1,1)\x{\rm SO}(2)\bigr)\,,&\cr\cr
&\txH={\rm SO}(1,1)\x{\rm SO}(2)\,,\qquad\qquad\hgt\equiv\gt{so}(1,1)\oplus\gt{so}(2)=\corr{J_{01}}\oplus\corr{J_{23}}\,,&
\qqq
with the body
\qq\nn
|{\rm s}\bigl({\rm AdS}_2\x\bS^2\bigr)|={\rm SO}(1,2)/{\rm SO}(1,1)\x{\rm SO}(3)/{\rm SO}(2)\equiv{\rm AdS}_2\x\bS^2\,,
\qqq 
and the HP one(s):
\bit
\item[{\bf \ref{eg:Zhou1}.1.}] the super-1-brane entirely in/over $\,{\rm AdS}_2$
\qq\nn
&{\rm SU}(1,1\,|\,2)_2/\bigl({\rm SO}(1,1)\x{\rm SO}(2)\bigr)\,,&\cr\cr
&\txH_{\rm vac}={\rm SO}(1,1)\x{\rm SO}(2)\,,\qquad\qquad\hgt_{\rm vac}=\corr{J_{01}}\oplus\corr{J_{23}}\,,&\cr\cr
&\dgt=\brd0\,,&\cr\cr
&\tgt_{\rm vac}^{(0)}=\corr{P_0,P_1}\,,\qquad\qquad\dgt^{-1}\tgt_{\rm vac}^{(0)}=\brd0\,;&
\qqq
\item[{\bf \ref{eg:Zhou1}.2.}] the super-1-brane astride
\qq\nn
&{\rm SU}(1,1\,|\,2)_2\,,&\cr\cr
&\txH_{\rm vac}=\bd1\,,\qquad\qquad\hgt_{\rm vac}=\brd0\,,&\cr\cr
&\dgt=\corr{J_{01},J_{23}}\,,&\cr\cr
&\tgt_{\rm vac}^{(0)}=\corr{P_0,P_2}\,,\qquad\qquad\dgt^{-1}\tgt_{\rm vac}^{(0)}=\corr{P_1,P_3}\equiv\egt^{(0)}\,;&
\qqq
\eit
\item[$\bullet$] The exponential superparametrisation(s):
\bit
\item[{\bf \ref{eg:Zhou1}.1.}] the super-1-brane entirely in/over $\,{\rm AdS}_2$
\qq\nn
\si_0^{\rm vac}\bigl(\theta^{\a'\a''I},x^a\bigr)=\ee^{\theta^{\a'\a''I}\ox Q_{\a'\a''I}}\cdot\ee^{x^a\ox P_a}\,;
\qqq
\item[{\bf \ref{eg:Zhou1}.2.}] the super-1-brane astride
\qq\nn
\si_0^{\rm vac}\bigl(\theta^{\a'\a''I},x^a,\phi^{01},\phi^{23}\bigr)=\ee^{\theta^{\a'\a''I}\ox Q_{\a'\a''I}}\cdot\ee^{x^a\ox P_a}\cdot\ee^{\phi^{01}\ox J_{01}+\phi^{23}\ox J_{23}}\,;
\qqq
\eit
\item[$\bullet$] The superbackgrounds: the NG one
\qq\nn
\sgt\Bgt_1^{{\rm (NG)}}&=&\bigl({\rm s}\bigl({\rm AdS}_2\x\bS^2\bigr),\eta_{ab}\,\theta^a_{\rm L}\ox\theta_{\rm L}^b,\Si_{\rm L}\wedge\bigl(\unl C\unl\g{}_a\ox\si_3\bigr)\,\Si_{\rm L}\wedge\theta_{\rm L}^a\equiv\underset{\tx{\ciut{(3)}}}{\chi}\hspace{-1pt}{}^{\rm Zh}\bigr)\,,\cr\cr
&&\theta_{\rm L}=\Si_{\rm L}^{\a'\a''I}\ox Q_{\a'\a''I}+\theta_{\rm L}^a\ox P_a+\theta^{01}_{\rm L}\ox J_{01}+\theta^{23}_{\rm L}\ox J_{23}
\qqq
with a supersymmetric global predecessor of the curving 
\qq\nn
\underset{\tx{\ciut{(2)}}}{\b}\hspace{-1pt}{}^{\rm Zh}=\Si_{\rm L}\wedge\bigl(\unl C\,\widetilde\g_3'\ox\si_1\bigr)\,\Si_{\rm L} 
\qqq
on $\,{\rm SU}(1,1\,|\,2)_2\,$ that descends to $\,{\rm s}({\rm AdS}_2\x\bS^2)$,\ and the HP one(s):
\bit
\item[{\bf \ref{eg:Zhou1}.1.}] the super-1-brane entirely in/over $\,{\rm AdS}_2\,$ (with $\,\ep_{01}=-1$)
\qq\nn
\sgt\Bgt_{1,\la_1}^{{\rm (HP)}}&=&\bigl({\rm SU}(1,1\,|\,2)_2/({\rm SO}(1,1)\x{\rm SO}(2)),\underset{\tx{\ciut{(3)}}}{\chi}\hspace{-1pt}{}^{\rm Zh}-\la_1\,\ep_{a'b'}\,\Si_{\rm L}\wedge\bigl(\unl C\,\unl\g{}^{a'}\ox\bd1_2\bigr)\,\Si_{\rm L}\wedge\theta_{\rm L}^{b'}\equiv\underset{\tx{\ciut{(3)}}}{\widehat\chi}\hspace{-1pt}{}^{\rm Zh}_{(1)}\bigr)\,,\cr\cr
&&\corr{P_0,P_1}\perp_\eta\corr{P_2,P_3}\,;
\qqq
\item[{\bf \ref{eg:Zhou1}.2.}] the super-1-brane astride (with $\,\ep_{02}=-1$)
\qq\nn
\sgt\Bgt_{1,\la_1}^{{\rm (HP)}}&=&\bigl({\rm SU}(1,1\,|\,2)_2,\underset{\tx{\ciut{(3)}}}{\chi}^{\rm Zh}-\la_1\,\bigl(\ep_{a'b''}\,\Si_{\rm L}\wedge\bigl(\unl C\,\unl\g{}^{a'}\ox\bd1_2\bigr)\,\Si_{\rm L}\wedge\theta_{\rm L}^{b'}-\theta_{\rm L}^1\wedge\theta_{\rm L}^2\wedge\theta_{\rm L}^{01}-\theta_{\rm L}^0\wedge\theta_{\rm L}^3\wedge\theta_{\rm L}^{23}\bigr)\equiv\underset{\tx{\ciut{(3)}}}{\widehat\chi}\hspace{-1pt}{}^{\rm Zh}_{(12)}\bigr)\,,\cr\cr
&&\corr{P_0,P_2}\perp_\eta\corr{P_1,P_3}\,;
\qqq
\eit
\item[$\bullet$] The Body-Localisation Constraints:
 \bit
\item[{\bf \ref{eg:Zhou1}.1.}] the super-1-brane entirely in/over $\,{\rm AdS}_2$:
\qq\nn
\theta_{\rm L}^{a''}\approx 0\,,\qquad a''\in\{2,3\}\,;
\qqq
\item[{\bf \ref{eg:Zhou1}.2.}] the super-1-brane astride: none.
\eit
\eit
\eeg

\beg\textbf{The Park--Rey super-1-brane in $\,{\rm s}({\rm AdS}_3\x\bS^3)$.}\label{eg:ParkRey1}
\bit
\item[$\bullet$] The mother super-Harish--Chandra pair:
\qq\nn
\bigl({\rm SU}(1,1\,|\,2)\x{\rm SU}(1,1\,|\,2)\bigr)_2\equiv\bigl({\rm SO}(2,2)\x{\rm SO}(4),\bigl(\gt{su}(1,1\,|\,2)\oplus\gt{su}(1,1\,|\,2)\bigr)_2\bigr)
\qqq
consisting of the product Lie group $\,{\rm SO}(2,2)\x{\rm SO}(4)\,$ and the Lie superalgebra
\qq\nn
\bigl(\gt{su}(1,1\,|\,2)\oplus\gt{su}(1,1\,|\,2)\bigr)_2&=&\bigoplus_{(\a',\a'',\a''', I)\in\{1,2\}^{\x 4}}\,\corr{Q_{\a'\a''\a''' I}}\oplus\bigoplus_{a'\in\{0,1,2\}}\,\corr{P_{a'}}\oplus\bigoplus_{a''\in\{3,4,5\}}\,\corr{P_{a''}}\cr\cr
&&\oplus\bigoplus_{a',b'\in\{0,1,2\}}\,\corr{J_{a'b'}=-J_{b'a'}}\oplus\bigoplus_{a'',b''\in\{3,4,5\}}\,\corr{J_{a''b''}=-J_{b''a''}}
\qqq
with the structure relations
\qq
\{Q_{\a'\a''\a'''I},Q_{\b'\b''\b'''J}\}=2\bigl(\unl C\,\unl\g^{a'}\cdot\bigl(\bd1_4\ox\si_2\bigr)\ox\bd1_2\bigr)_{\a'\a''\a'''I\b'\b''\b'''J}\,P_{a'}\cr\cr
-2\bigl(\unl C\,\unl\g^{a''}\unl\g{}_7\cdot\bigl(\bd1_4\ox\si_2\bigr)\ox\bd1_2\bigr)_{\a'\a''\a'''I\b'\b''\b'''J}\,P_{a''}\cr\cr
-\sfi\,\bigl(\unl C\,\unl\g^{a'b'}\unl\g{}_7\ox\si_3\bigr)_{\a'\a''\a'''I\b'\b''\b'''J}\,J_{a'b'}+\sfi\,\bigl(\unl C\,\unl\g^{a''b''}\unl\g{}_7\ox\si_3\bigr)_{\a'\a''\a'''I\b'\b''\b'''J}\,J_{a''b''}\,,\cr\cr\cr
[P_a,P_b]=\vep_{ab}\,J_{ab}\,,\qquad\vep_{ab}=\left\{ \barr{cl} +1 & \tx{if}\ a,b\in\{0,1,2\} \\
-1 & \tx{if}\ a,b\in\{3,4,5\} \\
0 & \tx{otherwise}\earr\right.\,,\cr\cr\cr
[J_{ab},J_{cd}]=\eta_{ad}\,J_{bc}-\eta_{ac}\,J_{bd}+\eta_{bc}\,J_{ad}-\eta_{bd}\,J_{ac}\,,\qquad\qquad[J_{ab},P_c]=\eta_{bc}\,P_a-\eta_{ac}\,P_b\,,\label{eq:AdSS3}\\ \cr\cr
[P_{a'},Q_{\a'\a''\a''' I}]=\tfrac{\sfi}{2}\,\bigl(\unl\g{}_{a'}\unl\g{}_7\cdot\bigl(\bd1_4\ox\si_2\bigr)\ox\si_3\bigr)^{\b'\b''\b'''J}_{\ \ \a'\a''\a'''I}\,Q_{\b'\b''\b''' J}\,,\cr\cr
[P_{a''},Q_{\a'\a''\a''' I}]=-\tfrac{\sfi}{2}\,\bigl(\unl\g{}_{a''}\cdot\bigl(\bd1_4\ox\si_2\bigr)\ox\si_3\bigr)^{\b'\b''\b'''J}_{\ \ \a'\a''\a'''I}\,Q_{\b'\b''\b''' J}\,,\cr\cr\cr
[J_{ab},Q_{\a'\a''\a''' I}]=\tfrac{1}{2}\,\bigl(\unl\g{}_{ab}\ox\bd1_2\bigr)^{\b'\b''\b'''J}_{\ \ \a'\a''\a'''I}\,Q_{\b'\b''\b'''J}\,,\nn
\qqq
expressed in terms of the Pauli matrices 
\qq\nn
\si_1=\left(\barr{cc} 0 & 1 \\ 1 & 0 \earr\right)\,,\qquad\qquad\si_2=\left(\barr{cc} 0 & -\sfi \\ \sfi & 0 \earr\right)\,,\qquad\qquad\si_3=\left(\barr{cc} 1 & 0 \\ 0 & -1 \earr\right)\,,
\qqq
the generators 
\qq\nn
\bigl\{\unl\g^{a'}\equiv\G^{a'}\ox\bd1_2\ox\si_1,\unl\g^{b''}\equiv\bd1_2\ox\G^{b''}\ox\si_2\bigr\}^{(a',b'')\in\{0,1,2\}\x\{3,4,5\}}
\qqq
of the Clifford algebra $\,{\rm Cliff}(\bR^{5,1})\,$ of the Minkowski space $\,\bR^{5,1}\,$ with the metric $\,\eta=\diag(-1,+1,$ $+1,+1,+1,+1)\,$ and their commutators
\qq\nn
\unl\g^{ab}=\tfrac{1}{2}\,\bigl[\unl\g^a,\unl\g^b\bigr]\,,\qquad a,b\in\ovl{0,5}
\qqq
built of the generators 
\qq\nn
\{\G^{a'}\}^{a'\in\{0,1,2\}}
\qqq
of the Clifford algebra $\,{\rm Cliff}(\bR^{2,1})\,$ (in the 2-dimensional spinor representation) and the generators 
\qq\nn
\{\G^{a''}\}^{a''\in\{3,4,5\}}
\qqq
of the Clifford algebra $\,{\rm Cliff}(\bR^{3,0})\,$ (also in the 2-dimensional spinor representation), and in terms of the chirality operator 
\qq\nn
\unl\g{}_7\equiv-\bd1_4\ox\si_3
\qqq
of $\,{\rm Cliff}(\bR^{5,1})$,\ as well as of the charge conjugation matrix 
\qq\nn
\unl C=C'\ox C''\ox\si_1=\unl C^{\rm T}
\qqq
given as the product of the charge conjugation matrices 
\qq\nn
C'=-C'{}^{\rm T}
\qqq
of $\,{\rm Cliff}(\bR^{2,1})\,$ and 
\qq\nn
C''=-C''{}^{\rm T}
\qqq
of $\,{\rm Cliff}(\bR^{3,0})$,\ such that 
\qq\nn
&C'\,\G^{a'}=\bigl(C'\,\G^{a'}\bigr)^{\rm T}\,,\qquad\qquad C''\,\G^{a''}=\bigl(C''\,\G^{a''}\bigr)^{\rm T}\,,&\cr\cr
&C'\,\G^{b'c'}=\bigl(C'\,\G^{b'c'}\bigr)^{\rm T}\,,\qquad\qquad C''\,\G^{d''e''}=\bigl(C''\,\G^{d''e''}\bigr)^{\rm T}\,,&
\qqq
so that -- in particular -- 
\qq\nn
\unl C\,\unl\g^a\,\unl C^{-1}=-\unl\g^{a\,{\rm T}}\,;
\qqq
\item[$\bullet$] The $\sfT_e\Ad_{{\rm SO}(2,1)\x{\rm SO}(3)}$-invariance of the vacuum splitting -- obvious;
\item[$\bullet$] The homogeneous spaces: the NG one
\qq\nn
&{\rm s}\bigl({\rm AdS}_3\x\bS^3\bigr)=\bigl({\rm SU}(1,1\,|\,2)\x{\rm SU}(1,1\,|\,2)\bigr)_2/\bigl({\rm SO}(2,1)\x{\rm SO}(3)\bigr)\,,&\cr\cr
&\txH={\rm SO}(2,1)\x{\rm SO}(3)\,,\qquad\qquad\hgt\equiv\gt{so}(2,1)\oplus\gt{so}(3)=\bigoplus_{a',b'\in\{0,1,2\}}\,\corr{J_{a'b'}}\oplus\bigoplus_{a'',b''\in\{3,4,5\}}\,\corr{J_{a''b''}}\,,&
\qqq
with the body
\qq\nn
|{\rm s}\bigl({\rm AdS}_3\x\bS^3\bigr)|={\rm SO}(2,2)/{\rm SO}(2,1)\x{\rm SO}(4)/{\rm SO}(3)\equiv{\rm AdS}_3\x\bS^3\,,
\qqq 
and the HP one(s):
\bit
\item[{\bf \ref{eg:ParkRey1}.1.}] the super-1-brane entirely in/over $\,{\rm AdS}_3$
\qq\nn
&\bigl({\rm SU}(1,1\,|\,2)\x{\rm SU}(1,1\,|\,2)\bigr)_2/\bigl({\rm SO}(1,1)\x{\rm SO}(3)\bigr)\,,&\cr\cr
&\txH_{\rm vac}={\rm SO}(1,1)\x{\rm SO}(3)\,,\qquad\qquad\hgt_{\rm vac}=\corr{J_{01}}\oplus\bigoplus_{a'',b''\in\{3,4,5\}}\,\corr{J_{a''b''}}\,,&\cr\cr
&\dgt=\corr{J_{02},J_{12}}\,,&\cr\cr
&\tgt_{\rm vac}^{(0)}=\corr{P_0,P_1}\,,\qquad\qquad\dgt^{-1}\tgt_{\rm vac}^{(0)}=\corr{P_2}\subsetneq\egt^{(0)}\,;&
\qqq
\item[{\bf \ref{eg:ParkRey1}.2.}] the super-1-brane astride
\qq\nn
&\bigl({\rm SU}(1,1\,|\,2)\x{\rm SU}(1,1\,|\,2)\bigr)_2/\bigl({\rm SO}(2)\x{\rm SO}(2)\bigr)\,,&\cr\cr
&\txH_{\rm vac}={\rm SO}(2)\x{\rm SO}(2)\,,\qquad\qquad\hgt_{\rm vac}=\corr{J_{12},J_{45}}\,,&\cr\cr
&\dgt=\corr{J_{01},J_{02},J_{34},J_{35}}\,,&\cr\cr
&\tgt_{\rm vac}^{(0)}=\corr{P_0,P_3}\,,\qquad\qquad\dgt^{-1}\tgt_{\rm vac}^{(0)}=\corr{P_1,P_2,P_4,P_5}\equiv\egt^{(0)}\,;&
\qqq
\eit
\item[$\bullet$] The exponential superparametrisation(s):
\bit
\item[{\bf \ref{eg:ParkRey1}.1.}] the super-1-brane entirely in/over $\,{\rm AdS}_3$
\qq\nn
\si_0^{\rm vac}\bigl(\theta^{\a'\a''\a'''I},x^a,\phi^{02},\phi^{12}\bigr)=\ee^{\theta^{\a'\a''\a'''I}\ox Q_{\a'\a''\a'''I}}\cdot\ee^{x^a\ox P_a}\cdot\ee^{\phi^{02}\ox J_{02}+\phi^{12}\ox J_{12}}\,;
\qqq
\item[{\bf \ref{eg:ParkRey1}.2.}] the super-1-brane astride
\qq\nn
\si_0^{\rm vac}\bigl(\theta^{\a'\a''\a'''I},x^a,\phi^{01},\phi^{02},\phi^{34},\phi^{35}\bigr)=\ee^{\theta^{\a'\a''\a'''I}\ox Q_{\a'\a''\a'''I}}\cdot\ee^{x^a\ox P_a}\cdot\ee^{\phi^{01}\ox J_{01}+\phi^{02}\ox J_{02}+\phi^{34}\ox J_{34}+\phi^{35}\ox J_{35}}\,;
\qqq
\eit
\item[$\bullet$] The superbackgrounds: the NG one
\qq\nn
\sgt\Bgt_1^{{\rm (NG)}}&=&\bigl({\rm s}\bigl({\rm AdS}_3\x\bS^3\bigr),\eta_{ab}\,\theta^a_{\rm L}\ox\theta_{\rm L}^b,\cr\cr
&&\Si_{\rm L}\wedge\bigl(\unl C\unl\g{}_{a'}\cdot\bigl(\bd1_4\ox\si_2\bigr)\ox\si_1\bigr)\,\Si_{\rm L}\wedge\theta_{\rm L}^{a'}-\Si_{\rm L}\wedge\bigl(\unl C\unl\g{}_{a''}\unl\g{}_7\cdot\bigl(\bd1_4\ox\si_2\bigr)\ox\si_1\bigr)\,\Si_{\rm L}\wedge\theta_{\rm L}^{a''}\equiv\underset{\tx{\ciut{(3)}}}{\chi}\hspace{-1pt}{}^{\rm PR}\bigr)\,,\cr\cr
&&\theta_{\rm L}=\Si_{\rm L}^{\a'\a''\a'''I}\ox Q_{\a'\a''\a'''I}+\theta_{\rm L}^a\ox P_a+\theta^{a'b'}_{\rm L}\ox J_{a'<b'}+\theta^{a''b''}_{\rm L}\ox J_{a''<b''}
\qqq
with a supersymmetric global predecessor of the curving 
\qq\nn
\underset{\tx{\ciut{(2)}}}{\b}\hspace{-1pt}{}^{\rm PR}=-\sfi\,\Si_{\rm L}\wedge\bigl(\unl C\,\unl\g{}_7\ox\si_2\bigr)\,\Si_{\rm L} 
\qqq
on $\,({\rm SU}(1,1\,|\,2)\x{\rm SU}(1,1\,|\,2))_2\,$ that descends to $\,{\rm s}({\rm AdS}_3\x\bS^3)$,\ and the HP one(s):
\bit
\item[{\bf \ref{eg:ParkRey1}.1.}] the super-1-brane entirely in/over $\,{\rm AdS}_3\,$ (with $\,\ep_{01}=-1$)
\qq\nn
\sgt\Bgt_{1,\la_1}^{{\rm (HP)}}&=&\bigl(\bigl({\rm SU}(1,1\,|\,2)\x{\rm SU}(1,1\,|\,2)\bigr)_2/\bigl({\rm SO}(1,1)\x{\rm SO}(3)\bigr),\cr\cr
&&\underset{\tx{\ciut{(3)}}}{\chi}\hspace{-1pt}{}^{\rm PR}-\la_1\,\ep_{a'b'}\,\bigl(\Si_{\rm L}\wedge\bigl(\unl C\,\unl\g{}^{a'}\cdot\bigl(\bd1_4\ox\si_2\bigr)\ox\bd1_2\bigr)\,\Si_{\rm L}\wedge\theta_{\rm L}^{b'}-\theta^{a'}_{\rm L}\wedge\theta^2_{\rm L}\wedge\theta_{\rm L}^{b'2}\bigr)\equiv\underset{\tx{\ciut{(3)}}}{\widehat\chi}\hspace{-1pt}{}^{\rm PR}_{(1)}\bigr)\,,\cr\cr
&&\corr{P_0,P_1}\perp_\eta\corr{P_2}\,;
\qqq
\item[{\bf \ref{eg:ParkRey1}.2.}] the super-1-brane astride
\qq\nn
\sgt\Bgt_{1,\la_1}^{{\rm (HP)}}&=&\bigl(\bigl({\rm SU}(1,1\,|\,2)\x{\rm SU}(1,1\,|\,2)\bigr)_2/\bigl({\rm SO}(2)\x{\rm SO}(2)\bigr)\cr\cr
&&\underset{\tx{\ciut{(3)}}}{\chi}\hspace{-1pt}{}^{\rm PR}+\la_1\,\bigl(\Si_{\rm L}\wedge\bigl(\unl C\,\unl\g{}^0\cdot\bigl(\bd1_4\ox\si_2\bigr)\ox\bd1_2\bigr)\,\Si_{\rm L}\wedge\theta_{\rm L}^3+\Si_{\rm L}\wedge\bigl(\unl C\,\unl\g{}^3\unl\g{}_7\cdot\bigl(\bd1_4\ox\si_2\bigr)\ox\bd1_2\bigr)\,\Si_{\rm L}\wedge\theta_{\rm L}^0\cr\cr
&&-\d_{a'b'}\,\theta^{a'}_{\rm L}\wedge\theta^3_{\rm L}\wedge\theta_{\rm L}^{0b'}-\d_{a''b''}\,\theta^0_{\rm L}\wedge\theta^{a''}_{\rm L}\wedge\theta_{\rm L}^{3b''}\bigr)\equiv\underset{\tx{\ciut{(3)}}}{\widehat\chi}\hspace{-1pt}{}^{\rm PR}_{(12)}\bigr)\,,\cr\cr
&&\corr{P_0,P_3}\perp_\eta\corr{P_1,P_2,P_4,P_5}\,;
\qqq
\eit
\item[$\bullet$] The Body-Localisation Constraints:
 \bit
\item[{\bf \ref{eg:ParkRey1}.1.}] the super-1-brane entirely in/over $\,{\rm AdS}_3$:
\qq\nn
\theta_{\rm L}^{a''}\approx 0\,,\qquad a''\in\{3,4,5\}\,;
\qqq
\item[{\bf \ref{eg:ParkRey1}.2.}] the super-1-brane astride: none.
\eit
\eit
\eeg

\beg\textbf{The Metsaev--Tseytlin super-1-brane in $\,{\rm s}({\rm AdS}_5\x\bS^5)$.}\label{eg:MT1}
\bit
\item[$\bullet$] The mother super-Harish--Chandra pair: 
\qq\nn
{\rm SU}(2,2\,|\,4)\equiv\bigl({\rm SO}(4,2)\x{\rm SO}(6),\gt{su}(2,2\,|\,4)\bigr)\,,
\qqq
consisting of the product Lie group $\,{\rm SO}(4,2)\x{\rm SO}(6)\,$ and the Lie superalgebra
\qq\nn
\gt{su}(2,2\,\vert\,4)&=&\bigoplus_{(\a',\a'', I)\in\ovl{1,4}\x\ovl{1,4}\x\{1,2\}}\,\corr{Q_{\a'\a'' I}}\oplus\bigoplus_{a'=0}^4\,\corr{P_{a'}}\oplus\bigoplus_{a''=5}^9\,\corr{P_{a''}}\cr\cr
&&\oplus\bigoplus_{a',b'=0}^4\,\corr{J_{a'b'}=-J_{b'a'}}\oplus\bigoplus_{a'',b''=5}^9\,\corr{J_{a''b''}=-J_{b''a''}}
\qqq
with the structure relations ($a,b\in\ovl{0,9}$)
\qq
\{Q_{\a'\a'' I},Q_{\b'\b'' J}\}=2\sfi\,\bigl(\bigl(\unl C\,\unl\g^{a'}\unl\g{}_{11}\bigr)_{\a'\a''I\b'\b''J}\,P_{a'}-\bigl(\unl C\,\unl\g^{a''}\bigr)_{\a'\a''I\b'\b''J}\,P_{a''}\bigr)\cr\cr
+\vep_{ab}\,\bigl(\unl C\,\unl\g^{ab}\bigr)_{\a'\a''I\b'\b''J}\,J_{ab}\,,\cr\cr\cr
[P_a,P_b]=\vep_{ab}\,J_{ab}\,,\qquad\vep_{ab}=\left\{ \barr{cl} +1 & \tx{if}\ a,b\in\ovl{0,4} \\
-1 & \tx{if}\ a,b\in\ovl{5,9} \\
0 & \tx{otherwise}\earr\right.\,,\cr\cr\cr
[J_{ab},J_{cd}]=\eta_{ad}\,J_{bc}-\eta_{ac}\,J_{bd}+\eta_{bc}\,J_{ad}-\eta_{bd}\,J_{ac}\,,\qquad\qquad[P_a,J_{bc}]=\eta_{ab}\,P_c-\eta_{ac}\,P_b\,,\label{eq:AdSS5}\\ \cr\cr
[P_{a'},Q_{\a'\a'' I}]=-\tfrac{\sfi}{2}\,\bigl(\unl\g{}_{a'}\unl\g{}_{11}\bigr)^{\b'\b''J}_{\ \ \a'\a''I}\,Q_{\b'\b'' J}\,,\qquad\qquad [P_{a''},Q_{\a'\a'' I}]=\tfrac{\sfi}{2}\,\bigl(\unl\g{}_{a''}\bigr)^{\b'\b''J}_{\ \ \a'\a''I}\,Q_{\b'\b'' J}\,,\cr\cr\cr
[J_{ab},Q_{\a'\a'' I}]=\tfrac{1}{2}\,\bigl(\unl\g{}_{ab}\bigr)^{\b'\b''J}_{\ \ \a'\a''I}\,Q_{\b'\b''J}\,.\nn
\qqq
expressed in terms of the Pauli matrices as above, the generators 
\qq\nn
\bigl\{\unl\g^{a'}\equiv\G^{a'}\ox\bd1_4\ox\si_1,\unl\g^{b''}\equiv\bd1_4\ox\G^{b''}\ox\si_2\bigr\}^{(a',b'')\in\ovl{0,4}\x\ovl{5,9}}
\qqq
of the Clifford algebra $\,{\rm Cliff}(\bR^{9,1})\,$ of the Minkowski space $\,\bR^{9,1}\,$ with the metric $\,\eta=\diag(-1,+1,$ $+1,+1,+1,+1,+1,+1,+1,+1)\,$ and their commutators
\qq\nn
\unl\g^{ab}=\tfrac{1}{2}\,\bigl[\unl\g^a,\unl\g^b\bigr]\,,\qquad a,b\in\ovl{0,9}
\qqq
built of the generators 
\qq\nn
\{\G^{a'}\}^{a'\in\ovl{0,4}}
\qqq
of the Clifford algebra $\,{\rm Cliff}(\bR^{4,1})\,$ (in the 4-dimensional spinor representation, in which they are traceless) and the generators 
\qq\nn
\{\G^{a''}\}^{a''\in\ovl{5,9}}
\qqq
of the Clifford algebra $\,{\rm Cliff}(\bR^{5,0})\,$ (also in the 4-dimensional spinor representation, in which they are traceless), and in terms of the chirality operator 
\qq\nn
\unl\g{}_{11}\equiv-\bd1_{16}\ox\si_3
\qqq
of $\,{\rm Cliff}(\bR^{9,1})$,\ as well as of  the charge conjugation matrix 
\qq\nn
\unl C=C'\ox C''\ox\sfi\,\si_2=-\unl C^{\rm T}
\qqq
of $\,{\rm Cliff}(\bR^{9,1})\,$ given as the product of the charge conjugation matrices 
\qq\nn
C'=-C'{}^{\rm T}
\qqq
of $\,{\rm Cliff}(\bR^{4,1})\,$ and 
\qq\nn
C''=-C''{}^{\rm T}
\qqq
of $\,{\rm Cliff}(\bR^{5,0})\,$ (as well as the Pauli matrix $\,\si_2$),\ such that 
\qq\nn
C'\,\G^{a'}=-\bigl(C'\,\G^{a'}\bigr)^{\rm T}\,,\qquad\qquad C''\,\G^{a''}=-\bigl(C''\,\G^{a''}\bigr)^{\rm T}\,,
\qqq
and
\qq\nn
C'\,\G^{a'b'}=\bigl(C'\,\G^{b'a'}\bigr)^{\rm T}\,,\qquad\qquad C''\,\G^{a''b''}=\bigl(C''\,\G^{a''b''}\bigr)^{\rm T}\,,
\qqq
so that -- in particular -- 
\qq\nn
\unl C\,\unl\g^a\,\unl C^{-1}=-\unl\g^{a\,{\rm T}}\,;
\qqq
\item[$\bullet$] The $\sfT_e\Ad_{{\rm SO}(4,1)\x{\rm SO}(5)}$-invariance of the vacuum splitting -- obvious;
\item[$\bullet$] The homogeneous spaces: the NG one
\qq\nn
&{\rm s}\bigl({\rm AdS}_5\x\bS^5\bigr)={\rm SU}(2,2\,|\,4)/\bigl({\rm SO}(4,1)\x{\rm SO}(5)\bigr)\,,&\cr\cr
&\txH={\rm SO}(4,1)\x{\rm SO}(5)\,,\qquad\qquad\hgt\equiv\gt{so}(4,1)\oplus\gt{so}(5)=\bigoplus_{a',b'=0}^4\,\corr{J_{a'b'}}\oplus\bigoplus_{a'',b''=5}^9\,\corr{J_{a''b''}}\,,&
\qqq
with the body
\qq\nn
|{\rm s}\bigl({\rm AdS}_5\x\bS^5\bigr)|={\rm SO}(4,2)/{\rm SO}(4,1)\x{\rm SO}(6)/{\rm SO}(5)\equiv{\rm AdS}_5\x\bS^5\,,
\qqq
and the HP one(s)
\bit
\item[{\bf \ref{eg:MT1}.1}] the super-1-brane entirely in/over $\,{\rm AdS}_5$:
\qq\nn
&{\rm SU}(2,2\,|\,4)/\bigl({\rm SO}(1,1)\x{\rm SO}(3)\x{\rm SO}(5)\bigr)\,,&\cr\cr
&\txH_{\rm vac}={\rm SO}(1,1)\x{\rm SO}(3)\x{\rm SO}(5)\,,\qquad\qquad\hgt_{\rm vac}=\corr{J_{01}}\oplus\bigoplus_{a',b'\in\{2,3,4\}}\,\corr{J_{a'b'}=-J_{b'a'}}\oplus\bigoplus_{a'',b''=5}^9\,\corr{J_{a''b''}=-J_{b''a''}}\,,&\cr\cr
&\dgt=\corr{J_{02},J_{03},J_{04},J_{12},J_{13},J_{14}}\,,&\cr\cr
&\tgt_{\rm vac}^{(0)}=\corr{P_0,P_1}\,,\qquad\qquad\dgt^{-1}\tgt_{\rm vac}^{(0)}=\corr{P_2,P_3,P_4}\subsetneq\egt^{(0)}\,;&
\qqq
\item[{\bf \ref{eg:MT1}.2}] the super-1-brane astride:
\qq\nn
&{\rm SU}(2,2\,|\,4)/\bigl({\rm SO}(4)\x{\rm SO}(4)\bigr)\,,&\cr\cr
&\txH_{\rm vac}={\rm SO}(4)\x{\rm SO}(4)\,,\qquad\qquad\hgt_{\rm vac}=\bigoplus_{a',b'=1}^4\,\corr{J_{a'b'}=-J_{b'a'}}\oplus\bigoplus_{a'',b''=6}^9\,\corr{J_{a''b''}=-J_{b''a''}}\,,&\cr\cr
&\dgt=\bigoplus_{a'=1}^4\,\corr{J_{0a'}}\oplus\bigoplus_{b''=6}^9\,\corr{J_{5b''}}\,,&\cr\cr
&\tgt_{\rm vac}^{(0)}=\corr{P_0,P_5}\,,\qquad\qquad\dgt^{-1}\tgt_{\rm vac}^{(0)}=\bigoplus_{a'=1}^4\,\corr{P_{a'}}\oplus\bigoplus_{b''=6}^9\,\corr{P_{b''}}\equiv\egt^{(0)}\,;&
\qqq
\eit
\item[$\bullet$] The exponential superparametrisation(s):
\bit
\item[{\bf \ref{eg:MT1}.1}] the super-1-brane entirely in/over $\,{\rm AdS}_5\,$ ($(\widehat b',\widehat c')\in\{2,3,4\}$):
\qq\nn
\si_0^{\rm vac}\bigl(\theta^{\a'\a''I},x^a,\phi^{0\widehat b'},\phi^{1\widehat c'}\bigr)=\ee^{\theta^{\a'\a''I}\ox Q_{\a'\a''I}}\cdot\ee^{x^a\ox P_a}\cdot\ee^{\phi^{0\widehat b'}\ox J_{0\widehat b'}+\phi^{1\widehat c'}\ox J_{0\widehat c'}}\,;
\qqq
\item[{\bf \ref{eg:MT1}.2}] the super-1-brane astride ($(\widehat b',\widehat c'')\in\ovl{1,4}\x\ovl{6,9}$):
\qq\nn
\si_0^{\rm vac}\bigl(\theta^{\a'\a''I},x^a,\phi^{0b'},\phi^{5c''}\bigr)=\ee^{\theta^{\a'\a''I}\ox Q_{\a'\a''I}}\cdot\ee^{x^a\ox P_a}\cdot\ee^{\phi^{0\widehat b'}\ox J_{0\widehat b'}+\phi^{5\widehat c''}\ox J_{5\widehat c''}}\,;
\qqq
\eit
\item[$\bullet$] The superbackgrounds: the NG one
\qq\nn
\sgt\Bgt_1^{{\rm (NG)}}&=&\bigl({\rm s}\bigl({\rm AdS}_5\x\bS^5\bigr),\eta_{ab}\,\theta^a_{\rm L}\ox\theta_{\rm L}^b,\sfi\,\Si_{\rm L}\wedge\unl C\,\unl\g{}_{a'}\,\Si_{\rm L}\wedge\theta^{a'}_{\rm L}-\sfi\,\Si_{\rm L}\wedge\unl C\,\unl\g{}_{a''}\unl\g{}_{11}\,\Si_{\rm L}\wedge\theta^{a''}_{\rm L}\equiv\underset{\tx{\ciut{(3)}}}{\chi}\hspace{-1pt}{}^{\rm MT}\bigr)\,,\cr\cr
&&\theta_{\rm L}=\Si_{\rm L}^{\a'\a''I}\ox Q_{\a'\a''I}+\theta_{\rm L}^a\ox P_a+\theta^{a'b'}_{\rm L}\ox J_{a'<b'}+\theta^{a''b''}_{\rm L}\ox J_{a''<b''}
\qqq
with a supersymmetric global predecessor of the curving 
\qq\nn
\underset{\tx{\ciut{(2)}}}{\b}\hspace{-1pt}{}^{\rm MT}=-\Si_{\rm L}\wedge\unl C\,\unl\g{}_{11}\,\Si_{\rm L} 
\qqq
on $\,{\rm SU}(2,2\,|\,4)\,$ that descends to $\,{\rm s}({\rm AdS}_5\x\bS^5)$,\ and the HP one(s):
\bit
\item[{\bf \ref{eg:MT1}.1}] the super-1-brane entirely in/over $\,{\rm AdS}_5\,$ (with $\,\ep_{01}=-1$):
\qq\nn
\sgt\Bgt_{1,\la_1}^{{\rm (HP)}}&=&\bigl({\rm SU}(2,2\,|\,4)/\bigl({\rm SO}(1,1)\x{\rm SO}(3)\x{\rm SO}(5)\bigr)
,\cr\cr
&&\underset{\tx{\ciut{(3)}}}{\chi}\hspace{-1pt}{}^{\rm MT}-\la_1\,\ep_{\unl a'\unl b'}\,\bigl(\sfi\,\Si_{\rm L}\wedge\unl C\,\unl\g{}^{\unl a'}\,\unl\g{}_{11}\,\Si_{\rm L}\wedge\theta_{\rm L}^{\unl b'}-\d_{\widehat c'\widehat d'}\,\theta_{\rm L}^{\unl a'}\wedge\theta_{\rm L}^{\widehat c'}\wedge\theta_{\rm L}^{\unl b'\widehat d'}\bigr)\equiv\underset{\tx{\ciut{(3)}}}{\widehat\chi}\hspace{-1pt}{}^{\rm MT}_{(1)}\bigr)\,,\cr\cr
&&\corr{P_0,P_1}\perp_\eta\bigoplus_{\widehat a=2}^9\,\corr{P_{\widehat a}}\,;
\qqq
\item[{\bf \ref{eg:MT1}.2}] the super-1-brane astride:
\qq\nn
\sgt\Bgt_{1,\la_1}^{{\rm (HP)}}&=&\bigl({\rm SU}(2,2\,|\,4)/\bigl({\rm SO}(4)\x{\rm SO}(4)\bigr),\cr\cr
&&\underset{\tx{\ciut{(3)}}}{\chi}\hspace{-1pt}{}^{\rm MT}+\la_1\,\bigl(\sfi\,\Si_{\rm L}\wedge\unl C\,\unl\g{}^0\,\unl\g{}_{11}\,\Si_{\rm L}\wedge\theta_{\rm L}^5+\sfi\,\Si_{\rm L}\wedge\unl C\,\unl\g{}^5\,\Si_{\rm L}\wedge\theta_{\rm L}^0-\d_{\widehat a'\widehat b'}\,\theta^{\widehat a'}_{\rm L}\wedge\theta^5_{\rm L}\wedge\theta_{\rm L}^{0\widehat b'}-\d_{\widehat a''\widehat b''}\,\theta^0_{\rm L}\wedge\theta^{\widehat a''}_{\rm L}\wedge\theta_{\rm L}^{5\widehat b''}\bigr)\equiv\underset{\tx{\ciut{(3)}}}{\widehat\chi}\hspace{-1pt}{}^{\rm MT}_{(12)}\bigr)\,,\cr\cr
&&\corr{P_0,P_5}\perp_\eta\bigoplus_{\widehat a'=1}^4\,\corr{P_{\widehat a'}}\oplus\bigoplus_{\widehat b''=6}^9\,\corr{P_{\widehat b''}}\,;
\qqq
\eit
\item[$\bullet$] The Body-Localisation Constraints:
 \bit
\item[{\bf \ref{eg:MT1}.1.}] the super-1-brane entirely in/over $\,{\rm AdS}_5$:
\qq\nn
\theta_{\rm L}^{a''}\approx 0\,,\qquad a''\in\ovl{5,9}\,;
\qqq
\item[{\bf \ref{eg:MT1}.2.}] the super-1-brane astride: none.
\eit
\eit
\eeg

\brem
In all the examples considered above, $\,\hgt_{\rm vac}\,$ is spanned on the generators $\,J_{ab}\,$ of $\,\hgt\,$ with both indices taken from the same subset (vacuum-vacuum resp.\ transverse-transverse), {\it i.e.}, those of the type $\,J_{\unl a\unl b}\,$ resp.\ $\,J_{\widehat a\widehat b}$,\ whereas $\,\dgt\,$ is spanned on the generators with mixed indices (vacuum-transverse), {\it i.e.}, those of the type $\,J_{\unl a\widehat b}$.\ Accordingly, the  Descendability Constraint is universally satisfied in consequence of the relations
\qq\nn
[J_{\unl a\unl b},J_{\unl c\widehat d}]=\eta_{\unl b\unl c}\,J_{\unl a\widehat d}-\eta_{\unl a\unl c}\,J_{\unl b\widehat d}\,,\qquad\qquad[J_{\widehat a\widehat b},J_{\unl c\widehat d}]=\d_{\widehat b\widehat d}\,J_{\unl c\widehat a}-\d_{\widehat a\widehat d}\,J_{\unl c\widehat b}\,.
\qqq
\erem

\noindent We are now ready to discuss, in all generality, the supersymmetry of the super-$\si$-models introduced above.

\section{The vacuum of the super-$\si$-model}\label{sec:vac}

Having defined the two classes of supersymmetric field theories of interest to us and stated a correspondence between them, we may, next, derive the dynamics of the super-$p$-brane in either one of the dual pair of models and study its (super)symmetries. As we ultimately intend to elucidate the (higher-)geometric nature of these symmetries, with emphasis on the local (or gauge) ones that only preserve a suitably relatively normalised \emph{combination} of the two terms in the action functional, we choose to work in the Hughes--Polchinski formulation of Def.\,\ref{def:GSinHP} in which both terms are of the same topological nature. The advantage of working in this formulation, to become apparent presently, is an intrinsically geometric form of the ensuing Euler--Lagrange equations and a neat geometric interpretation of the infinitesimal symmetries. Their derivation, as well as the symmetry analysis of the model, become tractable upon imposing a few additional constraints on the supersymmetry Lie superalgebra. These we discover one by one in the explicit calculations that follow.

We begin by writing out the logarithmic variation of the DF amplitude of the GS super-$\si$-model in the HP formulation engendered by an arbitrary section $\,\d\widehat\xi\in[\Om_p,\cT\cM_{\txH_{\rm vac}}]\,$ (in the $\cS$-point picture). As the amplitude is a pure differential character, {\it cp} \Rxcite{Sec.\,2}{Suszek:2019cum} and Sec.\,\ref{sec:hgeomphys}, we obtain
\qq\nn
-\sfi\,\d_{\d\widehat\xi}\log\,\cA_{\rm DF}^{{\rm (HP)},p,\la_p}[\widehat\xi]=\int_{\Om_p}\,\widehat\xi^*\bigl(\d\widehat\xi\con\underset{\tx{\ciut{(p+2)}}}{\widehat\txH}\hspace{-7pt}{}^{\rm vac}\bigr)\,,
\qqq
where
\qq\nn
\underset{\tx{\ciut{(p+2)}}}{\widehat\txH}\hspace{-7pt}{}^{\rm vac}\equiv\underset{\tx{\ciut{(p+2)}}}{\txH}\hspace{-7pt}{}^{\rm vac}+\la_p\,\sfd\underset{\tx{\ciut{(p+1)}}}{\txB}\hspace{-7pt}{}^{\rm (HP)}\,.
\qqq
In order to cast the result in a form amenable to further analysis, we use the fact that the tensorial components of the HP superbackground are pulled back from the mother supergroup $\,\txG\,$ and push forward the variation field $\,\d\widehat\xi\,$ to the sub-supermanifolds $\,\cV_i\,$ of \Reqref{eq:secHP} over which we contract the pushforwards with the $\txH_{\rm vac}$-basic super-$(p+2)$-form 
\qq\nn
\underset{\tx{\ciut{(p+2)}}}{\widehat\chi}\equiv\pi_{\txG/\txH_{\rm vac}}^*\underset{\tx{\ciut{(p+2)}}}{\widehat\txH}\hspace{-7pt}{}^{\rm vac}\,.
\qqq
The pushforward decomposes in the basis of the tangent sheaf $\,\cT\Si^{\rm HP}\,$ described in Prop.\,\ref{prop:bastanshHP} as
\qq\nn
\sfT_{\unl\chi{}_i}\si_i^{\rm vac}\d\widehat\xi(\unl\chi{}_i)&=:&\d\theta_i^\a\,\cT_{\a\,i}\bigl(\si_i^{\rm vac}(\unl\chi{}_i)\bigr)+\d x_i^a\,\cT_{a\,i}\bigl(\si_i^{\rm vac}(\unl\chi{}_i)\bigr)+\d\phi_i^{\widehat S}\,\cT_{\widehat S\,i}\bigl(\si_i^{\rm vac}(\unl\chi{}_i)\bigr)\cr\cr
&\equiv&\bigl(\d\widehat\xi^\z_i\,L_\z+\D^{\unl T}_i\,L_{\unl T}\bigr)\bigl(\si_i^{\rm vac}(\unl\chi{}_i)\bigr)\,,
\qqq
with the ($\txH_{\rm vac}$-)vertical correction $\,\D^{\unl T}_i\,L_{\unl T}\,$ ensuring that the pushforward is actually tangent to $\,\cV_i\subset\txG$,\ {\it cp} Prop.\,\ref{prop:bastanshHP}. The latter is linear in each of the independent coordinate variations $\,\d\theta_i^\a,\d x_i^a\,$ and $\,\d\phi_i^{\widehat S}\,$ and in the kernel of the super-$(p+2)$-form with which we contract the variation below, and so its presence does not affect linear independence of the coordinate variations. Thus, we find the local formula
\qq\nn
-\sfi\,\d_{\d\widehat\xi}\log\,\cA_{\rm DF}^{{\rm (HP)},p,\la_p}[\widehat\xi]=\sum_{\t\in\Tgt_{p+1}}\,\int_\t\,\bigl(\si_{\imath_\t}^{\rm vac}\circ\widehat\xi_\t\bigr)^*\bigl(\d\widehat\xi^\z_{\imath_\t}\,L_\z\con\underset{\tx{\ciut{(p+2)}}}{\widehat\chi}\bigr)\,.
\qqq
Taking into account The Dimensional Constraint and the structure of the HP super-$(p+1)$-form (alongside the super-Maurer--Cartan equations), that altogether imply
\qq\nn
\underset{\tx{\ciut{(p+2)}}}{\widehat\chi}&\equiv&\tfrac{1}{2p!}\,\bigl(\chi_{\a\b a_1 a_2\ldots a_p}\,\Si_{\rm L}^\a\wedge\Si_{\rm L}^\b\wedge\theta_{\rm L}^{a_1}\wedge\theta_{\rm L}^{a_2}\wedge\cdots\wedge\theta_{\rm L}^{a_p}\cr\cr
&&\hspace{.5cm}+(-1)^{|A|\cdot|B|+1}\,\la_p\,f_{AB}^{\ \ \ \unl a{}_0}\,\ep_{\unl a{}_0\unl a{}_1\unl a{}_2\ldots\unl a{}_p}\,\theta_{\rm L}^A\wedge\theta_{\rm L}^B\wedge\theta^{\unl a{}_1}_{\rm L}\wedge\theta^{\unl a{}_2}_{\rm L}\wedge\cdots\wedge\theta^{\unl a{}_p}_{\rm L}\bigr)\,,
\qqq
we establish the identity
\qq\nn
-\sfi\,\d_{\d\widehat\xi}\log\,\cA_{\rm DF}^{{\rm (HP)},p,\la_p}[\widehat\xi]&=&\sum_{\t\in\Tgt_{p+1}}\,\int_\t\,\bigl(\si_{\imath_\t}^{\rm vac}\circ\widehat\xi_\t\bigr)^*\bigl[\tfrac{1}{p!}\,\d\theta_{\imath_\t}^\a\,\bigl(\chi_{\a\b a_1 a_2\ldots a_p}\,\Si_{\rm L}^\b\wedge\theta_{\rm L}^{a_1}\wedge\theta_{\rm L}^{a_2}\wedge\cdots\wedge\theta_{\rm L}^{a_p}\cr\cr
&&\hspace{3.2cm}+\la_p\,f_{\a\b}^{\ \ \ \unl a{}_0}\,\ep_{\unl a{}_0\unl a{}_1\unl a{}_2\ldots\unl a{}_p}\,\Si_{\rm L}^\b\wedge\theta^{\unl a{}_1}_{\rm L}\wedge\theta^{\unl a{}_2}_{\rm L}\wedge\cdots\wedge\theta^{\unl a{}_p}_{\rm L}\bigr)\cr\cr
&&\hspace{2.5cm}+\tfrac{1}{2(p-1)!}\,\d x^{\unl a}_{\imath_\t}\,\bigl(\chi_{\a\b\unl a a_1 a_2\ldots a_{p-1}}\,\Si_{\rm L}^\a\wedge\Si_{\rm L}^\b\wedge\theta_{\rm L}^{a_1}\wedge\theta_{\rm L}^{a_2}\wedge\cdots\wedge\theta_{\rm L}^{a_{p-1}}\cr\cr
&&\hspace{2.1cm}+(-1)^{|A|\cdot|B|}\,\la_p\,f_{AB}^{\ \ \ \unl b}\,\ep_{\unl a\unl b\unl a{}_1\unl a{}_2\ldots\unl a{}_{p-1}}\,\theta_{\rm L}^A\wedge\theta_{\rm L}^B\wedge\theta^{\unl a{}_1}_{\rm L}\wedge\theta^{\unl a{}_2}_{\rm L}\wedge\cdots\wedge\theta^{\unl a{}_{p-1}}_{\rm L}\cr\cr
&&\hspace{3.2cm}-\tfrac{2\la_p}{p}\,f_{\unl a B}^{\ \ \ \unl b}\,\ep_{\unl b\unl a{}_1\unl a{}_2\ldots\unl a{}_p}\,\theta_{\rm L}^B\wedge\theta^{\unl a{}_1}_{\rm L}\wedge\theta^{\unl a{}_2}_{\rm L}\wedge\cdots\wedge\theta^{\unl a{}_p}_{\rm L}\bigr)\cr\cr
&&\hspace{2.5cm}+\tfrac{1}{2(p-1)!}\,\d x^{\widehat a}_{\imath_\t}\,\bigl(\chi_{\a\b\widehat a a_1 a_2\ldots a_{p-1}}\,\Si_{\rm L}^\a\wedge\Si_{\rm L}^\b\wedge\theta_{\rm L}^{a_1}\wedge\theta_{\rm L}^{a_2}\wedge\cdots\wedge\theta_{\rm L}^{a_{p-1}}\cr\cr
&&\hspace{3.2cm}-\tfrac{2\la_p}{p}\,f_{\widehat a B}^{\ \ \ \unl b}\,\ep_{\unl b\unl a{}_1\unl a{}_2\ldots\unl a{}_p}\,\theta_{\rm L}^B\wedge\theta^{\unl a{}_1}_{\rm L}\wedge\theta^{\unl a{}_2}_{\rm L}\wedge\cdots\wedge\theta^{\unl a{}_p}_{\rm L}\bigr)\cr\cr
&&\hspace{3.2cm}-\tfrac{\la_p}{p!}\,\d\phi_{\imath_\t}^{\widehat S}\,f_{\widehat S B}^{\ \ \ \unl b}\,\ep_{\unl b\unl a{}_1\unl a{}_2\ldots\unl a{}_p}\,\theta_{\rm L}^B\wedge\theta^{\unl a{}_1}_{\rm L}\wedge\theta^{\unl a{}_2}_{\rm L}\wedge\cdots\wedge\theta^{\unl a{}_p}_{\rm L}\bigr]\,.
\qqq
We shall now systematically derive from it the Euler--Lagrange equations of the super-$\si$-model in hand.

We begin with those implied by the requirement of the vanishing of the above variation for $\,\d\theta_i^\a=0=\d x_i^a\,$ and $\,\d\phi_i^{\widehat S}\neq 0$.\ The relevant variation now reduces to 
\qq\nn
-\sfi\,\d_{\d\widehat\xi}\log\,\cA_{\rm DF}^{{\rm (HP)},p,\la_p}[\widehat\xi]=-\tfrac{\la_p}{p!}\,\sum_{\t\in\Tgt_{p+1}}\,\int_\t\,\bigl(\si_{\imath_\t}^{\rm vac}\circ\widehat\xi_\t\bigr)^*\d\phi_{\imath_\t}^{\widehat S}\,f_{\widehat S\widehat a}^{\ \ \ \unl b}\,\ep_{\unl b\unl a{}_1\unl a{}_2\ldots\unl a{}_p}\,\theta_{\rm L}^{\widehat a}\wedge\theta^{\unl a{}_1}_{\rm L}\wedge\theta^{\unl a{}_2}_{\rm L}\wedge\cdots\wedge\theta^{\unl a{}_p}_{\rm L}
\qqq
and infers, in consequence of the identification of the HP super-$(p+1)$-form with the volume element on the vacuum ({\it i.e.}, on the embedded super-$p$-brane) that implies $\,\underset{\tx{\ciut{(p+1)}}}{\b}\hspace{-7pt}{}^{\rm (HP)}\neq 0$,\ the first subset of the Euler--Lagrange equations:
\qq\label{eq:EM1}
\theta_{\rm L}^{\widehat a}\approx 0\,,\qquad\widehat a\in\ovl{p+1,d-L}\,.
\qqq
In these, we readily recognise the Inverse Higgs Constraints (IHC) of Thm.\,\ref{thm:corrNGHP}, to be augmented with the Body-Localisation Constraints (BLC)
\qq\label{eq:BLC}
\theta_{\rm L}^{\widehat a}\approx 0\,,\qquad\widehat a\in\ovl{d-L+1,d}\,.
\qqq
Just to reiterate and -- in so doing -- explain the notation that we shall be using throughout, we extract from our previous discussion the following\medskip 

\noindent\textbf{Notational Convention:} 
The inscription 
\qq\nn
\theta_{\rm L}^\mu\approx 0\,,
\qqq 
written for some $\,\mu\in\ovl{0,\unl\d}$,\ means that the vacuum embedding $\,\widehat\xi\in[\Om_p,\cM_{\txH_{\rm vac}}]\,$ of the GS super-$\si$-model in the HP formulation is such that the image of the tangent of $\,\si_i^{\rm vac}\circ\widehat\xi,\ i\in I_{\txH_{\rm vac}}\,$ lies in the superdistribution $\,{\rm Ker}\,\theta_{\rm L}^\mu\rstr_{\cT\Si^{\rm HP}}\,$ within the tangent sheaf $\,\cT\Si^{\rm HP}\,$ of the HP section.\medskip
 
The first set of field equations, taken together with the BLC, has a fundamentally different status from that of the remaining field equations to be derived below -- indeed, it enforces correspondence with the original NG formulation. Their imposition reduces the variation to the form
\qq
-\sfi\,\d_{\d\widehat\xi}\log\,\cA_{\rm DF}^{{\rm (HP)},p,\la_p}[\widehat\xi]&=&\sum_{\t\in\Tgt_{p+1}}\,\int_\t\,\bigl(\si_{\imath_\t}^{\rm vac}\circ\widehat\xi_\t\bigr)^*\bigl[\tfrac{1}{p!}\,\d\theta_{\imath_\t}^\a\,\bigl(\chi_{\a\b\unl a{}_1\unl a{}_2\ldots\unl a{}_p}\,\Si_{\rm L}^\b\wedge\theta_{\rm L}^{\unl a{}_1}\wedge\theta_{\rm L}^{\unl a{}_2}\wedge\cdots\wedge\theta_{\rm L}^{\unl a{}_p}\cr\cr
&&\hspace{3.2cm}+\la_p\,f_{\a\b}^{\ \ \ \unl a{}_0}\,\ep_{\unl a{}_0\unl a{}_1\unl a{}_2\ldots\unl a{}_p}\,\Si_{\rm L}^\b\wedge\theta^{\unl a{}_1}_{\rm L}\wedge\theta^{\unl a{}_2}_{\rm L}\wedge\cdots\wedge\theta^{\unl a{}_p}_{\rm L}\bigr)\cr\cr
&&\hspace{2.5cm}+\tfrac{1}{2(p-1)!}\,\d x^{\unl a}_{\imath_\t}\,\bigl(\chi_{\a\b\unl a\unl a{}_1\unl a{}_2\ldots\unl a{}_{p-1}}\,\Si_{\rm L}^\a\wedge\Si_{\rm L}^\b\wedge\theta_{\rm L}^{\unl a{}_1}\wedge\theta_{\rm L}^{\unl a{}_2}\wedge\cdots\wedge\theta_{\rm L}^{\unl a{}_{p-1}}\cr\cr
&&\hspace{2.1cm}+(-1)^{|A|\cdot|B|}\,\la_p\,f_{AB}^{\ \ \ \unl b}\,\ep_{\unl a\unl b\unl a{}_1\unl a{}_2\ldots\unl a{}_{p-1}}\,\theta_{\rm L}^A\wedge\theta_{\rm L}^B\wedge\theta^{\unl a{}_1}_{\rm L}\wedge\theta^{\unl a{}_2}_{\rm L}\wedge\cdots\wedge\theta^{\unl a{}_{p-1}}_{\rm L}\cr\cr \label{eq:logvarIHC}
&&\hspace{3.2cm}-\tfrac{2\la_p}{p}\,f_{\unl a\unl c}^{\ \ \ \unl b}\,\ep_{\unl b\unl a{}_1\unl a{}_2\ldots\unl a{}_p}\,\theta_{\rm L}^{\unl c}\wedge\theta^{\unl a{}_1}_{\rm L}\wedge\theta^{\unl a{}_2}_{\rm L}\wedge\cdots\wedge\theta^{\unl a{}_p}_{\rm L}\\ \cr
&&\hspace{3.2cm}-\tfrac{2\la_p}{p}\,f_{\unl a\unl S}^{\ \ \ \unl b}\,\ep_{\unl b\unl a{}_1\unl a{}_2\ldots\unl a{}_p}\,\theta_{\rm L}^{\unl S}\wedge\theta^{\unl a{}_1}_{\rm L}\wedge\theta^{\unl a{}_2}_{\rm L}\wedge\cdots\wedge\theta^{\unl a{}_p}_{\rm L}\bigr)\cr\cr
&&\hspace{2.5cm}+\tfrac{1}{2(p-1)!}\,\d x^{\widehat a}_{\imath_\t}\,\bigl(\chi_{\a\b\widehat a\unl a{}_1\unl a{}_2\ldots \unl a{}_{p-1}}\,\Si_{\rm L}^\a\wedge\Si_{\rm L}^\b\wedge\theta_{\rm L}^{\unl a{}_1}\wedge\theta_{\rm L}^{\unl a{}_2}\wedge\cdots\wedge\theta_{\rm L}^{\unl a{}_{p-1}}\cr\cr
&&\hspace{3.2cm}-\tfrac{2\la_p}{p}\,f_{\widehat a\unl c}^{\ \ \ \unl b}\,\ep_{\unl b\unl a{}_1\unl a{}_2\ldots\unl a{}_p}\,\theta_{\rm L}^{\unl c}\wedge\theta^{\unl a{}_1}_{\rm L}\wedge\theta^{\unl a{}_2}_{\rm L}\wedge\cdots\wedge\theta^{\unl a{}_p}_{\rm L}\cr\cr
&&\hspace{3.2cm}-\tfrac{2\la_p}{p}\,f_{\widehat a\widehat S}^{\ \ \ \unl b}\,\ep_{\unl b\unl a{}_1\unl a{}_2\ldots\unl a{}_p}\,\theta_{\rm L}^{\widehat S}\wedge\theta^{\unl a{}_1}_{\rm L}\wedge\theta^{\unl a{}_2}_{\rm L}\wedge\cdots\wedge\theta^{\unl a{}_p}_{\rm L}\bigr)\bigr]\nn
\qqq
from which we readily extract another set of Euler--Lagrange equations by setting $\,\d x^a_i=0=\d x^{\widehat a}_i\,$ and $\,\d\theta^\a_i\neq 0\,$ and taking into account the non-singularity of the $\,\theta_{\rm L}^{\unl a}$,\ to wit,
\qq\label{eq:spinEL0}\hspace{3cm}
\bigl(\chi_{\a\b\unl a{}_1\unl a{}_2\ldots\unl a{}_p}+\la_p\,f_{\a\b}^{\ \ \ \unl a{}_0}\,\ep_{\unl a{}_0\unl a{}_1\unl a{}_2\ldots\unl a{}_p}\bigr)\,\Si_{\rm L}^\b\approx 0\,,\qquad\a\in\ovl{1,\d-d}\,,\quad\unl a{}_1,\unl a{}_2,\ldots,\unl a{}_p\in\ovl{0,p}\,.
\qqq
The latter result begs elucidation as it does not, on the face of it, have an obvious geometric meaning. Therefore, we interrupt our derivation to discuss the anticipated geometric form of the field equations in the formulation chosen, and the inevitable breakdown of translational symmetries effected by their solution, whereby we discover a path to its straightforward interpretation.

It is natural to expect,\label{kConsDisc} based -- in general -- on the geometric nature of the field theories under consideration and -- quite concretely -- on the hitherto results ({\it cp} also the remaining set of field equations, \Reqref{eq:rotEL}), that the Euler--Lagrange equations determine the body of the embedded worldvolume of the super-$p$-brane (up to the choice of the initial condition) covariantly through identification of its tangent within the tangent sheaf of the HP section as the intersection of kernels of the components of the Maurer--Cartan super-1-form along the direct-sum completion of the vacuum subspace $\,\tgt_{\rm vac}^{(0)}$.\ By supersymmetry, this scheme should have its Gra\ss mann-odd counterpart. Indeed, the freezing of the Gra\ss mann-even degrees of freedom transverse to the vacuum ({\it i.e.}, in particular, those along $\,\egt^{(0)}$) results in a spontaneous breakdown of supersymmetry which is transmitted, {\it via} the anticommutator
\qq\nn
\{\tgt^{(1)},\tgt^{(1)}\}\subset\tgt^{(0)}\oplus\hgt\,,
\qqq
to the Gra\ss mann-odd sector, and so the field equations in the odd sector ought to effect a `localisation' of the embedded worldvolume\footnote{One must keep in mind that the very definition of the super-$\si$-model embeddings presupposes a Gra\ss mann-odd extension of the Gra\ss mann-even worldvolume, as recalled in the opening paragraphs of Sec.\,\ref{sec:physmod}.} in what we presciently distinguished as the vacuum supspace $\,\tgt_{\rm vac}^{(1)}\equiv{\rm Im}\,\sfP^{(1)}\,$ on p.\,\pageref{ref:vacpresc}. Accordingly, we put forward
\medskip

\noindent\textbf{The $\k$-Symmetry Constraints:} We assume the Gra\ss mann-odd component of the Euler--Lagrange equations of the super-$\si$-model to define the corresponding sector of the vacuum of the field theory by (co)identifying its normal in the tangent sheaf of the HP section, with the algebraic model $\,\tgt_{\rm vac}^{(1)}$.\ To this end, we demand that the identity
\qq
\chi_{\a\b\unl a{}_1\unl a{}_2\ldots\unl a{}_p}+\la_p\,f_{\a\b}^{\ \ \ \unl a{}_0}\,\ep_{\unl a{}_0\unl a{}_1\unl a{}_2\ldots\unl a{}_p}\must\bigl(\bd1_{\d-d}-\sfP^{(1)}\bigr)^\g_{\ \a}\,\D_{\b\g\unl a{}_1\unl a{}_2\ldots\unl a{}_p}\equiv\bigl(\bd1_{\d-d}-\sfP^{(1)}\bigr)^\g_{\ \b}\,\D_{\a\g\unl a{}_1\unl a{}_2\ldots\unl a{}_p}\cr \label{eq:kConstr1}
\qqq
be satisfied for some non-singular tensor $\,\D_{\b\g\unl a{}_1\unl a{}_2\ldots\unl a{}_p}$,\ {\it i.e.} such that for any $p$-tuple $\,(\unl a{}_1\unl a{}_2\ldots\unl a{}_p)\in\ovl{0,p}\,$ the endomorphism 
\qq\nn
\D_{\unl a{}_1\unl a{}_2\ldots\unl a{}_p}=\D_{\a\b\unl a{}_1\unl a{}_2\ldots\unl a{}_p}\,\d^{\b\g}\,q^\a\ox Q_\g\in{\rm End}\bigl(\tgt^{(1)}\bigr)
\qqq
(written in the notation of \Reqref{eq:dualQ}) is invertible, and for a unique (up to a sign) value $\,\la_p\in\bR^\x$.\ The reduction of the odd degrees of freedom thus effected must be \emph{strictly} ({\it i.e.}, supersymmetrically) correlated with the similar reduction in the even sector encoded in Eqs.\,\eqref{eq:EM1}-\eqref{eq:BLC}, and so we further require
\qq\label{eq:kConstr2}
\{\tgt_{\rm vac}^{(1)},\tgt_{\rm vac}^{(1)}\}\overset{!}{\subset}\tgt_{\rm vac}^{(0)}\oplus\hgt\,.
\qqq

\noindent Clearly, whenever a projector can be read off from \Reqref{eq:spinEL0}, the latter field equations can be rewritten in the compact form
\qq\label{eq:spinEL}
\bigl(\bd1_{\d-d}-\sfP^{(1)}\bigr)^\a_{\ \b}\,\Si^\b_{\rm L}\approx 0\,,\qquad\a\in\ovl{1,\d-d}\,.
\qqq
\medskip

At this stage, in order to be able to proceed with our analysis, we make yet another simplifying assumption (intuited from inspection of the best known examples), which we term, by structural analogy with the standard differential calculus on metric manifolds,\medskip

\noindent\textbf{The No-Curvature and No-Torsion Constraints:} We assume the vacuum embedding to be (locally) flat and torsion-free up to a gauge transformation, in the sense expressed by the following identities
\qq\label{eq:FlatnoTor1}
[\tgt^{(0)}_{\rm vac},\tgt^{(0)}_{\rm vac}]\stackrel{!}{\subset}\hgt_{\rm vac}\stackrel{!}{\supset}[\egt^{(0)},\egt^{(0)}]\,.
\qqq
\medskip

\brem
As a consistency condition, implied by the Jacobi identity for triples from $\,\tgt_{\rm vac}^{(0)}\x\egt^{(0)}\x\dgt$,\ we derive from the the above the additional constraint:
\qq\label{eq:FlatnoTor2}
[\tgt_{\rm vac}^{(0)},\egt^{(0)}]\stackrel{!}{\subset}\dgt\,.
\qqq
\erem

\noindent These, in conjunction with the formerly derived field equations, lead to a much simplified expression for the variation:
\qq\nn
-\sfi\,\d_{\d\widehat\xi}\log\,\cA_{\rm DF}^{{\rm (HP)},p,\la_p}[\widehat\xi]&=&\sum_{\t\in\Tgt_{p+1}}\,\int_\t\,\bigl(\si_{\imath_\t}^{\rm vac}\circ\widehat\xi_\t\bigr)^*\bigl[\tfrac{\la_p}{p!}\,\d x^{\unl a}_{\imath_\t}\,\bigl(p\,f_{\unl S\unl c}^{\ \ \ \unl b}\,\ep_{\unl a\unl b\unl a{}_1\unl a{}_2\ldots\unl a{}_{p-1}}\,\theta_{\rm L}^{\unl S}\wedge\theta_{\rm L}^{\unl c}\wedge\theta^{\unl a{}_1}_{\rm L}\wedge\theta^{\unl a{}_2}_{\rm L}\wedge\cdots\wedge\theta^{\unl a{}_{p-1}}_{\rm L}\cr\cr
&&\hspace{3.2cm}-f_{\unl a\unl S}^{\ \ \ \unl b}\,\ep_{\unl b\unl a{}_1\unl a{}_2\ldots\unl a{}_p}\,\theta_{\rm L}^{\unl S}\wedge\theta^{\unl a{}_1}_{\rm L}\wedge\theta^{\unl a{}_2}_{\rm L}\wedge\cdots\wedge\theta^{\unl a{}_p}_{\rm L}\bigr)\cr\cr
&&\hspace{2.5cm}+\tfrac{1}{2(p-1)!}\,\d x^{\widehat a}_{\imath_\t}\,\bigl(\chi_{\a\b\widehat a\unl a{}_1\unl a{}_2\ldots \unl a{}_{p-1}}\,\Si_{\rm L}^\a\wedge\Si_{\rm L}^\b\wedge\theta_{\rm L}^{\unl a{}_1}\wedge\theta_{\rm L}^{\unl a{}_2}\wedge\cdots\wedge\theta_{\rm L}^{\unl a{}_{p-1}}\cr\cr
&&\hspace{3.2cm}-\tfrac{2\la_p}{p}\,f_{\widehat a\widehat S}^{\ \ \ \unl b}\,\ep_{\unl b\unl a{}_1\unl a{}_2\ldots\unl a{}_p}\,\theta_{\rm L}^{\widehat S}\wedge\theta^{\unl a{}_1}_{\rm L}\wedge\theta^{\unl a{}_2}_{\rm L}\wedge\cdots\wedge\theta^{\unl a{}_p}_{\rm L}\bigr)\bigr]\,,
\qqq
which reduces further upon taking into account the assumed unimodularity of the adjoint action of $\,\txH_{\rm vac}\,$ on $\,\tgt_{\rm vac}^{(0)}\,$ (as expressed in \Reqref{eq:unimodvac}),
\qq\nn
-\sfi\,\d_{\d\widehat\xi}\log\,\cA_{\rm DF}^{{\rm (HP)},p,\la_p}[\widehat\xi]&=&\sum_{\t\in\Tgt_{p+1}}\,\int_\t\,\bigl(\si_{\imath_\t}^{\rm vac}\circ\widehat\xi_\t\bigr)^*\bigl[\tfrac{1}{2(p-1)!}\,\d x^{\widehat a}_{\imath_\t}\,\bigl(\chi_{\a\b\widehat a\unl a{}_1\unl a{}_2\ldots \unl a{}_{p-1}}\,\Si_{\rm L}^\a\wedge\Si_{\rm L}^\b\wedge\theta_{\rm L}^{\unl a{}_1}\wedge\theta_{\rm L}^{\unl a{}_2}\wedge\cdots\wedge\theta_{\rm L}^{\unl a{}_{p-1}}\cr\cr
&&\hspace{3.2cm}-\tfrac{2\la_p}{p}\,f_{\widehat a\widehat S}^{\ \ \ \unl b}\,\ep_{\unl b\unl a{}_1\unl a{}_2\ldots\unl a{}_p}\,\theta_{\rm L}^{\widehat S}\wedge\theta^{\unl a{}_1}_{\rm L}\wedge\theta^{\unl a{}_2}_{\rm L}\wedge\cdots\wedge\theta^{\unl a{}_p}_{\rm L}\bigr)\bigr]\,.
\qqq
Note that the variation along $\,\tgt_{\rm vac}^{(0)}\,$ has vanished identically, and so we see that translations along $\,\tgt_{\rm vac}^{(0)}\,$ preserve the DF amplitudes after partial reduction (namely, after imposition of the IHC, of the BLC and of the spinorial field equations \eqref{eq:spinEL0}).

At this point, there remains one last condition to be imposed to conclude our derivation. As the only source of the spinorial indices carried by the tensors $\,\chi_{\a\b\widehat a\unl a{}_1\unl a{}_2\ldots \unl a{}_{p-1}}\,$ \emph{and} by the structure constants $\,f_{\a\b}^{\ \ \ \widehat a}\,$ are the generators of the relevant Clifford algebra (and the charge conjugation matrix), we are rather naturally led to impose on the GS super-$(p+2)$-cocycle\medskip

\noindent\textbf{The $\G$-Constraints:} We assume the following identities
\qq\label{eq:GamConstr}
\chi_{\a\g\widehat a\unl a{}_1\unl a{}_2\ldots \unl a{}_{p-1}}\,\sfP^{(1)\,\g}_{\ \ \ \ \ \b}=\chi_{\g\b\widehat a\unl a{}_1\unl a{}_2\ldots \unl a{}_{p-1}}\,\bigl(\bd1_{\d-d}-\sfP^{(1)}\bigr)^\g_{\ \a}
\qqq
to hold true.
\medskip

\noindent When used together with the Euler--Lagrange equations \eqref{eq:spinEL}, this yields
\qq\nn
-\sfi\,\d_{\d\widehat\xi}\log\,\cA_{\rm DF}^{{\rm (HP)},p,\la_p}[\widehat\xi]&=&\tfrac{\la_p}{p!}\,\sum_{\t\in\Tgt_{p+1}}\,\int_\t\,\bigl(\si_{\imath_\t}^{\rm vac}\circ\widehat\xi_\t\bigr)^*\d x^{\widehat a}_{\imath_\t}\,f_{\widehat S\widehat a}^{\ \ \ \unl b}\,\ep_{\unl b\unl a{}_1\unl a{}_2\ldots\unl a{}_p}\,\theta_{\rm L}^{\widehat S}\wedge\theta^{\unl a{}_1}_{\rm L}\wedge\theta^{\unl a{}_2}_{\rm L}\wedge\cdots\wedge\theta^{\unl a{}_p}_{\rm L}\,,
\qqq
and so, upon recalling the interpretation of the volume form in the vacuum, we arrive at the last set of field equations:
\qq\label{eq:rotEL}
\theta_{\rm L}^{\widehat S}\approx 0\,,\qquad\widehat S\in\ovl{D-\unl\d+1,D-\d}\,.
\qqq

\noindent We summarise our findings in 
\berop\label{prop:HPELeqs}
Adopt the assumptions and the notation of Def.\,\ref{def:GSinHP}. If the Hughes--Polchinski superbackground $\,\sgt\Bgt^{{\rm (HP)}}_{p,\la_p}\,$ satisfies the Even Effective-Mixing Constraints \eqref{eq:EMC0}-\eqref{eq:EMC3}, the Dimensional Constraint \eqref{eq:DimConstr}, the Descendability Constraint \eqref{eq:DesConstr}, the No-Curvature and No-Torsion Constraints \eqref{eq:FlatnoTor1} and \eqref{eq:FlatnoTor2}, and the $\G$-Constraints \eqref{eq:GamConstr}, then the Euler--Lagrange equations of the corresponding Green--Schwarz super-$\si$-model in the Hughes--Polchinski formulation for the super-$p$-brane in $\,\sgt\Bgt^{{\rm (HP)}}_{p,\la_p}\,$ restricted to field configurations subject to the Body-Localisation Constraints \eqref{eq:EM1} read
\qq\nn
\bigl(\theta_{\rm L}^{\widehat a},\bigl(\chi_{\a\b\unl a{}_1\unl a{}_2\ldots\unl a{}_p}+\la_p\,f_{\a\b}^{\ \ \ \unl a{}_0}\,\ep_{\unl a{}_0\unl a{}_1\unl a{}_2\ldots\unl a{}_p}\bigr)\,\Si_{\rm L}^\b,\theta_{\rm L}^{\widehat S}\bigr)\approx 0\,,\qquad\bigl(\widehat a,\a,\unl a{}_k,\widehat S\bigr)\in\ovl{p+1,d}\x\ovl{1,\d-d}\x\ovl{0,p}\x\ovl{D-\unl\d+1,D-\d}\,.
\qqq
If also the $\k$-Symmetry Constraints \eqref{eq:kConstr1} and \eqref{eq:kConstr2} are satisfied, the equations can be cast in the form
\qq 
\sfP^\ggt_{\ \egt\oplus\dgt}\circ\theta_{\rm L}\equiv\bigl(\theta_{\rm L}^{\widehat a},\bigl(\bd1_{\d-d}-\sfP^{(1)}\bigr)^\a_{\ \b}\,\Si^\b_{\rm L},\theta_{\rm L}^{\widehat S}\bigr)\approx 0\,,\qquad\bigl(\widehat a,\a,\widehat S\bigr)\in\ovl{p+1,d}\x\ovl{1,\d-d}\x\ovl{D-\unl\d+1,D-\d}\,, \cr \label{eq:ELeqssigmod}
\qqq
where 
\qq\nn
\sfP^\ggt_{\ \egt\oplus\dgt}\ :\ \ggt\circlearrowleft
\qqq 
is the projector onto $\,\egt\oplus\dgt\,$ with the kernel
\qq\nn
{\rm Ker}\,\sfP^\ggt_{\ \egt\oplus\dgt}=\tgt_{\rm vac}\oplus\hgt_{\rm vac}\,.
\qqq
\eerop
\beroof
Given above.
\eroof

\noindent We are thus led to
\bedef\label{def:vacHPsdistro}
Adopt the hitherto notation and assume that the Hughes--Polchinski superbackground $\,\sgt\Bgt^{{\rm (HP)}}_{p,\la_p}\,$ satisfies \emph{all} the Constraints listed in Prop.\,\ref{prop:HPELeqs}. The \textbf{Hughes--Polchinski vacuum superdistribution of} $\,\sgt\Bgt^{{\rm (HP)}}_{p,\la_p}\,$ is the sub-superdistribution of the correspondence superdistribution $\,{\rm Corr}_{\rm HP/NG}(\sgt\Bgt_{p,\la_p}^{\rm (HP)})\,$ of Def.\,\ref{def:Corrsdistro} given by
\qq\label{eq:Vacsdistro}
{\rm Vac}\bigl(\sgt\Bgt_{p,\la_p}^{\rm (HP)}\bigr):={\rm Ker}\,\bigl(\sfP^\ggt_{\ \egt\oplus\dgt}\circ\theta_{\rm L}\rstr_{\cT\Si^{\rm HP}}\bigr)\subset{\rm Corr}_{\rm HP/NG}\bigl(\sgt\Bgt_{p,\la_p}^{\rm (HP)}\bigr)\subset\cT\Si^{\rm HP}
\qqq
and modelled on $\,\tgt_{\rm vac}$.
\exdef

\brem
In order to neatly describe the HP vacuum superdistribution and related objects, we shall work with an eigenbasis of the projector $\,\sfP^{(1)}$.\ Elements of a basis of $\,{\rm Im}\,\sfP^{(1)}\equiv\tgt_{\rm vac}^{(1)}\,$ shall be denoted as 
\qq\nn
\unl Q{}_{\unl\a}\equiv\unl\La{}_{\ \unl\a}^\b\,Q_\b\,,\qquad\unl\a\in\ovl{1,q}\,,
\qqq
and those of a basis of $\,{\rm Ker}\,\sfP^{(1)}\equiv\tgt_{\rm vac}^{(1)}\,$ as
\qq\nn
\widehat Q_{\widehat\a}\equiv\widehat\La{}_{\ \widehat\a}^\b\,Q_\b\,,\qquad\widehat\a\in\ovl{q+1,\d-d}\,,
\qqq
where the $\,\unl\La{}_{\ \unl\a}^\b\,$ and the $\,\widehat\La{}_{\ \widehat\a}^\b\,$ are suitable numerical coefficiens.
\erem

There are three physically motivated regularity criteria\label{page:regularcrit} that we are compelled to invoke with regard to $\,{\rm Vac}(\sgt\Bgt_{p,\la_p}^{\rm (HP)})\,$ at this stage. We discuss at length the first two of them, postponing the last one to Sec.\,\ref{sec:susy} in which we shall have gathered the requisite formal and conceptual tools to make its analysis structural. The first of these is $\txH_{\rm vac}$-descendability which is indispensable if we wish to have access to vacua stretching across several trivialising patches $\,\cU_i^{\rm vac}\,$ of an \emph{a priori} nontrivial principal $\txH_{\rm vac}$-bundle \eqref{eq:homasprinc}, and even over a single patch it appears necessary to preserve the status of $\,\txH_{\rm vac}\,$ as the model of a gauge symmetry of the \emph{vacuum}. We have
\berop\label{prop:descvacsdistrolin}
Adopt the notation of Def.\,\ref{def:vacHPsdistro}. If the Hughes--Polchinski vacuum superdistribution $\,{\rm Vac}(\sgt\Bgt_{p,\la_p}^{\rm (HP)})\,$ is $\txH_{\rm vac}$-descendable in the sense of Def.\,\ref{def:Hvacdescsdistro}, then
\qq\nn
[\hgt_{\rm vac},\tgt^{(1)}_{\rm vac}]\subset\tgt^{(1)}_{\rm vac}\,.
\qqq
\eerop
\beroof
Obvious.
\eroof

\noindent The second natural criterion is \textbf{involutivity}, which for an arbitrary superdistribution $\,\cD\subset\cT\Si^{\rm HP}\,$ is expressed by the relation 
\qq\nn
[\cD,\cD\}\subset\cD\,.
\qqq
In the light of the supergeometric variant of The Frobenius Theorem ({\it cp} \Rxcite{Thms.\,6.1.12 \& 6.2.1}{Carmeli:2011}), this property ensures existence of a foliation of the HP section by the integral sub-supermanifolds of $\,\cD$.\ From the point of view of the underlying field theory, this is to be understood as a foliation by vacua corresponding to different initial conditions. We put it in a separate
\bedef\label{def:HPvacfol}
Adopt the hitherto notation. The foliation of the Hughes--Polchinski section $\,\Si^{\rm HP}\,$ by the integral leaves of the Hughes--Polchinski vacuum superdistribution $\,{\rm Vac}(\sgt\Bgt_{p,\la_p}^{\rm (HP)})$,\ whenever it exists, shall be termed the \textbf{Hughes--Polchinski vacuum foliation of} $\,\Si^{\rm HP}\,$ and denoted as $\,\Si^{\rm HP}_{\rm vac}\,$ (the disjoint union of its leaves), with the embedding
\qq\label{eq:HPvacsec}
\iota_{\rm vac}\ :\ \Si^{\rm HP}_{\rm vac}\emb\Si^{\rm HP}\,.
\qqq
\exdef
\noindent We have
\berop\label{prop:vacsalg}
Adopt the notation of Def.\,\ref{def:vacHPsdistro}. The Hughes--Polchinski vacuum superdistribution $\,{\rm Vac}(\sgt\Bgt_{p,\la_p}^{\rm (HP)})\,$ is involutive and hence determines the Hughes--Polchinski vacuum foliation of Def.\,\ref{def:HPvacfol} iff the following relations -- to be termed the \textbf{Vacuum-Superalgebra Constraints} henceforth -- are simultaneously satisfied:
\qq\nn
\{\tgt^{(1)}_{\rm vac},\tgt^{(1)}_{\rm vac}\}\subset\tgt_{\rm vac}^{(0)}\oplus\hgt_{\rm vac}\,,\qquad\qquad[\tgt_{\rm vac}^{(0)},\tgt^{(1)}_{\rm vac}]\subset\tgt^{(1)}_{\rm vac}\,,\qquad\qquad[\hgt_{\rm vac},\tgt^{(1)}_{\rm vac}]\subset\tgt^{(1)}_{\rm vac}\,.
\qqq
When put in conjunction with the Constraints listed in Prop.\,\ref{prop:HPELeqs} and assumed herein, they endow the \textbf{vacuum supervector space of} $\,\sgt\Bgt^{{\rm (HP)}}_{p,\la_p}\,$ defined as
\qq\nn
\gt{vac}\bigl(\sgt\Bgt_{p,\la_p}^{\rm (HP)}\bigr)\equiv\tgt_{\rm vac}\oplus\hgt_{\rm vac}\subset\ggt
\qqq
with the structure of a Lie sub-superalgebra, to be referred to as the \textbf{vacuum superalgebra of} $\,\sgt\Bgt^{{\rm (HP)}}_{p,\la_p}$,\ whenever it exists.
\eerop
\beroof
The vacuum superdistribution is a sub-superdistribution of the manifestly involutive superdistribution $\,\cT\Si^{\rm HP}\,$ locally generated (over $\,\cO_\txG(\cV_i)$) by the vector fields $\,\cT_{\mu\,i}\,$ of Prop.\,\ref{prop:bastanshHP}. Each of the latter is identified uniquely by its horizontal component $\,L_\mu\rstr_{\cV_i}\,$ with $\,\mu\in\ovl{0,\unl\d}$,\ and so it suffices to check which generators of $\,\fgt\,$ appear in the supercommutators $\,[\tgt_{\rm vac},\tgt_{\rm vac}\}\,$ (corresponding to the supercommutators $\,[L_{\unl{\unl A}},L_{\unl{\unl B}}\}\,$ of the horizontal components) and the commutators $\,[\hgt_{\rm vac},\tgt_{\rm vac}]\,$ (corresponding to the commutators $\,[L_{\unl S},L_{\unl{\unl A}}]\,$ of a horizontal component with a vertical correction). This yields the first statement of the proposition. The second part is obvious.
\eroof

\brem
The Vacuum-Superalgebra Constraints shall be encountered again when we come to investigate a peculiar odd local supersymmetry of the super-$\si$-model in the HP formulation that arises in the correspondence sector of its configuration space. Meanwhile, we indicate a simple way of realising the above (and previous) constraints, suggested by the analysis of known examples -- it consists in imposing the constraints
\qq
&\sfP^{(1)\,\g}_{\ \ \ \ \ \a}\,f_{\g\b}^{\ \ \ \unl a}=f_{\a\g}^{\ \ \ \unl a}\,\sfP^{(1)\,\g}_{\ \b}\,,\qquad\qquad\sfP^{(1)\,\g}_{\ \ \ \ \ \a}\,f_{\g\b}^{\ \ \ \widehat a}=f_{\a\g}^{\ \ \ \widehat a}\,\bigl(\bd1_{\d-d}-\sfP^{(1)}\bigr)^\g_{\ \b}\,,&\label{eq:kConstr4}\\ \cr
&\sfP^{(1)\,\g}_{\ \ \ \ \ \a}\,f_{\g\b}^{\ \ \ \unl S}=f_{\a\g}^{\ \ \ \unl S}\,\sfP^{(1)\,\g}_{\ \b}\,,\qquad\qquad\sfP^{(1)\,\g}_{\ \ \ \ \ \a}\,f_{\g\b}^{\ \ \ \widehat S}=f_{\a\g}^{\ \ \ \widehat S}\,\bigl(\bd1_{\d-d}-\sfP^{(1)}\bigr)^\g_{\ \b}\,,&\label{eq:kConstr5}\\ \cr
&\sfP^{(1)\,\g}_{\ \ \ \ \ \a}\,f_{\unl S\g}^{\ \ \ \b}=f_{\unl S\a}^{\ \ \ \g}\,\sfP^{(1)\,\b}_{\ \ \ \ \ \g}\,,\qquad\qquad\sfP^{(1)\,\g}_{\ \ \ \ \ \a}\,f_{\widehat S\g}^{\ \ \ \b}=f_{\widehat S\a}^{\ \ \ \g}\,\bigl(\bd1_{\d-d}-\sfP^{(1)}\bigr)^\b_{\ \g}\,,& \label{eq:kConstr6}\\ \cr
&\sfP^{(1)\,\g}_{\ \ \ \ \ \a}\,f_{\unl a\g}^{\ \ \ \b}=f_{\unl a\a}^{\ \ \ \g}\,\sfP^{(1)\,\b}_{\ \ \ \ \ \g}\,.&\label{eq:kConstr7}
\qqq
\erem

In the remainder of the present paper, we regard the regular case in which the HP vacuum foliation is \emph{both} $\txH_{\rm vac}$-descendable and involutive as the physically most appealing and natural one, hoping to return to the less obvious irregular case in a future study. In order to be able to quantify departures from regularity observed in some of the superbackgrounds listed in Sec.\,\ref{sec:physmod} that we analyse one by one at the end of this section, we need an adaptation, to the present supergeometric context, of the differential-geometric concepts that we introduce below after \Rcite{Tanaka:1970}. This extra conceptual investment will pay back in the analysis of local supersymmetries of the super-$\si$-model in the next section.
\bedef
Let $\,\cM\,$ be a smooth manifold and $\,\cD\subset\cT\cM\,$ a distribution over it. The \textbf{weak derived flag of} $\,\cD\,$ is the filtration 
\qq\nn
\cD^\bullet\ :\ \cD\equiv\cD^{-1}\subset\cD^{-2}\subset\ldots\subset\cD^{-j}\subset\ldots
\qqq
of $\,\cT\cM\,$ with components defined recursively as
\qq\nn
\cD^{-j}:=\cD^{-j+1}+[\cD,\cD^{-j+1}]\,,\qquad j>1\,.
\qqq
Given $\,x\in\cM$,\ denote by $\,\cD^{-j}(x)\,$ the $\bR$-linear span of all iterated Lie brackets, of length not greater than $\,j$,\ of the (local) generators of $\,\cD\,$ evaluated at $\,x$.\ A distribution $\,\cD\,$ is called \textbf{regular} if 
\qq\nn
\forall_{j\in\bN^\x}\ \forall_{x,y\in\cM}\ :\ \dim_\bR\,\cD^{-j}(x)=\dim_\bR\,\cD^{-j}(y)\,.
\qqq

The bounded function 
\qq\nn
\mu\ :\ \cM\too\bN^\x\ :\ x\longmapsto\min\{\ j\in\bN^\x \quad\vert\quad \cD^{-j-1}(x)=\cD^{-j}(x)\ \}
\qqq
shall be termed the \textbf{height of} $\,\cD^\bullet$.\ A distribution $\,\cD\,$ is called \textbf{bracket-generating} if \qq\nn
\forall_{x\in\cM}\ :\ \cD^{-\mu(x)}(x)=\sfT_x\cM\,.
\qqq
For any regular bracket-generating distribution $\,\cD$,\ and any $\,x\in\cM$,\ write
\qq\nn
\gtq^{-1}(x):=\cD^{-1}(x)\qquad\,,\qquad\gtq^{-j}(x):=\cD^{-j}(x)/\cD^{-j+1}(x)\,,\qquad j>1
\qqq
and
\qq\nn
\mgt(x):=\bigoplus_{j=1}^{\mu(x)}\,\gtq^{-j}(x)\,,
\qqq
endowing that latter vector space with the Lie bracket 
\qq\nn
[\cdot,\cdot]_{\mgt(x)}\ :\ \mgt(x)\x\mgt(x)\too\mgt(x)
\qqq
with restrictions
\qq\nn
[\cdot,\cdot]_{\mgt(x)}\rstr_{\gtq^{-j_1}(x)\x\gtq^{-j_2}(x)}\ &:&\ \gtq^{-j_1}(x)\x\gtq^{-j_2}(x)\too\gtq^{-j_1-j_2}(x)\cr\cr
&:&\ \bigl(X^{(-j_1)}_1(x)+\cD^{-j_1+1}(x),X^{(-j_2)}_2(x)+\cD^{-j_2+1}(x)\bigr)\cr\cr
&&\hspace{1cm}\longmapsto[X^{(-j_1)}_1,X^{(-j_2)}_2](x)+\cD^{-j_1-j_2+1}(x)\,,
\qqq
written in terms of (local) sections $\,X^{(-j_A)}_A\in\G_{\rm loc}(\cD^{-j_A}),\ A\in\{1,2\}$.\ The graded nilpotent Lie algebra 
\qq\nn
\bigl(\mgt(x),[\cdot,\cdot]_{\mgt(x)}\bigr)\,,
\qqq
generated by its subspace $\,\gtq^{-1}(x)\,$ (and hence termed \textbf{fundamental}), is called the \textbf{symbol of the distribution $\,\cD\,$ at the point $\,x$}. It is customary to choose at a given $\,x\in\cM\,$ a local basis $\,\cB_x\,$ of $\,\cT_x\cM\,$ adapted to the weak derived flag of $\,\cD\,$ evaluated at that point, {\it i.e.}, to form the corresponding filtration of local bases
\qq\nn
\cB_x^{-1}\subset\cB_x^{-2}\subset\ldots\subset\cB_x\,,\qquad\qquad\cD^{-j}(x)=\corr{\cB_x^{-j}}_\bR\,,
\qqq
and subsequently present $\,\gtq^{-j}(x)\,$ as 
\qq\nn
\gtq^{-j}(x)\equiv\corr{\cB_x^{-j}\setminus\cB_x^{-j+1}}_\bR\,.
\qqq
We adopt this convention in what follows.

Whenever there exists a fundamental graded nilpotent Lie algebra (of minimal degree $\,-\mu\in\bZ_{<0}$)
\qq\nn
\bigl(\mgt\equiv\bigoplus_{j=1}^\mu\,\gtq^{-j},[\cdot,\cdot]_\mgt\bigr)
\qqq
with the property 
\qq\nn
\forall_{x\in\cM}\ :\ \mgt(x)\cong\mgt\,,
\qqq
written in the category of graded nilpotent Lie algebras, we call $\,\cD\,$ a \textbf{distribution of constant symbol} $\,\mgt$.
\exdef

\noindent Thus, we shall be interested in -- among other things -- the \textbf{limit}
\qq\nn
{\rm Vac}^{-\infty}\bigl(\sgt\Bgt_{p,\la_p}^{\rm (HP)}\bigr)\equiv{\rm Vac}^{-\mu}\bigl(\sgt\Bgt_{p,\la_p}^{\rm (HP)}\bigr)\subset\cT\Si^{\rm HP}
\qqq
of the weak derived (super)flag $\,{\rm Vac}^\bullet(\sgt\Bgt_{p,\la_p}^{\rm (HP)})\,$ of the regular superdistribution $\,{\rm Vac}(\sgt\Bgt_{p,\la_p}^{\rm (HP)})\,$ and its relation to the mother tangent sheaf $\,\cT\Si^{\rm HP}\,$ in the familiar examples. Invariably, the regular vacuum superdistribution itself and the various components of its weak derived flag are sewn from the respective restrictions to the superdomains $\,\cV_i$,\ differing solely in the index $\,i\in I_{\txH_{\rm vac}}\,$ carried by the local generators. Therefore, whenever disclosing the anatomy of the breakdown of integrability, we give the local structure (in the form of an $\cO_\txG(\cV_i)$-linear span) exclusively.\medskip

\beg\textbf{The vacuum superdistribution for the Green--Schwarz super-0-brane in $\,{\rm sISO}(9,1\,|\,32)/{\rm SO}(9)$.}\label{eg:GSs0Vac}
The identity 
\qq\nn
\chi_{\a\b}+\la_0\,f_{\a\b}^{\ \ \ 0}=2\ovl\G_{11\,\a\b}+\la_0\,\ovl\G{}^0_{\a\b}\equiv 4\ovl\G{}^0_{\a\g}\,\left(\tfrac{\tfrac{\la_0}{2}-\G^0\,\G_{11}}{2}\right)^\g_{\ \b}\,,
\qqq
yields -- for $\,\la_0\in\{-2,2\}\,$ -- the possible projectors
\qq\nn
\sfP^{(1)}_{\pm 2}=\tfrac{\bd1_{32}\pm\G^0\,\G_{11}}{2}
\qqq
with the properties
\qq\nn
\G^0\,\sfP^{(1)}_{\pm 2}=\bigl(\bd1_{32}-\sfP^{(1)}_{\pm 2}\bigr)\,\G^0\,,\qquad\qquad\G^{\widehat a}\,\sfP^{(1)}_{\pm 2}=\sfP^{(1)}_{\pm 2}\,\G^{\widehat a}\,,\qquad\qquad C\,\sfP^{(1)}_{\pm 2}\,C^{-1}=\bigl(\bd1_{32}-\sfP^{(1)}_{\pm 2}\bigr)^{\rm T}\,,
\qqq
and so upon choosing, for the sake of concreteness,
\qq\nn
\sfP^{(1)}\equiv\sfP^{(1)}_{+2}=\tfrac{\bd1_{32}+\G^0\,\G_{11}}{2}\,,
\qqq
we readily verify 
\bit
\item the $\k$-Symmetry Constraints,
\qq\nn
\{\sfP^{(1)\,\g}_{\ \ \ \ \ \a}\,Q_\g,\sfP^{(1)\,\d}_{\ \ \ \ \ \b}\,Q_\d\}=\sfP^{(1)\,\g}_{\ \ \ \ \ \b}\,\ovl\G{}^0_{\a\g}\,P_0\in\corr{P_0}\oplus\bigoplus_{a,b=0}^9\,\corr{J_{ab}}\,;
\qqq
\item the Even Effective-Mixing Constraints, 
\qq\nn
[J_{\widehat a\widehat b},P_0]=0\in\corr{P_0}\ni\d_{\widehat a\widehat b}\,P_0=[J_{0\widehat a},P_{\widehat b}]\,,\qquad\qquad[J_{\widehat a\widehat b},P_{\widehat c}]=\d_{\widehat b\widehat c}\,P_{\widehat a}-\d_{\widehat a\widehat c}\,P_{\widehat b}\in\bigoplus_{\widehat d=1}^9\,\corr{P_{\widehat d}}\ni P_{\widehat a}=[J_{0\widehat a},P_0]\,;
\qqq
\item the No-Curvature and No-Torsion Constraints -- trivial;
\item the $\G$-Constraints -- trivial;
\item the Vacuum-Superalgebra Constraints, with
\qq\nn
\{\sfP^{(1)\,\g}_{\ \ \ \ \ \a}\,Q_\g,\sfP^{(1)\,\d}_{\ \ \ \ \ \b}\,Q_\d\}\in\corr{P_0}\oplus\bigoplus_{\widehat a,\widehat b=1}^9\,\corr{J_{\widehat a\widehat b}}
\qqq
and
\qq\nn
[J_{\widehat a\widehat b},\sfP^{(1)\,\b}_{\ \ \ \ \ \a}\,Q_\b]&=&\tfrac{1}{2}\,\G_{\widehat a\widehat b\,\a}^\b\,\sfP^{(1)\,\g}_{\ \ \ \ \ \b}\,Q_\g\in{\rm Im}\,\sfP^{(1)}\,,\cr\cr
[P_0,\sfP^{(1)\,\b}_{\ \ \ \ \ \a}\,Q_\b]&=&0\in{\rm Im}\,\sfP^{(1)}\,;
\qqq
\item ${\rm SO}(9)$-descendability.
\eit
Consequently, the HP vacuum superdistribution with restrictions
\qq\nn
{\rm Vac}\bigl({\rm sISO}(9,1\,|\,32)/{\rm SO}(9),\underset{\tx{\ciut{(2)}}}{\widehat\chi}\hspace{-1pt}{}^{\rm GS}\bigr)\rstr_{\cV_i}=\bigoplus_{\unl\a=1}^{16}\,\corr{\cT_{\unl\a\,i}}\oplus\corr{\cT_{0\,i}}
\qqq 
is an ${\rm SO}(9)$-descendable integrable superdistribution associated with the Lie superalgebra
\qq\nn
\gt{vac}\bigl({\rm sISO}(9,1\,|\,32)/{\rm SO}(9),\underset{\tx{\ciut{(2)}}}{\widehat\chi}\hspace{-1pt}{}^{\rm GS}\bigr)=\bigoplus_{\unl\a=1}^{16}\,\corr{\unl Q{}_{\unl\a}}\oplus\corr{P_0}\oplus\bigoplus_{\widehat a,\widehat b=1}^9\,\corr{J_{\widehat a\widehat b}}\,.
\qqq 
\eeg

\beg\textbf{The vacuum superdistribution for the Green--Schwarz super-$(4k+1)$-brane in $\,{\rm sISO}(d,1\,|\,D_{d,1})/({\rm SO}(4k+1,1)\x{\rm SO}(d-4k-1))\,$ for $\,k\in\{0,1,2\}$.} 
The identity 
\qq\nn
\chi_{\a\b\unl a{}_1\unl a{}_2\ldots\unl a{}_{4k+1}}+\la_{4k+1}\,f_{\a\b}^{\ \ \ \unl b}\,\ep_{\unl b\unl a{}_1\unl a{}_2\ldots\unl a{}_{4k+1}}=4(4k+1)!\,\ep_{\unl b\unl a{}_1\unl a{}_2\ldots\unl a{}_{4k+1}}\,\ovl\G{}^{\unl b}_{\a\g}\,\left(\tfrac{\tfrac{\la_{4k+1}}{2(4k+1)!}\,\bd1_{D_{d,1}}-\G^0\,\G^1\,\cdots\,\G^{4k+1}}{2}\right)^\g_{\ \b}\,,
\qqq
yields -- for $\,\la_{4k+1}\in\{-2(4k+1)!\equiv\la_{4k+1}^-,2(4k+1)!\equiv\la_{4k+1}^+\}\,$ -- the possible projectors
\qq\nn
\sfP^{(1)}_{\la_{4k+1}^\pm}=\tfrac{\bd1_{D_{d,1}}\pm\,\G^0\,\G^1\,\cdots\,\G^{4k+1}}{2}
\qqq
with properties
\qq\nn
\G^{\unl a}\,\sfP^{(1)}_{\la_{4k+1}^\pm}=\bigl(\bd1_{D_{d,1}}-\sfP^{(1)}_{\la_{4k+1}^\pm}\bigr)\,\G^{\unl a}\,,\qquad\qquad\G^{\widehat a}\,\sfP^{(1)}_{\la_{4k+1}^\pm}=\sfP^{(1)}_{\la_{4k+1}^\pm}\,\G^{\widehat a}\,,\qquad\qquad C\,\sfP^{(1)}_{\la_{4k+1}^\pm}\,C^{-1}=\bigl(\bd1_{D_{d,1}}-\sfP^{(1)}_{\la_{4k+1}^\pm}\bigr)^{\rm T}\,,
\qqq
and so upon choosing
\qq\nn
\sfP^{(1)}\equiv\sfP^{(1)}_{\la_{4k+1}^+}=\tfrac{\bd1_{D_{d,1}}+\G^0\,\G^1\,\cdots\,\G^{4k+1}}{2}\,,
\qqq
we check
\bit
\item the $\k$-Symmetry Constraints,
\qq\nn
\{\sfP^{(1)\,\g}_{\ \ \ \ \ \a}\,Q_\g,\sfP^{(1)\,\d}_{\ \ \ \ \ \b}\,Q_\d\}=\sfP^{(1)\,\g}_{\ \ \ \ \ \b}\,\ovl\G{}_{\a\g}^{\unl a}\,P_{\unl a}\in\bigoplus_{\unl a=0}^{4k+1}\,\corr{P_{\unl a}}\oplus\bigoplus_{a,b=0}^d\,\corr{J_{ab}}\,;
\qqq
\item the Even Effective-Mixing Constraints, 
\qq\nn
&[J_{\unl a\unl b},P_{\unl c}]=\eta_{\unl b\unl c}\,P_{\unl a}-\eta_{\unl a\unl c}\,P_{\unl b}\in\bigoplus_{\unl d=0}^{4k+1}\,\corr{P_{\unl d}}\ni 0=[J_{\widehat a\widehat b},P_{\unl c}]\,,\qquad\qquad[J_{\unl a\widehat b},P_{\widehat c}]=\d_{\widehat b\widehat c}\,P_{\unl a}\in\bigoplus_{\unl d=0}^{4k+1}\,\corr{P_{\unl d}}\,,&\cr\cr
&[J_{\unl a\unl b},P_{\widehat c}]=0\in\bigoplus_{\widehat d=4k+2}^d\,\corr{P_{\widehat d}}\ni\d_{\widehat b\widehat c}\,P_{\widehat a}-\d_{\widehat a\widehat c}\,P_{\widehat b}=[J_{\widehat a\widehat b},P_{\widehat c}]\,,\qquad\qquad[J_{\unl a\widehat b},P_{\unl c}]=-\eta_{\unl a\unl c}\,P_{\widehat b}\in\bigoplus_{\widehat d=4k+2}^d\,\corr{P_{\widehat d}}\,;&
\qqq
\item the No-Curvature and No-Torsion Constraints -- trivial;
\item the $\G$-Constraints,  
\qq\nn
\bigl(C\,\G_{\widehat a}\,\G_{\unl a{}_1}\,\G_{\unl a{}_2}\cdots\G_{\unl a{}_{4k}}\bigr)_{\a\g}\,\sfP^{(1)\,\g}_{\ \ \ \ \ \b}&=&\bigl(C\,\sfP^{(1)}\,\G_{\widehat a}\,\G_{\unl a{}_1}\,\G_{\unl a{}_2}\cdots\G_{\unl a{}_{4k}}\bigr)_{\a\b}=\bigl(\bigl(\bd1_{D_{d,1}}-\sfP^{(1)}\bigr)^{\rm T}\,C\,\G_{\widehat a}\,\G_{\unl a{}_1}\,\G_{\unl a{}_2}\cdots\G_{\unl a{}_{4k}}\bigr)_{\a\b}\cr\cr
&\equiv&\bigl(C\,\G_{\widehat a}\,\G_{\unl a{}_1}\,\G_{\unl a{}_2}\cdots\G_{\unl a{}_{4k}}\bigr)_{\g\b}\,\bigl(\bd1_{D_{d,1}}-\sfP^{(1)}\bigr)^\g_{\ \a}\,; 
\qqq
\item the Vacuum-Superalgebra Constraints, with
\qq\nn
\{\sfP^{(1)\,\g}_{\ \ \ \ \ \a}\,Q_\g,\sfP^{(1)\,\d}_{\ \ \ \ \ \b}\,Q_\d\}\in\bigoplus_{\unl a=0}^{4k+1}\,\corr{P_{\unl a}}\oplus\bigoplus_{\unl a,\unl b=0}^{4k+1}\,\corr{J_{\unl a\unl b}}\oplus\bigoplus_{\widehat a,\widehat b=4k+2}^d\,\corr{J_{\widehat a\widehat b}}\,,
\qqq
and
\qq\nn
&[J_{\unl a\unl b},\sfP^{(1)\,\b}_{\ \ \ \ \ \a}\,Q_\b]=\tfrac{1}{2}\,\G_{\unl a\unl b\,\a}^\b\,\sfP^{(1)\,\g}_{\ \ \ \ \ \b}\,Q_\g\in{\rm Im}\,\sfP^{(1)}\,,&\cr\cr
&[J_{\widehat a\widehat b},\sfP^{(1)\,\b}_{\ \ \ \ \ \a}\,Q_\b]=\tfrac{1}{2}\,\G_{\widehat a\widehat b\,\a}^\b\,\sfP^{(1)\,\g}_{\ \ \ \ \ \b}\,Q_\g\in{\rm Im}\,\sfP^{(1)}&
\qqq
and
\qq\nn
[P_{\unl a},\sfP^{(1)\,\b}_{\ \ \ \ \ \a}\,Q_\b]=0\in{\rm Im}\,\sfP^{(1)}\,;
\qqq
\item $({\rm SO}(4k+1,1)\x{\rm SO}(d-4k-1))$-descendability.
\eit
Consequently, the HP vacuum superdistribution with restrictions
\qq\nn
{\rm Vac}\bigl({\rm sISO}(d,1\,|\,D_{d,1})/({\rm SO}(4k+1,1)\x{\rm SO}(d-4k-1)),\underset{\tx{\ciut{(p+2)}}}{\widehat\chi}\hspace{-6pt}{}^{\rm GS}\bigr)\rstr_{\cV_i}=\bigoplus_{\unl\a=1}^{\frac{D_{d,1}}{2}}\,\corr{\cT_{\unl\a\,i}}\oplus\bigoplus_{\unl a=0}^{4k+1}\,\corr{\cT_{\unl a\,i}}
\qqq 
is an $({\rm SO}(4k+1,1)\x{\rm SO}(d-4k-1))$-descendable integrable superdistribution associated with the Lie superalgebra 
\qq\nn
&&\gt{vac}\bigl({\rm sISO}(d,1\,|\,D_{d,1})/({\rm SO}(4k+1,1)\x{\rm SO}(d-4k-1)),\underset{\tx{\ciut{(p+2)}}}{\widehat\chi}\hspace{-6pt}{}^{\rm GS}\bigr)\cr\cr
&=&\bigoplus_{\unl\a=1}^{\frac{D_{d,1}}{2}}\,\corr{\unl Q{}_{\unl\a}}\oplus\bigoplus_{\unl a=0}^{4k+1}\,\corr{P_{\unl a}}\oplus\bigoplus_{\unl a,\unl b=0}^{4k+1}\,\corr{J_{\unl a\unl b}}\oplus\bigoplus_{\widehat a,\widehat b=4k+2}^d\,\corr{J_{\widehat a\widehat b}}\,.
\qqq 
\eeg

\beg\textbf{The vacuum superdistribution for the Green--Schwarz super-$(4k+2)$-brane in $\,{\rm sISO}(d,1\,|\,D_{d,1})/({\rm SO}(4k+2,1)\x{\rm SO}(d-4k-2))\,$ for $\,k\in\{0,1\}$.} 
The identity 
\qq\nn
\chi_{\a\b\unl a{}_1\unl a{}_2\ldots\unl a{}_{4k+2}}+\la_{4k+2}\,f_{\a\b}^{\ \ \ \unl b}\,\ep_{\unl b\unl a{}_1\unl a{}_2\ldots\unl a{}_{4k+2}}=4(4k+2)!\,\ep_{\unl b\unl a{}_1\unl a{}_2\ldots\unl a{}_{4k+2}}\,\ovl\G{}^{\unl b}_{\a\g}\,\left(\tfrac{\tfrac{\la_{4k+2}}{2(4k+2)!}\,\bd1_{D_{d,1}}-\G^0\,\G^1\,\cdots\,\G^{4k+2}}{2}\right)^\g_{\ \b}\,,
\qqq
yields -- for $\,\la_{4k+2}\in\{-2(4k+2)!\equiv\la_{4k+2}^-,2(4k+2)!\equiv\la_{4k+2}^+\}\,$ -- the possible projectors
\qq\nn
\sfP^{(1)}_{\la_{4k+2}^\pm}=\tfrac{\bd1_{D_{d,1}}\pm\G^0\,\G^1\,\cdots\,\G^{4k+2}}{2}
\qqq
with properties
\qq\nn
\G^{\unl a}\,\sfP^{(1)}_{\la_{4k+2}^\pm}=\sfP^{(1)}_{\la_{4k+2}^\pm}\,\G^{\unl a}\,,\qquad\qquad\G^{\widehat a}\,\sfP^{(1)}_{\la_{4k+2}^\pm}=\bigl(\bd1_{D_{d,1}}-\sfP^{(1)}_{\la_{4k+2}^\pm}\bigr)\,\G^{\widehat a}\,,\qquad\qquad C\,\sfP^{(1)}_{\la_{4k+2}^\pm}\,C^{-1}=\sfP^{(1)\,{\rm T}}_{\la_{4k+2}^\pm}\,,
\qqq
and so upon choosing
\qq\nn
\sfP^{(1)}\equiv\sfP^{(1)}_{\la_{4k+2}^+}=\tfrac{\bd1_{D_{d,1}}+\,\G^0\,\G^1\,\cdots\,\G^{4k+2}}{2}\,,
\qqq
we check
\bit
\item the $\k$-Symmetry Constraints,
\qq\nn
\{\sfP^{(1)\,\g}_{\ \ \ \ \ \a}\,Q_\g,\sfP^{(1)\,\d}_{\ \ \ \ \ \b}\,Q_\d\}=\sfP^{(1)\,\g}_{\ \ \ \ \ \b}\,\ovl\G{}_{\a\g}^{\unl a}\,P_{\unl a}\in\bigoplus_{\unl a=0}^{4k+2}\,\corr{P_{\unl a}}\oplus\bigoplus_{a,b=0}^d\,\corr{J_{ab}}\,;
\qqq
\item the Even Effective-Mixing Constraints, 
\qq\nn
&[J_{\unl a\unl b},P_{\unl c}]=\eta_{\unl b\unl c}\,P_{\unl a}-\eta_{\unl a\unl c}\,P_{\unl b}\in\bigoplus_{\unl d=0}^{4k+2}\,\corr{P_{\unl d}}\ni 0=[J_{\widehat a\widehat b},P_{\unl c}]\,,\qquad\qquad[J_{\unl a\widehat b},P_{\widehat c}]=\d_{\widehat b\widehat c}\,P_{\unl a}\in\bigoplus_{\unl d=0}^{4k+2}\,\corr{P_{\unl d}}\,,&\cr\cr
&[J_{\unl a\unl b},P_{\widehat c}]=0\in\bigoplus_{\widehat d=4k+3}^d\,\corr{P_{\widehat d}}\ni\d_{\widehat b\widehat c}\,P_{\widehat a}-\d_{\widehat a\widehat c}\,P_{\widehat b}=[J_{\widehat a\widehat b},P_{\widehat c}]\,,\qquad\qquad[J_{\unl a\widehat b},P_{\unl c}]=-\eta_{\unl a\unl c}\,P_{\widehat b}\in\bigoplus_{\widehat d=4k+3}^d\,\corr{P_{\widehat d}}\,;&
\qqq
\item the No-Curvature and No-Torsion Constraints -- trivial;
\item the $\G$-Constraints,  
\qq\nn
\bigl(C\,\G_{\widehat a}\,\G_{\unl a{}_1}\,\G_{\unl a{}_2}\cdots\G_{\unl a{}_{4k+1}}\bigr)_{\a\g}\,\sfP^{(1)\,\g}_{\ \ \ \ \ \b}&=&\bigl(C\,\bigl(\bd1_{D_{d,1}}-\sfP^{(1)}\bigr)\,\G_{\widehat a}\,\G_{\unl a{}_1}\,\G_{\unl a{}_2}\cdots\G_{\unl a{}_{4k+1}}\bigr)_{\a\b}\cr\cr
&=&\bigl(\bigl(\bd1_{D_{d,1}}-\sfP^{(1)}\bigr)^{\rm T}\,C\,\G_{\widehat a}\,\G_{\unl a{}_1}\,\G_{\unl a{}_2}\cdots\G_{\unl a{}_{4k+1}}\bigr)_{\a\b}\cr\cr
&\equiv&\bigl(C\,\G_{\widehat a}\,\G_{\unl a{}_1}\,\G_{\unl a{}_2}\cdots\G_{\unl a{}_{4k+1}}\bigr)_{\g\b}\,\bigl(\bd1_{D_{d,1}}-\sfP^{(1)}\bigr)^\g_{\ \a}\,; 
\qqq
\item the Vacuum-Superalgebra Constraints, with
\qq\nn
\{\sfP^{(1)\,\g}_{\ \ \ \ \ \a}\,Q_\g,\sfP^{(1)\,\d}_{\ \ \ \ \ \b}\,Q_\d\}\in\bigoplus_{\unl a=0}^{4k+2}\,\corr{P_{\unl a}}\oplus\bigoplus_{\unl a,\unl b=0}^{4k+2}\,\corr{J_{\unl a\unl b}}\oplus\bigoplus_{\widehat a,\widehat b=4k+3}^d\,\corr{J_{\widehat a\widehat b}}\,,
\qqq
and
\qq\nn
&[J_{\unl a\unl b},\sfP^{(1)\,\b}_{\ \ \ \ \ \a}\,Q_\b]=\tfrac{1}{2}\,\G_{\unl a\unl b\,\a}^\b\,\sfP^{(1)\,\g}_{\ \ \ \ \ \b}\,Q_\g\in{\rm Im}\,\sfP^{(1)}\,,&\cr\cr
&[J_{\widehat a\widehat b},\sfP^{(1)\,\b}_{\ \ \ \ \ \a}\,Q_\b]=\tfrac{1}{2}\,\G_{\widehat a\widehat b\,\a}^\b\,\sfP^{(1)\,\g}_{\ \ \ \ \ \b}\,Q_\g\in{\rm Im}\,\sfP^{(1)}&
\qqq
and
\qq\nn
[P_{\unl a},\sfP^{(1)\,\b}_{\ \ \ \ \ \a}\,Q_\b]=0\in{\rm Im}\,\sfP^{(1)}\,;
\qqq
\item $({\rm SO}(4k+2,1)\x{\rm SO}(d-4k-2))$-descendability.
\eit
Consequently, the HP vacuum superdistribution with restrictions
\qq\nn
{\rm Vac}\bigl({\rm sISO}(d,1\,|\,D_{d,1})/({\rm SO}(4k+2,1)\x{\rm SO}(d-4k-2)),\underset{\tx{\ciut{(p+2)}}}{\widehat\chi}\hspace{-6pt}{}^{\rm GS}\bigr)\rstr_{\cV_i}=\bigoplus_{\unl\a=1}^{\frac{D_{d,1}}{2}}\,\corr{\cT_{\unl\a\,i}}\oplus\bigoplus_{\unl a=0}^{4k+2}\,\corr{\cT_{\unl a\,i}}
\qqq 
is an $({\rm SO}(4k+2,1)\x{\rm SO}(d-4k-2))$-descendable integrable superdistribution associated with the Lie superalgebra 
\qq\nn
&&\gt{vac}\bigl({\rm sISO}(d,1\,|\,D_{d,1})/({\rm SO}(4k+2,1)\x{\rm SO}(d-4k-2)),\underset{\tx{\ciut{(p+2)}}}{\widehat\chi}\hspace{-6pt}{}^{\rm GS}\bigr)\cr\cr
&=&\bigoplus_{\unl\a=1}^{\frac{D_{d,1}}{2}}\,\corr{\unl Q{}_{\unl\a}}\oplus\bigoplus_{\unl a=0}^{4k+2}\,\corr{P_{\unl a}}\oplus\bigoplus_{\unl a,\unl b=0}^{4k+2}\,\corr{J_{\unl a\unl b}}\oplus\bigoplus_{\widehat a,\widehat b=4k+3}^d\,\corr{J_{\widehat a\widehat b}}\,.
\qqq 
\eeg

\beg\textbf{The vacuum superdistribution for the Zhou super-1-brane in \linebreak $\,{\rm SU}(1,1\,|\,2)_2/({\rm SO}(1,1)\x{\rm SO}(2))$.}
The identity 
\qq\nn
\chi_{\a'\a'' I\b'\b'' J a'}+\la_1\,f_{\a'\a'' I\b'\b'' J}^{\ \ \ \ \ \ \ \ \ \ \ \ \ b'}\,\ep_{b'a'}=4\bigl(\unl C\,\unl\g^{b'}\ox\bd1_2\bigr)_{\a'\a''I\g'\g''K}\,\ep_{b'a'}\,\left(\tfrac{\la_1\,\bd1_8+\unl\g^0\,\unl\g^1\ox\si_3}{2}\right)^{\g'\g''K}_{\ \ \b'\b''J}\,,
\qqq
yields -- for $\,\la_0\in\{-1,1\}\,$ -- the possible projectors
\qq\nn
\sfP^{(1)}_{\mp 1}=\tfrac{\bd1_8\pm\unl\g^0\,\unl\g^1\ox\si_3}{2}
\qqq
with properties
\qq\nn
&\bigl(\unl\g^{\unl a}\ox\bd1_2\bigr)\,\sfP^{(1)}_{\pm 1}=\bigl(\bd1_8-\sfP^{(1)}_{\pm 1}\bigr)\,\bigl(\unl\g^{\unl a}\ox\bd1_2\bigr)\,,\qquad\qquad\bigl(\unl\g^{\widehat a}\ox\bd1_2\bigr)\,\sfP^{(1)}_{\pm 1}=\sfP^{(1)}_{\pm 1}\,\bigl(\unl\g^{\widehat a}\ox\bd1_2\bigr)\,,&\cr\cr 
&\bigl(\unl C\ox\bd1_2\bigr)\,\sfP^{(1)}_{\pm 1}\,\bigl(\unl C\ox\bd1_2\bigr)^{-1}=\bigl(\bd1_8-\sfP^{(1)}_{\pm 1}\bigr)^{\rm T}\,,&
\qqq
and so upon choosing
\qq\nn
\sfP^{(1)}\equiv\sfP^{(1)}_{-1}=\tfrac{\bd1_8+\unl\g^0\,\unl\g^1\ox\si_3}{2}\,,
\qqq
we check
\bit
\item the $\k$-Symmetry Constraints, 
\qq\nn
&&\{\sfP^{(1)\,\g'\g''K}_{\ \ \ \ \ \a'\a''I}\,Q_{\g'\g''K},\sfP^{(1)\,\d\d'L}_{\ \ \ \ \ \b'\b''J}\,Q_{\d\d'L}\}\cr\cr
&=&2\sfP^{(1)\,\g'\g''K}_{\ \ \ \ \ \b'\b''J}\,\bigl(\bigl(\unl C\,\unl\g^{\unl a}\ox\bd1_2\bigr)_{\a'\a''I\g'\g''K}\,P_{\unl a}-\sfi\,\bigl(\unl C\ox\si_2\bigr)_{\a'\a''I\g'\g''K}\,J_{01}-\sfi\,\bigl(\unl C\,\unl\g{}_5\ox\si_2\bigr)_{\a'\a''I\g'\g''K}\,J_{23}\bigr)\cr\cr
&\in&\corr{P_0,P_1}\oplus\corr{J_{01},J_{23}}\,;
\qqq
\item the Even Effective-Mixing Constraints,  
\qq\nn
&[J_{01},P_{\unl a}]=\d_{\unl a 1}\,P_0-\eta_{\unl a 0}\,P_1\in\corr{P_0,P_1}\in 0=[J_{23},P_{\unl a}]\,,&\cr\cr
&[J_{01},P_{\widehat a}]=0\in\corr{P_2,P_3}\in\d_{\widehat a 3}\,P_2-\d_{\widehat a 2}\,P_3=[J_{23},P_{\widehat a}]\,;&
\qqq
\item the No-Curvature and No-Torsion Constraints, with
\qq\nn
[P_0,P_1]=J_{01}\in\corr{J_{01},J_{23}}\ni-J_{23}=[P_2,P_3]\,;
\qqq
\item the $\G$-Constraints,  
\qq\nn
\bigl(\unl C\,\unl\g{}_{\widehat a}\ox\si_3\bigr)_{\a'\a''I\g'\g''K}\,\sfP^{(1)\,\g'\g''K}_{\ \ \ \ \ \b'\b''J}&=&\bigl(\bigl(\unl C\ox\bd1_2\bigr)\,\sfP^{(1)}\,\unl\g{}_{\widehat a}\ox\si_3\bigr)_{\a'\a''I\b'\b''J}\cr\cr
&=&\bigl(\bigl(\bd1_8-\sfP^{(1)}\bigr)^{\rm T}\,\bigl(\unl C\,\unl\g{}_{\widehat a}\ox\si_3\bigr)\bigr)_{\a'\a''I\b'\b''J}\cr\cr
&\equiv&\bigl(\unl C\,\unl\g{}_{\widehat a}\ox\si_3\bigr)_{\g'\g''K\b'\b''J}\,\bigl(\bd1_8-\sfP^{(1)}\bigr)^{\g'\g''K}_{\ \a'\a''I}\,; 
\qqq
\item the Vacuum-Superalgebra Constraints, with
\qq\nn
\{\sfP^{(1)\,\g'\g''K}_{\ \ \ \ \ \a'\a''I}\,Q_{\g'\g''K},\sfP^{(1)\,\d'\d''L}_{\ \ \ \ \ \b'\b''J}\,Q_{\d'\d''L}\}\in\corr{P_0,P_1}\oplus\corr{J_{01},J_{23}}
\qqq
and
\qq\nn
[J_{01},\sfP^{(1)\,\b'\b''J}_{\ \ \ \ \ \a'\a''I}\,Q_{\b'\b''J}]&=&\tfrac{1}{2}\,\bigl(\unl\g{}_0\unl\g{}_1\ox\bd1_2\bigr)^{\b'\b''J}_{\ \a'\a''I}\,\sfP^{(1)\,\g'\g''K}_{\ \ \ \ \ \b'\b''J}\,Q_{\g'\g''K}\in{\rm Im}\,\sfP^{(1)}\,,\cr\cr\cr
[J_{23},\sfP^{(1)\,\b'\b''J}_{\ \ \ \ \ \a'\a''I}\,Q_{\b'\b''J}]&=&\tfrac{1}{2}\,\bigl(\unl\g{}_2\unl\g{}_3\ox\bd1_2\bigr)^{\b'\b''J}_{\ \a'\a''I}\,\sfP^{(1)\,\g'\g''K}_{\ \ \ \ \ \b'\b''J}\,Q_{\g'\g''K}\in{\rm Im}\,\sfP^{(1)}\,,
\qqq
and
\qq\nn
[P_{\unl a},\sfP^{(1)\,\b'\b''J}_{\ \ \ \ \ \a'\a''I}\,Q_{\b'\b''J}]=\tfrac{\sfi}{2}\,\bigl(\unl\g{}_0\,\unl\g{}_1\,\unl\g{}_{\unl a}\ox\si_2\bigr)^{\g'\g''K}_{\ \a'\a''I}\,\sfP^{(1)\,\b'\b''J}_{\ \ \ \ \ \g'\g''K}\,Q_{\b'\b''J}\in{\rm Im}\,\sfP^{(1)}\,;
\qqq
\item $({\rm SO}(1,1)\x{\rm SO}(2))$-descendability.
\eit
Consequently, the HP vacuum superdistribution with restrictions
\qq\nn
{\rm Vac}\bigl({\rm SU}(1,1\,|\,2)_2/({\rm SO}(1,1)\x{\rm SO}(2)),\underset{\tx{\ciut{(3)}}}{\widehat\chi}\hspace{-1pt}{}^{\rm Zh}_{(1)}\bigr)\rstr_{\cV_i}=\bigoplus_{\unl\a=1}^4\,\corr{\cT_{\unl\a\,i}}\oplus\corr{\cT_{0\,i},\cT_{1\,i}}
\qqq 
is an $({\rm SO}(1,1)\x{\rm SO}(2))$-descendable integrable superdistribution associated with the Lie superalgebra
\qq\nn
\gt{vac}\bigl({\rm SU}(1,1\,|\,2)_2/({\rm SO}(1,1)\x{\rm SO}(2)),\underset{\tx{\ciut{(3)}}}{\widehat\chi}\hspace{-1pt}{}^{\rm Zh}_{(1)}\bigr)=\bigoplus_{\unl\a=1}^4\,\corr{\unl Q{}_{\unl\a}}\oplus\corr{P_0,P_1}\oplus\corr{J_{01},J_{23}}\,.
\qqq
\eeg

\beg\textbf{The vacuum superdistribution for the Zhou super-1-brane in $\,{\rm SU}(1,1\,|\,2)_2$.}
The identity 
\qq\nn
\chi_{\a'\a'' I\b'\b'' J\unl a}+\la_1\,f_{\a'\a'' I\b'\b'' J}^{\ \ \ \ \ \ \ \ \ \ \ \ \ \unl b}\,\ep_{\unl b\unl a}&=&2\bigl(\unl C\,\unl\g{}_{\unl a}\ox\si_3\bigr)_{\a'\a'' I\b'\b'' J}+2\la_1\,\bigl(\unl C\,\unl\G^{\unl b}\ox\bd1_2\bigr)_{\a'\a'' I\b'\b'' J}\,\ep_{\unl b\unl a}\cr\cr
&\equiv&4\bigl(\unl C\,\unl\g^{\unl b}\ox\bd1_2\bigr)_{\a'\a''I\g'\g''K}\,\ep_{\unl b\unl a}\,\left(\tfrac{\la_1\,\bd1_8+\unl\g^0\,\unl\g^2\ox\si_3}{2}\right)^{\g'\g''K}_{\ \ \b'\b''J}\,,
\qqq
yields -- for $\,\la_0\in\{-1,1\}\,$ -- the possible projectors
\qq\nn
\sfP^{(1)}_{\mp 1}=\tfrac{\bd1_8\pm\unl\g^0\,\unl\g^2\ox\si_3}{2}
\qqq
with properties
\qq\nn
&\bigl(\unl\g^{\unl a}\ox\bd1_2\bigr)\,\sfP^{(1)}_{\pm 1}=\bigl(\bd1_8-\sfP^{(1)}_{\pm 1}\bigr)\,\bigl(\unl\g^{\unl a}\ox\bd1_2\bigr)\,,\qquad\qquad\bigl(\unl\g^{\widehat a}\ox\bd1_2\bigr)\,\sfP^{(1)}_{\pm 1}=\sfP^{(1)}_{\pm 1}\,\bigl(\unl\g^{\widehat a}\ox\bd1_2\bigr)\,,&\cr\cr 
&\bigl(\unl C\ox\bd1_2\bigr)\,\sfP^{(1)}_{\pm 1}\,\bigl(\unl C\ox\bd1_2\bigr)^{-1}=\bigl(\bd1_8-\sfP^{(1)}_{\pm 1}\bigr)^{\rm T}\,,&
\qqq
and so upon choosing
\qq\nn
\sfP^{(1)}\equiv\sfP^{(1)}_{-1}=\tfrac{\bd1_8+\unl\g^0\,\unl\g^2\ox\si_3}{2}\,,
\qqq
we check
\bit
\item the $\k$-Symmetry Constraints,
\qq\nn
\{\sfP^{(1)\,\g'\g''K}_{\ \ \ \ \ \a'\a''I}\,Q_{\g'\g''K},\sfP^{(1)\,\d\d'L}_{\ \ \ \ \ \b'\b''J}\,Q_{\d\d'L}\}&=&2\sfP^{(1)\,\g'\g''K}_{\ \ \ \ \ \b'\b''J}\,\bigl(\bigl(\unl C\,\unl\g^{\unl a}\ox\bd1_2\bigr)_{\a'\a''I\g'\g''K}\,P_{\unl a}\cr\cr
&&-\sfi\,\bigl(\unl C\ox\si_2\bigr)_{\a'\a''I\g'\g''K}\,J_{01}-\sfi\,\bigl(\unl C\,\unl\g{}_5\ox\si_2\bigr)_{\a'\a''I\g'\g''K}\,J_{23}\bigr)\cr\cr
&\in&\corr{P_0,P_2}\oplus\corr{J_{01},J_{23}}\,;
\qqq
\item the Even Effective-Mixing Constraints 
\qq\nn
&[J_{01},P_0]=P_1\in\corr{P_1,P_3}\ni 0=[J_{23},P_0]\,,\qquad\qquad[J_{01},P_2]=0\in\corr{P_1,P_3}\ni -P_3=[J_{23},P_2]\,,&\cr\cr
&[J_{01},P_1]=P_0\in\corr{P_0,P_2}\ni 0=[J_{23},P_1]\,,\qquad\qquad[J_{01},P_3]=0\in\corr{P_0,P_2}\ni P_2=[J_{23},P_3]\,;&
\qqq
\item the No-Curvature and No-Torsion Constraints, with
\qq\nn
[P_0,P_2]=0\in\brd0\ni[P_1,P_3]\,;
\qqq
\item the $\G$-Constraints,  
\qq\nn
\bigl(\unl C\,\unl\g{}_{\widehat a}\ox\si_3\bigr)_{\a'\a''I\g'\g''K}\,\sfP^{(1)\,\g'\g''K}_{\ \ \ \ \ \b'\b''J}&=&\bigl(\bigl(\unl C\ox\bd1_2\bigr)\,\sfP^{(1)}\,\unl\g{}_{\widehat a}\ox\si_3\bigr)_{\a'\a''I\b'\b''J}\cr\cr
&=&\bigl(\bigl(\bd1_8-\sfP^{(1)}\bigr)^{\rm T}\,\bigl(\unl C\,\unl\g{}_{\widehat a}\ox\si_3\bigr)\bigr)_{\a'\a''I\b'\b''J}\cr\cr
&\equiv&\bigl(\unl C\,\unl\g{}_{\widehat a}\ox\si_3\bigr)_{\g'\g''K\b'\b''J}\,\bigl(\bd1_8-\sfP^{(1)}\bigr)^{\g'\g''K}_{\ \a'\a''I}\,.
\qqq
\eit
The Vacuum-Superalgebra Constraints, on the other hand, are \emph{not} satisfied in view of the above, 
\qq\nn
\{\sfP^{(1)\,\g'\g''K}_{\ \ \ \ \ \a'\a''I}\,Q_{\g'\g''K},\sfP^{(1)\,\d'\d''L}_{\ \ \ \ \ \b'\b''J}\,Q_{\d'\d''L}\}\notin\corr{P_0,P_2}\,,
\qqq
with 
\qq\nn
[P_{\unl a},\sfP^{(1)\,\b'\b''J}_{\ \ \ \ \ \a'\a''I}\,Q_{\b'\b''J}]=\tfrac{\sfi}{2}\,\bigl(\unl\g{}_0\,\unl\g{}_1\,\unl\g{}_{\unl a}\ox\si_2\bigr)^{\g'\g''K}_{\ \a'\a''I}\,\bigl(\bd1_8-\sfP^{(1)}\bigr)^{\b'\b''J}_{\ \g'\g''K}\,Q_{\b'\b''J}\in{\rm Im}\,\bigl(\bd1_8-\sfP^{(1)}\bigr)\,,
\qqq
and so the HP vacuum superdistribution with restrictions
\qq\nn
{\rm Vac}\bigl({\rm SU}(1,1\,|\,2)_2,\underset{\tx{\ciut{(3)}}}{\widehat\chi}\hspace{-1pt}{}^{\rm Zh}_{(12)}\bigr)\rstr_{\cV_i}=\bigoplus_{\unl\a=1}^4\,\corr{\cT_{\unl\a\,i}}\oplus\corr{\cT_{0\,i},\cT_{2\,i}}
\qqq
is nonintegrable. Its weak derived flag with (local) components of the symbol 
\qq\nn
&\gtq^{-1}_{(i)}\equiv{\rm Vac}\bigl({\rm SU}(1,1\,|\,2)_2,\underset{\tx{\ciut{(3)}}}{\widehat\chi}\hspace{-1pt}{}^{\rm Zh}_{(12)}\bigr)\rstr_{\cV_i}\,,&\cr\cr
&\gtq^{-2}_{(i)}=\bigoplus_{\widehat\a=5}^8\,\corr{\cT_{\widehat\a\,i}}\oplus\corr{\cT_{01\,i},\cT_{23\,i}}\,,\qquad\qquad\gtq^{-3}_{(i)}=\corr{\cT_{1\,i},\cT_{3\,i}}&
\qqq
is bracket-generating for the tangent sheaf of the HP section,
\qq\nn
{\rm Vac}^{-\infty}\bigl({\rm SU}(1,1\,|\,2)_2,\underset{\tx{\ciut{(3)}}}{\widehat\chi}\hspace{-1pt}{}^{\rm Zh}_{(12)}\bigr)=\cT\Si^{\rm HP}\,.
\qqq
\eeg

\beg\textbf{The vacuum superdistribution for the Park--Rey super-1-brane in $\,({\rm SU}(1,1\,|\,2)\x{\rm SU}(1,1\,|\,2))_2/({\rm SO}(1,1)\x{\rm SO}(3))$.}
The identity 
\qq\nn
&&\chi_{\a'\a''\a'''I\b'\b''\b'''J a'}+\la_1\,f_{\a'\a''\a'''I\b'\b''\b'''J}^{\ \ \ \ \ \ \ \ \ \ \ \ \ \ \ \ \ \ \ b'}\,\ep_{b'a'}\cr\cr
&=&4\bigl(\unl C\,\unl\g^{b'}\cdot\bigl(\bd1_4\ox\si_2\bigr)\ox\bd1_2\bigr)_{\a'\a''\a'''I\g'\g''\g'''K}\,\ep_{b'a'}\,\left(\tfrac{\la_1\,\bd1_{16}-\unl\g^0\,\unl\g^1\ox\si_1}{2}\right)^{\g'\g''\g'''K}_{\ \ \b'\b''\b'''J}\,,
\qqq
yields -- for $\,\la_0\in\{-1,1\}\,$ -- the possible projectors
\qq\nn
\sfP^{(1)}_{\pm 1}=\tfrac{\bd1_{16}\pm\unl\g^0\,\unl\g^1\ox\si_1}{2}
\qqq
with properties
\qq\nn
&\bigl(\unl\g^{\unl a}\ox\bd1_2\bigr)\,\sfP^{(1)}_{\pm 1}=\bigl(\bd1_{16}-\sfP^{(1)}_{\pm 1}\bigr)\,\bigl(\unl\g^{\unl a}\ox\bd1_2\bigr)\,,\qquad\qquad\bigl(\unl\g^{\widehat a}\ox\bd1_2\bigr)\,\sfP^{(1)}_{\pm 1}=\sfP^{(1)}_{\pm 1}\,\bigl(\unl\g^{\widehat a}\ox\bd1_2\bigr)\,,&\cr\cr 
&\bigl(\unl C\ox\bd1_2\bigr)\,\sfP^{(1)}_{\pm 1}\,\bigl(\unl C\ox\bd1_2\bigr)^{-1}=\bigl(\bd1_{16}-\sfP^{(1)}_{\pm 1}\bigr)^{\rm T}\,,&
\qqq
and so upon choosing
\qq\nn
\sfP^{(1)}\equiv\sfP^{(1)}_{+1}=\tfrac{\bd1_{16}+\unl\g^0\,\unl\g^1\ox\si_1}{2}\,,
\qqq
we check
\bit
\item the $\k$-Symmetry Constraints,
\qq\nn
\{\sfP^{(1)\,\g'\g''\g'''K}_{\ \ \ \ \ \a'\a''\a'''I}\,Q_{\g'\g''\g'''K},\sfP^{(1)\,\d\d'\d'''L}_{\ \ \ \ \ \b'\b''\b'''J}\,Q_{\d\d'\d'''L}\}&=&\sfP^{(1)\,\g'\g''\g'''K}_{\ \ \ \ \ \b'\b''\b'''J}\,\bigl(2\bigl(\unl C\,\unl\g^{\unl a}\cdot\bigl(\bd1_4\ox\si_2\bigr)\ox\bd1_2\bigr)_{\a'\a''\a'''I\g'\g''\g'''K}\,P_{\unl a}\cr\cr
&&+2\sfi\,\bigl(\unl C\,\unl\g^{01}\unl\g{}_7\ox\si_3\bigr)_{\a'\a''\a'''I\g'\g''\g'''K}\,J_{01}\cr\cr
&&-\sfi\,\bigl(\unl C\,\unl\g^{a''b''}\unl\g{}_7\ox\si_3\bigr)_{\a'\a''\a'''I\g'\g''\g'''K}\,J_{a''b''}\bigr)\cr\cr
&\in&\corr{P_0,P_1}\oplus\bigoplus_{a',b'=0}^2\,\corr{J_{a'b'}}\oplus\bigoplus_{a'',b''=3}^5\,\corr{J_{a''b''}}\,;
\qqq
\item the Even Effective-Mixing Constraints, 
\qq\nn
&[J_{01},P_{\unl a}]=\d_{\unl a 1}\,P_0-\eta_{\unl a 0}\,P_1\in\corr{P_0,P_1}\ni 0=[J_{a''b''},P_{\unl c}]\,,\qquad\qquad[J_{\unl a 2},P_{\widehat b}]=\d_{\widehat b 2}\,P_{\unl a}\in\corr{P_0,P_1}\ni\d_{\widehat b\widehat c}\,P_{\unl a} =[J_{\unl a\widehat b},P_{\widehat c}]\,,&\cr\cr
&[J_{01},P_{\widehat a}]=0\in\bigoplus_{\widehat d=2}^5\,\corr{P_{\widehat d}}\ni\d_{b''\widehat c}\,P_{a''}-\d_{a''\widehat c}\,P_{b''}=[J_{a''b''},P_{\widehat c}]\,,&\cr\cr
&[J_{\unl a 2},P_{\unl b}]=-\d_{\unl a\unl b}\,P_2\in\bigoplus_{\widehat d=2}^5\,\corr{P_{\widehat d}}\ni-\d_{\unl a\unl c}\,P_{\widehat b} =[J_{\unl a\widehat b},P_{\unl c}]\,;&
\qqq
\item the No-Curvature and No-Torsion Constraints, with
\qq\nn
[P_0,P_1]=J_{01}\in\corr{J_{01}}\oplus\bigoplus_{b'',c''=3}^5\,\corr{J_{b''c''}}\ni 0=[P_2,P_{a''}]\,,\qquad\qquad[P_{a''},P_{b''}]=-J_{a''b''}\in\corr{J_{01}}\oplus\bigoplus_{c'',d''=3}^5\,\corr{J_{c''d''}}\,;
\qqq
\item the $\G$-Constraints,  
\qq\nn
&&\bigl(\unl C\,\unl\g{}_2\cdot\bigl(\bd1_4\ox\si_2\bigr)\ox\si_1\bigr)_{\a'\a''\a'''I\g'\g''\g'''K}\,\sfP^{(1)\,\g'\g''\g'''K}_{\ \ \ \ \ \b'\b''\b'''J}\cr\cr
&=&\bigl(\bigl(\unl C\ox\bd1_2\bigr)\,\sfP^{(1)}\,\bigl(\unl\g{}_2\cdot\bigl(\bd1_4\ox\si_2\bigr)\ox\si_1\bigr)\bigr)_{\a'\a''\a'''I\b'\b''\b'''J}\cr\cr
&=&\bigl(\bigl(\bd1_{16}-\sfP^{(1)}\bigr)^{\rm T}\,\bigl(\unl C\,\unl\g{}_2\cdot\bigl(\bd1_4\ox\si_2\bigr)\ox\si_1\bigr)\bigr)_{\a'\a''\a'''I\b'\b''\b'''J}\cr\cr
&\equiv&\bigl(\unl C\,\unl\g{}_2\cdot\bigl(\bd1_4\ox\si_2\bigr)\ox\si_1\bigr)_{\g'\g''\g'''K\b'\b''\b'''J}\,\bigl(\bd1_{16}-\sfP^{(1)}\bigr)^{\g'\g''\g'''K}_{\ \a'\a''\a'''I}\,,\cr\cr\cr
&&-\bigl(\unl C\,\unl\g{}_{a''}\,\unl\g{}_7\cdot\bigl(\bd1_4\ox\si_2\bigr)\ox\si_1\bigr)_{\a'\a''\a'''I\g'\g''\g'''K}\,\sfP^{(1)\,\g'\g''\g'''K}_{\ \ \ \ \ \b'\b''\b'''J}\cr\cr
&=&-\bigl(\bigl(\unl C\ox\bd1_2\bigr)\,\sfP^{(1)}\,\bigl(\unl\g{}_{a''}\,\unl\g{}_7\cdot\bigl(\bd1_4\ox\si_2\bigr)\ox\si_1\bigr)\bigr)_{\a'\a''\a'''I\b'\b''\b'''J}\cr\cr
&=&-\bigl(\bigl(\bd1_{16}-\sfP^{(1)}\bigr)^{\rm T}\,\bigl(\unl C\,\unl\g{}_{a''}\,\unl\g{}_7\cdot\bigl(\bd1_4\ox\si_2\bigr)\ox\si_1\bigr)\bigr)_{\a'\a''\a'''I\b'\b''\b'''J}\cr\cr
&\equiv&-\bigl(\unl C\,\unl\g{}_{a''}\,\unl\g{}_7\cdot\bigl(\bd1_4\ox\si_2\bigr)\ox\si_1\bigr)_{\g'\g''\g'''K\b'\b''\b'''J}\,\bigl(\bd1_{16}-\sfP^{(1)}\bigr)^{\g'\g''\g'''K}_{\ \a'\a''\a'''I}\,;
\qqq
\item the Vacuum-Superalgebra Constraints, with 
\qq\nn
\{\sfP^{(1)\,\g'\g''\g'''K}_{\ \ \ \ \ \a'\a''\a'''I}\,Q_{\g'\g''\g'''K},\sfP^{(1)\,\d\d'\d'''L}_{\ \ \ \ \ \b'\b''\b'''J}\,Q_{\d\d'\d'''L}\}\in\corr{P_0,P_1}\oplus\corr{J_{01}}\oplus\bigoplus_{a'',b''=3}^5\,\corr{J_{a''b''}}
\qqq
and 
\qq\nn
[J_{01},\sfP^{(1)\,\b'\b''\b'''J}_{\ \ \ \ \ \a'\a''\a'''I}\,Q_{\b'\b''\b'''J}]&=&\tfrac{1}{2}\,\bigl(\unl\g{}_0\unl\g{}_1\ox\bd1_2\bigr)^{\b'\b''\b'''J}_{\ \a'\a''\a'''I}\,\sfP^{(1)\,\g'\g''\g'''K}_{\ \ \ \ \ \b'\b''\b'''J}\,Q_{\g'\g''\g'''K}\in{\rm Im}\,\sfP^{(1)}\,,\cr\cr
[J_{a''b''},\sfP^{(1)\,\b'\b''\b'''J}_{\ \ \ \ \ \a'\a''\a'''I}\,Q_{\b'\b''\b'''J}]&=&\tfrac{1}{2}\,\bigl(\unl\g{}_{a''}\unl\g{}_{b''}\ox\bd1_2\bigr)^{\b'\b''\b'''J}_{\ \a'\a''\a'''I}\,\sfP^{(1)\,\g'\g''\g'''K}_{\ \ \ \ \ \b'\b''\b'''J}\,Q_{\g'\g''\g'''K}\in{\rm Im}\,\sfP^{(1)}\,,
\qqq
and 
\qq\nn
[P_{\unl a},\sfP^{(1)\,\b'\b''\b'''J}_{\ \ \ \ \ \a'\a''\a'''I}\,Q_{\b'\b''\b'''J}]=\tfrac{\sfi}{2}\,\bigl(\unl\g{}_{\unl a}\,\unl\g{}_7\cdot\bigl(\bd1_4\ox\si_2\bigr)\ox\si_3\bigr)^{\g'\g''\g'''K}_{\ \a'\a''\a'''I}\,\sfP^{(1)\,\b'\b''\b'''J}_{\ \ \ \ \ \g'\g''\g'''K}\,Q_{\b'\b''\b'''J}\in{\rm Im}\,\sfP^{(1)}\,;
\qqq
\item $({\rm SO}(1,1)\x{\rm SO}(3))$-descendability.
\eit
Consequently, the HP vacuum superdistribution with restrictions
\qq\nn
{\rm Vac}\bigl(\bigl({\rm SU}(1,1\,|\,2)\x{\rm SU}(1,1\,|\,2)\bigr)_2/({\rm SO}(1,1)\x{\rm SO}(3)),\underset{\tx{\ciut{(3)}}}{\widehat\chi}\hspace{-1pt}{}^{\rm PR}_{(1)}\bigr)\rstr_{\cV_i}=\bigoplus_{\unl\a=1}^8\,\corr{\cT_{\unl\a\,i}}\oplus\corr{\cT_{0\,i},\cT_{1\,i}}
\qqq 
is an $({\rm SO}(1,1)\x{\rm SO}(3))$-descendable integrable superdistribution associated with the Lie superalgebra
\qq\nn
&&\gt{vac}\bigl(\bigl({\rm SU}(1,1\,|\,2)\x{\rm SU}(1,1\,|\,2)\bigr)_2/({\rm SO}(1,1)\x{\rm SO}(3)),\underset{\tx{\ciut{(3)}}}{\widehat\chi}\hspace{-1pt}{}^{\rm PR}_{(1)}\bigr)\cr\cr
&=&\bigoplus_{\unl\a=1}^8\,\corr{\unl Q{}_{\unl\a}}\oplus\corr{P_0,P_1}\oplus\corr{J_{01}}\oplus\bigoplus_{a'',b''=3}^5\,\corr{J_{a''b''}}\,.
\qqq
\eeg

\beg\textbf{The vacuum superdistribution for the Park--Rey super-1-brane in $\,({\rm SU}(1,1\,|\,2)\x{\rm SU}(1,1\,|\,2))_2/({\rm SO}(2)\x{\rm SO}(2))$.}
The identities
\qq\nn
&&\chi_{\a'\a''\a'''I\b'\b''\b'''J 0}-\la_1\,f_{\a'\a''\a'''I\b'\b''\b'''J}^{\ \ \ \ \ \ \ \ \ \ \ \ \ \ \ \ \ \ \ 3}\cr\cr
&=&4\bigl(\unl C\,\unl\g^3\,\unl\g{}_7\cdot\bigl(\bd1_4\ox\si_2\bigr)\ox\bd1_2\bigr)_{\a'\a''\a'''I\g'\g''\g'''K}\,\left(\tfrac{\la_1\,\bd1_{16}+\unl\g^0\,\unl\g^3\,\unl\g{}_7\ox\si_1}{2}\right)^{\g'\g''\g'''K}_{\ \ \b'\b''\b'''J}\,,\cr\cr\cr
&&\chi_{\a'\a''\a'''I\b'\b''\b'''J 3}+\la_1\,f_{\a'\a''\a'''I\b'\b''\b'''J}^{\ \ \ \ \ \ \ \ \ \ \ \ \ \ \ \ \ \ \ 0}\cr\cr
&=&4\bigl(\unl C\,\unl\g^0\cdot\bigl(\bd1_4\ox\si_2\bigr)\ox\bd1_2\bigr)_{\a'\a''\a'''I\g'\g''\g'''K}\,\left(\tfrac{\la_1\,\bd1_{16}+\unl\g^0\,\unl\g^3\,\unl\g{}_7\ox\si_1}{2}\right)^{\g'\g''\g'''K}_{\ \ \b'\b''\b'''J}\,,
\qqq
yield -- for $\,\la_0\in\{-1,1\}\,$ -- the possible projectors
\qq\nn
\sfP^{(1)}_{\mp 1}=\tfrac{\bd1_{16}\pm\unl\g^0\,\unl\g^3\,\unl\g{}_7\ox\si_1}{2}
\qqq
with properties
\qq\nn
&\bigl(\unl\g^{\unl a}\ox\bd1_2\bigr)\,\sfP^{(1)}_{\pm 1}=\sfP^{(1)}_{\pm 1}\,\bigl(\unl\g^{\unl a}\ox\bd1_2\bigr)\,,\qquad\qquad\bigl(\unl\g^{\widehat a}\ox\bd1_2\bigr)\,\sfP^{(1)}_{\pm 1}=\bigl(\bd1_{16}-\sfP^{(1)}_{\pm 1}\bigr)\,\bigl(\unl\g^{\widehat a}\ox\bd1_2\bigr)\,,&\cr\cr 
&\bigl(\unl C\ox\bd1_2\bigr)\,\sfP^{(1)}_{\pm 1}\,\bigl(\unl C\ox\bd1_2\bigr)^{-1}=\sfP^{(1)\,{\rm T}}_{\pm 1}\,,&
\qqq
and so upon choosing
\qq\nn
\sfP^{(1)}\equiv\sfP^{(1)}_{+1}=\tfrac{\bd1_{16}+\unl\g^0\,\unl\g^3\,\unl\g{}_7\ox\si_1}{2}\,,
\qqq
we check
\bit
\item the $\k$-Symmetry Constraints, 
\qq\nn
&&\{\sfP^{(1)\,\g'\g''\g'''K}_{\ \ \ \ \ \a'\a''\a'''I}\,Q_{\g'\g''\g'''K},\sfP^{(1)\,\d'\d''\d'''L}_{\ \ \ \ \ \b'\b''\b'''J}\,Q_{\d'\d''\d'''L}\}\cr\cr
&=&2\sfP^{(1)\,\g'\g''\g'''K}_{\ \ \ \ \ \b'\b''\b'''J}\,\bigl(\bigl(\unl C\,\unl\g^0\cdot\bigl(\bd1_4\ox\si_2\bigr)\ox\bd1_2\bigr)_{\a'\a''\a'''I\g'\g''\g'''K}\,P_0-\bigl(\unl C\,\unl\g^3\unl\g{}_7\cdot\bigl(\bd1_4\ox\si_2\bigr)\ox\bd1_2\bigr)_{\a'\a''\a'''I\g'\g''\g'''K}\,P_3\cr\cr
&&-\sfi\,\bigl(\unl C\,\unl\g^{0a'}\unl\g{}_7\ox\si_3\bigr)_{\a'\a''\a'''I\g'\g''\g'''K}\,J_{0a'}+\sfi\,\bigl(\unl C\,\unl\g^{3a''}\unl\g{}_7\ox\si_3\bigr)_{\a'\a''\a'''I\g'\g''\g'''K}\,J_{3a''}\bigr)\cr\cr
&\in&\corr{P_0,P_3}\oplus\bigoplus_{a',b'=0}^{2}\,\corr{J_{a',b'}}\oplus\bigoplus_{a'',b''=3}^5\,\corr{J_{a''b''}}\,;
\qqq
\item the Even Effective-Mixing Constraints, 
\qq\nn
&[J_{12},P_{\unl a}]=0\in\corr{P_0,P_3}\in 0=[J_{45},P_{\unl a}]\,,\qquad\qquad[J_{0a'},P_{\widehat b}]=\d_{a'\widehat b}\,P_0\in\corr{P_0,P_3}\ni\d_{a''\widehat b}\,P_3=[J_{3a''},P_{\widehat b}]\,,&\cr\cr
&[J_{12},P_{\widehat a}]=\d_{\widehat a 2}\,P_1-\d_{\widehat a 1}\,P_2\in\corr{P_1,P_2,P_4,P_5}\ni\d_{\widehat a 5}\,P_4-\d_{\widehat a 4}\,P_5=[J_{45},P_{\widehat a}]\,,&\cr\cr
&[J_{0a'},P_{\unl b}]=-\eta_{\unl b 0}\,P_{a'}\in\corr{P_1,P_2,P_4,P_5}\ni-\d_{\unl b 3}\,P_{a''}=[J_{3a''},P_{\unl b}]\,;&
\qqq
\item the No-Curvature and No-Torsion Constraints, with
\qq\nn
&[P_0,P_3]=0\in\corr{J_{12},J_{45}}\ni J_{12}=[P_1,P_2]\,,&\cr\cr
&[P_1,P_4]=[P_1,P_5]=[P_2,P_4]=[P_2,P_5]=0\in\corr{J_{12},J_{45}}\ni-J_{45}=[P_4,P_5]\,;&
\qqq
\item the $\G$-Constraints,  
\qq\nn
&&\bigl(\unl C\,\unl\g{}_{\widehat a'}\cdot\bigl(\bd1_4\ox\si_2\bigr)\ox\si_1\bigr)_{\a'\a''\a'''I\g'\g''\g'''K}\,\sfP^{(1)\,\g'\g''\g'''K}_{\ \ \ \ \ \b'\b''\b'''J}\cr\cr
&=&\bigl(\bigl(\unl C\ox\bd1_2\bigr)\,\bigl(\bd1_{16}-\sfP^{(1)}\bigr)\,\bigl(\unl\g{}_{\widehat a'}\cdot\bigl(\bd1_4\ox\si_2\bigr)\ox\si_1\bigr)\bigr)_{\a'\a''\a'''I\b'\b''\b'''J}\cr\cr
&=&\bigl(\bigl(\bd1_{16}-\sfP^{(1)}\bigr)^{\rm T}\,\bigl(\unl C\,\unl\g{}_{\widehat a'}\cdot\bigl(\bd1_4\ox\si_2\bigr)\ox\si_1\bigr)\bigr)_{\a'\a''\a'''I\b'\b''\b'''J}\cr\cr
&\equiv&\bigl(\unl C\,\unl\g{}_{\widehat a'}\cdot\bigl(\bd1_4\ox\si_2\bigr)\ox\si_1\bigr)_{\g'\g''\g'''K\b'\b''\b'''J}\,\bigl(\bd1_{16}-\sfP^{(1)}\bigr)^{\g'\g''\g'''K}_{\ \a'\a''\a'''I}\,,\qquad\widehat a'\in\{1,2\}\,,\cr\cr\cr
&&-\bigl(\unl C\,\unl\g{}_{\widehat a''}\,\unl\g{}_7\cdot\bigl(\bd1_4\ox\si_2\bigr)\ox\si_1\bigr)_{\a'\a''\a'''I\g'\g''\g'''K}\,\sfP^{(1)\,\g'\g''\g'''K}_{\ \ \ \ \ \b'\b''\b'''J}\cr\cr
&=&-\bigl(\bigl(\unl C\ox\bd1_2\bigr)\,\bigl(\bd1_{16}-\sfP^{(1)}\bigr)\,\bigl(\unl\g{}_{\widehat a''}\,\unl\g{}_7\cdot\bigl(\bd1_4\ox\si_2\bigr)\ox\si_1\bigr)\bigr)_{\a'\a''\a'''I\b'\b''\b'''J}\cr\cr
&=&-\bigl(\bigl(\bd1_{16}-\sfP^{(1)}\bigr)^{\rm T}\,\bigl(\unl C\,\unl\g{}_{\widehat a''}\,\unl\g{}_7\cdot\bigl(\bd1_4\ox\si_2\bigr)\ox\si_1\bigr)\bigr)_{\a'\a''\a'''I\b'\b''\b'''J}\cr\cr
&\equiv&-\bigl(\unl C\,\unl\g{}_{\widehat a''}\,\unl\g{}_7\cdot\bigl(\bd1_4\ox\si_2\bigr)\ox\si_1\bigr)_{\g'\g''\g'''K\b'\b''\b'''J}\,\bigl(\bd1_{16}-\sfP^{(1)}\bigr)^{\g'\g''\g'''K}_{\ \a'\a''\a'''I}\,,\qquad\widehat a''\in\{4,5\}\,;
\qqq
\item[$\bullet$] $({\rm SO}(2)\x{\rm SO}(2))$-descendability.
\eit
The Vacuum-Superalgebra Constraints, on the other hand, are \emph{not} satisfied in view of the above, 
\qq\nn
\{\sfP^{(1)\,\g'\g''\g'''K}_{\ \ \ \ \ \a'\a''\a'''I}\,Q_{\g'\g''\g'''K},\sfP^{(1)\,\d'\d''\d'''L}_{\ \ \ \ \ \b'\b''\b'''J}\,Q_{\d'\d''\d'''L}\}\notin\corr{P_0,P_3}\oplus\corr{J_{12},J_{45}}\,,
\qqq
with 
\qq\nn
[J_{12},\sfP^{(1)\,\b'\b''\b'''J}_{\ \ \ \ \ \a'\a''I}\,Q_{\b'\b''J}]&=&\tfrac{1}{2}\,\bigl(\unl\g{}_1\,\unl\g{}_2\ox\bd1_2\bigr)^{\g'\g''\g'''K}_{\ \a'\a''\a'''I}\,\sfP^{(1)\,\b'\b''\b'''J}_{\ \ \ \ \ \g'\g''\g'''K}\,Q_{\b'\b''\b'''J}\in{\rm Im}\,\sfP^{(1)}\,,\cr\cr
[J_{45},\sfP^{(1)\,\b'\b''\b'''J}_{\ \ \ \ \ \a'\a''I}\,Q_{\b'\b''J}]&=&\tfrac{1}{2}\,\bigl(\unl\g{}_4\,\unl\g{}_5\ox\bd1_2\bigr)^{\g'\g''\g'''K}_{\ \a'\a''\a'''I}\,\sfP^{(1)\,\b'\b''\b'''J}_{\ \ \ \ \ \g'\g''\g'''K}\,Q_{\b'\b''\b'''J}\in{\rm Im}\,\sfP^{(1)}
\qqq
but also
\qq\nn
[P_0,\sfP^{(1)\,\b'\b''\b'''J}_{\ \ \ \ \ \a'\a''I}\,Q_{\b'\b''J}]&=&\tfrac{\sfi}{2}\,\bigl(\unl\g{}_0\,\unl\g{}_7\,\bigl(\bd1_4\ox\si_2\bigr)\ox\si_3\bigr)^{\g'\g''\g'''K}_{\ \a'\a''\a'''I}\,\bigl(\bd1_{16}-\sfP^{(1)}\bigr)^{\b'\b''\b'''J}_{\ \g'\g''\g'''K}\,Q_{\b'\b''\b'''J}\in{\rm Im}\,\bigl(\bd1_{16}-\sfP^{(1)}\bigr)\,,\cr\cr
[P_3,\sfP^{(1)\,\b'\b''\b'''J}_{\ \ \ \ \ \a'\a''I}\,Q_{\b'\b''J}]&=&-\tfrac{\sfi}{2}\,\bigl(\unl\g{}_3\,\bigl(\bd1_4\ox\si_2\bigr)\ox\si_3\bigr)^{\g'\g''\g'''K}_{\ \a'\a''\a'''I}\,\bigl(\bd1_{16}-\sfP^{(1)}\bigr)^{\b'\b''\b'''J}_{\ \g'\g''\g'''K}\,Q_{\b'\b''\b'''J}\in{\rm Im}\,\bigl(\bd1_{16}-\sfP^{(1)}\bigr)\,,
\qqq
and so the HP vacuum superdistribution with restrictions
\qq\nn
{\rm Vac}\bigl(\bigl({\rm SU}(1,1\,|\,2)\x{\rm SU}(1,1\,|\,2)\bigr)_2/({\rm SO}(2)\x{\rm SO}(2)),\underset{\tx{\ciut{(3)}}}{\widehat\chi}\hspace{-1pt}{}^{\rm PR}_{(12)}\bigr)\rstr_{\cV_i}=\bigoplus_{\unl\a=1}^8\,\corr{\cT_{\unl\a\,i}}\oplus\corr{\cT_{0\,i},\cT_{3\,i}}
\qqq
is $({\rm SO}(2)\x{\rm SO}(2))$-descendable but not integrable. Its weak derived flag with (local) components of the symbol 
\qq\nn
&\gtq^{-1}_{(i)}\equiv{\rm Vac}\bigl(\bigl({\rm SU}(1,1\,|\,2)\x{\rm SU}(1,1\,|\,2)\bigr)_2/({\rm SO}(2)\x{\rm SO}(2)),\underset{\tx{\ciut{(3)}}}{\widehat\chi}\hspace{-1pt}{}^{\rm PR}_{(12)}\bigr)\rstr_{\cV_i}\,,&\cr\cr
&\gtq^{-2}_{(i)}=\bigoplus_{\widehat\a=9}^{16}\,\corr{\cT_{\widehat\a\,i}}\oplus\corr{\cT_{01\,i},\cT_{02\,i},\cT_{34\,i},\cT_{35\,i}}\,,\qquad\qquad\gtq^{-3}_{(i)}=\corr{\cT_{1\,i},\cT_{2\,i},\cT_{4\,i},\cT_{5\,i}}&
\qqq
is bracket-generating for the tangent sheaf of the HP section,
\qq\nn
{\rm Vac}^{-\infty}\bigl(\bigl({\rm SU}(1,1\,|\,2)\x{\rm SU}(1,1\,|\,2)\bigr)_2/({\rm SO}(2)\x{\rm SO}(2)),\underset{\tx{\ciut{(3)}}}{\widehat\chi}\hspace{-1pt}{}^{\rm PR}_{(12)}\bigr)=\cT\Si^{\rm HP}\,.
\qqq
\eeg

\beg\textbf{The vacuum superdistribution for the Metsaev--Tseytlin super-1-brane in $\,{\rm SU}(2,2\,|\,4)/({\rm SO}(1,1)\x{\rm SO}(3)\x{\rm SO}(5))$.}
The identity 
\qq\nn
\chi_{\a'\a''I\b'\b''J a'}+\la_1\,f_{\a'\a''I\b'\b''J}^{\ \ \ \ \ \ \ \ \ \ \ \ \ b'}\,\ep_{b'a'}=4\sfi\,\bigl(\unl C\,\unl\g^{b'}\unl\g{}_{11}\bigr)_{\a'\a''I\g'\g''K}\,\ep_{b'a'}\,\left(\tfrac{\la_1\,\bd1_{32}-\unl\g^0\,\unl\g^1\,\unl\g{}_{11}}{2}\right)^{\g'\g''K}_{\ \ \b'\b''J}\,,
\qqq
yields -- for $\,\la_0\in\{-1,1\}\,$ -- the possible projectors
\qq\nn
\sfP^{(1)}_{\pm 1}=\tfrac{\bd1_{32}\pm\unl\g^0\,\unl\g^1\,\unl\g{}_{11}}{2}
\qqq
with properties
\qq\nn
\unl\g^{\unl a}\,\sfP^{(1)}_{\pm 1}=\sfP^{(1)}_{\pm 1}\,\unl\g^{\unl a}\,,\qquad\qquad\unl\g^{\widehat a}\,\sfP^{(1)}_{\pm 1}=\bigl(\bd1_{32}-\sfP^{(1)}_{\pm 1}\bigr)\,\unl\g^{\widehat a}\,,\qquad\qquad\unl C\,\sfP^{(1)}_{\pm 1}\,\unl C^{-1}=\sfP^{(1)\,{\rm T}}_{\pm 1}\,,
\qqq
and so upon choosing
\qq\nn
\sfP^{(1)}\equiv\sfP^{(1)}_{+1}=\tfrac{\bd1_{32}+\unl\g^0\,\unl\g^1\,\unl\g{}_{11}}{2}\,,
\qqq
we check
\bit
\item the $\k$-Symmetry Constraints, 
\qq\nn
\{\sfP^{(1)\,\g'\g''K}_{\ \ \ \ \ \a'\a''I}\,Q_{\g'\g''K},\sfP^{(1)\,\d\d'L}_{\ \ \ \ \ \b'\b''J}\,Q_{\d\d'L}\}&=&2\sfP^{(1)\,\g'\g''K}_{\ \ \ \ \ \b'\b''J}\,\bigl(\sfi\,\bigl(\unl C\,\unl\g^{\unl a}\unl\g{}_{11}\bigr)_{\a'\a''I\g'\g''K}\,P_{\unl a}+\bigl(\unl C\,\unl\g^{01}\bigr)_{\a'\a''I\g'\g''K}\,J_{01}\bigr)\cr\cr
&\in&\corr{P_0,P_1}\oplus\bigoplus_{a',b'=0}^4\,\corr{J_{a'b'}}\oplus\bigoplus_{a'',b''=5}^9\,\corr{J_{a''b''}}\,;
\qqq
\item the Even Effective-Mixing Constraints,  
\qq\nn
&[J_{01},P_{\unl a}]=\d_{\unl a 1}\,P_0-\eta_{\unl a 0}\,P_1\in\corr{P_0,P_1}\ni 0=[J_{\widehat a\widehat b},P_{\unl c}]\,,\qquad\qquad[J_{\unl a\widehat b},P_{\widehat c}]=\d_{\widehat b\widehat c}\,P_{\unl a}\in\corr{P_0,P_1}\,,&\cr\cr
&[J_{01},P_{\widehat a}]=0\in\bigoplus_{\widehat d=2}^9\,\corr{P_{\widehat d}}\ni\d_{\widehat b\widehat c}\,P_{\widehat a}-\d_{\widehat a\widehat c}\,P_{\widehat b}=[J_{\widehat a\widehat b},P_{\widehat c}]\,,\qquad\qquad[J_{\unl a\widehat b},P_{\unl c}]=-\eta_{\unl a\unl c}\,P_{\widehat b}\in\bigoplus_{\widehat d=2}^9\,\corr{P_{\widehat d}}\,;&
\qqq
\item the No-Curvature and No-Torsion Constraints, with
\qq\nn
&[P_0,P_1]=J_{01}\in\corr{J_{01}}\oplus\bigoplus_{a',b'\in\{2,3,4\}}\,\corr{J_{a'b'}}\oplus\bigoplus_{b'',c''=5}^9\,\corr{J_{b''c''}}\ni 0=[P_2,P_{a''}]=[P_3,P_{a''}]=[P_4,P_{a''}]\,,&\cr\cr
&[P_2,P_3]=J_{23}\in\corr{J_{01}}\oplus\bigoplus_{a',b'\in\{2,3,4\}}\,\corr{J_{a'b'}}\oplus\bigoplus_{a'',b''=5}^9\,\corr{J_{a''b''}}\ni J_{24}=[P_2,P_4]\,,&\cr\cr
&[P_3,P_4]=J_{34}\in\corr{J_{01}}\oplus\bigoplus_{a',b'\in\{2,3,4\}}\,\corr{J_{a'b'}}\oplus\bigoplus_{c'',d''=5}^9\,\corr{J_{c''d''}}\ni-J_{a''b''}=[P_{a''},P_{b''}]\,;&
\qqq
\item the $\G$-Constraints,  
\qq\nn
&&\bigl(\unl C\,\unl\g{}_{\widehat a'}\bigr)_{\a'\a''I\g'\g''K}\,\sfP^{(1)\,\g'\g''K}_{\ \ \ \ \ \b'\b''J}=\bigl(\unl C\,\bigl(\bd1_{32}-\sfP^{(1)}\bigr)\,\unl\g{}_{\widehat a'}\bigr)_{\a'\a''I\b'\b''J}=\bigl(\bigl(\bd1_{32}-\sfP^{(1)}\bigr)^{\rm T}\,\bigl(\unl C\,\unl\g{}_{\widehat a'}\bigr)\bigr)_{\a'\a''I\b'\b''J}\cr\cr
&\equiv&\bigl(\unl C\,\unl\g{}_{\widehat a'}\bigr)_{\g'\g''K\b'\b''J}\,\bigl(\bd1_{32}-\sfP^{(1)}\bigr)^{\g'\g''K}_{\ \a'\a''I}\,,\qquad\widehat a'\in\{2,3,4\}\,,\cr\cr\cr
&&-\bigl(\unl C\,\unl\g{}_{a''}\,\unl\g{}_{11}\bigr)_{\a'\a''I\g'\g''K}\,\sfP^{(1)\,\g'\g''K}_{\ \ \ \ \ \b'\b''J}=-\bigl(\unl C\,\bigl(\bd1_{32}-\sfP^{(1)}\bigr)\,\unl\g{}_{a''}\,\unl\g{}_{11}\bigr)_{\a'\a''I\b'\b''J}\cr\cr
&=&-\bigl(\bigl(\bd1_{32}-\sfP^{(1)}\bigr)^{\rm T}\,\bigl(\unl C\,\unl\g{}_{a''}\,\unl\g{}_{11}\bigr)\bigr)_{\a'\a''I\b'\b''J}\equiv-\bigl(\unl C\,\unl\g{}_{a''}\,\unl\g{}_{11}\bigr)_{\g'\g''K\b'\b''J}\,\bigl(\bd1_{32}-\sfP^{(1)}\bigr)^{\g'\g''K}_{\ \a'\a''I}\,;
\qqq
\item the Vacuum-Superalgebra Constraints, with 
\qq\nn
\{\sfP^{(1)\,\g'\g''K}_{\ \ \ \ \ \a'\a''I}\,Q_{\g'\g''K},\sfP^{(1)\,\d\d'L}_{\ \ \ \ \ \b'\b''J}\,Q_{\d\d'L}\}\in\corr{P_0,P_1}\oplus\corr{J_{01}}\oplus\bigoplus_{a',b'\in\{2,3,4\}}\,\corr{J_{a'b'}}\oplus\bigoplus_{a'',b''=5}^9\,\corr{J_{a''b''}}
\qqq
and 
\qq\nn
[J_{01},\sfP^{(1)\,\b'\b''J}_{\ \ \ \ \ \a'\a''I}\,Q_{\b'\b''J}]&=&\tfrac{1}{2}\,\bigl(\unl\g{}_{01}\bigr)^{\b'\b''J}_{\ \a'\a''I}\,\sfP^{(1)\,\g'\g''K}_{\ \ \ \ \ \b'\b''J}\,Q_{\g'\g''K}\in{\rm Im}\,\sfP^{(1)}\,,\cr\cr
[J_{\widehat a\widehat b},\sfP^{(1)\,\b'\b''J}_{\ \ \ \ \ \a'\a''I}\,Q_{\b'\b''J}]&=&\tfrac{1}{2}\,\bigl(\unl\g{}_{\widehat a\widehat b}\bigr)^{\b'\b''J}_{\ \a'\a''I}\,\sfP^{(1)\,\g'\g''K}_{\ \ \ \ \ \b'\b''J}\,Q_{\g'\g''K}\in{\rm Im}\,\sfP^{(1)}\,,
\qqq
and 
\qq\nn
[P_{\unl a},\sfP^{(1)\,\b'\b''J}_{\ \ \ \ \ \a'\a''I}\,Q_{\b'\b''J}]=-\tfrac{\sfi}{2}\,\bigl(\unl\g{}_{\unl a}\,\unl\g{}_{11}\bigr)^{\g'\g''K}_{\ \a'\a''I}\,\sfP^{(1)\,\b'\b''J}_{\ \ \ \ \ \g'\g''K}\,Q_{\b'\b''J}\in{\rm Im}\,\sfP^{(1)}\,;
\qqq
\item $({\rm SO}(1,1)\x{\rm SO}(3)\x{\rm SO}(5))$-descendability.
\eit
Consequently, the HP vacuum superdistribution with restrictions
\qq\nn
{\rm Vac}\bigl({\rm SU}(2,2\,|\,4)/({\rm SO}(1,1)\x{\rm SO}(3)\x{\rm SO}(5)),\underset{\tx{\ciut{(3)}}}{\widehat\chi}\hspace{-1pt}{}^{\rm MT}_{(1)}\bigr)\rstr_{\cV_i}=\bigoplus_{\unl\a=1}^{16}\,\corr{\cT_{\unl\a\,i}}\oplus\corr{\cT_{0\,i},\cT_{1\,i}}
\qqq 
is an $({\rm SO}(1,1)\x{\rm SO}(3)\x{\rm SO}(5))$-descendable integrable superdistribution associated with the Lie superalgebra
\qq\nn
&&\gt{vac}\bigl({\rm SU}(2,2\,|\,4)/({\rm SO}(1,1)\x{\rm SO}(3)\x{\rm SO}(5)),\underset{\tx{\ciut{(3)}}}{\widehat\chi}\hspace{-1pt}{}^{\rm MT}_{(1)}\bigr)\cr\cr
&=&\bigoplus_{\unl\a=1}^{16}\,\corr{\unl Q{}_{\unl\a}}\oplus\corr{P_0,P_1}\oplus\corr{J_{01}}\oplus\bigoplus_{a',b'\in\{2,3,4\}}\,\corr{J_{a'b'}}\oplus\bigoplus_{a'',b''=5}^9\,\corr{J_{a''b''}}\,.
\qqq
\eeg

\beg\textbf{The vacuum superdistribution for the Metsaev--Tseytlin super-1-brane in $\,{\rm SU}(2,2\,|\,4)/({\rm SO}(4)\x{\rm SO}(4))$.}\label{eg:MTs1Vac}
The identities
\qq\nn
\chi_{\a'\a''I\b'\b''J 0}-\la_1\,f_{\a'\a''I\b'\b''J}^{\ \ \ \ \ \ \ \ \ \ \ \ 5}&=&4\sfi\,\bigl(\unl C\,\unl\g^5\bigr)_{\a'\a''I\g'\g''K}\,\left(\tfrac{\la_1\,\bd1_{32}+\unl\g^0\,\unl\g^5}{2}\right)^{\g'\g''K}_{\ \ \b'\b''J}\,,\cr\cr
\chi_{\a'\a''I\b'\b''J 3}+\la_1\,f_{\a'\a''I\b'\b''J}^{\ \ \ \ \ \ \ \ \ \ \ \ 0}&=&4\sfi\,\bigl(\unl C\,\unl\g^0\,\unl\g{}_{11}\bigr)_{\a'\a''I\g'\g''K}\,\left(\tfrac{\la_1\,\bd1_{32}+\unl\g^0\,\unl\g^5}{2}\right)^{\g'\g''K}_{\ \ \b'\b''J}\,,
\qqq
yield -- for $\,\la_0\in\{-1,1\}\,$ -- the possible projectors
\qq\nn
\sfP^{(1)}_{\mp 1}=\tfrac{\bd1_{32}\pm\unl\g^0\,\unl\g^5}{2}
\qqq
with properties
\qq\nn
\unl\g^{\unl a}\,\sfP^{(1)}_{\pm 1}=\bigl(\bd1_{32}-\sfP^{(1)}_{\pm 1}\bigr)\,\unl\g^{\unl a}\,,\qquad\qquad\unl\g^{\widehat a}\,\sfP^{(1)}_{\pm 1}=\sfP^{(1)}_{\pm 1}\,\unl\g^{\widehat a}\,,\qquad\qquad\unl C\,\sfP^{(1)}_{\pm 1}\,\unl C^{-1}=\bigl(\bd1_{32}-\sfP^{(1)}_{\pm 1}\bigr)^{\rm T}\,,
\qqq
and so upon choosing
\qq\nn
\sfP^{(1)}\equiv\sfP^{(1)}_{-1}=\tfrac{\bd1_{32}+\unl\g^0\,\unl\g^5}{2}\,,
\qqq
we check
\bit
\item the $\k$-Symmetry Constraints, 
\qq\nn
\{\sfP^{(1)\,\g'\g''K}_{\ \ \ \ \ \a'\a''I}\,Q_{\g'\g''K},\sfP^{(1)\,\d'\d''L}_{\ \ \ \ \ \b'\b''J}\,Q_{\d'\d''L}\}&=&2\sfP^{(1)\,\g'\g''K}_{\ \ \ \ \ \b'\b''J}\,\bigl(\sfi\,\bigl(\unl C\,\unl\g^0\,\unl\g{}_{11}\bigr)_{\a'\a''I\g'\g''K}\,P_0-\sfi\,\bigl(\unl C\,\unl\g^5\bigr)_{\a'\a''I\g'\g''K}\,P_5\cr\cr
&&+\bigl(\unl C\,\unl\g^{0a'}\bigr)_{\a'\a''I\g'\g''K}\,J_{0a'}-\bigl(\unl C\,\unl\g^{5a''}\bigr)_{\a'\a''I\g'\g''K}\,J_{5a''}\bigr)\cr\cr
&\in&\corr{P_0,P_5}\oplus\bigoplus_{a',b'=0}^{4}\,\corr{J_{a',b'}}\oplus\bigoplus_{a'',b''=5}^9\,\corr{J_{a''b''}}\,;
\qqq
\item the Even Effective-Mixing Constraints, 
\qq\nn
&[J_{\widehat a\widehat b},P_{\unl c}]=0\in\corr{P_0,P_5}\,,\qquad\qquad[J_{\unl a\widehat b},P_{\widehat c}]=\d_{\widehat b\widehat c}\,P_{\unl a}\in\corr{P_0,P_5}\,,&\cr\cr
&[J_{\widehat a\widehat b},P_{\widehat c}]=\d_{\widehat b\widehat c}\,P_{\widehat a}-\d_{\widehat a\widehat c}\,P_{\widehat b}\in\bigoplus_{a'=1}^{4}\,\corr{P_{a'}}\oplus\bigoplus_{b''=6}^9\,\corr{P_{b''}}\,,&\cr\cr
&[J_{\unl a\widehat b},P_{\unl c}]=-\eta_{\unl a\unl c}\,P_{\widehat b}\in\bigoplus_{a'=1}^{4}\,\corr{P_{a'}}\oplus\bigoplus_{b''=6}^9\,\corr{P_{b''}}\,;&
\qqq
\item the No-Curvature and No-Torsion Constraints, with
\qq\nn
[P_0,P_5]=0\in\bigoplus_{c',d'=1}^{4}\,\corr{J_{c'd'}}\oplus\bigoplus_{c'',d''=6}^9\,\corr{J_{c''d''}}\ni\d_{\widehat a}^{\ a'}\,\d_{\widehat b}^{\ b'}\,J_{a'b'}-\d_{\widehat a}^{\ a''}\,\d_{\widehat b}^{\ b''}\,J_{a''b''}=[P_{\widehat a},P_{\widehat b}]\,;
\qqq
\item the $\G$-Constraints, (written for $\,\widehat a'\in\{1,2,3,4\}\,$ and $\,\widehat a''\in\{6,7,8,9\}$)
\qq\nn
&&\bigl(\unl C\,\unl\g{}_{\widehat a'}\bigr)_{\a'\a''I\g'\g''K}\,\sfP^{(1)\,\g'\g''K}_{\ \ \ \ \ \b'\b''J}=\bigl(\unl C\,\sfP^{(1)}\,\unl\g{}_{\widehat a'}\bigr)_{\a'\a''I\b'\b''J}=\bigl(\bigl(\bd1_{32}-\sfP^{(1)}\bigr)^{\rm T}\,\bigl(\unl C\,\unl\g{}_{\widehat a'}\bigr)\bigr)_{\a'\a''I\b'\b''J}\cr\cr
&\equiv&\bigl(\unl C\,\unl\g{}_{\widehat a'}\bigr)_{\g'\g''K\b'\b''J}\,\bigl(\bd1_{32}-\sfP^{(1)}\bigr)^{\g'\g''K}_{\ \a'\a''I}\,,\cr\cr\cr
&&-\bigl(\unl C\,\unl\g{}_{\widehat a''}\,\unl\g{}_{11}\bigr)_{\a'\a''I\g'\g''K}\,\sfP^{(1)\,\g'\g''K}_{\ \ \ \ \ \b'\b''J}=-\bigl(\unl C\,\sfP^{(1)}\,\unl\g{}_{\widehat a''}\,\unl\g{}_{11}\bigr)_{\a'\a''I\b'\b''J}\cr\cr
&=&-\bigl(\bigl(\bd1_{32}-\sfP^{(1)}\bigr)^{\rm T}\,\bigl(\unl C\,\unl\g{}_{\widehat a''}\,\unl\g{}_{11}\bigr)\bigr)_{\a'\a''I\b'\b''J}\equiv-\bigl(\unl C\,\unl\g{}_{\widehat a''}\,\unl\g{}_{11}\bigr)_{\g'\g''K\b'\b''J}\,\bigl(\bd1_{32}-\sfP^{(1)}\bigr)^{\g'\g''K}_{\ \a'\a''I}\,;
\qqq
\item[$\bullet$] $({\rm SO}(4)\x{\rm SO}(4))$-descendability.
\eit
The Vacuum-Superalgebra Constraints, on the other hand, are \emph{not} satisfied in view of the above, 
\qq\nn
\{\sfP^{(1)\,\g'\g''K}_{\ \ \ \ \ \a'\a''I}\,Q_{\g'\g''K},\sfP^{(1)\,\d'\d''L}_{\ \ \ \ \ \b'\b''J}\,Q_{\d'\d''L}\}\notin\corr{P_0,P_5}\oplus\bigoplus_{a',b'=1}^{4}\,\corr{J_{a'b'}}\oplus\bigoplus_{a'',b''=6}^9\,\corr{J_{a''b''}}\,,
\qqq
with 
\qq\nn
[J_{\widehat a\widehat b},\sfP^{(1)\,\b'\b''J}_{\ \ \ \ \ \a'\a''I}\,Q_{\b'\b''J}]&=&\tfrac{1}{2}\,\bigl(\unl\g{}_{\widehat a}\,\unl\g{}_{\widehat b}\bigr)^{\g'\g''K}_{\ \a'\a''I}\,\sfP^{(1)\,\b'\b''J}_{\ \ \ \ \ \g'\g''K}\,Q_{\b'\b''J}\in{\rm Im}\,\sfP^{(1)}
\qqq
but also
\qq\nn
[P_0,\sfP^{(1)\,\b'\b''J}_{\ \ \ \ \ \a'\a''I}\,Q_{\b'\b''J}]&=&-\tfrac{\sfi}{2}\,\bigl(\unl\g{}_0\,\unl\g{}_{11}\bigr)^{\g'\g''K}_{\ \a'\a''I}\,\bigl(\bd1_{32}-\sfP^{(1)}\bigr)^{\b'\b''J}_{\ \g'\g''K}\,Q_{\b'\b''J}\in{\rm Im}\,\bigl(\bd1_{32}-\sfP^{(1)}\bigr)\,,\cr\cr
[P_5,\sfP^{(1)\,\b'\b''J}_{\ \ \ \ \ \a'\a''I}\,Q_{\b'\b''J}]&=&\tfrac{\sfi}{2}\,\bigl(\unl\g{}_5\bigr)^{\g'\g''K}_{\ \a'\a''I}\,\bigl(\bd1_{32}-\sfP^{(1)}\bigr)^{\b'\b''J}_{\ \g'\g''K}\,Q_{\b'\b''J}\in{\rm Im}\,\bigl(\bd1_{32}-\sfP^{(1)}\bigr)\,,
\qqq
and so the HP vacuum superdistribution with restrictions
\qq\nn
{\rm Vac}\bigl({\rm SU}(2,2\,|\,4)/({\rm SO}(4)\x{\rm SO}(4)),\underset{\tx{\ciut{(3)}}}{\widehat\chi}\hspace{-1pt}{}^{\rm MT}_{(12)}\bigr)\rstr_{\cV_i}=\bigoplus_{\unl\a=1}^{16}\,\corr{\cT_{\unl\a\,i}}\oplus\corr{\cT_{0\,i},\cT_{5\,i}}
\qqq
is nonintegrable. Its weak derived flag with (local) components of the symbol 
\qq\nn
&\gtq^{-1}_{(i)}\equiv{\rm Vac}\bigl({\rm SU}(2,2\,|\,4)/({\rm SO}(4)\x{\rm SO}(4)),\underset{\tx{\ciut{(3)}}}{\widehat\chi}\hspace{-1pt}{}^{\rm MT}_{(12)}\bigr)\rstr_{\cV_i}\,,&\cr\cr
&\gtq^{-2}_{(i)}=\bigoplus_{\widehat\a=17}^{32}\,\corr{\cT_{\widehat\a\,i}}\oplus\bigoplus_{a'=1}^{4}\,\corr{\cT_{0a'\,i}}\oplus\bigoplus_{b''=6}^9\,\corr{\cT_{5b''\,i}}\,,\qquad\qquad\gtq^{-3}_{(i)}=\bigoplus_{a'=1}^{4}\,\corr{\cT_{a'\,i}}\oplus\bigoplus_{b''=6}^9\,\corr{\cT_{b''\,i}}&
\qqq
is bracket-generating for the tangent sheaf of the HP section,
\qq\nn
{\rm Vac}^{-\infty}\bigl({\rm SU}(2,2\,|\,4)/({\rm SO}(4)\x{\rm SO}(4)),\underset{\tx{\ciut{(3)}}}{\widehat\chi}\hspace{-1pt}{}^{\rm MT}_{(12)}\bigr)=\cT\Si^{\rm HP}\,.
\qqq
\eeg

\section{Supersymmetries of the super-$\si$-model and of its vacuum}\label{sec:susy}

In the present section, we investigate in full detail (super)symmetries of the GS super-$\si$-model, with view to their geometrisation and subsequent lifting to the higher-geometric structures underlying the topological term in the DF amplitude that we recall in Sec.\,\ref{sec:hgeomphys}. Particular emphasis shall be laid on the peculiar local (or gauged) supersymmetry that appears -- through a universal mechanism of enhancement of gauge symmetry -- in the HP/NG correspondence sector and restores balance between bosonic and fermionic degrees of freedom of the vacuum of the super-$\si$-model, as first observed by de Azc\'arraga and Lukierski in \Rcite{deAzcarraga:1982njd}, later discussed by Siegel in Refs.\,\cite{Siegel:1983hh,Siegel:1983ke}, and recently elaborated by McArthur in \Rcite{McArthur:1999dy} and West {\it et al.} in Refs.\,\cite{Gomis:2006wu} from the more geometric perspective\footnote{{\it Cp} \Rcite{Suszek:2019cum} for a fairly complete list of references.}. The symmetry shall be demonstrated to geometrise in the HP formulation, but that in a rather non-trivial manner, to wit, as a distinguished (Gra\ss mann-)odd-generated superdistribution contained properly in the HP vacuum superdistribution $\,{\rm Vac}(\sgt\Bgt_{p,\la_p}^{\rm (HP)})\,$ over the HP section. In general, the superdistribution is \emph{neither} $\txH_{\rm vac}$-descendable \emph{nor} integrable but in the physically preferred circumstances in which $\,{\rm Vac}(\sgt\Bgt_{p,\la_p}^{\rm (HP)})\,$ defines a vacuum foliation of the HP section that descends to the supertarget $\,\cM_{\txH_{\rm vac}}$,\ the symmetry superdistribution is seen to bracket-generate\footnote{That is the limit of the weak derived flag of the superdistribution coincides with the vacuum superdistribution.} $\,{\rm Vac}(\sgt\Bgt_{p,\la_p}^{\rm (HP)})\,$ and -- in consequence -- envelop the vacuum, all that in a manner manifestly compatible with the (linearised) global supersymmetry.

\subsection{The global-supersymmetry group}\label{sec:globsusy}

The very formulation of the super-$\si$-model in terms of tensor products of the components of the left-invariant Maurer--Cartan super-1-form along $\,\fgt\,$ contracted with $\txH_{\rm vac}$-invariant tensors encodes the global supersymmetry of the field theory modelled by the Lie supergroup $\,\txG\,$ acting by left translations $\,[\ell]_\cdot\equiv[\ell]^{\txH_{\rm vac}}_\cdot$ as in \Reqref{eq:leftrans}. Indeed, there exist local lifts of $\,[\ell]_\cdot\,$ to $\,\Si^{\rm HP}\,$ determined, up to the action of the local gauge group $\,[\cU_i^{\txH_{\rm vac}},\txH_{\rm vac}]$,\ by the condition
\qq\label{eq:GonSiHPvac}
g\cdot\si_i^{\rm vac}(\unl\chi{}_i)\cdot\unl h_{ij}(\unl\chi{}_i;g)=\si_j^{\rm vac}\bigl(\unl\chi{}_j\bigl(g[\lact]\unl\chi{}_i\bigr)\bigr)
\qqq
written, in the $\cS$-point picture, for arbitrary $\,g\in\txG\,$ (an $\cS$-point) and $\,i\in I_{\txH_{\rm vac}}$,\ and for $\,j\in I_{\txH_{\rm vac}}\,$ such that $\,\pi_{\txG/\txH_{\rm vac}}(g\cdot\si_i^{\rm vac}(\unl\chi{}_i))\in\cU_j^{\txH_{\rm vac}}$,\ and with $\,[\lact]\,$ representing the induced action (in that picture), so that for any $\,k\in I_{\txH_{\rm vac}}\,$ with the same property we find
\qq\nn
\unl h_{ik}(\unl\chi{}_i;g)=\unl h_{ij}(\unl\chi{}_i;g)\cdot h_{jk}\bigl(\unl\chi{}_j\bigl(g[\lact]\unl\chi{}_i\bigr)\bigr)\,,
\qqq
where the $\,h_{jk}\,$ are the gluing mappings of the principal $\txH_{\rm vac}$-bundle \eqref{eq:homasprinc} (with $\,\txK=\txH_{\rm vac}$). The integrands in the action functional of the super-$\si$-model being (right-)$\txH_{\rm vac}$-basic (so that they descend to the homogeneous space) and -- by assumption -- \textbf{quasi-supersymmetric}, that is left-invariant up to a total external derivative in such a manner that the action functional for $\,\Om_p\,$ closed is left-$\txG$-invariant, the induced left action of $\,\txG\,$ on the supertarget $\,\cM_{\txH_{\rm vac}}\,$ preserves the DF amplitude. We shall discuss the deeper higher-geometric meaning of this global supersymmetry of the DF amplitudes in Sec.\,\ref{sec:hgeomphys}. Meanwhile, note that the non-linear realisation of the global-supersymmetry group on the HP section $\,\Si^{\rm HP}\,$ admits an `infinitesimal' presentation in terms of the corresponding `fundamental' vector fields, specified in
\berop\label{prop:leftglobfund}
In the hitherto notation, let the $\,\unl\cK{}_A,\ A\in\ovl{0,D}\,$ be the fundamental vector fields of the induced $\txG$-action $\,[\ell]_\cdot\,$ of $\,\txG\,$ on $\,\cM_{\txH_{\rm vac}}$.\ The local tangent lift
\qq\label{eq:fundtofundt}
\cK_{A\,i}\bigl(\si_i^{\rm vac}(\unl\chi{}_i)\bigr)\equiv\sfT_{\unl\chi{}_i}\si_i^{\rm vac}\bigl(\unl\cK{}_A(\unl\chi{}_i)\bigr)
\qqq
of the fundamental vector field $\,\unl\cK{}_A\,$ to $\,\cV_i,\ i\in I_{\txH_{\rm vac}}\,$ takes the form
\qq\label{eq:refform}
\cK_{A\,i}=R_A\rstr_{\cV_i}+\Xi_{A\,i}^{\ \ \ \unl S}\,L_{\unl S}\,,\qquad A\in\ovl{0,D}\,,
\qqq
with the sections $\,\Xi_{A\,i}^{\ \ \ \unl S}\in\cO_\txG(\cV_i)\,$ given by the formul\ae
\qq\nn
\Xi_{A\,i}^{\ \ \ \unl S}\bigl(\si_i^{\rm vac}(\unl\chi{}_i)\bigr)=H_A^{\ \nu}\bigl(\si_i^{\rm vac}(\unl\chi{}_i)\bigr)\,\bigl(\unl E(\unl\chi{}_i)^{-1}\bigr)_\nu^{\ \mu}\,E^{\ \unl S}_\mu(\unl\chi{}_i)-H_A^{\ \unl S}\bigl(\si_i^{\rm vac}(\unl\chi{}_i)\bigr)
\qqq
in terms of the change-of-basis sections $\,H_A^{\ B}\in\cO_\txG(\cV_i)\,$ defined as
\qq\nn
R_A=:H_A^{\ B}\,L_B\,.
\qqq
We shall call the $\bR$-linear span of the vector fields 
\qq\nn
\cK_A\in\G\bigl(\cT\Si^{\rm HP}\bigr)\,,\qquad\qquad\cK_A\rstr_{\cV_i}=\cK_{A\,i}
\qqq
the \textbf{global-supersymmetry subspace of} $\,\cT\Si^{\rm HP}\,$ and denote it as
\qq\nn
S^{\rm HP}_\txG=\corr{\ \cK_A\ \vert\ A\in\ovl{0,D}\ }\subset\G\bigl(\cT\Si^{\rm HP}\bigr)\,.
\qqq
\eerop
\beroof
The general structure of the reference formula \eqref{eq:refform} follows from Eqs.\,\eqref{eq:LRasfund} and \eqref{eq:GonSiHPvac}. The proof develops along similar lines as that of Prop.\,\ref{prop:bastanshHP}. Thus, we write
\qq\nn
\unl\cK{}_A(\unl\chi{}_i)=\D^\mu_{A\,i}(\unl\chi{}_i)\,\tfrac{\vec\p\ }{\p\unl\chi{}_i^\mu}
\qqq
and proceed with a calculation similar to the one leading to \Reqref{eq:hatlasrho}, 
whereby we arrive at the identity
\qq\nn
\D^\mu_{A\,i}\,E^{\ B}_\mu(\unl\chi{}_i)\equiv\unl\cK{}_A\con\si_i^{{\rm vac}\,*}\theta_{\rm L}^B(\unl\chi{}_i)=\cK_{A\,i}\con\theta_{\rm L}^B\bigl(\si_i^{\rm vac}(\unl\chi{}_i)\bigr)\equiv H_A^{\ B}\bigl(\si_i^{\rm vac}(\unl\chi{}_i)\bigr)+\Xi_{A\,i}^{\ \ \ \unl S}\bigl(\si_i^{\rm vac}(\unl\chi{}_i)\bigr)\,\d_{\unl S}^{\ B}\,,
\qqq
valid for any $\,B\in\ovl{0,D}$.\ Setting $\,B\equiv\nu\in\ovl{0,\unl\d}$,\ we obtain 
\qq\nn
\D^\mu_{A\,i}\,E^{\ \nu}_\mu(\unl\chi{}_i)=H_A^{\ \nu}\bigl(\si_i^{\rm vac}(\unl\chi{}_i)\bigr)\,.
\qqq 
\eroof

It is to be emphasised that while the maps $\,\ell_\cdot\,$ and $\,[\ell]_\cdot\,$ are left actions in the standard sense, the local lifts of the latter to the HP section do not compose, in general, a \emph{bona fide} action of the supersymmetry group due to the inherent ambiguity in the definition of the target index $\,j\,$ in \Reqref{eq:GonSiHPvac}. What \emph{does} survive the physically motivated restriction to $\,\Si^{\rm HP}\,$ is a \emph{linearised} realisation of the supersymmetry group on $\,\Si^{\rm HP}\,$ derived above. It is this linearised realisation that we shall work with when discussing the global supersymmetry of the various physically distinguished superdistributions over $\,\Si^{\rm HP}$.\ For that, however, we first need to formalise meaningfully the notion of a linearised global (super)symmetry in the case of a (super)distribution, which we do in  
\bedef\label{def:globlinGsdistro}
Let $\,\cM\,$ be a supermanifold, and let $\,S\subset\G(\cT\cM)\,$ be a Lie superalgebra.  A superdistribution $\,\cD\subset\cT\cM\,$ shall be called \textbf{$S$-symmetric} if the following identities are satisfied:
\qq\nn
\forall_{\cV\in S}\ :\ [\cV,\cD\}\subset\cD\,,
\qqq
so that the flows of the vector fields spanning $\,S\,$ preserve $\,\cD$.\ In particular, let $\,\txG\,$ be a Lie supergroup (with the tangent Lie superalgebra $\,\ggt$) that acts on $\,\cM$,\ inducing the fundamental vector fields $\,\cK^\la_X,\ X\in\ggt\,$ of \Reqref{eq:fundvecfields} that compose the $\bR$-linear subspace
\qq\nn
S_\txG:=\{\ \cK^\la_X \quad\vert\quad X\in\ggt\ \}\subset\G(\cT\cM)
\qqq
closed under the supercommutator and modelled on $\,\ggt$.\ An $S_\txG$-symmetric superdistribution $\,\cD\subset\cT\cM\,$ shall be termed \textbf{globally linearised-$\txG$-symmetric}, or \textbf{$\ggt$-invariant}.
\exdef
\noindent Taking into accout the correspondence between the `fundamental' vector fields $\,\cK_A\,$ on $\,\Si^{\rm HP}\,$ and the proper fundamental vector fields $\,\unl\cK{}_A\,$ on $\,\cM_{\txH_{\rm vac}}$,\ we adapt the above standard definition to the present situation as follows.
\bedef\label{def:globlinsusysdistro}
Adopt the notation of Props.\,\ref{prop:bastanshHP} and \ref{prop:leftglobfund} and let $\,S_\txG^{\rm HP}\subset\G(\cT\Si^{\rm HP})\,$ be the global-supersymmetry subspace of Prop.\,\ref{prop:leftglobfund}. A superdistribution $\,\cD\subset\cT\Si^{\rm HP}\,$ shall be called \textbf{globally linearised-supersymmetric} if $\,\cD\,$ is $S_\txG^{\rm HP}$-symmetric in the sense of Def.\,\ref{def:globlinGsdistro}.
\exdef
\noindent The above definition enables us to formulate and study the third and last from the list of regularity criteria, first mentioned on p.\,\pageref{page:regularcrit}, that can and should be applied to the HP vacuum superdistribution. This extra criterion is preservation of the vacuum superdistribution under global supersymmetry and existence of a residual global supersymmetry in the vacuum ({\it i.e.}, on any of the integral supermanifolds of that superdistribution). Below, we return to the discussion begun in the previous section.

The much reassuring general answer to the first of the two questions resulting from the global-supersymmetry criterion is given in
\berop\label{prop:descissusy}
The HP vacuum superdistribution $\,{\rm Vac}(\sgt\Bgt^{{\rm (HP)}}_{p,\la_p})\,$ of Def.\,\ref{def:vacHPsdistro} is globally linearised-supersymmetric in the sense of Def.\,\ref{def:globlinsusysdistro} iff 
\qq\nn
[\hgt_{\rm vac},\tgt^{(1)}_{\rm vac}]\subset\tgt^{(1)}_{\rm vac}\,.
\qqq
In particular, an $\txH_{\rm vac}$-descendable HP vacuum superdistribution is globally linearised-supersymmetric.
\eerop
\beroof
All vector fields from $\,S_\txG^{\rm HP}\,$ lie in $\,\cT\Si^{\rm HP}$,\ as do sections of $\,{\rm Vac}(\sgt\Bgt^{{\rm (HP)}}_{p,\la_p})$.\ Hence, taking into account Prop.\,\ref{prop:bastanshHP} and the standard (trivial) supercommutation relations of the right-invariant vector fields $\,R_B\,$ with the left invariant ones $\,L_C$,\ we conclude that the supercommutator of a `fundamental' vector field $\,\cK_{A\,i}\,$ of \Reqref{eq:refform} with a section $\,\tau_i^{\unl{\unl A}}\,\cT_{\unl{\unl A}\,i}$,\ written in terms of arbitrary $\,\t_i^{\unl{\unl A}}\in\cO_\txG(\cV_i)$,\ is determined uniquely by the horizontal components $\,L_\nu\rstr_{\cV_i}\,$ that can be obtained by supercommuting vertical components of $\,\cK_{A\,i}\,$ with the horizontal components $\,L_{\unl{\unl A}}\rstr_{\cV_i}\,$ of $\,\tau_i^{\unl{\unl A}}\,\cT_{\unl{\unl A}\,i}$.\ Thus, we arrive at the requirement
\qq\label{eq:hvacstabtvac}
[\hgt_{\rm vac},\tgt_{\rm vac}]\subset\tgt_{\rm vac}
\qqq
that boils down to the one from the claim of the proposition by the definition of $\,\hgt_{\rm vac}$.\ The second part of the claim follows directly from Prop.\,\ref{prop:descvacsdistrolin}.
\eroof

\noindent In the light of the last proposition, the vacuum foliation, whenever it exists, is preserved \emph{as a whole} by the flow of global supersymmetry, that is, vacua are carried into one another. It remains to distinguish those of the flows that preserve a \emph{particular} vacuum. We do that it in
\berop\label{prop:resglobsusysub}
Adopt the hitherto notation, and in particular -- that of Def.\,\ref{def:vacHPsdistro} and Prop.\,\ref{prop:leftglobfund}. A tangent lift
\qq\nn
\cK_X\equiv X^A\,\cK_A 
\qqq
of the fundamental vector field $\,X^A\,\unl\cK{}_A$,\ engendered by the induced action $\,[\ell]_\cdot\,$ of $\,\txG\,$ on $\,\cM_{\txH_{\rm vac}}$,\ along the family $\,\{\si_i^{\rm vac}\}_{i\in I_{\txH_{}\rm vac}}\,$ is tangent to the vacuum foliation of the HP section $\,\Si^{\rm HP}\,$ (the latter being assumed to exist) iff $\,X\equiv X^A\,t_A\,$ is a \emph{constant} (over $\,\Si^{\rm HP}$) solution to the set of $\,\unl\d-p-q-1\,$ linear equations
\qq\nn
\sfP^\ggt_{\ \egt\oplus\dgt}\bigl(X\,H\bigl(\si_i^{\rm vac}(\unl\chi_i)\bigr)\bigr)=0\,,
\qqq
written in terms of the endomorphism 
\qq\nn
H\bigl(\si_i^{\rm vac}(\unl\chi_i)\bigr)=\t^A\ox H_A^{\ B}\bigl(\si_i^{\rm vac}(\unl\chi_i)\bigr)\,t_B\ :\ \ggt\circlearrowleft
\qqq
in which $\,\{\t^A\}^{A\in\ovl{0,D}}\,$ is the basis of $\,\ggt^*\,$ dual to $\,\{t_A\}_{A\in\ovl{0,D}}$,
\qq\nn
\t^A(t_B)=\d^A_{\ B}\,.
\qqq
We shall call the linear span of these vector fields the \textbf{residual global-supersymmetry subspace of} $\,{\rm Vac}(\sgt\Bgt^{{\rm (HP)}}_{p,\la_p})\,$ and denote it as
\qq\label{eq:resglobsusysubssp}
S^{\rm HP, vac}_\txG\subset\G\bigl({\rm Vac}\bigl(\sgt\Bgt^{{\rm (HP)}}_{p,\la_p}\bigr)\bigr)\,.
\qqq
The subspace is closed under the supercommutator, and so modelled on a Lie superalgebra, to be termed the \textbf{residual global-supersymmetry subalgebra of} $\,\ggt\,$ and denoted as
\qq\label{eq:resglobsusysubsalg}
\sgt_{\rm vac}\subset\ggt\,,
\qqq
\eerop
\beroof
Obvious.
\eroof

\noindent We shall, next, examine another superdistribution within $\,\cT\Si^{\rm HP}\,$ distinguished by field-theoretic considerations.

\subsection{The $\k$-symmetry superdistribution -- an odd resolution of the vacuum}\label{sub:locSUSYgeom} 

Our next objective is the identification and investigation of a regular ({\it i.e.}, non-coincidental) enhancement of the residual hidden gauge-symmetry, modelled on the Lie algebra $\,\hgt_{\rm vac}$,\ of the GS super-$\si$-model in the HP formulation that occurs upon restriction of field configurations $\,\widehat\xi\in[\Om_p,\cM_{\txH_{\rm vac}}]\,$ to the HP/NG correspondence sector and takes the form of a superdistribution 
\qq\nn
\cG\cS\bigl(\sgt\Bgt^{{\rm (HP)}}_{p,\la_p}\bigr)\subset{\rm Corr}_{\rm HP}\bigl(\sgt\Bgt^{{\rm (HP)}}_{p,\la_p}\bigr)
\qqq 
with the defining property: \emph{arbitrary} variations $\,\d\widehat\xi\in[\Om_p,\cG\cS(\sgt\Bgt^{{\rm (HP)}}_{p,\la_p})]\,$ leave the DF amplitude unchanged (to the linear order) \emph{for field configurations from the correspondence sector}. We shall call it the \textbf{enhanced gauge-symmetry superdistribution of} $\,\sgt\Bgt^{{\rm (HP)}}_{p,\la_p}\,$ in what follows and demand that it descend to $\,\cM_{\txH_{\rm vac}}\,$ and be globally linearised-supersymmetric. The obvious reason to perform partial reduction of the field theory under study through imposition of the IHC and the BLC and require $\txH_{\rm vac}$-descendability of the ensuing gauge symmetry is that we are ultimately interested in the physical gauge symmetry of the dual NG super-$\si$-model. Once identified, we shall subsequently intersect $\,\cG\cS(\sgt\Bgt^{{\rm (HP)}}_{p,\la_p})\,$ with the vacuum superdistribution $\,{\rm Vac}(\sgt\Bgt^{{\rm (HP)}}_{p,\la_p})$,\ obtaining an object which -- generically -- will turn out to be spanned on those of the generators $\,\cT_{\mu\,i}\,$ of $\,\cT\Si^{\rm HP}\,$ listed in Prop.\,\ref{prop:bastanshHP} that carry labels $\,\mu\,$ of $\,\tgt^{(1)}$,\ and in any event, it will contain a distinguished sub-superdistribution $\,\k(\sgt\Bgt^{{\rm (HP)}}_{p,\la_p})\subset{\rm Vac}(\sgt\Bgt^{{\rm (HP)}}_{p,\la_p})\,$ of this form. Its tentative physical interpretation as an infinitesimal \emph{odd} local symmetry of the vacuum of an effectively \emph{topological} field theory prompts three intertwined geometric and field-theoretic questions, and it is only an affirmative answer to all three of them that grants $\,\k(\sgt\Bgt^{{\rm (HP)}}_{p,\la_p})\,$ its status of a gauge supersymmetry of the field theory under study. First of all, we must enquire -- once again, but this time with regard to a proper substructure -- if $\,\k(\sgt\Bgt^{{\rm (HP)}}_{p,\la_p})\,$ descends to the physical supertarget $\,\cM_{\txH_{\rm vac}}$.\ If this is the case, we are bound to ask if it does so in a manner compatible with the global supersymmetry present, or -- in other words -- if it is globally linearised-supersymmetric. By the argument from the proof of Prop.\,\ref{prop:descissusy}, we know that a positive answer to the first question actually implies a positive answer to the second one. The last of the three questions regards the limit of its weak derived (super)flag,
\qq\nn
\k^{-\infty}\bigl(\sgt\Bgt^{{\rm (HP)}}_{p,\la_p}\bigr)\subseteq\cT\Si^{\rm HP}\,.
\qqq
In fact, it makes sense to speak of $\,\k(\sgt\Bgt^{{\rm (HP)}}_{p,\la_p})\,$ as a proper symmetry of the vacuum iff the limit stays within $\,{\rm Vac}(\sgt\Bgt^{{\rm (HP)}}_{p,\la_p})$,\ in which case it foliates the vacuum by gauge orbits. Under the physically preferred circumstances in which all three questions have been answered in the positive, we are dealing with a sub-superdistribution  
\qq\nn
\k\bigl(\sgt\Bgt^{{\rm (HP)}}_{p,\la_p}\bigr)\subseteq{\rm Vac}\bigl(\sgt\Bgt^{{\rm (HP)}}_{p,\la_p}\bigr)
\qqq
whose weak derived flag is closed under the supercommutator and hence modelled on a Lie sub-superalgebra 
\qq\nn
\ggt\sgt_{\rm vac}\bigl(\sgt\Bgt^{{\rm (HP)}}_{p,\la_p}\bigr)\subseteq\gt{vac}\bigl(\sgt\Bgt^{{\rm (HP)}}_{p,\la_p}\bigr)
\qqq
of the vacuum superalgebra of Prop.\,\ref{prop:vacsalg}. At this point, it seems fit to pause briefly in order to articulate an intuition drawn from experience with standard topological gauge field theories\footnote{The Reader is advised to consult Atiyah's foundational paper \cite{Atiyah:1989vu} for an axiomatic distillate obtained from the various {\it ad hoc} definitions of a topological field theory employed in the physics literature, with due emphasis on (and a rigorus rendering of) its constitutive properties that give a meaning to the name.} to which the GS super-$\si$-model in the HP formulation appears to bear structural affinity. The crucial observation is that in a topological field theory, in the absence of local degrees of freedom, propagation of configurations localised at Cauchy slices of the theory's spacetime is realised, or -- indeed -- replaced by (a class of) gauge transformations, {\it cp}, {\it e.g.}, the extensively studied three-dimensional Chern--Simons topological gauge field theory (of \Rcite{Freed:1992vw}) on a cylinder over a (punctured) Riemann surface for an explicit instantiation of this phenomenon. This leads us to anticipate that the ($\txH_{\rm vac}$-descendable and globally linearised-supersymmetric) superdistribution $\,\k(\sgt\Bgt^{{\rm (HP)}}_{p,\la_p})\,$ should fill up the vacuum superdistribution, {\it i.e.}, that it should be bracket-generating for $\,{\rm Vac}(\sgt\Bgt^{{\rm (HP)}}_{p,\la_p})$,\ so that the vector fields spanning the limit of its weak derived flag envelop the embedded worldvolume, winning it its name -- the \textbf{square root of the vacuum} -- that features in the title of the present paper. This may seem like a lot to expect (in particular, this expectation presupposes some sort of `completeness' of the Lie superalgebra $\,\ggt\sgt_{\rm vac}(\sgt\Bgt^{{\rm (HP)}}_{p,\la_p})\,$ as a model of residual gauge transformations preserving the vacuum) but explicit computations carried out for the superbackgrounds from those of Examples \ref{eq:s0gsMink}-\ref{eg:MT1} that possess a supersymmetric vacuum foliation actually confirm our intuition. With this reassuring note in mind, we now pass to the derivation of $\,\cG\cS(\sgt\Bgt^{{\rm (HP)}}_{p,\la_p})\,$ and $\,\k(\sgt\Bgt^{{\rm (HP)}}_{p,\la_p})$.

The point of departure of our analysis is formula \eqref{eq:logvarIHC} for an arbitrary variation of the DF amplitude in which on top of the restriction to the HP correspondence section, we also impose (on the supersymmetry algebra) the Even Effective-Mixing Constraints \eqref{eq:EMC0}-\eqref{eq:EMC3}, the Descendability Constraint \eqref{eq:DesConstr}, the No-Curvature and No-Torsion Constraints \eqref{eq:FlatnoTor1} and \eqref{eq:FlatnoTor2}, as well as the $\k$-Symmetry Constraints \eqref{eq:kConstr1} and \eqref{eq:kConstr2}, to the effect: 
\qq\nn
&&-\sfi\,\d_{\d\widehat\xi}\log\,\cA_{\rm DF}^{{\rm (HP)},p,\la_p}[\widehat\xi]\cr\cr
&=&\sum_{\t\in\Tgt_{p+1}}\,\int_\t\,\bigl(\si_{\imath_\t}^{\rm vac}\circ\widehat\xi_\t\bigr)^*\bigl[\tfrac{1}{p!}\,\d\theta_{\imath_\t}^\a\,\bigl(\bd1_{\d-d}-\sfP^{(1)}\bigr)^\g_{\ \a}\,\D_{\b\g\unl a{}_1\unl a{}_2\ldots\unl a{}_p}\,\Si_{\rm L}^\b\wedge\theta_{\rm L}^{\unl a{}_1}\wedge\theta_{\rm L}^{\unl a{}_2}\wedge\cdots\wedge\theta_{\rm L}^{\unl a{}_p}\cr\cr
&&\hspace{2.5cm}+\tfrac{1}{2(p-1)!}\,\d x^{\unl a}_{\imath_\t}\,\bigl(\bigl(\bd1_{\d-d}-\sfP^{(1)}\bigr)^\g_{\ \b}\,\D_{\a\g\unl a\unl a{}_2\unl a{}_3\ldots\unl a{}_p}\,\Si_{\rm L}^\a\wedge\Si_{\rm L}^\b\wedge\theta_{\rm L}^{\unl a{}_1}\wedge\theta_{\rm L}^{\unl a{}_2}\wedge\cdots\wedge\theta_{\rm L}^{\unl a{}_{p-1}}\bigr)\cr\cr
&&\hspace{2.5cm}+\tfrac{1}{2(p-1)!}\,\d x^{\widehat a}_{\imath_\t}\,\bigl(\chi_{\a\b\widehat a\unl a{}_1\unl a{}_2\ldots \unl a{}_{p-1}}\,\Si_{\rm L}^\a\wedge\Si_{\rm L}^\b\wedge\theta_{\rm L}^{\unl a{}_1}\wedge\theta_{\rm L}^{\unl a{}_2}\wedge\cdots\wedge\theta_{\rm L}^{\unl a{}_{p-1}}\cr\cr
&&\hspace{3.2cm}-\tfrac{2\la_p}{p}\,f_{\widehat a\widehat S}^{\ \ \ \unl b}\,\ep_{\unl b\unl a{}_1\unl a{}_2\ldots\unl a{}_p}\,\theta_{\rm L}^{\widehat S}\wedge\theta^{\unl a{}_1}_{\rm L}\wedge\theta^{\unl a{}_2}_{\rm L}\wedge\cdots\wedge\theta^{\unl a{}_p}_{\rm L}\bigr)\bigr]\,.
\qqq
The sole subtlety in the above expression lurks in its second line in which for any $(p-1)$-tuple of (even) vacuum indices $\,(\unl a{}_2,\unl a{}_3,\ldots,\unl a{}_p)\in\ovl{0,p}\,$ we have a \emph{pair} $\,(\unl a{}_0,\unl a{}_1)\in\ovl{0,p}\,$ of complementary indices, with, say, $\,\ep_{\unl a{}_0\unl a{}_1\unl a{}_2\ldots\unl a{}_p}=1$.\ In the light of the assumptions made with regard to the tensors $\,\D_{\a\b\unl b{}_0\unl b{}_1\ldots\unl b{}_p}$,\ we may write, for a $(p+1)$-tuple as above,
\qq\nn
&&\d x^{\unl a}_{\imath_\t}\,\bigl(\bd1_{\d-d}-\sfP^{(1)}\bigr)^\g_{\ \b}\,\D_{\a\g\unl a\unl a{}_2\unl a{}_3\ldots\unl a{}_p}=\bigl(\d x^{\unl a{}_0}_{\imath_\t}\,\D_{\a\g\unl a{}_0\unl a{}_2\unl a{}_3\ldots\unl a{}_p}+\d x^{\unl a{}_1}_{\imath_\t}\,\D_{\a\g\unl a{}_1\unl a{}_2\unl a{}_3\ldots\unl a{}_p}\bigr)\,\bigl(\bd1_{\d-d}-\sfP^{(1)}\bigr)^\g_{\ \b}\cr\cr
&=&\D_{\a\g\unl a{}_1\unl a{}_2\unl a{}_3\ldots\unl a{}_p}\,\bigl(\d x^{\unl a{}_1}_{\imath_\t}\,\d^\g_{\ \vep}+\d x^{\unl a{}_0}_{\imath_\t}\,\bigl(\D^{-1}_{\unl a{}_1\unl a{}_2\unl a{}_3\ldots\unl a{}_p}\bigr)^{\g\d}\,\D_{\d\vep\unl a{}_0\unl a{}_2\unl a{}_3\ldots\unl a{}_p}\bigr)\,\bigl(\bd1_{\d-d}-\sfP^{(1)}\bigr)^\vep_{\ \b}
\qqq
(\emph{no} summation over the repeated indices $\,\unl a{}_0\,$ and $\,\unl a{}_1$), and so the possibility arises of having a `chiral' gauge symmetry with
\qq\nn
\d x^{\unl a{}_1}_{\imath_\t}=-\tfrac{\tr_{\tgt^{(1)}}\bigl(\D^{-1}_{\unl a{}_1\unl a{}_2\unl a{}_3\ldots\unl a{}_p}\,\D_{\unl a{}_0\unl a{}_2\unl a{}_3\ldots\unl a{}_p}\,\bigl(\bd1_{\d-d}-\sfP^{(1)}\bigr)\bigr)}{\d-d-q}\,\d x^{\unl a{}_0}_{\imath_\t}
\qqq
(recall that we assumed $\,\sfP^{(1)}\neq\bd1_{\d-d}$), generated by vector fields
\qq\nn
\cT_{\unl a{}_0\unl a{}_1}=\cT_{\unl a{}_0}-\tfrac{\tr_{\tgt^{(1)}}\bigl(\D^{-1}_{\unl a{}_1\unl a{}_2\unl a{}_3\ldots\unl a{}_p}\,\D_{\unl a{}_0\unl a{}_2\unl a{}_3\ldots\unl a{}_p}\,\bigl(\bd1_{\d-d}-\sfP^{(1)}\bigr)\bigr)}{\d-d-q}\,\cT_{\unl a{}_1}\,.
\qqq
Above, we identify the matrices $\,\D_{\a\b\unl b{}_0\unl b{}_1\ldots\unl b{}_p}\,$ with the corresponding endomorphisms of $\,\tgt^{(1)}$.\ Inspection of the examples leads us to think of such symmetries as exceptional and treat them separately, which justifies imposition of\medskip

\noindent\textbf{The Even Achirality Constraints:} We assume, in the hitherto notation, the following identities
\qq\label{eq:EvAchir}
\Pi_{(\unl a{}_0,\unl a{}_1|\unl a{}_2,\unl a{}_3,\ldots,\unl a{}_p)}\equiv\tr_{\tgt^{(1)}}\bigl(\D^{-1}_{\unl a{}_0\unl a{}_2\unl a{}_3\ldots\unl a{}_p}\,\D_{\unl a{}_1\unl a{}_2\unl a{}_3\ldots\unl a{}_p}\,\bigl(\bd1_{\d-d}-\sfP^{(1)}\bigr)\bigr)\must 0
\qqq
to hold true for any $(p+1)$-tuple $\,(\unl a{}_0,\unl a{}_1,\ldots,\unl a{}_p)\in\ovl{0,p}$.
\medskip

\noindent With the Constraints in force, we readily infer from the above that, in a generic situation, the enhanced gauge-symmetry superdistribution is sewn from its local restrictions of the form
\qq\label{eq:GSsdistrogen}
\cG\cS\bigl(\sgt\Bgt^{{\rm (HP)}}_{p,\la_p}\bigr)\rstr_{\cV_i}=\bigoplus_{\unl\a=1}^q\,\corr{\cT_{\unl\a\,i}}\oplus\bigoplus_{\widehat S=D-\unl\d+1}^{D-\d}\,\corr{\cT_{\widehat S\,i}}
\qqq
confirmed -- without a single exception -- through scrutiny of Examples \ref{eq:s0gsMink}-\ref{eg:MT1}. Thus, we recover the \emph{full} hidden gauge symmetry of the dual formulation (the vertical component being realised trivially) augmented with translations along the Gra\ss mann-odd generators of the vacuum superdistribution. Our identification of the invariances as infinitesimal gauge symmetries stems from the following observation: Their very derivation implies that they can be regarded as `infinitesimal' \emph{right} translations of field configurations from the HP/NG correspondence sector in the directions of the supervector subspace 
\qq\label{eq:GSmodalg}
\ggt\sgt\bigl(\sgt\Bgt^{{\rm (HP)}}_{p,\la_p}\bigr)\equiv\tgt_{\rm vac}^{(1)}\oplus\dgt\subset\ggt\,.
\qqq 
While left translations do \emph{not} depend on the choice of the local sections of the principal $\txH_{\rm vac}$-bundle \eqref{eq:homasprinc} (with $\,\txK=\txH_{\rm vac}$) and, consequently, descend to the homogeneous space along the canonical projection $\,\pi_{\cM_{\txH_{\rm vac}}}$,\ the right ones \emph{do} depend on the precise form of the $\,\si_i^{\rm vac}$,\ and so the inherent \emph{locality} of these transformations is an irremovable consequence of the hidden gauge freedom of the super-$\si$-model ({\it cp} also \Rxcite{Sec.\,3}{McArthur:1999dy}). In view of the peculiar relation between the lagrangean density and the presymplectic form of the GS super-$\si$-model in the topological HP formulation, this identification is in keeping with the standard definition of gauge symmetries as generators of the kernel of the presymplectic form of the field theory, {\it cp} \Rcite{Gawedzki:1972ms}. 

Upon taking into account the structure of the vacuum superdistribution, we are led to 
\bedef\label{def:ksymmsdistro}
Adopt the hitherto notation, and in particular that of Prop.\,\ref{prop:bastanshHP}. The superdistribution 
\qq\label{eq:ksymsdistro}
\k\bigl(\sgt\Bgt^{{\rm (HP)}}_{p,\la_p}\bigr)\subset{\rm Vac}\bigl(\sgt\Bgt^{{\rm (HP)}}_{p,\la_p}\bigr)\,,\qquad\qquad
\k\bigl(\sgt\Bgt^{{\rm (HP)}}_{p,\la_p}\bigr)\rstr_{\cV_i}=\bigoplus_{\unl\a=1}^q\,\corr{\cT_{\unl\a\,i}}\,,
\qqq
spanned on generators of Gra\ss mann-odd local symmetries of the HP/NG correspondence sector of the (generic) Green--Schwarz super-$\si$-model in the Hughes--Polchinski formulation of Def.\,\ref{def:GSinHP}, shall be called the \textbf{$\k$-symmetry superdistribution of} $\,\sgt\Bgt^{{\rm (HP)}}_{p,\la_p}$.\ Whenever the limit of its weak derived flag is contained in the vacuum superdistribution, that is
\qq\nn
\k^{-\infty}\bigl(\sgt\Bgt^{{\rm (HP)}}_{p,\la_p}\bigr)\subseteq{\rm Vac}\bigl(\sgt\Bgt^{{\rm (HP)}}_{p,\la_p}\bigr)\,,
\qqq
we call the ensuing Lie superalgebra 
\qq\nn
\ggt\sgt_{\rm vac}\bigl(\sgt\Bgt^{{\rm (HP)}}_{p,\la_p}\bigr)\subseteq\gt{vac}\bigl(\sgt\Bgt^{{\rm (HP)}}_{p,\la_p}\bigr)
\qqq
the \textbf{vacuum gauge-symmetry superalgebra}, or -- for historical reasons -- the \textbf{$\k$-symmetry superalgebra}. 
\exdef
\brem 
Parenthetically, let us note that vacua with $\,q\equiv\dim\,\tgt_{\rm vac}^{(1)}\,$ out of the $\,D_{d,1}\equiv\dim\,\tgt^{(1)}\,$ supercharges (or `supersymmetries') of the theory left unbroken are usually referred to as \textbf{$\frac{q}{D_{d,1}}$-BPS states} in the physics parlance. Hence, in the regular ({\it i.e.}, $\txH_{\rm vac}$-descendable and integrable) case, we might speak of the \textbf{BPS fraction of the vacuum of} $\,\sgt\Bgt^{{\rm (HP)}}_{p,\la_p}$,
\qq\nn
{\rm BPS}\bigl(\sgt\Bgt^{{\rm (HP)}}_{p,\la_p}\bigr):=\tfrac{q}{D_{d,1}}\equiv\tfrac{1}{D_{d,1}}\,\tr_{\tgt^{(1)}}\,\sfP^{(1)}\,.
\qqq
\erem

\noindent Our derivation of the $\k$-symmetry superdistribution paves the way to further analysis, subordinated to the primary goal of answering the three basic structural questions formulated above. Upon invoking our former considerations and results, we readily arrive at
\berop\label{prop:kglobsusy}
In the hitherto notation and under the assumption that the Even Effective-Mixing Constraints \eqref{eq:EMC0}-\eqref{eq:EMC3}, the Descendability Constraint \eqref{eq:DesConstr}, the No-Curvature and No-Torsion Constraints \eqref{eq:FlatnoTor1} and \eqref{eq:FlatnoTor2}, the $\k$-Symmetry Constraints \eqref{eq:kConstr1} and \eqref{eq:kConstr2}, as well as the Even Achirality Constraints \eqref{eq:EvAchir} are satisfied, the $\k$-symmetry superdistribution of the HP superbackground $\,\sgt\Bgt^{{\rm (HP)}}_{p,\la_p}\,$ given in \Reqref{eq:ksymsdistro} is globally linearised-supersymmetric in the sense of Def.\,\ref{def:globlinsusysdistro} iff 
\qq\nn
[\hgt_{\rm vac},\tgt^{(1)}_{\rm vac}]\subset\tgt^{(1)}_{\rm vac}\,.
\qqq
Thus, $\,\k(\sgt\Bgt^{{\rm (HP)}}_{p,\la_p})\,$ is globally linearised-supersymmetric iff $\,{\rm Vac}(\Bgt^{{\rm (HP)}}_{p,\la_p})\,$ is. In particular, an $\txH_{\rm vac}$-descendable $\k$-symmetry superdistribution is globally linearised-supersymmetric.
\eerop
\beroof
Follows directly from Prop.\,\ref{prop:descissusy} and its proof.
\eroof

\noindent We also have
\berop\label{prop:kwdfinvac}
In the hitherto notation and under the assumption that the Even Effective-Mixing Constraints \eqref{eq:EMC0}-\eqref{eq:EMC3}, the Descendability Constraint \eqref{eq:DesConstr}, the No-Curvature and No-Torsion Constraints \eqref{eq:FlatnoTor1} and \eqref{eq:FlatnoTor2}, the $\k$-Symmetry Constraints \eqref{eq:kConstr1} and \eqref{eq:kConstr2}, as well as the Even Achirality Constraints \eqref{eq:EvAchir} are satisfied, the limit $\,\k^{-\infty}(\sgt\Bgt^{{\rm (HP)}}_{p,\la_p})\,$ of the weak derived flag $\,\k^\bullet(\sgt\Bgt^{{\rm (HP)}}_{p,\la_p})\,$ of the $\k$-symmetry superdistribution $\,\k(\sgt\Bgt^{{\rm (HP)}}_{p,\la_p})\,$ of the HP superbackground $\,\sgt\Bgt^{{\rm (HP)}}_{p,\la_p}$,\ given in \Reqref{eq:ksymsdistro}, is contained in the HP vacuum superdistribution $\,{\rm Vac}(\sgt\Bgt^{{\rm (HP)}}_{p,\la_p})\,$ of Def.\,\ref{def:vacHPsdistro},
\qq\nn
\k^{-\infty}\bigl(\sgt\Bgt^{{\rm (HP)}}_{p,\la_p}\bigr)\subseteq{\rm Vac}\bigl(\sgt\Bgt^{{\rm (HP)}}_{p,\la_p}\bigr)\,,
\qqq
if the Vacuum-Superalgebra Constraints of Prop.\,\ref{prop:vacsalg} are satisfied. If, in addition, the endomorphisms
\qq\nn
f_{\unl a}\equiv f_{\a\b}^{\ \ \ \unl a}\,\d^{\b\g}\,q^\a\ox Q_\g\in{\rm End}\bigl(\tgt^{(1)}\bigr)\,,\qquad\unl a\in\ovl{0,p}
\qqq 
(expressed in the notation of \Reqref{eq:dualQ}) satisfy the \textbf{Odd Achirality Constraints} 
\qq\nn
&\forall_{\unl a\in\ovl{0,p}}\ \exists_{f_{\unl a}^{-1}\in{\rm End}(\tgt^{(1)})}\ :\ f_{\unl a}^{-1}\circ f_{\unl a}=\id_{\tgt^{(1)}}\,,&\cr\cr
&\forall_{\unl a,\unl b\in\ovl{0,p}}\ \exists_{\la_{\unl a}\in\bR^\x}\ :\ \Pi_{\unl a\unl b}\equiv\tr_{\tgt^{(1)}}\bigl(f_{\unl a}^{-1}\circ\sfP^{(1)\,{\rm T}}\circ f_{\unl b}\circ\sfP^{(1)}\bigr)\must\la_{\unl a}\,\d_{\unl a\unl b}\,,&
\qqq
$\,\k(\sgt\Bgt^{{\rm (HP)}}_{p,\la_p})\,$ is bracket-generating for $\,{\rm Vac}(\sgt\Bgt^{{\rm (HP)}}_{p,\la_p})$,
\qq\nn
\k^{-\infty}\bigl(\sgt\Bgt^{{\rm (HP)}}_{p,\la_p}\bigr)={\rm Vac}\bigl(\sgt\Bgt^{{\rm (HP)}}_{p,\la_p}\bigr)\,,
\qqq
and the $\k$-symmetry superalgebra of $\,\sgt\Bgt^{{\rm (HP)}}_{p,\la_p}\,$ coincides with the vacuum superalgebra of $\,\sgt\Bgt^{{\rm (HP)}}_{p,\la_p}$,\ as defined {\it ibidem},
\qq\nn
\ggt\sgt_{\rm vac}\bigl(\sgt\Bgt^{{\rm (HP)}}_{p,\la_p}\bigr)=\gt{vac}\bigl(\sgt\Bgt^{{\rm (HP)}}_{p,\la_p}\bigr)\,.
\qqq
\eerop
\beroof
As for the first part, it is fully analogous to that of Prop.\,\ref{prop:descissusy}.
\eroof

\noindent In view of the physical significance of the various results scattered in the propositions written out and cited heretofore, we summarise our findings in 
\bethe\label{thm:kdemyst}
Let $\,\txG=(|\txG|,\cO_\txG)\,$ be a Lie supergroup with the tangent Lie superalgebra $\,\ggt=\ggt^{(0)}\oplus\ggt^{(1)}$,\ and let $\,\txH\,$ and $\,\txH_{\rm vac}\subset\txH\,$ be two Lie subgroups of the body $\,|\txG|\,$ of $\,\txG\,$ that correspond to two reductive decompositions of $\,\ggt$,
\qq\nn
\ggt=\fgt\oplus\kgt\,,\qquad(\fgt,\kgt)\in\{(\tgt,\hgt),(\tgt\oplus\dgt,\hgt_{\rm vac})\}\,,
\qqq
in which $\,\hgt=\dgt\oplus\hgt_{\rm vac}\,$ and $\,\hgt_{\rm vac}\,$ are the tangent Lie algebras of $\,\txH\,$ and $\,\txH_{\rm vac}$,\ respectively, satisfying the relations
\qq\nn
[\hgt_{\rm vac},\hgt_{\rm vac}]\subset\hgt_{\rm vac}\,,\qquad\qquad[\hgt_{\rm vac},\dgt]\subset\dgt\,,\qquad\qquad[\dgt,\dgt]\subset\hgt_{\rm vac}\,,
\qqq
with the direct-sum complement $\,\tgt\,$ of $\,\hgt\,$ in $\,\ggt\,$ further decomposing as
\qq\nn
\tgt=\tgt_{\rm vac}\oplus\egt
\qqq
into supervector sub-spaces 
\qq\nn
\tgt_{\rm vac}=\tgt^{(0)}_{\rm vac}\oplus\tgt^{(1)}_{\rm vac}\,,\qquad\qquad\egt=\egt^{(0)}\oplus\egt^{(1)}
\qqq 
with the properties
\qq\nn
[\hgt_{\rm vac},\tgt_{\rm vac}^{(0)}]\subset\tgt_{\rm vac}^{(0)}\,,\qquad\qquad[\hgt_{\rm vac},\egt^{(0)}]\subset\egt^{(0)}\,,\qquad\qquad[\dgt,\tgt^{(0)}_{\rm vac}]\subset\egt^{(0)}\,,\qquad\qquad[\dgt,\egt^{(0)}]\subset\tgt_{\rm vac}^{(0)}\,,
\qqq
and such that the relations
\qq\nn
&[\tgt^{(0)}_{\rm vac},\tgt^{(0)}_{\rm vac}]\subset\hgt_{\rm vac}\supset[\egt^{(0)},\egt^{(0)}]\,,\qquad\qquad[\tgt_{\rm vac}^{(0)},\egt^{(0)}]\subset\dgt\,,&\cr\cr
&\{\tgt_{\rm vac}^{(1)},\tgt_{\rm vac}^{(1)}\}\subset\tgt_{\rm vac}^{(0)}\oplus\hgt&
\qqq
hold true. Consider the principal (super)bundles 
\qq\nn
\alxydim{@C=1.5cm@R=1.5cm}{ \txK \ar[r] & \txG \ar[d]^{\pi_{\txG/\txK}} \\ & \txG/\txK}\,,\qquad\txK\in\{\txH,\txH_{\rm vac}\}\,,
\qqq
coming with the respective families $\,\{\si_i^{\rm vac}\ :\ \cU_i^{\txH_{\rm vac}}\too\txG\}_{i\in I_{\txH_{\rm vac}}}\,$ and $\,\{\si_i^{\xcancel{\rm vac}}\ :\ \cU_i^\txH\too\txG\}_{i\in I_{\txH_{\rm vac}}}\,$ of sections associated with open covers $\,\{\cU_i^\txK\}_{i\in I_\txK}\,$ of the respective bases, defined on p.\,\pageref{page:sivacnovac}. The Green--Schwarz super-$\si$-model in the Hughes--Polchinski formulation of Def.\,\ref{def:GSinHP} for the superbackground $\,\sgt\Bgt^{{\rm (HP)}}_{p,\la_p}\,$ of \Reqref{eq:HPspbgrnd} satisfying the $\k$-Symmetry Constraints \eqref{eq:kConstr1} and \eqref{eq:kConstr2} as well as the $\G$-Constraints \eqref{eq:GamConstr} -- defined for objects of the mapping supermanifold $\,[\Om_p,\Si^{\rm HP}]\,$ of the $(p+1)$-dimensional worldvolume $\,\Om_p\,$ into the Hughes--Polchinski section
\qq\nn
\Si^{\rm HP}=\bigsqcup_{i\in I_{\txH_{\rm vac}}}\,\si_i^{\rm vac}\bigl(\cU_i^{\txH_{\rm vac}}\bigr)\subset\txG
\qqq
and equivalent to the Green--Schwarz super-$\si$-model in the Nambu--Goto formulation of Def.\,\ref{def:GSinNG} for the superbackground $\,\sgt\Bgt^{{\rm (GS)}}_p\,$ of \Reqref{eq:GSspbgrnd} in the HP/NG correspondence sector composed of mappings with tangents restricted to the correspondence superdistribution 
\qq\nn
{\rm Corr}_{\rm HP/NG}\bigl(\sgt\Bgt^{{\rm (HP)}}_{p,\la_p}\bigr)={\rm Ker}\,\bigl(\sfP^\ggt_{\ \egt^{(0)}}\circ\theta_{\rm L}\rstr_{\cT\Si^{\rm HP}}\bigr)\subset\cT\Si^{\rm HP}
\qqq
of Def.\,\ref{def:Corrsdistro} (upon postcomposition with tangents of the $\,\si_i^{\rm vac}$) -- determines the vacuum superdistribution 
\qq\nn
{\rm Vac}\bigl(\sgt\Bgt^{{\rm (HP)}}_{p,\la_p}\bigr)={\rm Ker}\,\bigl(\sfP^\ggt_{\ \egt\oplus\dgt}\circ\theta_{\rm L}\rstr_{\cT\Si^{\rm HP}}\bigr)\subset{\rm Corr}_{\rm HP/NG}\bigl(\sgt\Bgt^{{\rm (HP)}}_{p,\la_p}\bigr)
\qqq
of Def.\,\ref{def:vacHPsdistro} that is globally linearised-supersymmetric iff
\qq\nn
[\hgt_{\rm vac},\tgt_{\rm vac}^{(1)}]\subset\tgt_{\rm vac}^{(1)}\,,
\qqq
which is also a necessary condition for it to descend to $\,\cM_{\txH_{\rm vac}}$.\ The vacuum superdistribution is involutive, and hence defines the vacuum foliation of the Hughes--Polchinski section of Prop.\,\ref{prop:vacsalg} iff the additional conditions
\qq\nn
\{\tgt^{(1)}_{\rm vac},\tgt^{(1)}_{\rm vac}\}\subset\tgt_{\rm vac}^{(0)}\oplus\hgt_{\rm vac}\,,\qquad\qquad[\tgt_{\rm vac}^{(0)},\tgt^{(1)}_{\rm vac}]\subset\tgt^{(1)}_{\rm vac}\,,\qquad\qquad[\hgt_{\rm vac},\tgt^{(1)}_{\rm vac}]\subset\tgt^{(1)}_{\rm vac}
\qqq
are satisfied, making the vacuum supervector space
\qq\nn
\gt{vac}\bigl(\sgt\Bgt^{{\rm (HP)}}_{p,\la_p}\bigr)=\tgt_{\rm vac}\oplus\hgt_{\rm vac}\subset\ggt
\qqq
into the vacuum (Lie) superalgebra.

Furthermore, the Green--Schwarz super-$\si$-model in the Hughes--Polchinski formulation distinguishes the enhanced gauge-symmetry superdistribution 
\qq\nn
\cG\cS\bigl(\sgt\Bgt^{{\rm (HP)}}_{p,\la_p}\bigr)\subset{\rm Corr}_{\rm HP}\bigl(\sgt\Bgt^{{\rm (HP)}}_{p,\la_p}\bigr)
\qqq
of local symmetries of its HP/NG correspondence sector. The associated $\k$-symmetry superdistribution 
\qq\nn
\k\bigl(\sgt\Bgt^{{\rm (HP)}}_{p,\la_p}\bigr)={\rm Ker}\,\bigl(\bigl(\id_\ggt-\sfP^\ggt_{\ \tgt^{(1)}_{\rm vac}}\bigr)\circ\theta_{\rm L}\rstr_{\cT\Si^{\rm HP}}\bigr)
\qqq
of Def.\,\ref{def:ksymmsdistro} is globally linearised-supersymmetric iff 
\qq\nn
[\hgt_{\rm vac},\tgt_{\rm vac}^{(1)}]\subset\tgt_{\rm vac}^{(1)}\,,
\qqq
which is also a necessary condition for it to descend to $\,\cM_{\txH_{\rm vac}}$.\ The limit $\,\k^{-\infty}(\sgt\Bgt^{{\rm (HP)}}_{p,\la_p})\,$ of its weak derived flag coincides with the limit $\,{\rm Vac}^{-\infty}(\sgt\Bgt^{{\rm (HP)}}_{p,\la_p})\,$ of the weak derived flag of the vacuum superdistribution if the conditions 
\qq\nn
&\forall_{\unl a\in\ovl{0,p}}\ \exists_{f_{\unl a}^{-1}\in{\rm End}(\tgt^{(1)})}\ :\ f_{\unl a}^{-1}\circ f_{\unl a}=\id_{\tgt^{(1)}}\,,&\cr\cr
&\forall_{\unl a,\unl b\in\ovl{0,p}}\ \exists_{\la_{\unl a}\in\bR^\x}\ :\ \tr_{\tgt^{(1)}}\bigl(f_{\unl a}^{-1}\circ\sfP^{(1)\,{\rm T}}\circ f_{\unl b}\circ\sfP^{(1)}\bigr)=\la_{\unl a}\,\d_{\unl a\unl b}\,,&
\qqq
are satisfied, and so -- whenever this happens -- it lies in that superdistribution iff the latter is involutive, in which case the $\k$-symmetry superdistribution is bracket-generating for $\,{\rm Vac}(\sgt\Bgt^{{\rm (HP)}}_{p,\la_p})$,\ that is
\qq\label{eq:lwdfkisVac}
\k^{-\infty}\bigl(\sgt\Bgt^{{\rm (HP)}}_{p,\la_p}\bigr)={\rm Vac}\bigl(\sgt\Bgt^{{\rm (HP)}}_{p,\la_p}\bigr)\,,
\qqq 
and the $\k$-symmetry superalgebra $\,\ggt\sgt_{\rm vac}(\sgt\Bgt^{{\rm (HP)}}_{p,\la_p})\,$ of Def.\,\ref{def:ksymmsdistro} coincides with the vacuum superalgebra,
\qq\nn
\ggt\sgt_{\rm vac}\bigl(\sgt\Bgt^{{\rm (HP)}}_{p,\la_p}\bigr)=\gt{vac}\bigl(\sgt\Bgt^{{\rm (HP)}}_{p,\la_p}\bigr)\,.
\qqq 
\ethe

\noindent Altogether, we conclude that physical considerations favour -- as leading to a meaningful `localisation' of the vacuum within the supertarget which is compatible with the global supersymmetry present -- superbackgrounds with involutive (and $\txH_{\rm vac}$-descendable) vacuum superdistributions, and these are bracket-generated by their $\k$-symmetry sub-superdistributions that envelop the embedded vacua, in conformity with our TFT intuition. The vacua are expected, furthermore, to exhibit global (linearised) supersymmetry, realised by the residual global-supersymmetry subspace $\,S_\txG^{\rm HP, vac}\subset{\rm Vac}(\sgt\Bgt^{{\rm (HP)}}_{p,\la_p})$.\ Given the simple and highly constrained\footnote{{\it I.a.}, by the super-Jacobi identities of the mother Lie superalgebra $\,\ggt$.} (super)algebraic model of the $\k$-symmetry superdistribution and of its weak derived flag, it is completely straightforward to identify the sources of a potential obstruction against both: integrability of the vacuum superdistribution and identity \eqref{eq:lwdfkisVac}. The first of these anomalies is the projection
\qq\nn
\agt_{\rm int}:=\sfP^\ggt_{\ \egt^{(1)}}\bigl([\tgt_{\rm vac}^{(0)},\tgt_{\rm vac}^{(1)}]\bigr)
\qqq
that quantifies the violation of the second of the Vacuum-Superalgebra Constraints of Prop.\,\ref{prop:vacsalg}. Note that it also encodes the potential violation of the first of these Constraints as a non-zero projection
\qq\nn
\sfP^\ggt_{\ \dgt}\bigl(\{\tgt_{\rm vac}^{(1)},\tgt_{\rm vac}^{(1)}\}\bigr)\neq 0
\qqq 
of the anticommutator of the supercharges of the vacuum onto the vacuum-changing component $\,\dgt\,$ of the mother gauge-symmetry algebra $\,\hgt\,$ yields
\qq\nn
\sfP^\ggt_{\ \egt^{(0)}}\bigl([\{\tgt_{\rm vac}^{(1)},\tgt_{\rm vac}^{(1)}\},\tgt_{\rm vac}^{(0)}]\bigr)\neq 0
\qqq
in virtue of the Even Effective-Mixing Constraints \eqref{eq:EMC1}, so that if we assume 
\qq\nn
\agt_{\rm int}=0\,,
\qqq
we are immediately led -- {\it via} the super-Jacobi identity restricted to the triple $\,\tgt_{\rm vac}^{(1)}\x\tgt_{\rm vac}^{(1)}\x\tgt_{\rm vac}^{(0)}\,$ -- to the contradiction 
\qq\nn
\sfP^\ggt_{\ \egt^{(0)}}\bigl([\{\tgt_{\rm vac}^{(1)},\tgt_{\rm vac}^{(1)}\},\tgt_{\rm vac}^{(0)}]\bigr)\subset\sfP^\ggt_{\ \egt^{(0)}}\bigl(\{[\tgt_{\rm vac}^{(0)},\tgt_{\rm vac}^{(1)}],\tgt_{\rm vac}^{(1)}\}\bigr)\subset\sfP^\ggt_{\ \egt^{(0)}}\bigl(\{\tgt_{\rm vac}^{(1)},\tgt_{\rm vac}^{(1)}\}\bigr)\subset\sfP^\ggt_{\ \egt^{(0)}}\bigl(\tgt_{\rm vac}^{(0)}\oplus\hgt\bigr)=0\,.
\qqq
The last (and independent) anomaly is the projection 
\qq\nn
\agt_{\rm susy}:=\sfP^\ggt_{\ \egt^{(1)}}\bigl([\hgt_{\rm vac},\tgt_{\rm vac}^{(1)}]\bigr)
\qqq
that captures the potential obstruction against global (linearised) supersymmetry. Prior to establishing the first step towards a consistent higher-geometric lift of our symmetry analysis in the next section, we present below a complete list of the superalgebraic models and anomalies of the $\k$-symmetry superdistributions for the superbackgrounds from Examples \ref{eq:s0gsMink}-\ref{eg:MT1}, writing them out in the convention
\qq\nn
\agt_{\rm int}=\bigl(X^{\unl a}\,[P_{\unl a},\sfP^{(1)\,\b}_{\ \ \ \ \ \a}\,Q_\b]\bigr)\,,\qquad\qquad\agt_{\rm susy}=\bigl(\Phi^{\unl S}\,[J_{\unl S},\sfP^{(1)\,\b}_{\ \ \ \ \ \a}\,Q_\b]\bigr)\,,
\qqq
with $\,X^{\unl a}\,$ and $\,\Phi^{\unl S}\,$ arbitrary (real) parameters. Note that all regular cases have the BPS fraction $\,\frac{1}{2}$.
\medskip

\beg\textbf{The square root of the Green--Schwarz super-0-brane in $\,{\rm sISO}(9,1\,|\,32)/{\rm SO}(9)$.}\label{eg:GSs0k}
The tensor
\qq\nn
f_0\equiv\ovl\G{}^0
\qqq
satisfies the Odd Achirality Constraints
\qq\nn
\Pi_{00}=\tfrac{1}{2}\,\tr_{\tgt^{(1)}}\bigl(\G_0\,\G^0\,\bigl(\bd1_{32}+\G^0\,\G_{11}\bigr)\bigr)=\tfrac{1}{2}\,\tr_{\tgt^{(1)}}\bigl(\bd1_{32}+\G^0\,\G_{11}\bigr)=16\,.
\qqq
The Even Achirality Constraints are satisfied trivially. The $\k$-symmetry superdistribution with restrictions
\qq\nn
\k\bigl({\rm sISO}(9,1\,|\,32)/{\rm SO}(9),\underset{\tx{\ciut{(2)}}}{\widehat\chi}\hspace{-1pt}{}^{\rm GS}\bigr)\rstr_{\cV_i}=\bigoplus_{\unl\a=1}^{16}\,\corr{\cT_{\unl\a\,i}}
\qqq 
is an ${\rm SO}(9)$-descendable superdistribution with the limit of its weak derived flag with restrictions
\qq\nn
\k^{-\infty}\bigl({\rm sISO}(9,1\,|\,32)/{\rm SO}(9),\underset{\tx{\ciut{(2)}}}{\widehat\chi}\hspace{-1pt}{}^{\rm GS}\bigr)\rstr_{\cV_i}=\bigoplus_{\unl\a=1}^{16}\,\corr{\cT_{\unl\a\,i}}\oplus\corr{\cT_{0\,i}}\equiv{\rm Vac}\bigl({\rm sISO}(9,1\,|\,32)/{\rm SO}(9),\underset{\tx{\ciut{(2)}}}{\widehat\chi}\hspace{-1pt}{}^{\rm GS}\bigr)\rstr_{\cV_i}\,.
\qqq 
\eeg

\beg\textbf{The square root of the Green--Schwarz super-$(4k+1)$-brane in \linebreak $\,{\rm sISO}(d,1\,|\,D_{d,1})/({\rm SO}(4k+1,1)\x{\rm SO}(d-4k-1))\,$ for $\,k\in\{1,2\}$.} 
The tensors
\qq\nn
\D_{\unl a{}_1\unl a{}_2\ldots\unl a{}_{4k+1}}\equiv 4(4k+1)!\,\ep_{\unl a\unl a{}_1\unl a{}_2\ldots\unl a{}_{4k+1}}\,\ovl\G{}^{\unl a}\,,\qquad\qquad f_{\unl a}\equiv\ovl\G{}^{\unl a}
\qqq
satisfy the Even Achirality Constraints,
\qq\nn
\Pi_{(\unl a{}_0,\unl a{}_1|\unl a{}_2,\unl a{}_3,\ldots,\unl a{}_{4k+1})}&=&-\tfrac{1}{2}\,\tr_{\tgt^{(1)}}\bigl(\G_{\unl a{}_1}\,\G^{\unl a{}_0}\,\bigl(\bd1_{D_{d,1}}-\G^0\,\G^1\,\cdots\,\G^{4k+1}\bigr)\bigr)=\tfrac{1}{2}\,\eta_{\unl a{}_0\unl a{}_0}\,\tr_{\tgt^{(1)}}\bigl(\G^{\unl a{}_2}\,\G^{\unl a{}_2}\,\cdots\,\G^{\unl a{}_{4k+1}}\bigr)\cr\cr
&=&\tfrac{(-1)^{4k-1}}{2}\,\eta_{\unl a{}_0\unl a{}_0}\,\tr_{\tgt^{(1)}}\bigl(\G^{\unl a{}_3}\,\G^{\unl a{}_4}\,\cdots\,\G^{\unl a{}_{4k+1}}\,\G^{\unl a{}_2}\bigr)\equiv-\Pi_{(\unl a{}_0,\unl a{}_1|\unl a{}_2,\unl a{}_3,\ldots,\unl a{}_{4k+1})}=0\,,
\qqq
and the Odd Achirality Constraints,
\qq\nn
\Pi_{\unl a\unl b}=\tfrac{1}{2}\,\tr_{\tgt^{(1)}}\bigl(\G_{\unl a}\,\G^{\unl b}\,\bigl(\bd1_{D_{d,1}}+\G^0\,\G^1\,\cdots\,\G^{4k+1}\bigr)\bigr)=\d_{\unl a}^{\ \unl b}\,\tfrac{1}{2}\,\tr_{\tgt^{(1)}}\bigl(\bd1_{D_{d,1}}+\G^0\,\G^1\,\cdots\,\G^{4k+1}\bigr)=\tfrac{D_{d,1}}{2}\,\d_{\unl a\unl b}\,.
\qqq
The $\k$-symmetry superdistribution with restrictions
\qq\nn
\k\bigl({\rm sISO}(d,1\,|\,D_{d,1})/({\rm SO}(4k+1,1)\x{\rm SO}(d-4k-1)),\underset{\tx{\ciut{(p+2)}}}{\widehat\chi}\hspace{-6pt}{}^{\rm GS}\bigr)\rstr_{\cV_i}=\bigoplus_{\unl\a=1}^{\frac{D_{d,1}}{2}}\,\corr{\cT_{\unl\a\,i}}
\qqq 
is an $({\rm SO}(4k+1,1)\x{\rm SO}(d-4k-1))$-descendable superdistribution with the limit of its weak derived flag with restrictions
\qq\nn
&&\k^{-\infty}\bigl({\rm sISO}(d,1\,|\,D_{d,1})/({\rm SO}(4k+1,1)\x{\rm SO}(d-4k-1)),\underset{\tx{\ciut{(p+2)}}}{\widehat\chi}\hspace{-6pt}{}^{\rm GS}\bigr)\rstr_{\cV_i}=\bigoplus_{\unl\a=1}^{\frac{D_{d,1}}{2}}\,\corr{\cT_{\unl\a\,i}}\oplus\bigoplus_{\unl a=0}^{4k+1}\,\corr{\cT_{\unl a\,i}}\cr\cr
&\equiv&{\rm Vac}\bigl({\rm sISO}(d,1\,|\,D_{d,1})/({\rm SO}(4k+1,1)\x{\rm SO}(d-4k-1)),\underset{\tx{\ciut{(p+2)}}}{\widehat\chi}\hspace{-6pt}{}^{\rm GS}\bigr)\rstr_{\cV_i}\,.
\qqq 
\eeg

\beg\textbf{The square root of the Green--Schwarz super-$(4k+2)$-brane in \linebreak $\,{\rm sISO}(d,1\,|\,D_{d,1})/({\rm SO}(4k+2,1)\x{\rm SO}(d-4k-2))\,$ for $\,k\in\{0,1\}$.} 
The tensors
\qq\nn
\D_{\unl a{}_1\unl a{}_2\ldots\unl a{}_{4k+2}}\equiv 4(4k+2)!\,\ep_{\unl a\unl a{}_1\unl a{}_2\ldots\unl a{}_{4k+2}}\,\ovl\G{}^{\unl a}\,,\qquad\qquad f_{\unl a}\equiv\ovl\G{}^{\unl a}
\qqq
satisfy the Even Achirality Constraints,
\qq\nn
\Pi_{(\unl a{}_0,\unl a{}_1|\unl a{}_2,\unl a{}_3,\ldots,\unl a{}_{4k+2})}&=&-\tfrac{1}{2}\,\tr_{\tgt^{(1)}}\bigl(\G_{\unl a{}_1}\,\G^{\unl a{}_0}\,\bigl(\bd1_{D_{d,1}}-\G^0\,\G^1\,\cdots\,\G^{4k+2}\bigr)\bigr)=\tfrac{1}{2}\,\eta_{\unl a{}_0\unl a{}_0}\,\tr_{\tgt^{(1)}}\bigl(\G^{\unl a{}_2}\,\G^{\unl a{}_2}\,\cdots\,\G^{\unl a{}_{4k+2}}\bigr)\cr\cr
&=&\tfrac{1}{2}\,\eta_{\unl a{}_0\unl a{}_0}\,\tr_{\tgt^{(1)}}\bigl(\G^{\unl a{}_{4k+2}\,{\rm T}}\,\G^{\unl a{}_{4k}\,{\rm T}}\,\cdots\,\,\G^{\unl a{}_2\,{\rm T}}\bigr)=\tfrac{(-1)^{4k+1}}{2}\,\eta_{\unl a{}_0\unl a{}_0}\,\tr_{\tgt^{(1)}}\bigl(C\,\G^{\unl a{}_{4k+2}}\,\G^{\unl a{}_{4k}}\,\cdots\,\,\G^{\unl a{}_2}\,C^{-1}\bigr)\cr\cr
&=&\tfrac{(-1)^{4k+1+\frac{4k(4k+1)}{2}}}{2}\,\eta_{\unl a{}_0\unl a{}_0}\,\tr_{\tgt^{(1)}}\bigl(\G^{\unl a{}_2}\,\G^{\unl a{}_3}\,\cdots\,\,\G^{\unl a{}_{4k+2}}\bigr)\equiv-\Pi_{(\unl a{}_0,\unl a{}_1|\unl a{}_2,\unl a{}_3,\ldots,\unl a{}_{4k+2})}=0\,,
\qqq
and the Odd Achirality Constraints,
\qq\nn
\Pi_{\unl a\unl b}=\tfrac{1}{2}\,\tr_{\tgt^{(1)}}\bigl(\G_{\unl a}\,\G^{\unl b}\,\bigl(\bd1_{D_{d,1}}+\G^0\,\G^1\,\cdots\,\G^{4k+2}\bigr)\bigr)=\d_{\unl a}^{\ \unl b}\,\tfrac{1}{2}\,\tr_{\tgt^{(1)}}\bigl(\bd1_{D_{d,1}}+\G^0\,\G^1\,\cdots\,\G^{4k+2}\bigr)=\tfrac{D_{d,1}}{2}\,\d_{\unl a\unl b}\,.
\qqq
The $\k$-symmetry superdistribution with restrictions
\qq\nn
\k\bigl({\rm sISO}(d,1\,|\,D_{d,1})/({\rm SO}(4k+2,1)\x{\rm SO}(d-4k-2)),\underset{\tx{\ciut{(p+2)}}}{\widehat\chi}\hspace{-6pt}{}^{\rm GS}\bigr)\rstr_{\cV_i}=\bigoplus_{\unl\a=1}^{\frac{D_{d,1}}{2}}\,\corr{\cT_{\unl\a\,i}}
\qqq 
is an $({\rm SO}(4k+2,1)\x{\rm SO}(d-4k-2))$-descendable superdistribution with the limit of its weak derived flag with restrictions
\qq\nn
&&\k^{-\infty}\bigl({\rm sISO}(d,1\,|\,D_{d,1})/({\rm SO}(4k+2,1)\x{\rm SO}(d-4k-2)),\underset{\tx{\ciut{(p+2)}}}{\widehat\chi}\hspace{-6pt}{}^{\rm GS}\bigr)\rstr_{\cV_i}=\bigoplus_{\unl\a=1}^{\frac{D_{d,1}}{2}}\,\corr{\cT_{\unl\a\,i}}\oplus\bigoplus_{\unl a=0}^{4k+2}\,\corr{\cT_{\unl a\,i}}\cr\cr
&\equiv&{\rm Vac}\bigl({\rm sISO}(d,1\,|\,D_{d,1})/({\rm SO}(4k+2,1)\x{\rm SO}(d-4k-2)),\underset{\tx{\ciut{(p+2)}}}{\widehat\chi}\hspace{-6pt}{}^{\rm GS}\bigr)\rstr_{\cV_i}\,.
\qqq 
\eeg

\beg\textbf{The square root of the Zhou super-1-brane in $\,{\rm SU}(1,1\,|\,2)_2/({\rm SO}(1,1)\x{\rm SO}(2))$.}
The tensors
\qq\nn
\D_0\equiv 4\unl C\,\unl\g^1\ox\bd1_2\,,\qquad\D_1\equiv-4\unl C\,\unl\g^0\ox\bd1_2\,,\qquad\qquad f_{\unl a}\equiv 2\unl C\,\unl\g{}^{\unl a}\ox\bd1_2
\qqq
satisfy the Even Achirality Constraints,
\qq\nn
\Pi_{(01|-)}&=&-\tfrac{1}{2}\,\tr_{\tgt^{(1)}}\bigl(\bigl(\unl\g{}_1\,\unl\g^0\ox\bd1_2\bigr)\cdot\bigl(\bd1_8-\unl\g^0\,\unl\g^1\ox\si_3\bigr)\bigr)=-\tfrac{1}{2}\,\bigl(2\tr\,\bigl(\unl\g{}_1\,\unl\g^0\bigr)+4\tr\,\si_3\bigr)=0\,,
\qqq
and the Odd Achirality Constraints,
\qq\nn
\Pi_{\unl a\unl b}=\tfrac{1}{2}\,\tr_{\tgt^{(1)}}\bigl(\bigl(\unl\g{}_{\unl a}\,\unl\g^{\unl b}\ox\bd1_2\bigr)\,\bigl(\bd1_8+\unl\g^0\,\unl\g^1\ox\si_3\bigr)\bigr)=\d_{\unl a}^{\ \unl b}\,\tfrac{1}{2}\,\tr_{\tgt^{(1)}}\bigl(\bd1_8+\unl\g^0\,\unl\g^1\ox\si_3\bigr)=4\,\d_{\unl a\unl b}\,.
\qqq
The $\k$-symmetry superdistribution with restrictions
\qq\nn
\k\bigl({\rm SU}(1,1\,|\,2)_2/({\rm SO}(1,1)\x{\rm SO}(2)),\underset{\tx{\ciut{(3)}}}{\widehat\chi}\hspace{-1pt}{}^{\rm Zh}_{(1)}\bigr)\rstr_{\cV_i}=\bigoplus_{\unl\a=1}^4\,\corr{\cT_{\unl\a\,i}}
\qqq 
is an $({\rm SO}(1,1)\x{\rm SO}(2))$-descendable superdistribution with the limit of its weak derived flag with restrictions
\qq\nn
&&\k^{-\infty}\bigl({\rm SU}(1,1\,|\,2)_2/({\rm SO}(1,1)\x{\rm SO}(2)),\underset{\tx{\ciut{(3)}}}{\widehat\chi}\hspace{-1pt}{}^{\rm Zh}_{(1)}\bigr)\rstr_{\cV_i}=\bigoplus_{\unl\a=1}^4\,\corr{\cT_{\unl\a\,i}}\oplus\corr{\cT_{0\,i},\cT_{1\,i}}\cr\cr
&\equiv&{\rm Vac}\bigl({\rm SU}(1,1\,|\,2)_2/({\rm SO}(1,1)\x{\rm SO}(2)),\underset{\tx{\ciut{(3)}}}{\widehat\chi}\hspace{-1pt}{}^{\rm Zh}_{(1)}\bigr)\rstr_{\cV_i}\,.
\qqq 
\eeg

\beg\textbf{No square root for the Zhou super-1-brane in $\,{\rm SU}(1,1\,|\,2)_2$.}
The tensors
\qq\nn
\D_0\equiv 4\unl C\,\unl\g^2\ox\bd1_2\,,\qquad\D_2\equiv-4\unl C\,\unl\g^0\ox\bd1_2\,,\qquad\qquad f_{\unl a}\equiv 2\unl C\,\unl\g{}^{\unl a}\ox\bd1_2
\qqq
satisfy the Even Achirality Constraints,
\qq\nn
\Pi_{(02|-)}&=&-\tfrac{1}{2}\,\tr_{\tgt^{(1)}}\bigl(\bigl(\unl\g{}_2\,\unl\g^0\ox\bd1_2\bigr)\cdot\bigl(\bd1_8-\unl\g^0\,\unl\g^2\ox\si_3\bigr)\bigr)=-\tfrac{1}{2}\,\bigl(2\tr\,\bigl(\unl\g{}_2\,\unl\g^0\bigr)+4\tr\,\si_3\bigr)=0\,,
\qqq
and the Odd Achirality Constraints,
\qq\nn
\Pi_{\unl a\unl b}=\tfrac{1}{2}\,\tr_{\tgt^{(1)}}\bigl(\bigl(\unl\g{}_{\unl a}\,\unl\g^{\unl b}\ox\bd1_2\bigr)\,\bigl(\bd1_8+\unl\g^0\,\unl\g^2\ox\si_3\bigr)\bigr)=\d_{\unl a}^{\ \unl b}\,\tfrac{1}{2}\,\tr_{\tgt^{(1)}}\bigl(\bd1_8+\unl\g^0\,\unl\g^2\ox\si_3\bigr)=4\,\d_{\unl a\unl b}\,.
\qqq
The $\k$-symmetry superdistribution with restrictions
\qq\nn
\k\bigl({\rm SU}(1,1\,|\,2)_2,\underset{\tx{\ciut{(3)}}}{\widehat\chi}\hspace{-1pt}{}^{\rm Zh}_{(12)}\bigr)\rstr_{\cV_i}=\bigoplus_{\unl\a=1}^4\,\corr{\cT_{\unl\a\,i}}
\qqq
does not bracket-generate the HP vacuum superdistribution due to the anomaly 
\qq\nn
\agt_{\rm int}=\bigl(\tfrac{\sfi}{2}\,X^{\unl a}\,\bigl(\unl\g{}_0\,\unl\g{}_1\,\unl\g{}_{\unl a}\ox\si_2\bigr)^{\g'\g''K}_{\ \a'\a''I}\,\bigl(\bd1_8-\sfP^{(1)}\bigr)^{\b'\b''J}_{\ \g'\g''K}\,Q_{\b'\b''J}\bigr)\,,\qquad\qquad\agt_{\rm susy}=0\,.
\qqq
Instead, the limit of its weak derived flag envelops the HP section,
\qq\nn
\k^{-\infty}\bigl({\rm SU}(1,1\,|\,2)_2,\underset{\tx{\ciut{(3)}}}{\widehat\chi}\hspace{-1pt}{}^{\rm Zh}_{(12)}\bigr)=\cT\Si^{\rm HP}\,.
\qqq
\eeg

\beg\textbf{The square root of the Park--Rey super-1-brane in $\,({\rm SU}(1,1\,|\,2)\x$ \linebreak ${\rm SU}(1,1\,|\,2))_2/({\rm SO}(1,1)\x{\rm SO}(3))$.}
The tensors 
\qq\nn
\D_0\equiv 4\unl C\,\unl\g^1\cdot\bigl(\bd1_4\ox\si_2\bigr)\ox\bd1_2\,,\qquad\D_1\equiv-4\unl C\,\unl\g^0\cdot\bigl(\bd1_4\ox\si_2\bigr)\ox\bd1_2\,,\qquad\qquad f_{\unl a}\equiv 2\unl C\,\unl\g^{\unl a}\cdot\bigl(\bd1_4\ox\si_2\bigr)\ox\bd1_2
\qqq
satisfy the Even Achirality Constraints,
\qq\nn
\Pi_{(01|-)}&=&-\tfrac{1}{2}\,\tr_{\tgt^{(1)}}\bigl(\bigl(\bigl(\bd1_4\ox\si_2\bigr)\cdot\unl\g{}_1\,\unl\g^0\cdot\bigl(\bd1_4\ox\si_2\bigr)\ox\bd1_2\bigr)\cdot\bigl(\bd1_{16}-\unl\g^0\,\unl\g^1\ox\si_1\bigr)\bigr)\cr\cr
&=&-\tfrac{1}{2}\,\tr_{\tgt^{(1)}}\bigl(\bigl(\unl\g{}_1\,\unl\g^0\ox\bd1_2\bigr)\cdot\bigl(\bd1_{16}-\unl\g^0\,\unl\g^1\ox\si_1\bigr)\bigr)=-\tfrac{1}{2}\,\bigl(2\tr\,\bigl(\unl\g{}_1\,\unl\g^0\bigr)+8\tr\,\si_1\bigr)=0\,,
\qqq
and the Odd Achirality Constraints,
\qq\nn
\Pi_{\unl a\unl b}&=&\tfrac{1}{2}\,\tr_{\tgt^{(1)}}\bigl(\bigl(\bigl(\bd1_4\ox\si_2\bigr)\cdot\unl\g{}_{\unl a}\,\unl\g^{\unl b}\cdot\bigl(\bd1_4\ox\si_2\bigr)\ox\bd1_2\bigr)\cdot\bigl(\bd1_{16}+\unl\g^0\,\unl\g^1\ox\si_1\bigr)\bigr)\cr\cr
&=&\d_{\unl a}^{\ \unl b}\,\tfrac{1}{2}\,\tr_{\tgt^{(1)}}\bigl(\bd1_{16}+\unl\g^0\,\unl\g^1\ox\si_1\bigr)=8\,\d_{\unl a\unl b}\,.
\qqq
The $\k$-symmetry superdistribution with restrictions
\qq\nn
\k\bigl(\bigl({\rm SU}(1,1\,|\,2)\x{\rm SU}(1,1\,|\,2)\bigr)_2/({\rm SO}(1,1)\x{\rm SO}(3)),\underset{\tx{\ciut{(3)}}}{\widehat\chi}\hspace{-1pt}{}^{\rm PR}_{(1)}\bigr)\rstr_{\cV_i}=\bigoplus_{\unl\a=1}^8\,\corr{\cT_{\unl\a\,i}}\oplus\corr{\cT_{0\,i},\cT_{1\,i}}
\qqq 
is an $({\rm SO}(1,1)\x{\rm SO}(3))$-descendable superdistribution with the limit of its weak derived flag with restrictions
\qq\nn
&&\k^{-\infty}\bigl(\bigl({\rm SU}(1,1\,|\,2)\x{\rm SU}(1,1\,|\,2)\bigr)_2/({\rm SO}(1,1)\x{\rm SO}(3)),\underset{\tx{\ciut{(3)}}}{\widehat\chi}\hspace{-1pt}{}^{\rm PR}_{(1)}\bigr)\rstr_{\cV_i}=\bigoplus_{\unl\a=1}^8\,\corr{\cT_{\unl\a\,i}}\oplus\corr{\cT_{0\,i},\cT_{1\,i}}\cr\cr
&\equiv&{\rm Vac}\bigl(\bigl({\rm SU}(1,1\,|\,2)\x{\rm SU}(1,1\,|\,2)\bigr)_2/({\rm SO}(1,1)\x{\rm SO}(3)),\underset{\tx{\ciut{(3)}}}{\widehat\chi}\hspace{-1pt}{}^{\rm PR}_{(1)}\bigr)\rstr_{\cV_i}\,.
\qqq 
\eeg

\beg\textbf{No square root for the Park--Rey super-1-brane in $\,({\rm SU}(1,1\,|\,2)\x$ \linebreak ${\rm SU}(1,1\,|\,2))_2/({\rm SO}(2)\x{\rm SO}(2))$.}
The tensors 
\qq\nn
&\D_0\equiv 4\unl C\,\unl\g^3\,\unl\g{}_7\cdot\bigl(\bd1_4\ox\si_2\bigr)\ox\bd1_2\,,\qquad\D_3\equiv 4\unl C\,\unl\g^0\cdot\bigl(\bd1_4\ox\si_2\bigr)\ox\bd1_2\,,&\cr\cr
&f_0\equiv 2\unl C\,\unl\g^0\cdot\bigl(\bd1_4\ox\si_2\bigr)\ox\bd1_2\,,\qquad\qquad f_3\equiv-2\unl C\,\unl\g^3\,\unl\g{}_7\cdot\bigl(\bd1_4\ox\si_2\bigr)\ox\bd1_2&
\qqq
satisfy the Even Achirality Constraints,
\qq\nn
\Pi_{(03|-)}&=&\tfrac{1}{2}\,\tr_{\tgt^{(1)}}\bigl(\bigl(\bigl(\bd1_4\ox\si_2\bigr)\cdot\unl\g{}_7\,\unl\g{}_3\,\unl\g^0\cdot\bigl(\bd1_4\ox\si_2\bigr)\ox\bd1_2\bigr)\cdot\bigl(\bd1_{16}-\unl\g^0\,\unl\g^3\,\unl\g{}_7\ox\si_1\bigr)\bigr)\cr\cr
&=&\tfrac{1}{2}\,\tr_{\tgt^{(1)}}\bigl(\bigl(\unl\g{}_7\,\unl\g{}_3\,\unl\g^0\ox\bd1_2\bigr)\cdot\bigl(\bd1_{16}-\unl\g^0\,\unl\g^3\,\unl\g{}_7\ox\si_1\bigr)\bigr)=\tfrac{1}{2}\,\bigl(2\sfi\,\tr\,\bigl(\G^0\bigr)\,\tr\,\bigl(\G^3\bigr)+8\tr\,\si_1\bigr)\cr\cr
&=&\sfi\,\tr\,\bigl(\G^{0\,{\rm T}}\bigr)\,\tr\,\bigl(\G^3\bigr)=-\sfi\,\tr\,\bigl(C'\,\G^0\,C'{}^{-1}\bigr)\,\tr\,\bigl(\G^3\bigr)=-\sfi\,\tr\,\bigl(\G^0\bigr)\,\tr\,\bigl(\G^3\bigr)=0\,,
\qqq
and the Odd Achirality Constraints,
\qq\nn
\Pi_{03}&=&-\tfrac{1}{2}\,\tr_{\tgt^{(1)}}\bigl(\bigl(\bigl(\bd1_4\ox\si_2\bigr)\cdot\unl\g{}_0\,\unl\g^3\,\unl\g{}_7\cdot\bigl(\bd1_4\ox\si_2\bigr)\ox\bd1_2\bigr)\cdot\bigl(\bd1_{16}+\unl\g^0\,\unl\g^3\,\unl\g{}_7\ox\si_1\bigr)\bigr)\cr\cr
&=&-\tfrac{1}{2}\,\tr_{\tgt^{(1)}}\bigl(\bigl(\unl\g{}_0\,\unl\g^3\,\unl\g{}_7\ox\bd1_2\bigr)\cdot\bigl(\bd1_{16}+\unl\g^0\,\unl\g^3\,\unl\g{}_7\ox\si_1\bigr)\bigr)=-\tfrac{1}{2}\,\bigl(2\sfi\,\tr\,\bigl(\G^0\bigr)\,\tr\,\bigl(\G^3\bigr)+8\tr\,\si_1\bigr)=0\,,\cr\cr
\Pi_{\unl a\unl a}&\equiv&8\,.
\qqq
The $\k$-symmetry superdistribution with restrictions
\qq\nn
\k\bigl(\bigl({\rm SU}(1,1\,|\,2)\x{\rm SU}(1,1\,|\,2)\bigr)_2/({\rm SO}(2)\x{\rm SO}(2)),\underset{\tx{\ciut{(3)}}}{\widehat\chi}\hspace{-1pt}{}^{\rm PR}_{(12)}\bigr)\rstr_{\cV_i}=\bigoplus_{\unl\a=1}^8\,\corr{\cT_{\unl\a\,i}}\oplus\corr{\cT_{0\,i},\cT_{1\,i}}
\qqq
is $({\rm SO}(2)\x{\rm SO}(2))$-descendable but does not bracket-generate the HP vacuum superdistribution due to the anomaly 
\qq\nn
\agt_{\rm int}&=&\bigl(\tfrac{\sfi}{2}\,X^0\,\bigl(\unl\g{}_0\,\unl\g{}_7\,\bigl(\bd1_4\ox\si_2\bigr)\ox\si_3\bigr)^{\g'\g''\g'''K}_{\ \a'\a''\a'''I}\,\bigl(\bd1_{16}-\sfP^{(1)}\bigr)^{\b'\b''\b'''J}_{\ \g'\g''\g'''K}\,Q_{\b'\b''\b'''J},\cr\cr
&&-\tfrac{\sfi}{2}\,X^3\,\bigl(\unl\g{}_3\,\bigl(\bd1_4\ox\si_2\bigr)\ox\si_3\bigr)^{\g'\g''\g'''K}_{\ \a'\a''\a'''I}\,\bigl(\bd1_{16}-\sfP^{(1)}\bigr)^{\b'\b''\b'''J}_{\ \g'\g''\g'''K}\,Q_{\b'\b''\b'''J}\bigr)\,,\cr\cr
\agt_{\rm susy}&=&0\,.
\qqq
Instead, the limit of its weak derived flag envelops the HP section,
\qq\nn
\k^{-\infty}\bigl(\bigl({\rm SU}(1,1\,|\,2)\x{\rm SU}(1,1\,|\,2)\bigr)_2/({\rm SO}(2)\x{\rm SO}(2)),\underset{\tx{\ciut{(3)}}}{\widehat\chi}\hspace{-1pt}{}^{\rm PR}_{(12)}\bigr)=\cT\Si^{\rm HP}\,.
\qqq
\eeg

\beg\textbf{The square root of the Metsaev--Tseytlin super-1-brane in \linebreak $\,{\rm SU}(2,2\,|\,4)/({\rm SO}(1,1)\x{\rm SO}(3)\x{\rm SO}(5))$.}
The tensors 
\qq\nn
\D_0\equiv 4\sfi\,\unl C\,\unl\g^1\,\unl\g{}_{11}\,,\qquad\D_1\equiv-4\sfi\,\unl C\,\unl\g^0\,\unl\g{}_{11}\,,\qquad\qquad f_{\unl a}\equiv 2\sfi\,\unl C\,\unl\g^{\unl a}\,\unl\g{}_{11}
\qqq
satisfy the Even Achirality Constraints,
\qq\nn
\Pi_{(01|-)}&=&-\tfrac{1}{2}\,\tr_{\tgt^{(1)}}\bigl(\unl\g{}_{11}\,\unl\g{}_1\,\unl\g^0\,\unl\g{}_{11}\cdot\bigl(\bd1_{32}-\unl\g^0\,\unl\g^1\,\unl\g{}_{11}\bigr)\bigr)=-\tfrac{1}{2}\,\tr_{\tgt^{(1)}}\bigl(\unl\g{}_1\,\unl\g^0\cdot\bigl(\bd1_{32}-\unl\g^0\,\unl\g^1\,\unl\g{}_{11}\bigr)\bigr)\cr\cr
&=&-\tfrac{1}{2}\,\tr_{\tgt^{(1)}}\,\unl\g{}_{11}=8\,\tr\,\si_3=0\,,
\qqq
and the Odd Achirality Constraints,
\qq\nn
\Pi_{\unl a\unl b}=\tfrac{1}{2}\,\tr_{\tgt^{(1)}}\bigl(\unl\g{}_{11}\,\unl\g{}_{\unl a}\,\unl\g^{\unl b}\,\unl\g{}_{11}\cdot\bigl(\bd1_{32}+\unl\g^0\,\unl\g^1\,\unl\g{}_{11}\bigr)\bigr)=\d_{\unl a}^{\ \unl b}\,\tfrac{1}{2}\,\tr_{\tgt^{(1)}}\bigl(\bd1_{32}+\unl\g^0\,\unl\g^1\,\unl\g{}_{11}\bigr)=16\,\d_{\unl a\unl b}\,.
\qqq
The $\k$-symmetry superdistribution with restrictions
\qq\nn
\k\bigl({\rm SU}(2,2\,|\,4)/({\rm SO}(1,1)\x{\rm SO}(3)\x{\rm SO}(5)),\underset{\tx{\ciut{(3)}}}{\widehat\chi}\hspace{-1pt}{}^{\rm MT}_{(1)}\bigr)\rstr_{\cV_i}=\bigoplus_{\unl\a=1}^{16}\,\corr{\cT_{\unl\a\,i}}\oplus\corr{\cT_{0\,i},\cT_{1\,i}}
\qqq 
is an $({\rm SO}(1,1)\x{\rm SO}(3)\x{\rm SO}(5))$-descendable superdistribution with the limit of its weak derived flag with restrictions
\qq\nn
&&\k^{-\infty}\bigl({\rm SU}(2,2\,|\,4)/({\rm SO}(1,1)\x{\rm SO}(3)\x{\rm SO}(5)),\underset{\tx{\ciut{(3)}}}{\widehat\chi}\hspace{-1pt}{}^{\rm MT}_{(1)}\bigr)\rstr_{\cV_i}=\bigoplus_{\unl\a=1}^{16}\,\corr{\cT_{\unl\a\,i}}\oplus\corr{\cT_{0\,i},\cT_{1\,i}}\oplus\corr{\cT_{0\,i},\cT_{1\,i}}\cr\cr
&\equiv&{\rm Vac}\bigl({\rm SU}(2,2\,|\,4)/({\rm SO}(1,1)\x{\rm SO}(3)\x{\rm SO}(5)),\underset{\tx{\ciut{(3)}}}{\widehat\chi}\hspace{-1pt}{}^{\rm MT}_{(1)}\bigr)\rstr_{\cV_i}\,.
\qqq 
\eeg

\beg\textbf{No square root for the Metsaev--Tseytlin super-1-brane in $\,{\rm SU}(2,2\,|\,4)/({\rm SO}(4)\x{\rm SO}(4))$.}\label{eg:MTs1k}
The tensors 
\qq\nn
\D_0\equiv 4\sfi\,\unl C\,\unl\g^5\,,\qquad\D_5\equiv4\sfi\,\unl C\,\unl\g^0\,\unl\g{}_{11}\,,\qquad\qquad f_0\equiv 2\sfi\,\unl C\,\unl\g^0\,\unl\g{}_{11}\,,\qquad f_5\equiv-2\sfi\,\unl C\,\unl\g^5
\qqq
satisfy the Even Achirality Constraints,
\qq\nn
\Pi_{(05|-)}&=&\tfrac{1}{2}\,\tr_{\tgt^{(1)}}\bigl(\unl\g{}_5\,\unl\g^0\,\unl\g{}_{11}\cdot\bigl(\bd1_{32}-\unl\g^0\,\unl\g^5\bigr)\bigr)=\tfrac{1}{2}\,\bigl(2\sfi\,\tr\,\bigl(\G^0\bigr)\,\tr\,\bigl(\G^5\bigr)-16\tr\,\si_3\bigr)=0\,,
\qqq
and the Odd Achirality Constraints,
\qq\nn
\Pi_{05}&=&-\tfrac{1}{2}\,\tr_{\tgt^{(1)}}\bigl(\unl\g{}_{11}\,\unl\g{}_0\,\unl\g^5\cdot\bigl(\bd1_{32}+\unl\g^0\,\unl\g^5\bigr)\bigr)=-\tfrac{1}{2}\,\tr_{\tgt^{(1)}}\bigl(2\sfi\,\tr\,\bigl(\G^0\bigr)\,\tr\,\bigl(\G^5\bigr)+16\tr\,\si_3\bigr)=0\,,\cr\cr
\Pi_{\unl a\unl a}&=&16\,.
\qqq
The $\k$-symmetry superdistribution with restrictions
\qq\nn
\k\bigl({\rm SU}(2,2\,|\,4)/({\rm SO}(4)\x{\rm SO}(4)),\underset{\tx{\ciut{(3)}}}{\widehat\chi}\hspace{-1pt}{}^{\rm MT}_{(12)}\bigr)\rstr_{\cV_i}=\bigoplus_{\unl\a=1}^{16}\,\corr{\cT_{\unl\a\,i}}
\qqq
is $({\rm SO}(4)\x{\rm SO}(4))$-descendable but does not bracket-generate the HP vacuum superdistribution due to the anomaly 
\qq\nn
\agt_{\rm int}&=&\bigl(-\tfrac{\sfi}{2}\,X^0\,\bigl(\unl\g{}_0\,\unl\g{}_{11}\bigr)^{\g'\g''K}_{\ \a'\a''I}\,\bigl(\bd1_{32}-\sfP^{(1)}\bigr)^{\b'\b''J}_{\ \g'\g''K}\,Q_{\b'\b''J},\cr\cr
&&\tfrac{\sfi}{2}\,X^5\,\bigl(\unl\g{}_5\bigr)^{\g'\g''K}_{\ \a'\a''I}\,\bigl(\bd1_{32}-\sfP^{(1)}\bigr)^{\b'\b''J}_{\ \g'\g''K}\,Q_{\b'\b''J}\bigr)\,,\cr\cr
\agt_{\rm susy}&=&0\,.
\qqq
Instead, the limit of its weak derived flag envelops the HP section,
\qq\nn
\k^{-\infty}\bigl({\rm SU}(2,2\,|\,4)/({\rm SO}(4)\x{\rm SO}(4)),\underset{\tx{\ciut{(3)}}}{\widehat\chi}\hspace{-1pt}{}^{\rm MT}_{(12)}\bigr)=\cT\Si^{\rm HP}\,.
\qqq
\eeg
~\medskip

\noindent There is one super-$\si$-model with an integrable vacuum from the previous list of Examples \ref{eg:GSs0Vac}-\ref{eg:MTs1Vac} that we left out above as it fails to satisfy both Achirality Constraints. In view of its relevance, as a fundamental super-$\si$-model of the superstring, we discuss it separately below.\medskip 

\beg\textbf{The square root of the Green--Schwarz super-1-brane in \linebreak $\,{\rm sISO}(d,1\,|\,D_{d,1})/({\rm SO}(1,1)\x{\rm SO}(d-1))$.}\label{eg:sqroots1bsMink}
The tensors
\qq\nn
\D_0\equiv-4\ovl\G{}^1\,,\qquad\D_1\equiv 4\ovl\G{}^0\,,\qquad\qquad f_{\unl a}\equiv\ovl\G{}^{\unl a}
\qqq
satisfy the identities
\qq\nn
\Pi_{(0,1|-)}&=&-\tfrac{1}{2}\,\tr_{\tgt^{(1)}}\bigl(\G_1\,\G^0\,\bigl(\bd1_{D_{d,1}}-\G^0\,\G^1\bigr)\bigr)=-\tfrac{D_{d,1}}{2}
\qqq
and
\qq\nn
\Pi_{\unl a\unl b}=\tfrac{1}{2}\,\tr_{\tgt^{(1)}}\bigl(\G_{\unl a}\,\G^{\unl b}\,\bigl(\bd1_{D_{d,1}}+\G^0\,\G^1\bigr)\bigr)=\tfrac{D_{d,1}}{2}\,\bigl(\d_{\unl a}^{\ \unl b}-\eta_{\unl b\unl b}\,\ep_{\unl a\unl b}\bigr)\,,
\qqq
and so manifestly violate the Even and Odd Achirality Constraints. In consequence, the correspondence sector of the super-$\si$-model exhibits an additional Gra\ss mann-even symmetry generated by the vector fields $\,\cT_0+\cT_1$,\ whence the enhanced gauge-symmetry superdistribution in the exceptional form
\qq\nn
\cG\cS\bigl({\rm sISO}(d,1\,|\,D_{d,1})/({\rm SO}(1,1)\x{\rm SO}(d-1)),\underset{\tx{\ciut{(3)}}}{\widehat\chi}\hspace{-1pt}{}^{\rm GS}\bigr)=\bigoplus_{\unl\a=1}^{\frac{D_{d,1}}{2}}\,\corr{\cT_{\unl\a}}\oplus\corr{\cT_0+\cT_1}\oplus\bigoplus_{(\unl a,\widehat b)\in\{0,1\}\x\ovl{2,d}}\,\corr{\cT_{\unl a\widehat b}}
\qqq
whereas the limit of the weak derived flag of the $({\rm SO}(1,1)\x{\rm SO}(d-1))$-descendable $\k$-symmetry superdistribution
\qq\nn
\k\bigl({\rm sISO}(d,1\,|\,D_{d,1})/({\rm SO}(1,1)\x{\rm SO}(d-1)),\underset{\tx{\ciut{(3)}}}{\widehat\chi}\hspace{-1pt}{}^{\rm GS}\bigr)\equiv\bigoplus_{\unl\a=1}^{\frac{D_{d,1}}{2}}\,\corr{\cT_{\unl\a}}
\qqq
is given by
\qq\nn
&&\k^{-\infty}\bigl({\rm sISO}(d,1\,|\,D_{d,1})/({\rm SO}(1,1)\x{\rm SO}(d-1)),\underset{\tx{\ciut{(3)}}}{\widehat\chi}\hspace{-1pt}{}^{\rm GS}\bigr)=\bigoplus_{\unl\a=1}^{\frac{D_{d,1}}{2}}\,\corr{\cT_{\unl\a}}\oplus\corr{\cT_0-\cT_1}\cr\cr
&\subsetneq&{\rm Vac}\bigl({\rm sISO}(d,1\,|\,D_{d,1})/({\rm SO}(1,1)\x{\rm SO}(d-1)),\underset{\tx{\ciut{(3)}}}{\widehat\chi}\hspace{-1pt}{}^{\rm GS}\bigr)\,.
\qqq 
The two vector fields: 
\qq\nn
\cT_\pm:=\cT_0\pm\cT_1
\qqq
are complementary, and it is natural to think of them as target-superspace counterparts of the chiral worldsheet diffeomorphisms (in, say, the static gauge). Thus, $\,\k({\rm sISO}(d,1\,|\,D_{d,1})/({\rm SO}(1,1)\x{\rm SO}(d-1)),\underset{\tx{\ciut{(3)}}}{\widehat\chi}\hspace{-1pt}{}^{\rm GS})\,$ alone is seen to generate one chiral half of the integrable vacuum of the superstring. As noted already in \Rxcite{Remark 6.2}{Suszek:2019cum} (in a different description adopted {\it ibidem}), the appearance of the chiral field $\,\cT_-\,$ in the weak derived flag of the $\k$-symmetry superdistribution is to be understood in this context as a variant of the chiral Sugawara extension.

Clearly, upon extending the purely Gra\ss mann-odd $\k$-symmetry superdistribution by the span of $\,\cT_+$,\ that is by defining the \textbf{extended $\k$-symmetry superdistribution} 
\qq\nn
\k_{\rm ext}\bigl({\rm sISO}(d,1\,|\,D_{d,1})/({\rm SO}(1,1)\x{\rm SO}(d-1)),\underset{\tx{\ciut{(3)}}}{\widehat\chi}\hspace{-1pt}{}^{\rm GS}\bigr):=\bigoplus_{\unl\a=1}^{\frac{D_{d,1}}{2}}\,\corr{\cT_{\unl\a}}\oplus\corr{\cT_+}\,,
\qqq
we obtain the anticipated result
\qq\nn
\k_{\rm ext}^{-\infty}\bigl({\rm sISO}(d,1\,|\,D_{d,1})/({\rm SO}(1,1)\x{\rm SO}(d-1)),\underset{\tx{\ciut{(3)}}}{\widehat\chi}\hspace{-1pt}{}^{\rm GS}\bigr)\equiv{\rm Vac}\bigl({\rm sISO}(d,1\,|\,D_{d,1})/({\rm SO}(1,1)\x{\rm SO}(d-1)),\underset{\tx{\ciut{(3)}}}{\widehat\chi}\hspace{-1pt}{}^{\rm GS}\bigr)\,.
\qqq
\eeg

\section{The higher geometry behind the physics}\label{sec:hgeomphys}

The (super)background of a (super-)$\si$-model is but the lowest rung in a hierarchy of geometric structures over the (super)target co-defining the field theory. As we are about to recapitulate, the higher structures, associated with the (super)background's $(p+2)$-form component, encode essential information on a distinguished quantisation scheme for the (super-)$\si$-model, and so any discussion of a constitutive property of the latter is necessarily incomplete, and potentially even wrong, until we lift it to those higher structures. This is particularly true of classical symmetries whose consistent transposition to the quantum r\'egime is a key component of the quantisation scheme. Accordingly, in the last two sections of the present paper, we take the first step on the path towards a full-fledged realisation of the enhanced gauge symmetry of the GS super-$\si$-model in its HP formulation on objects that geometrise the super-$(p+2)$-cocycles of the HP superbackground through a construction which features a beautiful interplay between higher cohomology and differential geometry. Below, we merely recapitulate\footnote{The idea was reviewed at great length in \Rcite{Suszek:2017xlw}, where, moreover, a long bibliographical list was drawn.} those aspects thereof that are of immediate relevance to the intended symmetry analysis.

The higher cohomology behind the topological term in the Dirac--Feynman amplitude of the purely \emph{bosonic} (or Gra\ss mann-even) $\si$-model was originally identified by Alvarez in \Rcite{Alvarez:1984es} and, more structurally, by Gaw\c{e}dzki in \Rcite{Gawedzki:1987ak}, and later geometrised by Murray and Stevenson in Refs.\,\cite{Murray:1994db,Murray:1999ew} in a manner amenable to various subsequent generalisations and extensions. The point of departure is the interpretation of the said term in the Dirac--Feynman amplitude,
\qq\nn
&\cA_{\rm DF, top}^{{\rm NG},p}:=\ee^{\sfi\,(S_{\si,p}^{\rm (NG)}-S_{\si,p,{\rm metr}}^{\rm (NG)})}\ :\ [\Om_p,\unl\cM]\too\uj\,,&\cr\cr 
&S_{\si,p,{\rm metr}}^{\rm (NG)}[x]:=\int_{\Om_p}\,\sqrt{\det_{(p)}\,\bigl(x^*\unl\g\bigr)}\,,\qquad x\in[\Om_p,\unl\cM]\,,&
\qqq
in a background
\qq\nn
\Bgt_p=\bigl(\unl\cM,\unl\g,\underset{\tx{\ciut{(p+2)}}}{\unl{\chi}}\bigr)
\qqq
composed of a target space given by a standard metric ($C^\infty$-)manifold $\,(\unl\cM,\unl\g)\,$ and of a de Rham $(p+2)$-cocycle $\,\underset{\tx{\ciut{(p+2)}}}{\unl{\chi}}\in Z^{p+2}_{\rm dR}(\unl\cM)\,$ with periods 
\qq\nn
{\rm Per}\bigl(\underset{\tx{\ciut{(p+2)}}}{\unl{\chi}}\bigr)\subset 2\pi\bZ\,.
\qqq 
The term admits a local presentation
\qq\nn
\cA_{\rm DF, top}^{{\rm NG},p}[x]=\ee^{\sfi\,\int_{\Om_p}\,x^*\underset{\tx{\ciut{(p+1)}}}{\unl{\b}}}
\qqq
for $\,x\in[\Om_p,\cU]\subset[\Om_p,\unl\cM]$,\ where $\,\cU\subset\unl\cM\,$ is an open subset with the property
\qq\nn
\underset{\tx{\ciut{(p+2)}}}{\unl{\chi}}\rstr_\cU=\sfd\underset{\tx{\ciut{(p+1)}}}{\unl{\b}}
\qqq
for some $\,\underset{\tx{\ciut{(p+1)}}}{\unl{\b}}\in\Om^{p+1}(\unl\cM)$.\ It yields the so-called \textbf{$(p+1)$-volume holonomy}
\qq\nn
\cA_{\rm DF, top}^{{\rm NG},p}\equiv{\rm Hol}_{\cG^{(p)}}\,,
\qqq
over $\,x(\Om_p)\in Z_{p+1}(\unl\cM)$,\ of an \textbf{abelian $p$-gerbe} $\,\cG^{(p)}$,\ the latter being a geometrisation of (the cohomology class of) $\,\underset{\tx{\ciut{(p+2)}}}{\unl{\chi}}\,$ in the spirit of \Rcite{Gajer:1996}, inspired by the pioneering papers \cite{Murray:1994db,Murray:1999ew}. The holonomy is an example of a Cheeger--Simons differential character of degree $p+1$ modulo $\,2\pi\bZ\,$ ({\it cp} \Rcite{Cheeger:1985}),
\qq\nn
{\rm Hol}_{\cG^{(p)}}[x]=:h_{\cG^{(p)}}\bigl(x(\Om_p)\bigr)\,,\qquad\qquad h_{\cG^{(p)}}\in{\rm Hom}_{\rm {\bf AbGrp}}\bigl(Z_{p+1}(\unl\cM),\uj\bigr)
\qqq
with the property
\qq\nn
\forall_{\underset{\tx{\ciut{(p+2)}}}{c}\in C_{p+2}(\unl\cM)}\ :\ h_{\cG^{(p)}}\bigl(\p\underset{\tx{\ciut{(p+2)}}}{c}\bigr)=\vep_{(p)}\bigl(\underset{\tx{\ciut{(p+2)}}}{c}\bigr)\,,
\qqq
expressed in terms of the $(p+2)$-cochain 
\qq\nn
\vep_{(p)}\in{\rm Hom}_{\rm {\bf AbGrp}}\bigl(C_{p+2}(\unl\cM),\uj\bigr)
\qqq
given by
\qq\nn
\vep_{(p)}\equiv\ee^{\sfi\,\int_\cdot\,{\rm curv}(\cG^{(p)})}\ :\ C_{p+2}(\unl\cM)\too\uj\ :\ \underset{\tx{\ciut{(p+2)}}}{c}\longmapsto\ee^{\sfi\,\int_{\underset{\tx{\ciut{(p+2)}}}{c}}\,{\rm curv}(\cG^{(p)})}\,,
\qqq
for
\qq\nn
{\rm curv}\bigl(\cG^{(p)}\bigr)\equiv\underset{\tx{\ciut{(p+2)}}}{\unl{\chi}}
\qqq
the {\bf curvature} of $\,\cG^{(p)}$.\ A little more abstractly, the holonomy is the image 
\qq\nn
{\rm Hol}_{\cG^{(p)}}[x]\equiv\iota_p\bigl([x^*\cG^{(p)}]\bigr)
\qqq
of the isoclass of the flat gerbe\footnote{The notion of a pullback in the (higher) category of $p$-gerbes was also recalled in \Rcite{Suszek:2017xlw}.} $\,x^*\cG^{(p)}\,$ over $\,\Om_p\,$ under the canonical isomorphism
\qq\nn
\iota_p\ :\ \cW^{p+2}(\Om_p;0)\xrightarrow{\ \cong\ }\uj
\qqq
between the group $\,\cW^{p+2}(\Om_p;0)\,$ of isoclasses of flat abelian $p$-gerbes over $\,\Om_p\,$ and $\,\uj$.\ It may be given an entirely explicit form, though, in terms of a trivialisation of $\,{\rm curv}(\cG^{(p)})\,$ over some open cover $\,\{\cO_i\}_{i\in I}\equiv\cO_{\unl\cM}\,$ of $\,\unl\cM\,$ ({\it e.g.}, a \emph{good} one, {\it i.e.}, one with all non-empty multiple intersections $\,\cO_{i_1}\cap\cO_{i_2}\cap\cdots\cap\cO_{i_N}\equiv\cO_{i_1 i_2\ldots i_N},\ N\in\bN^\x\,$ contractible, which always exists on a $C^2$-manifold by The Weil--de Rham Theorem) that consists of sheaf-cohomological data of $\,\cG^{(p)}$.\ The relevant cohomology is the real Deligne--Beilinson hypercohomology, {\it i.e.}, the direct limit, over refinements of (good) open covers, of the cohomologies of the total complexes of the bicomplexes formed by an extension of the bounded Deligne complex $\,\cD(p+1)^\bullet$:
\qq\nn
\begin{tikzpicture}[descr/.style={fill=white,inner sep=1.5pt}]
        \matrix (m) [
            matrix of math nodes,
            row sep=1.cm,
            column sep=2.5cm,
            text height=1.5ex, text depth=0.25ex
        ]
        { \unl\bZ{}_{\unl\cM}\equiv\cD(p+1)^{-1} & \unl\bR{}_{\unl\cM}\equiv\cD(p+1)^0 & \unl{\Om^1(\unl\cM)}\equiv\cD(p+1)^1 \\ \unl{\Om^2(\unl\cM)}\equiv\cD(p+1)^2 & \cdots & \unl{\Om^{p+1}(\unl\cM)}\equiv\cD(p+1)^{p+1} \\
        };

        \path[overlay,->, font=\scriptsize,>=angle 45]
        (m-1-1) edge node[descr,yshift=2.3ex] {$2\pi\,\id_{\unl\bZ{}_{\unl\cM}}\equiv\sfd^{(-1)}$} (m-1-2)
        (m-1-2) edge node[descr,yshift=2.3ex] {$\ \sfd\equiv\sfd^{(0)}\ $} (m-1-3)
        (m-1-3) edge[out=355,in=175] node[descr,xshift=-15.ex,yshift=1.1ex] {$\sfd\equiv\sfd^{(1)}$} (m-2-1)
        (m-2-1) edge node[descr,yshift=-2.3ex] {$\ \sfd\equiv\sfd^{(2)}\ $} (m-2-2)
        (m-2-2) edge node[descr,yshift=-2.3ex] {$\ \sfd\equiv\sfd^{(p)}\ $} (m-2-3);
\end{tikzpicture}\,,
\qqq
of sheaves of local integer constants, locally smooth maps (containing the former as the image of the sheaf counterpart of the injection $\,\bZ\ni n\longmapsto 2\pi n\in\bR$) and $k$-forms (for $\,k\in\ovl{1,p+1}$) on $\,\unl\cM\,$ in the direction of the \Cv ech cohomology associated with a (fixed) good open cover $\,\cO_{\unl\cM}$,\ {\it cp.}\ \Rcite{Johnson:2003}. The latter cohomologies being defined for groups of $k$-cochains
\qq\nn
\textrm{\Cv D}^k\bigl(\cO_{\unl\cM},\cD(p+1)^\bullet\bigr)=\bigoplus_{\substack{(m,n)\in\bN\x(\{-1\}\cup\bN) \\ m+n=k}}\,\vC^m\bigl(\cO_{\unl\cM},\cD(p+1)^n\bigr)\,,
\qqq 
with 
\qq\nn
\vC^m\bigl(\cO_{\unl\cM},\cD(p+1)^n\bigr)=\left\{ \barr{lc} \bZ_{\bigsqcup_{i_0,i_1,\ldots,i_{m+1}\in I}\,\cO_{i_0 i_1\ldots i_{m+1}}} & \textrm{for}\ n=-1 \cr\cr
C^\infty\bigl(\bigsqcup_{i_0,i_1,\ldots,i_m\in I}\,\cO_{i_0 i_1\ldots i_m},\bR\bigr) & \textrm{for}\ n=0 \cr\cr
\Om^n\bigl(\bigsqcup_{i_0,i_1,\ldots,i_m\in I}\,\cO_{i_0 i_1\ldots i_m}\bigr) & \textrm{for}\ 0<n\leq p+1\cr\cr
0 & \textrm{for}\ n>p+1 \earr \right.\,,
\qqq
and the Deligne coboundary operators
\qq
&D^{(k)}\ :\ \textrm{\Cv D}^k\bigl(\cO_{\unl\cM},\cD(p+1)^\bullet\bigr)\too\textrm{\Cv D}^{k+1}\bigl(\cO_{\unl\cM},\cD(p+1)^\bullet\bigr)\,,&\cr &&\label{eq:Delignecob}\\
&D^{(k)}\rstr_{\vC^m(\cO_{\unl\cM},\cD(p+1)^n)}=\sfd^{(n)}+(-1)^{n+1}\,\vd^{(m)}\,,\qquad m+n=k\,,& \nn
\qqq
given in terms of the \Cv ech coboundary operators
\qq\nn
\vd^{(m)}\ &:&\ \vC^m\bigl(\cO_{\unl\cM},\cD(p+1)^n\bigr)\too\vC^{m+1}\bigl(\cO_{\unl\cM},\cD(p+1)^n\bigr)\cr\cr
&:&\ \bigl(\underset{\tx{\ciut{(n)}}}{\varpi}{}_{i_0 i_1\ldots i_m}\bigr)\longmapsto\bigl(\sum_{l=0}^{m+1}\,(-1)^l\,\underset{\tx{\ciut{(n)}}}{\varpi}{}_{\underset{\widehat{i_l}}{i_0 i_1\ldots i_m}}\rstr_{\cO_{i_0i_1\ldots i_{m+1}}}\equiv\bigl(\vd^{(m)}\underset{\tx{\ciut{(n)}}}{\varpi}\bigr)_{i_0i_1\ldots i_{m+1}}\bigr)\,.
\qqq
We have the useful
\berop\label{prop:LieD}
Adopt the above notation and define -- for a \Cv ech--Deligne $k$-cochain (for $\,k<p+1$)
\qq\nn
\underset{\tx{\ciut{(k)}}}{\unl{\xcA}}&\equiv&\bigl(\underset{\tx{\ciut{(k)}}}{\unl{\a}}\hspace{-5pt}{}_{i_0},\underset{\tx{\ciut{(k-1)}}}{\unl{\a}}{}_{i'_0 i'_1},\ldots,\underset{\tx{\ciut{(-1)}}}{\unl{\a}}{}_{i^{(k)}_0 i^{(k)}_1\ldots i^{(k)}_{k+1}}\bigr)\in\textrm{\Cv D}^k\bigl(\cO_{\unl\cM},\cD(p+1)^\bullet\bigr)\,.
\qqq
and a smooth vector field $\,\cV\in\G(\cT\unl\cM)\,$ -- the \Cv ech--Deligne $(k-1)$-cochain
\qq\nn
\cV\con\underset{\tx{\ciut{(k)}}}{\unl{\xcA}}:=\bigl(\cV\con\underset{\tx{\ciut{(k)}}}{\unl{\a}}\hspace{-5pt}{}_{i_0},\cV\con\underset{\tx{\ciut{(k-1)}}}{\unl{\a}}{}_{i'_0 i'_1},\ldots,\cV\con\underset{\tx{\ciut{(1)}}}{\unl{\a}}{}_{i^{(k-2)}_0 i^{(k-2)}_1\ldots i^{(k-2)}_{k-1}},0\bigr)\in\textrm{\Cv D}^{k-1}\bigl(\cO_{\unl\cM},\cD(p+1)^\bullet\bigr)\,.
\qqq
The $k$-cochain
\qq\nn
\pLie{\cV}\underset{\tx{\ciut{(k)}}}{\unl{\xcA}}:=\bigl(\pLie{\cV}\underset{\tx{\ciut{(k)}}}{\unl{\a}}\hspace{-5pt}{}_{i_0},\pLie{\cV}\underset{\tx{\ciut{(k-1)}}}{\unl{\a}}{}_{i'_0 i'_1},\ldots,\pLie{\cV}\underset{\tx{\ciut{(0)}}}{\unl{\a}}{}_{i^{(k-1)}_0 i^{(k-1)}_1\ldots i^{(k-1)}_k},0\bigr)\in\textrm{\Cv D}^k\bigl(\cO_{\unl\cM},\cD(p+1)^\bullet\bigr)\,.
\qqq
satisfies the identity
\qq\nn
\pLie{\cV}\underset{\tx{\ciut{(k)}}}{\unl{\xcA}}=D^{(k-1)}\bigl(\cV\con\underset{\tx{\ciut{(k)}}}{\unl{\xcA}}\bigr)+\cV\con D^{(k)}\underset{\tx{\ciut{(k)}}}{\unl{\xcA}}\,.
\qqq
\eerop
\beroof
Straightforward.
\eroof
\noindent In this setting, the local data of $\,\cG^{(p)}\,$ compose a $(p+1)$-cocycle 
\qq\nn
\underset{\tx{\ciut{(p+1)}}}{\unl{\xcB}}&\equiv&\bigl(\underset{\tx{\ciut{(p+1)}}}{\unl{\b}}\hspace{-5pt}{}_{i_0},\underset{\tx{\ciut{(p)}}}{\unl{\b}}{}_{i'_0 i'_1},\ldots,\underset{\tx{\ciut{(-1)}}}{\unl{\b}}{}_{i^{(p+1)}_0 i^{(p+1)}_1\ldots i^{(p+1)}_{p+2}}\bigr)\in{\rm Ker}\,D^{(p+1)}\subset\textrm{\Cv D}^{p+1}\bigl(\cO_{\unl\cM},\cD(p+1)^\bullet\bigr)\,.
\qqq
(1-)Isomorphic $p$-gerbes $\,\cG_A^{(p)},\ A\in\{1,2\}\,$ over a common base are described by (DB-)cohomologous $(p+1)$-cocycles $\,\underset{\tx{\ciut{(p+1)}}}{\unl{\xcB}}\hspace{-5pt}{}_A,\ A\in\{1,2\}\,$ (for a suitable choice of the cover),\ {\it i.e.}, there exists a \Cv ech--Deligne $p$-cochain
\qq\label{eq:CDpcoch}
\underset{\tx{\ciut{(p)}}}{\unl{\xcP}}{}^{(1)}\equiv\bigl(\underset{\tx{\ciut{(p)}}}{\unl{P}}{}_{i_0}^{(1)},\underset{\tx{\ciut{(p-1)}}}{\unl{P}}\hspace{-3pt}{}_{i'_0 i'_1}^{(1)},\ldots,\underset{\tx{\ciut{(-1)}}}{\unl{P}}{}_{i^{(p)}_0 i^{(p)}_1\ldots i^{(p)}_{p+1}}^{(1)}\bigr)\in\textrm{\Cv D}^p\bigl(\cO_{\unl\cM},\cD(p+1)^\bullet\bigr)
\qqq
satisfying the identity
\qq\nn
\underset{\tx{\ciut{(p+1)}}}{\unl{\xcB}}\hspace{-5pt}{}_1+D^{(p)}\underset{\tx{\ciut{(p)}}}{\unl{\xcP}}{}^{(1)}=\underset{\tx{\ciut{(p+1)}}}{\unl{\xcB}}\hspace{-5pt}{}_2\,.
\qqq
Similarly, (2-)isomorphic (1-)isomorphisms represented by $p$-cochains $\,\underset{\tx{\ciut{(p)}}}{\unl{\xcP}}{}_A^{(1)},\ A\in\{1,2\}\,$ as above are (DB-)cohomologous (for a suitable choice of the cover), {\it i.e.}, there exists a \Cv ech--Deligne $(p-1)$-cochain
\qq\nn
\underset{\tx{\ciut{(p-1)}}}{\unl{\xcP}}\hspace{-5pt}{}^{(2)}\equiv\bigl(\underset{\tx{\ciut{(p-1)}}}{\unl{P}}{}_{i_0}^{(2)},\underset{\tx{\ciut{(p-2)}}}{\unl{P}}\hspace{-3pt}{}_{i'_0 i'_1}^{(2)},\ldots,\underset{\tx{\ciut{(-1)}}}{\unl{P}}{}_{i^{(p)}_0 i^{(p)}_1\ldots i^{(p)}_p}^{(2)}\bigr)\in\textrm{\Cv D}^{p-1}\bigl(\cO_{\unl\cM},\cD(p+1)^\bullet\bigr)
\qqq
satisfying the identity
\qq\nn
\underset{\tx{\ciut{(p)}}}{\unl{\xcP}}^{(1)}\hspace{-9pt}{}_1\hspace{5pt}+D^{(p-1)}\underset{\tx{\ciut{(p-1)}}}{\unl{\xcP}}\hspace{-3pt}{}^{(2)}=\underset{\tx{\ciut{(p)}}}{\unl{\xcP}}^{(1)}\hspace{-9pt}{}_2\,,
\qqq
and this hierarchy continues all the way to $(p+1)$-isomorphisms for which equivalence means proper equality.

The fundamental advantage of the gerbe-theoretic description of the $\si$-model, justifying the introduction of the somewhat heavy cohomological and geometric formalism, is the canonical way to prequantisation that it paves. The way leads through the \textbf{transgression map}
\qq\nn
\t_p\ :\ \bH^{p+1}\left(\unl\cM,\cD(p+1)^\bullet\right)\too\bH^1\left(\xcC_p\unl\cM,\cD(1)^\bullet\right)\,,
\qqq
first noted and put to use for $\,p=1\,$ in \Rcite{Gawedzki:1987ak}, that assigns to the isoclass of the $p$-gerbe $\,\cG^{(p)}\,$ over $\,\unl\cM$,\ represented by a class in $\,\bH^{p+1}(\unl\cM,\cD(p+1)^\bullet)$,\ the isoclass of a principal $\bC^\x$-bundle 
\qq\label{eq:transbund}
\alxydim{@C=1cm@R=1cm}{\bC^\x \ar[r] & \xcL_{\cG^{(p)}} \ar[d]^{\pi_{\xcL_{\cG^{(p)}}}} \\ & \xcC_p\unl\cM }\,,
\qqq 
termed the {\bf transgression bundle}, over the configuration space 
\qq\nn
\xcC_p\unl\cM\equiv\bigl[\xcC_p,\unl\cM\bigr]
\qqq 
of the $\si$-model attached to a Cauchy hypersurface $\,\xcC_p\subset\Om_p$,\ with connection $\,\nabla_{\xcL_{\cG^{(p)}}}\,$ of curvature 
\qq\nn
\curv\bigl(\nabla_{\xcL_{\cG^{(p)}}}\bigr)=\int_{\xcC_p}\,\ev_p^*{\rm curv}\bigl(\cG^{(p)}\bigr)\,,
\qqq
written for 
\qq\nn
\ev_p\ :\ \xcC_p\unl\cM\x\xcC_p\too\unl\cM\ :\ (\unl x,c)\longmapsto\unl x(c)\,.
\qqq
The curvature of $\,\nabla_{\xcL_{\cG^{(p)}}}\,$ is to be compared with the (pre)symplectic form\footnote{The (pre)symplectic form for the lagrangean field theory of maps from $\,[\Om_p,\unl\cM]\,$ defined by $\,S_{\si,p}^{\rm (NG)}\,$ can be derived in the first-order formalism of Refs.\,\cite{Gawedzki:1972ms,Kijowski:1973gi,Kijowski:1974mp,Kijowski:1976ze,Szczyrba:1976,Kijowski:1979dj}, {\it cp} also \Rcite{Saunders:1989jet} for a modern treatment.} 
\qq\nn
\Om_\si=\d\vartheta_{\sfT^*\xcC_p\unl\cM}+\pi_{\sfT^*\xcC_p\unl\cM}^*\int_{\xcC_p}\,\ev_p^*{\rm curv}\bigl(\cG^{(p)}\bigr)
\qqq
of the $\si$-model over its space of states $\,\sfT^*\xcC_p\unl\cM\,$ (written out in its simplest form) canonically projecting to the configuration space, $\,\pi_{\sfT^*\xcC_p\unl\cM}\ :\ \sfT^*\xcC_p\unl\cM\too\xcC_p\unl\cM$,\ the 2-form being expressed in terms of the so-called (kinetic-action) 1-form $\,\vartheta_{\sfT^*\xcC_p\unl\cM}\in\Om^1(\sfT^*\xcC_p\unl\cM)\,$ with the familiar presentation 
\qq\nn
\vartheta_{\sfT^*\xcC_p\unl\cM}[\unl x,\txp]=\int_{\cC_p}\,{\rm Vol}(\cC_p)\wedge\txp_\mu(\cdot)\d\unl x^\mu(\cdot)
\qqq
in the coordinates $\,(\unl x^\mu,\txp_\nu)\,$ on $\,\sfT^*\xcC_p\unl\cM\,$ (here, $\,\txp\,$ is the \emph{kinetic}-momentum field over the Cauchy hypersurface). Suitably polarised sections of the line bundle associated\footnote{The principal $\bC^\x$-bundle $\,\pi_{\sfT^*\xcC_p\unl\cM}^*\xcL_{\cG^{(p)}}\,$ (with the principal $\bC^\x$-connection 1-form corrected by the pullback of the action 1-form $\,\vartheta_{\sfT^*\xcC_p\unl\cM}\,$ from its base) is to be understood as the frame bundle $\,\sfF\xcL_\si\,$ of the prequantum bundle $\,\xcL_\si\cong(\sfF\xcL_\si\x\bC)/\bC^\x$.} with the principal $\bC^\x$-bundle given by the pullback, along the bundle projection $\,\pi_{\sfT^*\xcC_p\unl\cM}$,\ of the transgression bundle to the classical space of states $\,\sfT^*\xcC_p\unl\cM\,$ define the Hilbert space of the $\si$-model.

The intrinsically quantum-mechanical nature of the geometrisation $\,\cG^{(p)}\,$ of the $(p+2)$-cocycle $\,\underset{\tx{\ciut{(p+2)}}}{\unl{\chi}}$,\ derived from Dirac's ingenious interpretation of the classical lagrangean density as a transport operator between quantum states (later elaborated by Feynman), makes the higher geometry associated with $\,\cG^{(p)}\,$ a natural arena for a rigorous discussion of (pre)quantisable symmetries of the $\si$-model. The latter are customarily grouped into two categories: 
\bit
\item \textbf{global symmetries} -- represented by those isometries of $\,(\unl\cM,\unl\g)$,\ composing a group $\,\txG_\si\,$ acting smoothly as
\qq\nn
\unl\la{}_\cdot\ :\ \txG_\si\x\unl\cM\too\unl\cM\ :\ (g,m)\longmapsto\la_g(m)\,,
\qqq 
that lift to $p$-gerbe (1-)isomorphisms\label{p:globsym}
\qq\label{eq:actinvGp}
\unl\Phi{}_g\ :\ \unl\la{}_g^*\cG^{(p)}\xrightarrow{\ \cong\ }\cG^{(p)}\,,\qquad g\in\txG_\si\,,
\qqq
with --  for an open cover $\,\cO_{\unl\cM}\,$ endowed with index maps 
\qq\nn
\imath_g\ :\ I\circlearrowleft
\qqq
such that
\qq\nn
\unl\la{}_g(\cO_i)\subset\cO_{\imath_g(i)}\,,
\qqq
{\it i.e.}, defining an extension 
\qq\nn
\widehat{\unl\la}{}_g^*\ &:&\ \textrm{\Cv D}^k\bigl(\cO_{\unl\cM},\cD(p+1)^\bullet\bigr)\circlearrowleft\cr\cr 
&:&\ \bigl(\underset{\tx{\ciut{(k)}}}{\unl{\varpi}}{}_{i_0},\underset{\tx{\ciut{(k-1)}}}{\unl{\varpi}}\hspace{-5pt}{}_{i'_0 i'_1},\ldots,\underset{\tx{\ciut{(-1)}}}{\unl{\varpi}}{}_{i^{(k)}_0 i^{(k)}_1\ldots i^{(k)}_{k+1}}\bigr)\longmapsto\bigl(\unl\la{}_g^*\underset{\tx{\ciut{(k)}}}{\unl{\varpi}}{}_{\imath_g(i_0)},\unl\la{}_g^*\underset{\tx{\ciut{(k-1)}}}{\unl{\varpi}}\hspace{-5pt}{}_{\imath_g(i'_0) \imath_g(i'_1)},\ldots,\unl\la{}_g^*\underset{\tx{\ciut{(-1)}}}{\unl{\varpi}}{}_{\imath_g(i^{(k)}_0) \imath_g(i^{(k)}_1)\ldots \imath_g(i^{(k)}_{k+1})}\bigr)
\qqq
of the manifold map $\,\unl\la_\cdot\,$ -- a sheaf-cohomological presentation furnished by a \Cv ech--Deligne $p$-cochain $\,\underset{\tx{\ciut{(p)}}}{\unl{\xcP}}{}_g^{(1)}\,$ as in \Reqref{eq:CDpcoch}, satisfying the identity
\qq\nn
\widehat{\unl\la}{}_g^*\underset{\tx{\ciut{(p+1)}}}{\unl{\xcB}}+D^{(p)}\underset{\tx{\ciut{(p)}}}{\unl{\xcP}}{}_g^{(1)}=\underset{\tx{\ciut{(p+1)}}}{\unl{\xcB}}\,;
\qqq
\item \textbf{local symmetries} \emph{induced from the global ones with a model $\,\txG_\si\,$} -- represented by $\txG_\si$-equi\-vari\-ant structures on $\,\cG^{(p)}\,$ that may be understood -- in the cohomological picture -- as completions of the $(p+1)$-cocycles $\,\underset{\tx{\ciut{(p+1)}}}{\unl{\xcB}}\,$ in the Deligne--Beilinson cohomology to those in its extension in the direction of group cohomology (for a suitable choice of the open cover $\,\cO_{\unl\cM}$,\ {\it cp} \Rxcite{App.\,I}{Gawedzki:2012fu}) that are based on the sheaf-cohomological data of the $p$-gerbe 1-isomorphism
\qq\label{eq:locsusyequiv1}
\Upsilon_p\equiv\Upsilon_p^{(1)}\ :\ \unl\la{}_\cdot^*\cG^{(p)}\xrightarrow{\ \cong\ }\pr_2^*\cG^{(p)}\ox\cI_{\underset{\tx{\ciut{(p+1)}}}{\varrho_{-\theta_{\rm L}}}}^{(p)}
\qqq
over the arrow manifold $\,\txG_\si\x\unl\cM\,$ of the action groupoid $\,\txG_\si\lx\unl\cM\,$ (with the source and target maps given by $\,\pr_2\,$ and $\,\unl\la{}_\cdot$,\ respectively, {\it cp} Def.\,\ref{def:symmequivstr}), where $\,\cI_{\underset{\tx{\ciut{(p+1)}}}{\varrho_{-\theta_{\rm L}}}}^{(p)}\,$ is the trivial $p$-gerbe with the global curving 
\qq\nn
{\underset{\tx{\ciut{(p+1)}}}{\varrho_{-\theta_{\rm L}}}}=\sum_{k=1}^{p+1}\,\tfrac{(-1)^{p-k}}{k!}\,\pr_2^*\underset{\tx{\ciut{(p+1-k)}}}{\a_{A_1 A_2\ldots A_k}}\wedge\pr_1^*\bigl(\theta^{A_1}_{\rm L}\wedge\theta^{A_2}_{\rm L}\wedge\cdots\wedge\theta^{A_k}_{\rm L}\bigr)\in\Om^{p+1}\bigl(\txG_\si\x M\bigr)
\qqq
written in terms of components $\,\theta_{\rm L}^A,\ A\in\ovl{1,\dim\,\txG_\si}\,$ of the left-invariant Maurer--Cartan 1-form 
\qq\nn
\theta_{\rm L}=\theta_{\rm L}^A\ox\t_A\in\Om^1(\txG_\si)\ox\ggt_\si
\qqq
corresponding to generators $\,\t_A\,$ of the Lie algebra $\,\ggt_\si\,$ of $\,\txG_\si$,\ as well as of the $(p-k)$-forms 
\qq\nn
\underset{\tx{\ciut{(p-k)}}}{\a_{A_1 A_2\ldots A_{k+1}}}=(-1)^{\frac{k(2p-k-1)}{2}}\,\cK^{\unl\la{}_\cdot}_{A_1}\con\cK^{\unl\la{}_\cdot}_{A_2}\con\cdots\con\cK^{\unl\la{}_\cdot}_{A_k}\con\underset{\tx{\ciut{(p)}}}{\k_{A_{k+1}}}\,,\qquad k\in\ovl{1,p}
\qqq
that are all determined by the fundamental vector fields $\,\cK^{\unl\la{}_\cdot}_A\equiv\cK^{\unl\la{}_\cdot}_{\t_A}\in\G(\sfT\unl\cM)\,$ for $\,\unl\la{}_\cdot\,$ associated with the $\,\t_A\,$ in the usual manner and by the $p$-forms $\,\underset{\tx{\ciut{(p)}}}{\k_A}\,$ (assumed to exist) satisfying the conditions
\qq\nn
\sfd\underset{\tx{\ciut{(p)}}}{\k_A}=-\cK^{\unl\la{}_\cdot}_A\con{\rm curv}\bigl(\cG^{(p)}\bigr)\,,\qquad A\in\ovl{1,\dim\,\txG_\si}\,,
\qqq
{\it cp} \Rxcite{Sec.\,2}{Suszek:2019cum} for details and examples.
\eit
\brem\label{rem:descequivstr}
It is important to note that $p$-gerbes admitting a $\txG_\si$-equivariant structure with a \emph{vanishing} curving,
\qq\nn
{\underset{\tx{\ciut{(p+1)}}}{\varrho_{-\theta_{\rm L}}}}\equiv 0\,,
\qqq
are precisely the ones that descend to the quotient manifold $\,\unl\cM/\txG_\si\,$ whenever the latter exists, that is every gerbe over $\,\unl\cM\,$ with that property is (then) a pullback of a $p$-gerbe over $\,\unl\cM/\txG_\si\,$ and {\it vice versa}, {\it cp} \Rxcite{Thm.\,5.3}{Gawedzki:2010rn} (and also \Rxcite{Thms.\,8.15 \& 8.17}{Gawedzki:2012fu}).
\erem
\noindent The above identification of the higher-geometric realisations of $\si$-model symmetries was confirmed in \Rcite{Suszek:2012ddg}, where it was demonstrated that the $p$-gerbe isomorphisms transgress to automorphisms of the prequantum bundle, and so (some of them) give rise to symmetries of the quantum theory, whereas equivariant structures provide target-space data for the gauge-symmetry defects of \Rcite{Suszek:2011,Suszek:2012ddg,Suszek:2013} that implement the gauging of the global symmetry $\,\txG_\si\,$ through a natural generalisation of the worldsheet-orbifold construction of Refs.\,\cite{Dixon:1985jw,Dixon:1986jc} and of its more recent application in the framework of the TFT quantisation of CFT in \Rcite{Frohlich:2009gb} ({\it cp} also \Rcite{Jureit:2006yf}). Equivariant structures themselves were first introduced in the context of the gauging of global $\si$-model symmetries in Refs.\,\cite{Gawedzki:2010rn,Gawedzki:2012fu}, where a mixture of geometric and categorial arguments enriched with field-theoretic considerations was invoked to prove that the structures do, indeed, ensure descent of the field theory from the original target space $\,\unl\cM\,$ to the orbit space $\,\unl\cM/\txG_\si\,$ or, whenever the latter is \emph{not} a smooth manifold, transform the original theory into a model of a field theory with $\,\unl\cM/\txG_\si\,$ as the target space through application of the Universal Gauge Principle worked out in those papers.\bigskip

All the field-theoretic, higher-geometric and -cohomological ideas brought up above converge naturally and become particularly tightly entangled in the setting of super-$\si$-models on homogeneous spaces of Lie supergroups, considered in the present paper. Indeed, the couplings of the background Green--Schwarz (super-)$(p+2)$-cocycles to the (super)charge currents sourced by the propagation of super-$p$-branes `in' the homogeneous spaces, defining the topological terms in the corresponding Dirac--Feynman amplitudes, call for a deep and precise understanding and a rigorous formal description just like their Gra\ss mann-even counterparts, and super{\it symmetry} -- both global and local -- is \emph{the} organising principle in the construction of the relevant field theories, constraining severly the admissible choices of a consistent backround and the topological charge of the super-$p$-brane, or -- more precisely -- the ratio of the gravitational and topological charge ({\it i.e.}, the relative normalisation of the two terms in the super-$\si$-model action functional). This is the basic rationale behind the postulate and a systematic advancement of the programme of a supersymmetry-equivariant geometrisation of the (super-)$(p+2)$-form fields and their topological couplings to `embedded' super-$p$-brane worldvolumes that was put forward in \Rcite{Suszek:2017xlw} and subsequently elaborated from a variety of angles in Refs.\,\cite{Suszek:2019cum,Suszek:2018bvx} and \Rcite{Suszek:2018ugf}. The programme combines the intuitions and techniques derived and borrowed from its predecessor -- the comprehensive study of the gerbe theory of the two-dimensional \emph{bosonic} $\si$-model with the Wess--Zumino term in the action functional\footnote{Of particular relevance are the $\si$-models with the background $\,(\txG,\k_\ggt\circ(\theta_{\rm L}\ox\theta_{\rm L}),\underset{\tx{\ciut{(3)}}}{\txH}=\k_\ggt\circ([\theta_{\rm L}\overset{\wedge}{,}\theta_{\rm L}]\wedge\theta_{\rm L}))\,$ given by a compact Lie group $\,\txG\,$ with the Cartan--Killing metric ($\k_\ggt\,$ is the Killing form on the Lie algebra $\,\ggt\,$ of $\,\txG$) and the canonical Cartan 3-form on it, {\it i.e.}, the so-called Wess--Zumino--Witten $\si$-models of \Rcite{Witten:1983ar}, {\it cp} also \Rcite{Gawedzki:1999bq} for an in-depth exposition.}, laying emphasis on the reformulation and elucidation of global symmetries and of the Universal Gauge Principle in these field theories in the higher-geometric (or gerbe-theoretic) language  -- with the Kleinian(-type) supergeometry of the said homogeneous spaces and the cohomology \emph{sensu largo} of Lie supergroups. The fundamental idea underlying the postulate is the physically motivated identification of the supersymmetry-invariant refinement of the standard de Rham cohomology as the appropriate cohomology in which to analyse and trivialise the Green--Schwarz (super-)$(p+2)$-cocycles, in conjunction with the interpretation of the discrepancy between that refinement and the original de Rham cohomology in terms of the nontrivial topology (or homology) of the Gra\ss mann-odd fibre of an orbifold of the supertarget space by the action of a discrete subgroup of the supersymmetry group generated by `integral' supertranslations\footnote{For an interpretation of this somewhat imprecise definition of the subgroup -- of the Kosteleck\'y--Rabin type ({\it cp} \Rcite{Kostelecky:1983qu}) -- in the $\cS$-point picture, we refer the Reader to \Rxcite{Rem.\,4.1}{Suszek:2017xlw}.}, the orbifold being understood as the implicit supertarget space of the field theory in question. The interpretation, originally advanced by Rabin in \Rcite{Rabin:1985tv} (on the basis of his earlier work with Crane on global aspects of supergeometry, reported in \Rcite{Rabin:1984rm}), and then rephrased and further corroborated by the Author, in \Rcite{Suszek:2019cum}, through the study of the wrapping(-charge) anomaly in the Poisson algebra of Noether charges of global supersymmetry in the canonical description of the super-$\si$-model in the presence of non-vanishing monodromies of (Cauchy) states along the Gra\ss mann-odd directions in the supertarget space, led to the construction of higher-supergeometric objects, termed super-$p$-gerbes, that are in exactly the same structural relation to the Green--Schwarz (super-)$(p+2)$-cocycles as the one between $p$-gerbes and de Rham $(p+2)$-cocycles in the purely Gra\ss mann-even setting, but with the supersymmetry-invariant cohomology replacing the standard de Rham cohomology. Thus, by a straightforward generalisation of \Rxcite{Def.\,5.21}{Suszek:2017xlw}, we arrive at the postulative
\bedef\label{def:spG}
Let $\,\txG\,$ be a Lie supergroup and $\,\cM\,$ a supermanifold endowed with an action 
\qq\nn
\la\ :\ \txG\x\cM\too\cM
\qqq
of $\,\txG$,\ in the sense of \Rxcite{Def.\,7.2.7}{Carmeli:2011}. A \textbf{$\txG$-invariant $p$-gerbe} is an abelian (bundle) $p$-gerbe, in the sense of \Rcite{Gajer:1996}, with total spaces of all surjective submersions entering its definition endowed with the respective (projective) lifts of $\,\la$,\ commuting with the defining $\bC^\x$-actions on the total spaces of all the principal $\bC^\x$-bundles present, with respect to which the submersions are equivariant, all connections are invariant and all (connection-preserving) principal $\bC^\x$-bundle isomorphisms are equivariant. \textbf{$\txG$-invariant $k$-isomorphisms between $\txG$-invariant $p$-gerbes} for $\,k\in\ovl{1,p+1}\,$ are defined analogously.

Whenever $\,\txG\,$ has a non-zero odd component of its super-dimension ({\it i.e.}, it is a proper Lie supergrup that is \emph{not} a Lie group), the corresponding $\txG$-invariant $p$-gerbes are called \textbf{super-$p$-gerbes}, or \textbf{supersymmetric $p$-gerbes}, and $\txG$-invariant $k$-isomorphisms between them are called \textbf{super-$p$-gerbe $k$-isomorphisms}, or \textbf{supersymmetric $p$-gerbe $k$-isomorphisms}.
\exdef
\brem
Using the structure of a $\txG$-invariant $p$-gerbe, we may readily construct $p$-gerbe 1-isomorphisms \eqref{eq:actinvGp}, lifting the action of the (super)symmetry group $\,\txG\,$ to the geometric object $\,\cG^{(p)}\,$ from its base. This fact was illustrated (on 0- and 1-gerbes) in \Rxcite{Secs.\,4.1 \& 4.2}{Suszek:2019cum}.
\erem
\noindent The general definition, which can readily be derived from the hitherto developments of the super\-sym\-metry-equivariant geometrisation programme\footnote{{\it Cp}, {\it e.g.}, the opening discussion on the notion of supersymmetry of a 0- and 1-gerbe over a supermanifold in \Rxcite{Sec.\,4.1}{Suszek:2019cum} and \Rxcite{Sec.\,4.2}{Suszek:2019cum}, respectively.}, has the following important specialisations that play a prime r\^ole in the analysis of concrete examples.
\bedef\label{def:s0g}
In the notation of Def.\,\ref{def:spG}, a \textbf{super-0-gerbe}, or a \textbf{supersymmetric 0-gerbe} over $\,\cM\,$ of curvature $\,\underset{\tx{\ciut{(2)}}}{\chi}\in Z^2(\cM)$,\ with $\,\txG\,$ as the supersymmetry group, is a principal $\bC^\x$-bundle $\,(\sfY\cM,\pi_{\sfY\cM},\underset{\tx{\ciut{(1)}}}{\cA})$,\ described by the diagram
\qq\nn
\alxydim{@C=1.5cm@R=1.5cm}{ \bC^\x \ar[r] & \sfY\cM,\underset{\tx{\ciut{(1)}}}{\cA} \ar[d]^{\pi_{\sfY\cM}} \\ & \cM,\underset{\tx{\ciut{(2)}}}{\chi}}
\qqq
in which $\,\pi_{\sfY\cM}\,$ is a surjective submersion and 
\qq\nn
\underset{\tx{\ciut{(1)}}}{\cA}\in\Om^1(\sfY\cM)
\qqq
is a principal $\bC^\x$-connection such that
\qq\nn
\sfd\underset{\tx{\ciut{(1)}}}{\cA}\in\pi_{\sfY\cM}^*\underset{\tx{\ciut{(2)}}}{\chi}
\qqq
(the pullback having the usual realisation in the local-coordinate ({\it i.e.}, $\cS$-point) picture), with the following properties:
\bit
\item the action $\,\la\,$ admits a lift $\,\sfY\la\,$ to the total space $\,\sfY\cM\,$ determined by the commutative diagram
\qq\label{diag:Yliftla}
\alxydim{@C=2cm@R=1.5cm}{ \txG\x\sfY\cM \ar[r]^{\quad\sfY\la} \ar[d]_{\id_\txG\x\pi_{\sfY\cM}} & \sfY\cM \ar[d]^{\pi_{\sfY\cM}} \\ \txG\x\cM \ar[r]_{\quad\la} & \cM}
\qqq
that commutes with the defining action
\qq\nn
r^{\sfY\cM}\ :\ \sfY\cM\x\bC^\x\too\sfY\cM
\qqq
of the structure group $\,\bC^\x\,$ on $\,\sfY\cM$,\ as expressed by the commutative diagram
\qq\nn
\alxydim{@C=2cm@R=1.5cm}{ \txG\x\sfY\cM\x\bC^\x \ar[d]_{\sfY\la\x\id_{\bC^\x}} \ar[r]^{\quad\id_\txG\x r^{\sfY\cM}} & \txG\x\sfY\cM \ar[d]^{\sfY\la} \\ \sfY\cM\x\bC^\x \ar[r]_{\quad r^{\sfY\cM}} & \sfY\cM}\,;
\qqq
\item the lift preserves the principal $\bC^\x$-connection, which reduces, in the local-coordinate ({\it i.e.}, $\cS$-point) picture, to the standard property of invariance of the latter super-1-form under pullback along the mapping, induced by $\,{\rm Yon}_\la(\cS)$,\ that effects the action of a \emph{fixed} $\,g\in{\rm Yon}_\txG(\cS)\,$ on $\,{\rm Yon}_\cM(\cS)\,$ (both in a local-coordinate description), to be written -- by a customary abuse of notation -- as 
\qq\nn
\sfY\la_g^*\underset{\tx{\ciut{(1)}}}{\cA}=\underset{\tx{\ciut{(1)}}}{\cA}\,,\qquad g\in{\rm Yon}_\txG(\cS)\,.
\qqq
\eit

An \textbf{isomorphism between super-0-gerbes} $\,(\sfY_A\cM,\pi_{\sfY_A\cM},\underset{\tx{\ciut{(1)}}}{\cA}{}_A),\ A\in\{1,2\}$,\ endowed with the respective lifts $\,\sfY_A\la\,$ of $\,\la$,\ is a connection-preserving isomorphism 
\qq\nn
\alxydim{@C=2cm@R=1.5cm}{ \sfY_1\cM \ar[r]^{\Phi} \ar[d]_{\pi_{\sfY_1\cM}} & \sfY_2\cM \ar[d]^{\pi_{\sfY_2\cM}} \\ \cM \ar@{=}[r]_{\quad\id_\cM} & \cM}
\qqq
between the principal $\bC^\x$-bundles that intertwines the lifts, {\it i.e.}, satisfies
\qq\nn
\alxydim{@C=2cm@R=1.5cm}{ \txG\x\sfY_1\cM \ar[r]^{\quad\sfY_1\la} \ar[d]_{\id_\txG\x\Phi} & \sfY_1\cM \ar[d]^{\Phi} \\ \txG\x\sfY_2\cM \ar[r]_{\quad\sfY_2\la} & \sfY_2\cM}\,.
\qqq
\exdef
\bedef\label{def:s1g}
In the notation of Def.\,\ref{def:spG}, a \textbf{super-1-gerbe}, or a \textbf{supersymmetric 1-gerbe} of curvature $\,\underset{\tx{\ciut{(3)}}}{\chi}\in Z^3(\cM)\,$ over $\,\cM$,\ with $\,\txG\,$ as the supersymmetry group, is an abelian (bundle) 1-gerbe $\,(\sfY\cM,\pi_{\sfY\cM},\underset{\tx{\ciut{(2)}}}{\cB},L,\pi_L,\underset{\tx{\ciut{(1)}}}{\cA}{}_L,\mu_L)$,\ described by the diagram
\qq\nn
\alxydim{@C=2cm@R=1.5cm}{ \mu_L\ :\ \pr_{1,2}^*L\ox\pr_{2,3}^*L\xrightarrow{\ \cong\ }\pr_{1,3}^*L \ar[d] & \bC^\x \ar[r] & L,\underset{\tx{\ciut{(1)}}}{\cA}{}_L \ar[d]^{\pi_L} & \\ \sfY^{[3]}\cM \ar@/^.75pc/[rr]^{\pr_{1,2}} \ar@<0ex>[rr]|-{\pr_{2,3}} \ar@/^-.75pc/[rr]_{\pr_{1,3}} & & \sfY^{[2]}\cM \ar@<-.75ex>[r]_{\pr_2} \ar@<.75ex>[r]^{\pr_1} & \sfY\cM,\underset{\tx{\ciut{(2)}}}{\cB} \ar[d]^{\pi_{\sfY\cM}} \\ & & & \cM,\underset{\tx{\ciut{(3)}}}{\chi} }
\qqq
in which $\,\pi_{\sfY\cM}\,$ is a surjective submersion, $\,\underset{\tx{\ciut{(2)}}}{\cB}\in\Om^2(\sfY\cM)\,$ is the curving such that
\qq\nn
\sfd\underset{\tx{\ciut{(2)}}}{\cB}=\pi_{\sfY\cM}^*\underset{\tx{\ciut{(3)}}}{\chi}\,,
\qqq
and $\,L\,$ is a principal $\bC^\x$-bundle over the fibred square\footnote{{\it Cp}, {\it e.g.}, \Rxcite{Def.\,B.1.11}{Carmeli:2011}.}, described by the commutative diagram
\qq\nn
\alxydim{@C=.75cm@R=1cm}{ & \sfY^{[2]}\cM \ar[rd]^{\pr_2} \ar[ld]_{\pr_1} & \\ \sfY\cM \ar[rd]_{\pi_{\sfY\cM}} & & \sfY\cM \ar[ld]^{\pi_{\sfY\cM}} \\ & \cM & }\,,
\qqq
and endowed with a principal $\bC^\x$-connection $\,\underset{\tx{\ciut{(1)}}}{\cA}{}_L\in\Om^1(L)\,$ of curvature $\,(\pr_2^*-\pr_1^*)\underset{\tx{\ciut{(2)}}}{\cB}$, 
\qq\nn
\sfd\underset{\tx{\ciut{(1)}}}{\cA}{}_L=\pi_L^*(\pr_2^*-\pr_1^*)\underset{\tx{\ciut{(2)}}}{\cB}\,,
\qqq
and a (connection-preserving) principal $\bC^\x$-bundle isomorphism $\,\mu_L\,$ of (the tensor product\footnote{The tensor product $\,L_1\ox L_2\,$ of principal $\bC^\x$-bundles $\,L_\a,\ \a\in\{1,2\}\,$ is defined, after \Rcite{Brylinski:1993ab}, as the (principal) bundle $\,(L_1\x L_2)/\bC^\x\,$ associated with $\,L_1\,$ through the defining $\bC^\x$-action on $\,L_2$,\ {\it cp} also \Rxcite{Rem.\,5.5}{Suszek:2017xlw} and \Rxcite{Def.\,6.2.1}{Kessler:2019bwp}.} of) the pullback principal $\bC^\x$-bundles
\qq\nn
\alxydim{@C=2cm@R=1.5cm}{ \pr_{i,j}^*L\equiv\sfY^{[3]}\cM\x_{\pr_{i,j}}L \ar[r]^{\qquad\qquad\pr_2} \ar[d]_{\pr_1} & L \ar[d]^{\pi_L} \\ \sfY^{[3]}\cM \ar[r]_{\pr_{i,j}\equiv(\pr_i,\pr_j)} & \sfY^{[2]}\cM}\,,\qquad (i,j)\in\{(1,2),(2,3),(1,3)\}
\qqq
over the fibred cube $\,\sfY^{[3]}\cM$,\ described by the commutative diagram
\qq\nn
\alxydim{@C=1.5cm@R=1cm}{ & \sfY^{[3]}\cM \ar[rd]^{\pr_3} \ar[d]_{\pr_2} \ar[ld]_{\pr_1} & \\ \sfY\cM \ar[rd]_{\pi_{\sfY\cM}} & \sfY\cM \ar[d]_{\pi_{\sfY\cM}} & \sfY\cM \ar[ld]^{\pi_{\sfY\cM}} \\ & \cM & }\,,
\qqq
that induces a groupoid structure on its fibres, being subject to the associativity constraint
\qq\nn
\pr_{1,2,4}^*\mu_L\circ\bigl(\id_{\pr_{1,2}^*L}\ox\pr_{2,3,4}^*\mu_L\bigr)=\pr_{1,3,4}^*\mu_L\circ\bigl(\pr_{1,2,3}^*\mu_L\ox \id_{\pr_{3,4}^*L}\bigr)
\qqq
over the fibred tetrahedron $\,\sfY^{[4]}\cM\,$ (with its canonical projections $\,\pr_{i,j,k}\equiv(\pr_i,\pr_j,\pr_k)\ :\ \sfY^{[4]}M\too \sfY^{[3]}M\,$ and $\,\pr_{i,j}\equiv(\pr_i,\pr_j)\ :\ \sfY^{[4]}M\too \sfY^{[2]}M,\ i,j\in\{1,2,3,4\}$), described by the commutative diagram
\qq\nn
\alxydim{@C=1.5cm@R=1cm}{ & & \sfY^{[4]}\cM \ar[rrd]^{\pr_4} \ar[rd]_{\pr_3} \ar[ld]^{\pr_2} \ar[lld]_{\pr_1} & & \\ \sfY\cM \ar[rrd]_{\pi_{\sfY\cM}} & \sfY\cM \ar[rd]^{\pi_{\sfY\cM}} & & \sfY\cM \ar[ld]_{\pi_{\sfY\cM}}  & \sfY\cM \ar[lld]^{\pi_{\sfY\cM}} \\ & & \cM & & }\,,
\qqq
with the following properties:
\bit
\item the action $\,\la\,$ admits a lift $\,\sfY\la\,$ to the total space $\,\sfY\cM\,$ determined by the commutative Diag.\,\eqref{diag:Yliftla}, and so also to the fibred products $\,\sfY^{[n]}\cM,\ n\in\{2,3,4\}$,
\qq\nn
\sfY^{[n]}\la\equiv\bigl(\sfY\la\circ\pr_{1,2},\sfY\la\circ\pr_{1,3},\ldots,\sfY\la\circ\pr_{1,n}\bigr)\ :\ \txG\x\sfY^{[n]}\cM\too\sfY^{[n]}\cM\,,
\qqq
and from $\,\sfY^{[2]}\cM\,$ to the total space $\,L$,
\qq\nn
\alxydim{@C=2cm@R=1.5cm}{ \txG\x L \ar[r]^{\quad L\la} \ar[d]_{\id_\txG\x\pi_L} & L \ar[d]^{\pi_L} \\ \txG\x\sfY^{[2]}\cM \ar[r]_{\quad\sfY^{[2]}\la} & \sfY^{[2]}\cM}\,,
\qqq
the latter lift commuting with the defining action
\qq\nn
r^L\ :\ L\x\bC^\x\too L
\qqq
of the structure group $\,\bC^\x\,$ on $\,L$,\ as expressed by the commutative diagram
\qq\nn
\alxydim{@C=2cm@R=1.5cm}{ \txG\x L\x\bC^\x \ar[d]_{L\la\x\id_{\bC^\x}} \ar[r]^{\quad\id_\txG\x r^L} & \txG\x L \ar[d]^{L\la} \\ L\x\bC^\x \ar[r]_{\quad r^L} & L}\,;
\qqq
\item the lifts $\,\sfY^{[3]}\la\,$ and $\,L\la\,$ induce, in turn, the canonical lifts 
\qq\nn
L_{i,j}\la\equiv\bigl(\sfY^{[3]}\la\circ(\pr_1\x\pr_2),L\la\circ(\pr_1\x\pr_3)\bigr)\ :\ \txG\x\pr_{i,j}^*L\equiv\txG\x\bigl(\sfY^{[3]}\cM\x_{\pr_{i,j}}L\bigr)\too\pr_{i,j}^*L
\qqq
of $\,\la\,$ to the respective pullback bundles, and so also a (diagonal) lift to the tensor-product bundle
\qq\nn
L_{1,2;2,3}\la\ :\ \txG\x\bigl(\pr_{1,2}^*L\ox\pr_{2,3}^*L\bigr)\too\pr_{1,2}^*L\ox\pr_{2,3}^*L\,;
\qqq
\item the lifts $\,\sfY\la\,$ and $\,L\la\,$ preserve the curving $\,\underset{\tx{\ciut{(2)}}}{\cB}\,$ and the principal $\bC^\x$-connection $\,\underset{\tx{\ciut{(1)}}}{\cA}{}_L$,\ respectively, which -- in the previously introduced notation -- may be written as
\qq\nn
\sfY\la_g^*\underset{\tx{\ciut{(2)}}}{\cB}=\underset{\tx{\ciut{(2)}}}{\cB}\qquad\land\qquad L\la_g^*\underset{\tx{\ciut{(1)}}}{\cA}{}_L=\underset{\tx{\ciut{(1)}}}{\cA}{}_L\,,\qquad g\in{\rm Yon}_\txG(\cS)\,;
\qqq
\item the groupoid structure $\,\mu_L\,$ is equivariant with respect to the lifted actions,
\qq\nn
\mu_L\circ L_{1,2;2,3}\la=L_{1,3}\la\circ(\id_\txG\x\mu_L)\,.
\qqq
\eit

An \textbf{isomorphism between super-1-gerbes} $\,(\sfY_A\cM,\pi_{\sfY_A\cM},\underset{\tx{\ciut{(2)}}}{\cB}{}_A,L_A,\pi_{L_A},\underset{\tx{\ciut{(1)}}}{\cA}{}_{L_A},\mu_{L_A}),\ A\in\{1,2\}$,\ endowed with the respective lifts $\,\sfY_A\la\,$ and $\,L_A\la$,\ is a gerbe 1-isomorphism $\,\Phi=(\sfY\sfY_{1,2}\cM,\pi_{\sfY\sfY_{1,2}\cM},E,\pi_E,$ $\underset{\tx{\ciut{(1)}}}{\cA}{}_E,\a_E)$,\ composed of a surjective submersion 
\qq\nn
\pi_{\sfY\sfY_{1,2}\cM}\ :\ \sfY\sfY_{1,2}\cM\too\sfY_{1,2}\cM
\qqq
over the fibred product 
\qq\nn
\alxydim{@C=.75cm@R=1cm}{ & \sfY_{1,2}\cM\equiv\sfY_1\cM\x_\cM\sfY_2\cM \ar[rd]^{\pr_2} \ar[ld]_{\pr_1} & \\ \sfY_1\cM \ar[rd]_{\pi_{\sfY_1\cM}} & & \sfY_2\cM \ar[ld]^{\pi_{\sfY_2\cM}} \\ & \cM & }\,,
\qqq
and a principal $\bC^\x$-bundle
\qq\nn
\alxydim{@C=1.5cm@R=1.5cm}{ \bC^\x \ar[r] & E,\underset{\tx{\ciut{(1)}}}{\cA}{}_E \ar[d]^{\pi_E} \\ & \sfY\sfY_{1,2}\cM}
\qqq
over its total space with a surjective submersion $\,\pi_E\,$ and a principal $\bC^\x$-connection 
\qq\nn
\underset{\tx{\ciut{(1)}}}{\cA}{}_E\in\Om^1(E)
\qqq
of curvature $\,\pr_2^*\underset{\tx{\ciut{(2)}}}{\cB}{}_2-\pr_1^*\underset{\tx{\ciut{(2)}}}{\cB}{}_1$,
\qq\nn
\sfd\underset{\tx{\ciut{(1)}}}{\cA}{}_E=\pi_E^*\bigl(\pr_2^*\underset{\tx{\ciut{(2)}}}{\cB}{}_2-\pr_1^*\underset{\tx{\ciut{(2)}}}{\cB}{}_1\bigr)\,,
\qqq
together with a connection-preserving principal $\bC^\x$-bundle isomorphism 
\qq\nn
\a_E\ :\ \bigl(\pi_{\sfY\sfY_{1,2}\cM}\x\pi_{\sfY\sfY_{1,2}\cM}\bigr)^*\pr_{1,3}^*L_1\ox\pr_2^*E\xrightarrow{\ \cong\ }\pr_1^*E\ox\bigl(\pi_{\sfY\sfY_{1,2}\cM}\x\pi_{\sfY\sfY_{1,2}\cM}\bigr)^*\pr_{2,4}^*L_2
\qqq
over the fibred square 
\qq\nn
\alxydim{@C=.75cm@R=1cm}{ & \sfY^{[2]}\sfY_{1,2}\cM\equiv\sfY\sfY_{1,2}\cM\x_\cM\sfY\sfY_{1,2}\cM \ar[rd]^{\pr_2} \ar[ld]_{\pr_1} & \\ \sfY\sfY_{1,2}\cM \ar[rd]_{\pi_{\sfY_1\cM}\circ\pr_1\circ\pi_{\sfY\sfY_{1,2}\cM}\qquad} & & \sfY\sfY_{1,2}\cM \ar[ld]^{\qquad\quad\pi_{\sfY_2\cM}\circ\pr_2\circ\pi_{\sfY\sfY_{1,2}\cM}} \\ & \cM & }
\qqq 
satisfying the identity
\qq\nn
&&\bigl(\id_{\pr_1^*E}\ox\pi_{1,2,3}^*\circ\pr_{2,4,6}^*\mu_{L_2}\bigr)\circ\bigl(\pr_{1,2}^*\a_E\ox\id_{\pi_{2,3}^*\circ\pr_{2,4}^*L_2}\bigr)\circ\bigl(\id_{\pi_{1,2}^*\circ\pr_{1,3}^*L_1}\ox\pr_{2,3}^*\a_E\bigr)\cr\cr
&=&\pr_{1,3}^*\a_E\circ\bigl(\pi_{1,2,3}^*\circ\pr_{1,3,5}^*\mu_{L_1}\ox\id_{\pr_3^*E}\bigr)
\qqq
over the fibred cube
\qq\nn
\alxydim{@C=3cm@R=1cm}{ & \sfY^{[3]}\sfY_{1,2}\cM\equiv\sfY\sfY_{1,2}\cM\x_\cM\sfY\sfY_{1,2}\cM\x_\cM\sfY\sfY_{1,2}\cM \ar[rd]^{\pr_3} \ar[d]_{\pr_2} \ar[ld]_{\pr_1} & \\ \sfY\sfY_{1,2}\cM \ar[rd]_{\pi_{\sfY_1\cM}\circ\pr_1\circ\pi_{\sfY\sfY_{1,2}\cM}\qquad} & \sfY\sfY_{1,2}\cM \ar[d]_{\pi_{\sfY_2\cM}\circ\pr_2\circ\pi_{\sfY\sfY_{1,2}\cM}} & \sfY\sfY_{1,2}\cM \ar[ld]^{\qquad\pi_{\sfY_1\cM}\circ\pr_1\circ\pi_{\sfY\sfY_{1,2}\cM}} \\ & \cM & }\,,
\qqq
equipped with the mappings
\qq\nn
&\pi_{i,j}=(\pi_{\sfY\sfY_{1,2}M}\x\pi_{\sfY\sfY_{1,2}M})\circ\pr_{i,j}\,,\quad(i,j)\in\{(1,2),(2,3),(1,3)\}\,,&\cr\cr &\pi_{1,2,3}=\pi_{\sfY\sfY_{1,2}M}\x\pi_{\sfY\sfY_{1,2}M}\x\pi_{\sfY\sfY_{1,2}M}\,,&
\qqq
the gerbe 1-isomorphism $\,\Phi\,$ having the following properties:
\bit
\item the action 
\qq\nn
\sfY_{1,2}\la\equiv\bigl(\sfY_1\la\circ\pr_{1,2},\sfY_2\la\circ\pr_{1,3}\bigr)\ :\ \txG\x\sfY_{1,2}\cM\too\sfY_{1,2}\cM\,,
\qqq 
induced canonically by the $\,\sfY_A\la\,$ admits a lift $\,\sfY\sfY_{1,2}\la\,$ to $\,\sfY\sfY_{1,2}\cM$,\ described by the commutative diagram
\qq\nn
\alxydim{@C=2cm@R=1.5cm}{ \txG\x\sfY\sfY_{1,2}\cM \ar[r]^{\quad\sfY\sfY_{1,2}\la} \ar[d]_{\id_\txG\x\pi_{\sfY\sfY_{1,2}\cM}} & \sfY\sfY_{1,2}\cM \ar[d]^{\pi_{\sfY\sfY_{1,2}\cM}} \\ \txG\x\sfY_{1,2}\cM \ar[r]_{\quad\sfY_{1,2}\la} & \sfY_{1,2}\cM}\,,
\qqq
and the latter further lifts to the total space $\,E$,
\qq\nn
\alxydim{@C=2cm@R=1.5cm}{ \txG\x E \ar[r]^{\quad E\la} \ar[d]_{\id_\txG\x\pi_E} & E \ar[d]^{\pi_E} \\ \txG\x\sfY\sfY_{1,2}\cM \ar[r]_{\quad\sfY\sfY_{1,2}\la} & \sfY\sfY_{1,2}\cM}\,,
\qqq
the resulting lift commuting with the defining action
\qq\nn
r^E\ :\ E\x\bC^\x\too E
\qqq
of the structure group $\,\bC^\x\,$ on $\,E$,\ as expressed by the commutative diagram
\qq\nn
\alxydim{@C=2cm@R=1.5cm}{ \txG\x E\x\bC^\x \ar[d]_{E\la\x\id_{\bC^\x}} \ar[r]^{\quad\id_\txG\x r^E} & \txG\x E \ar[d]^{E\la} \\ E\x\bC^\x \ar[r]_{\quad r^E} & E}\,;
\qqq
\item the lifts $\,\sfY\sfY_{1,2}\la\,$ and $\,E\la$,\ in conjunction with the formerly introduced ones, induce, in turn, the canonical lifts
\qq\nn
\sfY^{[2]}_{A,A+2}L_A\la&\equiv&\bigl(\sfY^{[2]}\sfY_{1,2}\la\circ\pr_{1,2},\bigl(\sfY_{1,2}\la\circ\pr_{1,2},\sfY_{1,2}\la\circ\pr_{1,3}\bigr)\circ\pr_{1,3},L_A\la\circ\pr_{1,4}\bigr)\cr\cr 
&:&\ \txG\x\bigl(\sfY^{[2]}\sfY_{1,2}\cM\x_{\pi_{\sfY\sfY_{1,2}M}\x\pi_{\sfY\sfY_{1,2}M}}\bigl(\sfY^{[2]}_{1,2}\cM\x_{\pr_{A,A+2}}L_A\bigr)\bigr)\cr\cr
&\too&\sfY^{[2]}\sfY_{1,2}\cM\x_{\pi_{\sfY\sfY_{1,2}M}\x\pi_{\sfY\sfY_{1,2}M}}\bigl(\sfY^{[2]}_{1,2}\cM\x_{\pr_{A,A+2}}L_A\bigr)\,,\cr\cr\cr
\sfY^{[2]}_A E\la&\equiv&\bigl(\sfY^{[2]}\sfY_{1,2}\la\circ\pr_{1,2},E\la\circ\pr_{1,3}\bigr)\ :\ \txG\x\bigl(\sfY^{[2]}\sfY_{1,2}\cM\x_{\pr_A}E\bigr)\too\sfY^{[2]}\sfY_{1,2}\cM\x_{\pr_A}E
\qqq
written for $\,A\in\{1,2\}$,\ for the fibred square
\qq\nn
\alxydim{@C=.75cm@R=1cm}{ & \sfY^{[2]}_{1,2}\cM\equiv\sfY_{1,2}\cM\x_\cM\sfY_{1,2}\cM \ar[rd]^{\pr_2} \ar[ld]_{\pr_1} & \\ \sfY_{1,2}\cM \ar[rd]_{\pi_{\sfY_1\cM}\circ\pr_1\quad} & & \sfY_{1,2}\cM \ar[ld]^{\quad\pi_{\sfY_2\cM}\circ\pr_2} \\ & \cM & }
\qqq
and for the canonical lift
\qq\nn
\sfY^{[2]}\sfY_{1,2}\la\equiv\bigl(\sfY\sfY_{1,2}\la\circ\pr_{1,2},\sfY\sfY_{1,2}\la\circ\pr_{1,3}\bigr)\ :\ \txG\x\sfY^{[2]}\sfY_{1,2}\cM\too\sfY^{[2]}\sfY_{1,2}\cM
\qqq
of $\,\sfY\sfY_{1,2}\la\,$ to $\,\sfY^{[2]}\sfY_{1,2}\cM$,\ and so also canonical (diagonal) lifts 
\qq\nn
\sfY^{[2]}_{1,3;2}L_1E\la\ &:&\ \txG\x\bigl(\pi_{\sfY\sfY_{1,2}\cM}\x\pi_{\sfY\sfY_{1,2}\cM}\bigr)^*\pr_{1,3}^*L_1\ox\pr_2^*E\too\bigl(\pi_{\sfY\sfY_{1,2}\cM}\x\pi_{\sfY\sfY_{1,2}\cM}\bigr)^*\pr_{1,3}^*L_1\ox\pr_2^*E\,,\cr\cr
\sfY^{[2]}_{1;1,3}EL_2\la\ &:&\ \txG\x\pr_1^*E\ox\bigl(\pi_{\sfY\sfY_{1,2}\cM}\x\pi_{\sfY\sfY_{1,2}\cM}\bigr)^*\pr_{2,4}^*L_2\too\pr_1^*E\ox\bigl(\pi_{\sfY\sfY_{1,2}\cM}\x\pi_{\sfY\sfY_{1,2}\cM}\bigr)^*\pr_{2,4}^*L_2
\qqq
to the tensor-product bundles are induced;
\item the lift $\,E\la\,$ preserves the principal $\bC^\x$-connection $\,\underset{\tx{\ciut{(1)}}}{\cA}{}_E$,\ which -- in the previously introduced notation -- may be written as
\qq\nn
E\la_g^*\underset{\tx{\ciut{(1)}}}{\cA}{}_E=\underset{\tx{\ciut{(1)}}}{\cA}{}_E\,,\qquad g\in{\rm Yon}_\txG(\cS)\,;
\qqq
\item the isomorphism $\,\a_E\,$ is equivariant with respect to the lifted actions,
\qq\nn
\a_E\circ\sfY^{[2]}_{1,3;2}L_1E\la=\sfY^{[2]}_{1;1,3}EL_2\la\circ(\id_\txG\x\a_E)\,.
\qqq 
\eit
\exdef
\noindent The above constitute a most natural abstraction of the concrete results of Refs.\,\cite{Suszek:2017xlw,Suszek:2019cum,Suszek:2018bvx,Suszek:2018ugf}, the latter being tailored to the peculiar supergeometric circumstances in which physically relevant models are realised (by construction). These are, invariably, the circumstances in which the $\txG$-supermanifold of a supersymmetry Lie supergroup $\,\txG\,$ of immediate interest is a homogeneous space $\,\txG/\txH\,$ relative to an isotropy Lie subgroup $\,\txH\subset|\txG|\equiv\cM_\txH\,$ of its body, the mother supergroup fibring principally over the base $\,\cM_\txH\,$ in the manner discussed at length in Sec.\,\ref{sec:physmod}. Furtermore, the corresponding super-$(p+2)$-cocycles $\,\underset{\tx{\ciut{(p+2)}}}{\txH}\,$ (descended from those on $\,\txG$,\ {\it cp} \Reqref{eq:Hachi}) define nontrivial classes in the supersymmetry-invariant refinement of the de Rham cohomology of $\,\cM_\txH$,\ at least as long as we insist -- in keeping with the principles which underlie the construction of the associated super-$\si$-models, as postulated in the original papers and subsequently recalled and adapted to the higher-geometric framework in \Rcite{Suszek:2018bvx} -- that the curvings on homogeneous spaces with a super-Minkowskian limit of the (homogeneous) \.In\"on\"u--Wigner blow-up should asymptote to their flat-superspacetime counterparts, {\it cp} \Rcite{Suszek:2018bvx} for details of the argument. As a consequence, the task of geometrising $\,\underset{\tx{\ciut{(p+2)}}}{\txH}\,$ relative to the \emph{supersymmetrically-invariant refinement} of the de Rham cohomology of $\,\cM_\txH$,
\qq\nn
H^\bullet_{\rm dR}(\cM_\txH,\bR)^\txG\,,
\qqq 
boils down to associating with the nontrivial class $\,[\underset{\tx{\ciut{(p+2)}}}{\chi}]_{{\rm CaE},\txH-{\rm basic}}\,$ of $\,\underset{\tx{\ciut{(p+2)}}}{\chi}\,$ in the (right-)$\txH$-basic\footnote{A (super-)$k$-form $\,\underset{\tx{\ciut{(k)}}}{\om}\,$ on an $\txH$-(super)manifold $\,\cM\,$ is termed $\txH$-basic if it is $\txH$-invariant and $\txH$-horizontal, the latter meaning that contractions of fundamental vector fields on $\,\cM\,$ (induced by the $\txH$-action) with $\,\underset{\tx{\ciut{(k)}}}{\om}\,$ vanish identically.} refinement of the Cartan--Eilenberg cohomology of $\,\txG$,\ that is the de Rham cohomology  
\qq\nn
{\rm CaE}^\bullet(\txG)_{\txH-{\rm basic}}\equiv H^\bullet_{\rm dR}(\txG,\bR)^\txG_{\txH-{\rm basic}}
\qqq
of (left-)$\txG$-invariant and $\txH$-basic superdifferential forms on $\,\txG$,\ a particular abelian super-$p$-gerbe over $\,\txG\,$ that descends (as a super-$p$-gerbe) to the quotient $\,\txG/\txH$.\ By a super-counterpart of the argument recalled in Rem.\,\ref{rem:descequivstr}, this requires that the super-$p$-gerbe over $\,\txG\,$ be equipped with an $\txH$-equivariant structure of a vanishing curving which should be taken supersymmetric ({\it i.e.}, compatible with the global supersymmetry present, essentially along the lines of \Rxcite{Sec.\,4}{Suszek:2019cum}). Being a resolution of $\,\underset{\tx{\ciut{(p+2)}}}{\chi}\,$ in the ($\txH$-basic) Cartan--Eilenberg cohomology, the mother super-$p$-gerbe over $\,\txG\,$ may acquire a Lie-superalgebraic description determined by the super-variant of -- on the one hand -- the classical equality between the Cartan--Eilenberg cohomology $\,{\rm CaE}^\bullet(\txG)\,$ and the Chevalley--Eilenberg cohomology 
\qq\nn
{\rm CE}^\bullet(\ggt)\equiv H^\bullet(\ggt;\bR)
\qqq
of the Lie superalgebra $\,\ggt\,$ with values in the trivial $\ggt$-module $\,\bR$,\ as derived in \Rxcite{App.\,C}{Suszek:2017xlw}, and -- on the other hand -- an interpretation of classes in $\,H^\bullet(\ggt;\bR)\,$ in terms of superalgebraic extensions of $\,\ggt$,\ further assumed integrable to surjective submersions over $\,\txG$.\ The latter interpretation has been considered in two as yet unrelated (in all generality) guises: 
\bit
\item the direct one-to-one correspondence between classes in $\,{\rm CE}^{p+2}(\ggt)\,$ and the peculiar $L_\infty$-su\-per\-al\-ge\-bras introduced by Baez and Huerta in Refs.\,\cite{Baez:2010ye,Huerta:2011ic} and termed \textbf{slim Lie $(p+1)$-superalgebras} ({\it cp} also \Rcite{Baez:2004hda6} for a non-$\bZ/2\bZ$-graded precursor of that correspondence) -- some preliminary ideas about integration of these Lie $(p+1)$-superalgebras to the so-called \textbf{Lie $(p+2)$-supergroups} were presented in \Rxcite{Chap.\,7}{Huerta:2011ic} but it seems that the range of applicability of the integration method is an open question and -- to the best of the Author's knowledge -- no \emph{concrete} examples of Lie $(p+2)$-supergroups of physical (superstring-theoretic) relevance have been constructed explicitly to date (let alone the sought-after super-$p$-gerbes);
\item a physics-guided construction of an extension $\,\widehat\txG\xrightarrow{\ \widehat\pi\ }\txG\,$ that integrates an extension $\,\widehat\ggt\xrightarrow{\ \widehat\sfP\ }\ggt\,$ parametrised by topological charges carried by the super-$p$-brane and has the fundamental property 
\qq\nn
[\widehat\pi^*\underset{\tx{\ciut{(p+2)}}}{\chi}]_{{\rm CaE},\txH-{\rm basic}}=0\in{\rm CaE}^{p+2}(\widehat\txG)_{\txH-{\rm basic}}
\qqq
that no proper sub-superalgebra of $\,\widehat\ggt\,$ shares with it -- the extension may admit a description in terms of a short exact sequence of Lie supergroups
\qq\nn
\alxydim{@C=1cm@R=1cm}{\bd1 \ar[r] & Z \ar[r]^{\jmath_Z} & \widehat\txG \ar[r]^{\widehat\pi} & \txG \ar[r] & \bd1}
\qqq
integrating a short-exact sequence of the corresponding Lie superalgebras
\qq\nn
\alxydim{@C=1cm@R=1cm}{\brd0 \ar[r] & \zgt \ar[r]^{\jmath_\zgt} & \widehat\ggt \ar[r]^{\widehat\sfP} & \ggt \ar[r] & \brd0}
\qqq
that results from a sequence of supercentral extensions induced, each, by a nontrivial class in the 2nd cohomology group $\,{\rm CE}^2(\ggt_{\rm int})_{\txH-{\rm basic}}\,$ of an intermediate Lie superalgebra $\,\ggt_{\rm int}\,$ ($\ggt\,$ being one of them) engendered by $\,\underset{\tx{\ciut{(p+2)}}}{\chi}\,$ in such a manner that every extension in the sequence yields a partial, or termwise, trivialisation of the pullback of $\,\underset{\tx{\ciut{(p+2)}}}{\chi}\,$ to the Lie supergroup $\,\txG_{\rm int}\,$ (to which $\,\ggt_{\rm int}\,$ integrates\footnote{Integrability is essentially controlled by the Tuynman--Wiegerinck criterion, {\it cp} \Rcite{Tuynman:1987ij}.}) in the corresponding $\txH$-basic Cartan--Eilenberg cohomology $\,{\rm CaE}^\bullet(\txG_{\rm int})_{\txH-{\rm basic}}\,$ (this is the construction, originally due to de Azc\'arraga, exploited and elaborated in the super-Minkowskian setting\footnote{The peculiarity of the super-Minkowskian setting is the existence of a Lie supergroup structure on the homogeneous space itself. As a consequence, one may seek to extend the supersymmetry Lie superalgebra and erect the super-$p$-gerbe directly over the supertarget. This idea was put to work in \Rcite{Suszek:2017xlw}.} of \Rcite{Suszek:2017xlw} where it led to the emergence of the so-called \textbf{extended superspacetimes}, {\it cp} \Rcite{Chryssomalakos:2000xd}, and in the case of Zhou's super-0-brane in $\,{\rm s}({\rm AdS}_2\x\bS^2)$,\ the latter being constrained severly by the Green--Schwarz limit of the \.In\"on\"u--Wigner blow-up $\,{\rm s}({\rm AdS}_2\x\bS^2)\too{\rm sMink}(3,1\,|\,D_{3,1})$), or -- alternatively, in the case of curved supertargets -- it may come from an enrichment of the original supersymmetry Lie superalgebra $\,\ggt\,$ by the said supercharges $\,\{Z_i\}_{i\in\ovl{1,N}}$,\ spanning a subspace $\,\zgt=\bigoplus_{i=1}^N\,\corr{Z_i}\subset\widehat\ggt\,$ with the property
\qq\nn
[\zgt,\zgt\}_{\widehat\ggt}\cap\ggt\neq 0\,,
\qqq  
as dictated by the physically motivated (\.In\"on\"u--Wigner) asymptotics of $\,\widehat\ggt\,$ (this seems to be the case for the supertargets $\,{\rm s}({\rm AdS}_3\x\bS^3)\,$ and $\,{\rm s}({\rm AdS}_5\x\bS^5)\,$ of super-1-brane (superstring) propagation, {\it cp} the tentative proposal of \Rxcite{Sec.\,9}{Suszek:2018bvx}).
\eit  
\brem
Whenever the homogeneous space $\,\cM_\txH\,$ is endowed with the structure of a Lie supergroup, we may try to restrict the extension procedure to it rather than going \emph{via} the mother Lie supergroup $\,\txG$.\ This was, indeed, the tactic successfully applied to the super-Minkowskian superbackgrounds treated at length in \Rcite{Suszek:2017xlw}, where the relevant surjective submersions came from Lie-supergroup extensions of the supertranslation group.
\erem
The above line of reasoning, based on Lie-supergroup/-superalgebra extensions, leads to a specialisation of the former definitions of a super-$p$-gerbe,
\bedef
A \textbf{Cartan--Eilenberg super-$p$-gerbe} is a super-$p$-gerbe object, as described in Def.\,\ref{def:spG}, in the category of Lie supergroups, with all constitutive surjective surjections corresponding to Lie-supergroup extensions.
\exdef
\noindent and, in particular, of a super-0- and a super-1-gerbe.
\bedef
A \textbf{Cartan--Eilenberg super-0-gerbe} is a super-0-gerbe in the sense of Def.\,\ref{def:s0g}, such that 
\bit
\item $\,\cM\equiv\txG$,\ taken with the left action induced by the supergroup multiplication in $\,\txG$;
\item $\,\sfY\cM\equiv\sfY\txG\,$ is endowed with the structure of a Lie supergroup that extends that on $\,\txG$;
\item $\,\pi_{\sfY\cM}\equiv\pi_{\sfY\txG}\,$ is a Lie-supergroup epimorphism.
\eit

An \textbf{isomorphism between Cartan--Eilenberg super-0-gerbes} $\,(\sfY_A\txG,\pi_{\sfY_A\txG},\underset{\tx{\ciut{(1)}}}{\cA}{}_A),\ A\in\{1,2\}\,$ is a super-0-gerbe isomorphism $\,\Phi\,$ as in Def.\,\ref{def:s0g} that is simultaneously a Lie-supergroup isomorphism.
\exdef
\bedef
A \textbf{Cartan--Eilenberg super-1-gerbe} is a super-1-gerbe in the sense of Def.\,\ref{def:s1g}, such that 
\bit
\item $\,\cM\equiv\txG$,\ taken with the left action induced by the supergroup multiplication in $\,\txG$;
\item $\,\sfY\cM\equiv\sfY\txG\,$ (and so also $\,\sfY^{[k]}\cM\equiv\sfY^{[k]}\txG,\ k\in\{2,3,4\}$) as well as $\,L\,$ (and so also the various pullbacks thereof along the canonical projections, and their tensor products) are endowed with the structure of a Lie supergroup extending that on $\,\txG$;
\item $\,\pi_{\sfY\cM}\equiv\pi_{\sfY\txG}\,$ and $\,\pi_L\,$ are Lie-supergroup epimorphisms, and $\,\mu_L\,$ is a Lie-supergroup isomorphism establishing an equivalence of the respective extensions.
\eit

An \textbf{isomorphism between Cartan--Eilenberg super-1-gerbes} $\,(\sfY_A\txG,\pi_{\sfY_A\txG},\underset{\tx{\ciut{(2)}}}{\cB}{}_A,L_A,\pi_{L_A},\underset{\tx{\ciut{(1)}}}{\cA}{}_{L_A},$ $\mu_{L_A})$ is a super-1-gerbe isomorphism $\,\Phi=(\sfY\sfY_{1,2}\txG,\pi_{\sfY\sfY_{1,2}\txG},E,\pi_E,\underset{\tx{\ciut{(1)}}}{\cA}{}_E,\a_E)\,$ as in Def.\,\ref{def:s1g} with the following properties:
\bit
\item $\,\sfY\sfY_{1,2}\txG\,$ is endowed with the structure of a Lie supergroup that extends that on $\,\sfY_{1,2}\txG\,$ (the product one);
\item $\,E\,$  is endowed with the structure of a Lie supergroup that extends that on $\,\sfY\sfY_{1,2}\txG$;
\item $\,\pi_{\sfY\sfY_{1,2}\txG}\,$ and $\,\pi_E\,$ are Lie-supergroup epimorphisms, and $\,\a_E\,$ is a Lie-supergroup isomorphism (for the induced Lie-supergroup structures induced in a canonical manner on its domain and codomain).
\eit
\exdef

While the geometrisation scheme based on Lie-supergroup extensions induced by the super-$p$-brane charge is largely non-algorithmic and oftentimes driven by physical intuition\footnote{{\it E.g.}, the asymptotic correspondence between the curved and flat superbackgrounds, as well as the canonical analysis of the field-theoretic realisation of the supersymmetry algebra, {\it cp} \Rxcite{Sec.\,3}{Suszek:2018bvx}.} rather than some obvious mathematical correspondence, we believe -- in view of the central r\^ole played by the said intuition in the development of the supersymmetric field theories of interest (starting with the pioneering work of Metsaev and Tseytlin on superstrings in $\,{\rm s}({\rm AdS}_5\x\bS^5)$) and their subsequent applications (in particular, in the celebrated but still formally inadequately understood AdS/CFT correspondence), and of the concreteness of the ensuing supergeometric objects, implying their amenability to a hands-on verification of such important properties as supersymmetry-equivariance and compatibility with a properly defined and understood $\k$-symmetry -- that it constitutes a structurally most natural and tractable proposal for a geometrisation of the physically motivated supersymmetry-invariant cohomology, developing in a close conceptual analogy with the by now much-advanced theory of bundle gerbes behind the bosonic $\si$-model and related topological field theories. Accordingly, we shall pursue this latter line of thinking in what follows, leaving the much interesting question regarding a precise relation between the two approaches to a future study.

There are currently several working examples of the abstract structures recalled above, to wit, the Cartan--Eilenberg super-$p$-gerbes for $\,p\in\{0,1,2\}\,$ of \Rxcite{Sec.\,5}{Suszek:2017xlw} over the super-Minkowski space, associated with the respective Green--Schwarz super-$(p+2)$-cocycles $\,\underset{\tx{\ciut{(p+2)}}}{\txH}\hspace{-6pt}^{\rm GS}\,$ (such that $\,\underset{\tx{\ciut{(p+2)}}}{\chi}\hspace{-7pt}^{\rm GS}=\pi_{{\rm sISO}(d,1\,\vert\,ND_{d,1})/{\rm SO}(d,1)}^*\underset{\tx{\ciut{(p+2)}}}{\txH}\hspace{-6pt}^{\rm GS}$,\ for $\,N\in\{1,2\}\,$ as in Examples \ref{eq:s0gsMink} and \ref{eq:spgsMink}), and the super-$0$-gerbe of \Rxcite{Sec.\,5}{Suszek:2018ugf} over $\,{\rm s}({\rm AdS}_2\x\bS^2)$,\ associated with the Zhou super-2-cocycle $\,\underset{\tx{\ciut{(2)}}}{\chi}\hspace{-1pt}^{\rm Zh}$.\ There is also the trivial Cartan--Eilenberg super-1-gerbe of \Rxcite{Sec.\,6}{Suszek:2018bvx}, associated with the Metsaev--Tseytlin super-3-cocycle $\,\underset{\tx{\ciut{(3)}}}{\chi}\hspace{-1pt}^{\rm MT}$,\ but the latter super-1-gerbe was demonstrated \emph{not} to asymptote to the Green--Schwarz super-1-gerbe over $\,{\rm sMink}(9,1\,\vert\,32)$,\ and so it was argued that the super-3-cocycle itself may have to be corrected for the sake of promoting the principle of \.In\"on\"u--Wigner correspondence between the curved and flat super-$\si$-models to the rank of a higher-geometric correspondence between the respective super-1-gerbes (a hint as to where to look for a suitable correction was given in \Rxcite{Sec.\,9}{Suszek:2018bvx}). It deserves to be noted at this point that the super-Minkowskian super-$p$-gerbes referred to above were proven, in \Rxcite{Sec.\,4}{Suszek:2019cum}, to carry a supersymmetric $\Ad$-equivariant structure in perfect conformity with the Gra\ss mann-even intuition.\bigskip

While the global supersymmetry of the super-$\si$-model, realised by the action of the Lie supergroup $\,\txG\,$ and preserving \emph{separately} each of the two factors in the DF amplitude: the metric one and the topological one, is accommodated directly in the construction of the super-$p$-gerbe, any attempt at `gerbification' -- in the form of a suitable equivariant structure -- of the peculiar local supersymmetry modelled on the supervector space $\,\ggt\sgt(\sgt\Bgt^{{\rm (HP)}}_{p,\la_p})\,$ of \Reqref{eq:GSmodalg} in the generic situation or by the $\k$-symmetry superalgebra $\,\ggt\sgt_{\rm vac}(\sgt\Bgt^{{\rm (HP)}}_{p,\la_p})\,$ in the regular case, and \emph{mixing} the two factors nontrivially, requires that the higher-geometric discussion be transcribed from the NG formulation over $\,\cM_\txH\,$ to the HP formulation over the HP section $\,\Si^{\rm HP}\,$ of \Reqref{eq:HPsec} resp.\ over the HP vacuum foliation $\,\Si^{\rm HP}_{\rm vac}\,$ of Def.\,\ref{def:HPvacfol}. That such a `gerbification' is altogether possible is a consequence of the purely topological nature of the DF amplitude in the HP formulation. Cohomological triviality (in the supersymmetric and $\txH_{\rm vac}$-basic refinement of the de Rham cohomology) of the new tensorial component $\,\sfd\underset{\tx{\ciut{(p+1)}}}{\b}\hspace{-7pt}{}^{\rm (HP)}\,$ of the superbackground that replaces the supersymmetric and $\txH$-basic metric actually renders the lift quite straightforward. The stage for a concretisation of these ideas is set, after \Rxcite{Sec.\,6.2}{Suszek:2019cum}, in the following
\bedef\label{def:restrextHPpgerb}
Let $\,\cG^{(p)}\,$ be the super-$p$-gerbe, as described in Def.\,\ref{def:spG}, over $\,\cM_\txH\equiv\txG/\txH\,$ with the curvature given by the descendant $\,\underset{\tx{\ciut{(p+1)}}}{\txH}\in Z^{p+2}_{\rm dR}(\cM_\txH)^\txG$,\ from the total space $\,\txG\,$ to the base $\,\cM_\txH\,$ of the principal $\txH$-bundle $\,\pi_{\txG/\txH}\ :\ \txG\too\txG/\txH\,$ of \Reqref{eq:homasprinc}, of the $\txH_{\rm vac}$-basic Green--Schwarz super-$(p+2)$-cocycle $\,\underset{\tx{\ciut{(p+1)}}}{\chi}\,$ defining a class $\,[\underset{\tx{\ciut{(p+1)}}}{\chi}]_{{\rm CaE},\txH_{\rm vac}-{\rm basic}}\in{\rm CaE}^{p+2}(\txG)_{\txH_{\rm vac}}\,$,\ and let $\,\underset{\tx{\ciut{(p+1)}}}{\b}\hspace{-7pt}{}^{\rm (HP)}\in\Om^{p+1}(\txG)^\txG\,$ be the Hughes--Polchinski super-$(p+1)$-form of \Reqref{eq:HPcurv} that enters definition \eqref{eq:HPspbgrnd} of the Hughes--Polchinski super-$p$-brane background and determines, in the usual manner, the trivial Cartan--Eilenberg super-$p$-gerbe over $\,\txG$,\ to be denoted as $\,\cI^{(p)}_{\underset{\tx{\ciut{(p+1)}}}{\b}\hspace{-7pt}{}^{\rm (HP)}}$.\ \textbf{The extended Hughes--Polchinski $p$-gerbe} over $\,\txG\,$ is the product bundle $p$-gerbe
\qq\nn
\widehat\cG{}^{(p)}:=\pi_{\txG/\txH}^*\cG^{(p)}\ox\cI^{(p)}_{\la_p\,\underset{\tx{\ciut{(p+1)}}}{\b}\hspace{-7pt}{}^{\rm (HP)}}\,.
\qqq
Its \textbf{vacuum restriction} is the $p$-gerbe 
\qq\nn
\iota_{\rm vac}^*\widehat\cG{}^{(p)}\equiv\widehat\cG{}^{(p)}\rstr_{\Si^{\rm HP}_{\rm vac}}
\qqq
obtained by pullback along the embedding \eqref{eq:HPvacsec} of the Hughes--Polchinski vacuum foliation $\,\Si^{\rm HP}_{\rm vac}\,$ of Def.\,\ref{def:HPvacfol} in the Hughes--Polchinski section $\,\Si^{\rm HP}\,$ of \Reqref{eq:HPsec}.
\exdef

\brem
As has already been mentioned in passing, the appearance of extensions of supersymmetry algebras, underlying the geometrisation scheme advanced herein, is entirely natural from the point of view of the field-theoretic realisation of supersymmetry in the GS super-$\si$-model. In the HP formulation, this can be seen as follows. The ability to explicitly model the infinitesimal action of the supersymmetry group on the HP section $\,\Si^{\rm HP}\,$ with the help of the vector fields $\,\cK_A\,$ paves the way to the canonical analysis of supersymmetry of the GS super-$\si$-model (to be understood in the very same functorial/$\cS$-point spirit as the DF amplitudes itself). Indeed, upon contracting (the trivial lifts, to the mapping supermanifold $\,[\xcC_p,\cM_{\txH_{\rm vac}}]$,\ of) the fundamental vector fields $\,\cK_X\equiv X^A\,\cK_A,\ X=X^A\,t_A\in\ggt\,$ with the presymplectic form (note that as a result of the topological character of the HP formulation the presymplectic form depends only on the configuration $\,\widehat\xi_{\xcC_p}\equiv\widehat\xi\rstr_{\xcC_p}$) 
\qq\nn
\Om_\si^{({\rm HP})}[\widehat\xi_{\xcC_p}]=\int_{\xcC_p}\,\widehat\xi_{\xcC_p}^*\bigl(\la_p\,\d\underset{\tx{\ciut{(p+1)}}}{\txB}\hspace{-7pt}{}^{\rm (HP)}+\underset{\tx{\ciut{(p+2)}}}{\txH}\hspace{-7pt}{}^{\rm vac}\bigr)\,,
\qqq
we readily derive the Noether supersymmetry charges ($\widehat\xi_{\t\cap\xcC_p}\equiv\widehat\xi\rstr_{\t\cap\xcC_p}$)
\qq\nn
h_X[\widehat\xi_{\xcC_p}]=\sum_{\t\in\Tgt_{p+1}}\,\int_{\t\cap\xcC_p}\,\bigl(\si_{\imath_\t}^{\rm vac}\circ\widehat\xi_{\t\cap\xcC_p}\bigr)^*\bigl(\la_p\,R_X\con\underset{\tx{\ciut{(p+1)}}}{\b}\hspace{-7pt}{}^{\rm (HP)}+\underset{\tx{\ciut{(p)}}}{\k}{}_X\bigr)\,,\qquad X\in\ggt\,,
\qqq
expressing them in terms of the $p$-forms $\,\underset{\tx{\ciut{(p)}}}{\k}{}_X\in\Om^p(\txG)\,$ defined as
\qq\nn
\cK_X\con\underset{\tx{\ciut{(p+2)}}}{\chi}\equiv R_X\con\underset{\tx{\ciut{(p+2)}}}{\chi}=:-\sfd\underset{\tx{\ciut{(p)}}}{\k}{}_X\,.
\qqq
Their existence follows from the assumed quasi-supersymmetry of $\,\underset{\tx{\ciut{(p+1)}}}{\b}$,\ or -- in other words -- from global supersymmetry of $\,\cG^{(p)}$.\ The Poisson bracket of the charges (associated with the vectors $\,X_1,X_2\in\ggt$), as determined by the presymplectic form $\,\Om_\si^{({\rm HP})}$,\ reads
\qq\nn
\{h_{X_1},h_{X_2}\}_{\Om_\si^{({\rm HP})}}[\widehat\xi_{\xcC_p}]=h_{[X_1,X_2\}}[\widehat\xi_{\xcC_p}]-\sum_{\t\in\Tgt_{p+1}}\,\int_{\t\cap\xcC_p}\,\bigl(\si_{\imath_\t}^{\rm vac}\circ\widehat\xi_{\t\cap\xcC_p}\bigr)^*\underset{\tx{\ciut{(p)}}}{\a_{X_1,X_2}}\,,
\qqq
with the integrand of the (classical) \textbf{wrapping anomaly} given by
\qq\nn
\underset{\tx{\ciut{(p)}}}{\a_{X_1,X_2}}=\pLie{R_{X_1}}\underset{\tx{\ciut{(p)}}}{\k}{}_{X_2}+\underset{\tx{\ciut{(p)}}}{\k}{}_{[X_1,X_2\}}\,.
\qqq
The reason why this field-theoretic departure from the underlying supersymmetry Lie superalgebra was dubbed thus in \Rxcite{Sec.\,3}{Suszek:2018bvx} (strictly speaking, the effect was investigated in the NG formulation) is that it is determined by a de Rham $p$-cocycle,
\qq\nn
\sfd\underset{\tx{\ciut{(p)}}}{\a_{X_1,X_2}}=-\pLie{R_{X_1}}\bigl(R_{X_2}\con\underset{\tx{\ciut{(p+2)}}}{\chi}\bigr)+\sfd\underset{\tx{\ciut{(p)}}}{\k}{}_{[X_1,X_2\}}=R_{[X_1,X_2\}}\con\underset{\tx{\ciut{(p+2)}}}{\chi}+\sfd\underset{\tx{\ciut{(p)}}}{\k}{}_{[X_1,X_2\}}=0\,,
\qqq
and so it is non-zero solely if the embedding $\,\widehat\xi\,$ has a nontrivial monodromy around the Cauchy hypersurface $\,\xcC_p$,\ or the embedded worldvolume contains -- in the functorial picture -- a non-contractible cycle. In \Rxcite{Sec.\,4 \& 5}{Suszek:2018bvx}, monodromies around (compactified) Gra\ss mann-odd cycles were postulated to encode the supercentral extensions of the supertarget $\,\txG\,$ (resp.\ $\,\cM_\txH\,$ whenever the latter is a Lie supergroup) engendered by the GS super-$(p+2)$-cocycle $\,\underset{\tx{\ciut{(p+2)}}}{\chi}\,$ that co-determines the super-$\si$-model.
\erem

The assumption of existence of a \emph{global} (super)symmetry in a field theory with some further \emph{gauged} (super)symmetry and of its lift to the geometrisation of the theory's topological content impose nontrivial constraints upon the equivariant structure on that geometrisation that incarnates the local (super)symmetry. Indeed, it is natural to demand that the equivariant structure be compatible with the global (super)symmetry in an obvious manner that we recall after \Rxcite{Sec.\,4}{Suszek:2019cum}, where it was described concisely by The Invariance Postulate and quantified -- in the \emph{super}symmetric setting in hand -- through
\bedef\label{def:symmequivstr}
Let $\,\cM\,$ be a (super)manifold with an abelian $p$-gerbe $\,\cG^{(p)}\,$ over it, as described before, and 
let $\,\txG\,$ and $\,\txG_{\rm loc}\,$ be two Lie (super)groups, with the respective binary operations $\,\mu\,$ and $\,\mu_{\rm loc}\,$ and the respective tangent Lie (super)algebras $\,\ggt\,$ and $\,\ggt_{\rm loc}$,\ acting on $\,\cM\,$ as
\qq\nn
\la\ :\ \txG\x\cM\too\cM\,,\qquad\qquad\la_{\rm loc}\ :\ \txG_{\rm loc}\x\cM\too\cM\,,
\qqq
the former to be thought of as the global (super)symmetry group and the latter as the local (super)symmetry group. Consider the action (super)groupoid\footnote{{\it Cp} \Rxcite{Def.\,8.6}{Suszek:2012ddg} for a definition of the notion and a comprehensive discussion of its naturalness in and relevance to the description of gauged symmetries of a field theory.}
\qq\nn
\txG_{\rm loc}\lx\cM\qquad :\qquad \alxydim{@C=2cm@R=1.5cm}{\txG_{\rm loc}\x\cM \ar@<.75ex>[r]^{\quad\la_{\rm loc}} \ar@<-.75ex>[r]_{\quad\pr_2} & \cM}
\qqq
and form the nerve $\,\sfN^\bullet(\txG_{\rm loc}\lx\cM)\,$ thereof,
\qq\label{eq:actgrpdsimpl}
\alxydim{@C=1.5cm@R=1.5cm}{ \cdots \ar@<.75ex>[r]^{d_\bullet^{(3)}\quad} \ar@<.25ex>[r] \ar@<-.25ex>[r]
\ar@<-.75ex>[r] & \txG_{\rm loc}^{\x 2}\x\cM \ar@<.5ex>[r]^{\ d_\bullet^{(2)}} \ar@<0.ex>[r]
\ar@<-.5ex>[r] & \txG_{\rm loc}\x\cM \ar@<.5ex>[r]^{\quad d_\bullet^{(1)}} \ar@<-.5ex>[r] & \cM}\,,
\qqq
with face maps
\qq\nn
d_l^{(m)}\ :\ \txG_{\rm loc}^{\x m}\x\cM\too\txG_{\rm loc}^{\x m-1}\x\cM\,,\qquad l\in\ovl{0,m}\,,\quad m\in\bN^\x
\qqq
given by the formul\ae
\qq\nn
&d_0^{(m)}=\pr_{2,3,\ldots,m,m+1}\,,\qquad\qquad d_m^{(m)}=\id_{\txG_{\rm loc}^{\x m-1}}\x\la_{\rm loc},\,,&\cr\cr
&d_i^{(m)}=\id_{\txG_{\rm loc}^{\x i-1}}\x\mu_{\rm loc}\x\id_{\txG_{\rm loc}^{\x m-1-i}\x\cM}\,,\qquad i\in\ovl{1,m-1}\,.&
\qqq
Assume the existence of a $\txG_{\rm loc}$-equivariant structure on $\,\cG^{(p)}$,\ understood as a collection of $p$-gerbe $k$-isomorphisms $\,\Upsilon_p^{(k)},\ k\in\ovl{1,p+1}\,$ over the respective components $\,\sfN^k(\txG_{\rm loc}\lx\cM)\equiv\txG_{\rm loc}^{\x k}\x\cM\,$ of the nerve, with $\,\Upsilon_p^{(p+1)}\,$ subject to a coherence condition over $\,\sfN^{p+2}(\txG_{\rm loc}\lx\cM)\equiv\txG_{\rm loc}^{\x p+2}\x\cM$,\ that altogether form a natural generalisation of the specific coherent (simplicial) objects explicited in \Rxcite{Sec.\,2}{Suszek:2019cum} for $\,p\in\{0,1\}\,$ ({\it cp} Sec.\,\ref{sec:susykappa} for an explicit sheaf-cohomological description of its linearisation) We say that the equivariant structure is a \textbf{$\txG$-invariantly $\txG_{\rm loc}$-equivariant structure} on $\,\cG^{(p)}\,$ if there exists an action
\qq\nn
\la^1\ :\ \txG\x\sfN^1(\txG_{\rm loc}\lx\cM)\too\sfN^1(\txG_{\rm loc}\lx\cM)
\qqq 
of $\,\txG\,$ that lifts $\,\la\,$ to $\,\sfN^1(\txG_{\rm loc}\lx\cM)\equiv\txG_{\rm loc}\x\cM\,$ (and so canonically induces lifts $\,\la^k\,$ to the remaining components $\,\sfN^k(\txG_{\rm loc}\lx\cM),\ k>1\,$ of $\,\sfN^\bullet(\txG_{\rm loc}\lx\cM)$) in a manner compatible with $\,\la$,\ as expressed by the commutative diagram 
\qq\label{eq:liftcompface}
\alxydim{@C=2cm@R=1.5cm}{ \txG\x\cM \ar[d]_{\la} & \txG\x(\txG_{\rm loc}\x\cM) \ar[d]_{\la^1} \ar[r]^{\qquad\id_\txG\x\pr_2} \ar[l]_{\id_\txG\x\la_{\rm loc}\quad} & \txG\x\cM \ar[d]^{\la} & \\ \cM & \txG_{\rm loc}\x\cM \ar[r]_{\qquad \pr_2} \ar[l]^{\la_{\rm loc}} & \cM}
\qqq
and such that each component $\,\Upsilon_p^{(k)}\,$ is $\txG$-invariant with respect to the corresponding action $\,\la^k\,$ in the sense of Def.\,\ref{def:spG}.
\exdef
\noindent Adaptation of the above structure to the situation of immediate interest, that is to the description of the enhanced gauge symmetry resp.\ the $\k$-symmetry of the globally supersymmetric super-$\si$-model in the HP formulation, requires a careful reworking of the original concepts, taking into account the peculiar nature of the local symmetry, the localisation of the vacuum itself within $\,\Si^{\rm HP}$,\ as well as the implicit requirement of descent of the higher-geometric structure to the homogeneous space $\,\cM_{\txH_{\rm vac}}$.\ Therefore, rather than unpacking the last definition in all generality, we leave it as it stands and, instead, pass directly to the study of a higher-geometric extension of the local supersymmetry of the super-$\si$-model that we are, at long last, ready to address.

\section{Supersymmetry-invariant gerbification of linearised $\k$-symmetry}\label{sec:susykappa}

In this closing section of the present paper, we seek to investigate a gerbe-theoretic realisation of the enhanced local supersymmetry of the GS super-$\si$-model in the topological HP formulation determined by the extended HP $p$-gerbe. The supersymmetry is invariably represented by a distinguished superdistribution in the tangent sheaf of the HP section $\,\Si^{\rm HP}\,$ -- be it the generic enhanced gauge-symmetry superdistribution $\,\cG\cS(\sgt\Bgt^{{\rm (HP)}}_{p,\la_p})\,$ for the correspondence sector of the field theory, or the limit $\,\k^{-\infty}(\sgt\Bgt^{{\rm (HP)}}_{p,\la_p})\,$ of the weak derived flag of the $\k$-symmetry superdistribution $\,\k(\sgt\Bgt^{{\rm (HP)}}_{p,\la_p})\,$ (resp.\ its extended variant in Example \ref{eg:sqroots1bsMink}) for its vacuum in the regular case of an integrable HP vacuum superdistribution $\,{\rm Vac}(\sgt\Bgt^{{\rm (HP)}}_{p,\la_p})$.\ Based on the intuition developed in the Gra\ss mann-even setting and recalled in the previous section, we anticipate the emergence of an equivariant structure of sorts with respect to the local-symmetry superdistribution on the extended HP $p$-gerbe. Its precise identification calls for a reformulation of Def.\,\ref{def:symmequivstr} of a supersymmetric equivariant structure that accomodates the following three facts established previously:
\bit
\item The local supersymmetry is realised \emph{linearly} on the HP/NG correspondence sector of the super-$\si$-model and in the case of the $\k$-symmetry superdistribution $\,\k(\sgt\Bgt^{{\rm (HP)}}_{p,\la_p})$,\ it restricts further to the HP vacuum foliation that it envelops, the foliation being engendered by the involutive HP vacuum superdistribution that $\,\k(\sgt\Bgt^{{\rm (HP)}}_{p,\la_p})\,$ bracket-generates. In either situation, integrating any component of the tangential structure to a gauge-symmetry Lie supergroup -- even if the corresponding model Lie superalgebra should allow it -- would \emph{a priori} be meaningless as the supergroup could \emph{not} be realised on $\,\Si^{\rm HP}\,$ or $\,\Si_{\rm vac}^{\rm HP}\,$ in the standard manner by the same argument\footnote{Of course, we might and -- indeed -- ought to try to descend the integrated symmetry to the homogeneous space for which, however, we should have to devise new tools. We hope to return to this idea in the future.} as the one invoked in Sec.\,\ref{sec:globsusy} for the action of the global-supersymmetry group $\,\txG$.
\item By the said argument, the global-supersymmetry group $\,\txG\,$ is \emph{not}, in general, realised on $\,\Si^{\rm HP}\,$ or $\,\Si_{\rm vac}^{\rm HP}\,$ globally, but does admit a \emph{linearised} realisation by the global-supersymmetry subspace $\,S_\txG^{\rm HP}\,$ of Prop.\,\ref{prop:leftglobfund}, with generators labelled by the Lie superalgebra $\,\ggt$,\ resp.\ by its vacuum-preserving subspace -- the residual global-supersymmetry subspace $\,S_\txG^{\rm HP, vac}\,$ of Prop.\,\ref{prop:resglobsusysub}, with generators labelled by the Lie sub-superalgebra $\,\sgt_{\rm vac}$.
\item Global structures that arise over the HP section $\,\Si^{\rm HP}$,\ such as, {\it e.g.}, the extended HP $p$-gerbe and any equivariant structure on it, implicitly model corresponding structures on the homogeneous space $\,\cM_{\txH_{\rm vac}}$. 
\eit 
Putting all these facts together, we conclude that the structure that we are after is a linearisation of a standard (super)group-equivariant structure on the extended HP $p$-gerbe, compatible with linearised global supersymmetry up to linearised hidden gauge transformations from $\,\hgt_{\rm vac}\,$ (a linearisation of the vacuum isotropy group $\,\txH_{\rm vac}$) -- all that in a manifestly $\txH_{\rm vac}$-descendable form. Generically, the linearisation is not expected to extend beyond the level of the relevant super-$p$-gerbe 1-isomorphism $\,\Upsilon_p^{(1)}\,$ of \Reqref{eq:locsusyequiv1} (or, more accurately, a linearised version thereof), representing the bare local-supersymmetry superdistribution and further assumed globally linearised(-$\txG$-)supersymmetric, but in the physically favoured regular situation in which the HP vacuum superdistribution is integrable, we should climb with the vacuum restriction $\,\iota_{\Si^{\rm HP}_{\rm vac}}^*\widehat\cG{}^{(p)}\,$ of the extended HP $p$-gerbe of Def.\,\ref{def:restrextHPpgerb} one level up in the hierarchy defining the standard (super)group-equivariant structure ({\it i.e.}, demand either coherence of $\,\Upsilon_0^{(1)}$,\ or existence of $\,\Upsilon_p^{(2)}\,$ for $\,p>0$) and subsequently perform linearisation and check compatibility with linearised global supersymmetry up to vacuum gauge symmetry for this extended structure. As we approach the task of constructing such a linearised equivariant structure in the intrinsically geometric formalism of gerbe theory, an important difference between the two superdistributions comes to the fore that effectively rules out the generic structure $\,\cG\cS(\sgt\Bgt^{{\rm (HP)}}_{p,\la_p})\,$ as a subject of gerbification. Indeed, while the $\k$-symmetry superdistribution $\,\k(\sgt\Bgt^{{\rm (HP)}}_{p,\la_p})\,$ gives rise to an involutive supersymmetry superdistribution $\,\k^{-\infty}(\sgt\Bgt^{{\rm (HP)}}_{p,\la_p})\,$ tangent to a family of \emph{sub}-supermanifolds embedded in $\,\Si^{\rm HP}$,\ and so can be meaningfully restricted to any one of these sub-supermanifolds ({\it i.e.}, the vacuum), as can be the $p$-gerbe, the enhanced gauge-symmetry superdistribution $\,\cG\cS(\sgt\Bgt^{{\rm (HP)}}_{p,\la_p})\,$ acquires its status upon restriction of field configurations to the non-integrable correspondence superdistribution, which it then generically leaves {\it via} its weak derived flag that itself does not asymptote to a larger gauge-symmetry superdistribution of the underlying field theory. There is -- on one hand -- no obvious \emph{geometric}\footnote{We might, in principle, try to impose the restriction in the sheaf-theoretic language but the lack of a (sub-)supermanifold structure behind the restriction would inevitably render such a construction non-canonical.} mechanism of imposing the field-theoretic restriction to the correspondence sector upon the extended HP $p$-gerbe, and -- on the other hand -- there is no reason to expect equivariance of the latter without that restriction or with respect to the integrable limit of the weak derived flag of $\,\cG\cS(\sgt\Bgt^{{\rm (HP)}}_{p,\la_p})\,$ on the (sub-)supermanifold of $\,\Si^{\rm HP}\,$ that it envelops. Thus, it appears, we are bound to study equivariance of the vacuum restriction $\,\iota_{\Si^{\rm HP}_{\rm vac}}^*\widehat\cG{}^{(p)}\,$ of the extended HP $p$-gerbe with respect to the vacuum-generating gauged supersymmetry realised by $\,\k^{-\infty}(\sgt\Bgt^{{\rm (HP)}}_{p,\la_p})\equiv{\rm Vac}(\sgt\Bgt^{{\rm (HP)}}_{p,\la_p})\,$ and modelled on the $\k$-symmetry superalgebra $\,\ggt\sgt_{\rm vac}(\sgt\Bgt^{{\rm (HP)}}_{p,\la_p})\equiv\gt{vac}(\sgt\Bgt^{{\rm (HP)}}_{p,\la_p})$.\ In the present section, we treat the lowest layer of such a structure that, as argued above, accounts for the linear (or supervector-space) structure on $\,\ggt\sgt_{\rm vac}(\sgt\Bgt^{{\rm (HP)}}_{p,\la_p})$,\ relegating the issue of a higher-geometric implementation of the superalgebra structure to a future study. 

We begin by deriving a linearised version of the constraint of compatibility up to gauge symmetry imposed upon $\,\la^1\,$ through Diag.\,\ref{eq:liftcompface}. We do that in local coordinates ({\it i.e.}, in the $\cS$-point picture) on the vacuum foliation 
\qq\nn
\Si^{\rm HP}_{\rm vac}=\bigsqcup_{i\in I_{\txH_{\rm vac}}}\,\bigsqcup_{\upsilon_i\in\Upsilon_i}\,\cV^{\rm vac}_{i,\upsilon_i}
\qqq 
which we write as the disjoint union of integral sub-supermanifolds (or vacua) $\,\cV^{\rm vac}_{i,\upsilon_i}\subset\cV_i\,$ of the HP vacuum superdistribution $\,{\rm Vac}(\sgt\Bgt^{{\rm (HP)}}_{p,\la_p})\,$ (locally labelled by sets $\,\Upsilon_i$), the latter being assumed integrable below. For the coordinates, we use the shorthand notation
\qq\nn
\unl\xi_i\equiv\si_i^{\rm vac}(\unl\chi{}_i)\,.
\qqq
In it, the aforementioned realisation, over $\,\cV^{\rm vac}_{i,\upsilon_i}\,$ of the residual global-supersymmetry algebra, with the homogeneous generators ($|S_{\breve A}|\equiv|\breve A|$)
\qq\nn
\sgt_{\rm vac}\equiv\bigoplus_{\breve A=1}^{S_{\rm vac}}\,\corr{S_{\breve A}}\subset\ggt\,,\qquad S_{\rm vac}\equiv\dim\,S^{\rm HP, vac}_\txG\,,
\qqq 
and of the local-supersymmetry supervector space (for which we introduce a new symbol to unclutter the notation)
\qq\nn
\sgt_{\rm loc}\equiv\ggt\sgt_{\rm vac}\bigl(\sgt\Bgt^{{\rm (HP)}}_{p,\la_p}\bigr)\equiv\bigoplus_{\widetilde A=0}^{p+q+D-\unl\d}\,\corr{V_{\widetilde A}}\subset\ggt\,,
\qqq 
with
\qq\nn
V_{\widetilde A}=\left\{ \barr{cl} P_{\widetilde A} & \tx{ if } \widetilde A\in\ovl{0,p} \cr\cr \unl Q_{\widetilde A-p} & \tx{ if } \widetilde A\in\ovl{p+1,p+q} \cr\cr J_{\widetilde A-p-q}  & \tx{ if } \widetilde A\in\ovl{p+q+1,p+q+D-\unl\d} \earr \right.\,,
\qqq
reads
\qq\nn
\sgt_{\rm vac}\x\cV^{\rm vac}_{i,\upsilon_i}\ni\bigl(X\equiv X^{\breve A}\,S_{\breve A},\unl\xi{}_i\bigr)\longmapsto X^{\breve A}\,\cK_{S_{\breve A}\,i}(\unl\xi{}_i)\in\cT\cV^{\rm vac}_{i,\upsilon_i}
\qqq
and
\qq\nn
\sgt_{\rm loc}\x\cV^{\rm vac}_{i,\upsilon_i}\ni\bigl(\G\equiv\G^{\widetilde A}\,V_{\widetilde A},\unl\xi{}_i)\longmapsto\G^{\widetilde A}\,\cT_{V_{\widetilde A}\,i}(\unl\xi{}_i\bigr)\equiv\G^{\widetilde A}\,\cT_{\widetilde A\,i}(\unl\xi{}_i)\in\cT\cV^{\rm vac}_{i,\upsilon_i}\,,
\qqq
respectively, the latter being trivial in the $\hgt_{\rm vac}$-sector, 
\qq\nn
\G^{\widetilde A}\,\cT_{\widetilde A\,i}\equiv\G^{\unl{\unl A}}\,\cT_{\unl{\unl A}\,i}\,.
\qqq
Note that the $\,X^{\breve A}\,$ and the $\,\G^{\widetilde A}\,$ appear here in the r\^ole of global coordinates on the superspaces $\,\sgt_{\rm vac}\,$ and $\,\sgt_{\rm loc}$,\ respectively, and so they carry Gra\ss mann parity according to the rule
\qq\nn
|X^{\breve A}|=|\breve A|\,,\qquad\qquad|\G^{\widetilde A}|=|\widetilde A|\,.
\qqq
The task in hand now boils down to finding a collection of (smooth) sections 
\qq\label{eq:rotsec}\qquad\qquad
\La_{\breve A\widetilde B\,i}^{\ \ \ \ \widetilde C}\in\cO_\txG\bigl(\cV^{\rm vac}_{i,\upsilon_i}\bigr)\,,\qquad(\breve A,\widetilde B,\widetilde C)\in\ovl{1,S_{\rm vac}}\x\ovl{0,p+q+D-\unl\d}^{\x 2}\,,\quad \upsilon_i\in\Upsilon_i\,,\quad i\in I_{\txH_{\rm vac}}
\qqq
satifying the identities which by a mild abuse of the notation may be written, up to terms quadratic in $\,X\,$ or $\,\G$,\ as
\qq\nn
&&\unl\xi{}_i+X^{\breve A}\,\cK_{S_{\breve A}\,i}\bigl(\unl\xi{}_i\bigr)\con\sfd\unl\xi{}_i+\G^{\widetilde A}\,\bigl(\d_{\widetilde A}^{\ \widetilde B}+X^C\,\La_{C\widetilde A\,i}^{\ \ \ \ \widetilde B}\bigl(\unl\xi{}_i\bigr)\bigr)\,\cT_{\widetilde B\,i}\bigl(\unl\xi{}_i+X^{\breve D}\,\cK_{S_{\breve D}\,i}\bigl(\unl\xi{}_i\bigr)\con\sfd\unl\xi{}_i\bigr)\con\sfd\unl\xi{}_i+\xcO(X^2,\G^2)\cr\cr
&=&\unl\xi{}_i+\G^{\widetilde A}\,\cT_{\widetilde A\,i}\bigl(\unl\xi{}_i\bigr)\con\sfd\unl\xi{}_i+X^{\breve C}\,\cK_{S_{\breve C}\,i}\bigl(\unl\xi{}_i+\G^{\widetilde A}\,\cT_{\widetilde A\,i}\bigl(\unl\xi{}_i\bigr)\con\sfd\unl\xi{}_i\bigr)\con\sfd\unl\xi{}_i+\G^{\widetilde A}\,X^{\breve C}\,\La_{\breve C\widetilde A\,i}^{\ \ \ \ p+q+\unl S}\bigl(\unl\xi{}_i\bigr)\,L_{\unl S}\bigl(\unl\xi{}_i\bigr)\,.
\qqq
Upon expanding the above (dropping terms of order $\,\xcO(X^2,\G^2)$) and removing the (arbitrary) coefficients $\,X^{\breve C}\,$ and $\,\G^{\widetilde A}$,\ we arrive at the desired compact equation
\qq\label{eq:lingloblocomp}
[\cK_{S_{\breve A}\,i},\cT_{\widetilde B\,i}\}\bigl(\unl\xi{}_i\bigr)=-\La_{\breve A\widetilde B\,i}^{\ \ \ \ \unl{\unl C}}\bigl(\unl\xi{}_i\bigr)\,\cT_{\unl{\unl C}\,i}\bigl(\unl\xi{}_i\bigr)-\La_{\breve A\widetilde B\,i}^{\ \ \ \ p+q+\unl S}\bigl(\unl\xi{}_i\bigr)\,L_{\unl S}\bigl(\unl\xi{}_i\bigr)\,,
\qqq
to be solved for the $\,\La_{\breve A\widetilde B\,i}^{\ \ \ \ \widetilde C}$.\ The existence and uniqueness of the solution follows from
\berop\label{prop:linrealsusyonkap}
Adopt the hitherto notation, and in particular that of Props.\,\ref{prop:bastanshHP} and \ref{prop:leftglobfund}. There exists a canonical realisation of the residual global-supersymmetry algebra $\,\sgt_{\rm vac}\,$ on $\,\sgt_{\rm loc}\x\Si^{\rm HP}_{\rm vac}$,\ 
\qq\nn
&&\sgt_{\rm vac}\x\bigl(\sgt_{\rm loc}\x\Si^{\rm HP}_{\rm vac}\bigr)\supset\ggt\x\bigl(\sgt_{\rm loc}\x\cV^{\rm vac}_{i,\upsilon_i}\bigr)\too\cT\bigl(\sgt_{\rm loc}\x\cV^{\rm vac}_{i,\upsilon_i}\bigr)\subset\cT\bigl(\sgt_{\rm loc}\x\Si^{\rm HP}_{\rm vac}\bigr)\cr\cr 
&:&\ \bigl(X\equiv X^{\breve A}\,S_{\breve A},\bigl(\G\equiv\G^{\widetilde A}\,t_{\widetilde A},\xi_i\bigr)\bigr)\longmapsto X^{\breve A}\,\cK_{S_{\breve A}\,i}\bigl(\unl\xi{}_i\bigr)+\G^{\widetilde B}\,X^{\breve C}\,\La_{\breve C\widetilde B\,i}^{\ \ \ \ \widetilde A}\bigl(\unl\xi{}_i\bigr)\,\tfrac{\vec\p\ \ }{\p\G^{\widetilde A}}\,,
\qqq
compatible with the realisation of the local-supersymmetry supervector space $\,\sgt_{\rm loc}\,$ on $\,\cV^{\rm vac}_{i,\upsilon_i}\,$ by the limit $\,\k^{-\infty}(\sgt\Bgt^{{\rm (HP)}}_{p,\la_p})\,$ of the weak derived flag of the $\k$-symmetry superdistribution $\,\k(\sgt\Bgt^{{\rm (HP)}}_{p,\la_p})\,$ and that of $\,\sgt_{\rm vac}\,$ by the residual global-supersymmetry subspace $\,S_\txG^{\rm HP, vac}\,$ with a trivial linearised correcting ($\hgt_{\rm vac}$-)gauge transformation. It is determined by the sections $\,\La_{A\widetilde B\,i}^{\ \ \ \ \widetilde C}\in\cO_\txG(\cV^{\rm vac}_{i,\upsilon_i})\,$ of \Reqref{eq:rotsec} of the form 
\qq\nn
\La_{\breve A\widetilde B\,i}^{\ \ \ \ \unl{\unl C}}=-[\cK_{S_{\breve A}\,i},\cT_{\widetilde B\,i}\}\con\theta_{\rm L}^{\unl{\unl C}}\equiv-\breve\Xi{}_{\breve A\,i}^{\ \ \ \unl S}\,f_{\unl S\widetilde B}^{\ \ \ \unl{\unl C}}\,,\qquad\qquad\La_{A\widetilde B\,i}^{\ \ \ \ p+q+\unl S}\equiv 0\,,
\qqq
in which $\,\unl S\in\ovl{1,D-\unl\d},\unl{\unl C}\in\ovl{0,p+q}\,$ and the sections $\,\breve\Xi{}_{A\,i}^{\ \ \ \unl S}\in\cO_\txG(\cV^{\rm vac}_{i,\upsilon_i})\,$ are uniquely defined -- in the spirit of Prop.\,\ref{prop:leftglobfund} -- by the formula
\qq\nn
\cK_{S_{\breve A}\,i}=R_{S_{\breve A}}\rstr_{\cV^{\rm vac}_{i,\upsilon_i}}+\breve\Xi{}_{\breve A\,i}^{\ \ \ \unl S}\,L_{\unl S}\,.
\qqq
\eerop
\beroof
Follows straightforwardly from an adaptation of the proof of Prop.\,\ref{prop:descissusy} to the integrable sub-superdistribution $\,{\rm Vac}(\sgt\Bgt^{{\rm (HP)}}_{p,\la_p})\subset\cT\Si^{\rm HP}$,\ and from the identity 
\qq\nn
\cT_{\unl{\unl C}\,i}\con\theta_{\rm L}^A\bigl(\unl\xi{}_i\bigr)=\d_{\unl{\unl C}}^{\ A}+T_{\unl{\unl C}\,i}^{\ \ \ \unl S}\,\d_{\unl S}^{\ A}\,,
\qqq
{\it cp} \Reqref{eq:TmuasL}, taken together with the calculation
\qq\nn
[\cK_{S_{\breve A}\,i},\cT_{\widetilde B\,i}\}&=&\bigl((-1)^{|\breve A|\cdot|\widetilde B|+1}\,L_{\widetilde B}\con\sfd\breve\Xi{}_{\breve A\,i}^{\ \ \ \unl S}+\breve\Xi{}_{\breve A\,i}^{\ \ \ \unl T}\,L_{\unl T}\con\sfd T_{\widetilde B\,i}^{\ \ \ \unl S}+(-1)^{|\breve A|\cdot|\widetilde B|+1}\,T_{\widetilde B\,i}^{\ \ \ \unl T}\,L_{\unl T}\con\sfd\breve\Xi{}_{\breve A\,i}^{\ \ \ \unl S}\cr\cr
&&+R_{S_{\breve A}}\con\sfd T_{\widetilde B\,i}^{\ \ \ \unl S}+\breve\Xi{}_{\breve A\,i}^{\ \ \ \unl U}\,T_{\widetilde B\,i}^{\ \ \ \unl V}\,f_{\unl U\unl V}^{\ \ \ \unl S}\bigr)\,L_{\unl S}+\breve\Xi{}_{\breve A\,i}^{\ \ \ \unl S}\,f_{\unl S\widetilde B}^{\ \ \ \widetilde C}\,L_{\widetilde C}
\qqq
in which relation \eqref{eq:hvacstabtvac} has been used.
\eroof
\noindent Prior to lifting the linearised realisation of the residual supersymmetry on $\,\sgt_{\rm loc}\x\Si^{\rm HP}_{\rm vac}\,$ to the extended HP $p$-gerbe, we pause to reformulate the fundamental identity \eqref{eq:lingloblocomp} in a manner reflecting its actual meaning (as a condition of compatibility of the two actions), and with view to its subsequent applications. Thus, we have
\berop\label{prop:commlocgloblin}
Adopt the notation of Prop.\,\ref{prop:linrealsusyonkap} and consider the fundamental vector fields for the linearised action of $\,\txG\,$ on $\,\sgt_{\rm loc}\x\Si^{\rm HP}_{\rm vac}$,\ with the local coordinate presentation 
\qq\nn
\breve\cK{}_{\breve A}(\G,\unl\xi{}_i):=\cK_{S_{\breve A}}(\unl\xi{}_i)+(-1)^{|\breve A|\cdot|\widetilde B|}\,\G^{\widetilde B}\,\La_{\breve A\widetilde B\,i}^{\ \ \ \ \widetilde C}\bigl(\unl\xi{}_i\bigr)\,\tfrac{\vec\p\ \ }{\p\G^{\widetilde C}}\,,
\qqq
spanning a subspace in the tangent sheaf $\,\cT(\sgt_{\rm loc}\x\Si^{\rm HP}_{\rm vac})\,$ to be denoted as
\qq\nn
\widehat S{}^{\rm HP, vac}_\txG\subset\G\bigl(\cT\bigl(\sgt_{\rm loc}\x\Si^{\rm HP}_{\rm vac}\bigr)\bigr)\,,
\qqq
alongside those generating the limit $\,\k^{-\infty}(\sgt\Bgt^{{\rm (HP)}}_{p,\la_p})\,$ of the weak derived flag of the $\k$-symmetry superdistribution $\,\k(\sgt\Bgt^{{\rm (HP)}}_{p,\la_p})$,\ now regarded as $\sgt_{\rm loc}$-linear vector fields on $\,\sgt_{\rm loc}\x\Si^{\rm HP}_{\rm vac}\,$ as {\it per}
\qq\label{eq:fundglocup}
\widehat\cT(\G,\unl\xi_i):=\G^{\widetilde A}\,\cT_{\widetilde A}(\unl\xi{}_i)\,.
\qqq
These satisfy the commutation relations
\qq\nn
[\breve\cK_{\breve A},\widehat\cT](\G,\unl\xi{}_i)=0+\xcO(\G^2)\,,
\qqq
expressing compatibility of the two actions (in the linear order).
\eerop
\beroof
Follows from Prop.\,\ref{prop:linrealsusyonkap}.
\eroof 

In the remainder of the present paper, we establish a higher-geometric lift of the structure identified above. To this end, we first derive, from a detailed analysis of the restriction of a standard group-equivariant structure (in the Gra\ss mann-even setting) to an infinitesimal vicinity of the group unit, the appropriate notion of a $\sgt_{\rm loc}$-equivariant structure, to be understood as a consistent realisation of the local-symmetry superdistribution $\,\k^{-\infty}(\sgt\Bgt^{{\rm (HP)}}_{p,\la_p})\,$ on the vacuum restriction of the extended HP $p$-gerbe. The underlying supergeometric setting will be the one delineated in Def.\,\ref{def:symmequivstr}. Given the infinitesimal and hence local nature of the realisation sought after, we are free to employ the sheaf-theoretic description of $p$-gerbes and the associated $k$-isomorphisms, which will prove particularly convenient.

Thus, consider a $p$-gerbe $\,\cG^{(p)}\,$ of curvature 
\qq\nn
{\rm curv}\bigl(\cG^{(p)}\bigr)\equiv\underset{\tx{\ciut{(p+2)}}}{\chi}
\qqq
represented by a $(p+1)$-cocycle $\,\underset{\tx{\ciut{(p+1)}}}{\xcB}\in{\rm Ker}\,D^{(p+1)}\subset\textrm{\Cv D}^{p+1}(\cO_\cM,\cD(p+1)^\bullet)$,
\qq\nn
D^{(p+1)}\underset{\tx{\ciut{(p+1)}}}{\xcB}=0\,.
\qqq
and assume existence of a \emph{descendable} $\txG_{\rm loc}$-equivariant structure on $\,\cG^{(p)}$,\ {\it i.e.}, one whose 1-isomorphism
\qq\nn
\Upsilon_p^{(1)}\ :\ \la_{\rm loc}^*\cG^{(p)}\xrightarrow{\ \cong\ }\pr_2^*\cG^{(p)}
\qqq
has 
\qq\nn
{\underset{\tx{\ciut{(p+1)}}}{\varrho_{-\theta_{\rm L}}}}\equiv 0\,.
\qqq
A necessary condition for that to be the case is the horizontality of the curvature with respect to the fundamental vector fields $\,\cK{}^{\la_{\rm loc}}_\cdot$,
\qq\label{eq:curvGphor}
\cK{}^{\la_{\rm loc}}_\G\con\underset{\tx{\ciut{(p+2)}}}{\chi}=0\,,\qquad\G\in\ggt_{\rm loc}\,.
\qqq
Note that the fundamental vector fields induce $\ggt_{\rm loc}$-linear vector fields on $\,\ggt_{\rm loc}\x\cM\,$ as {\it per}
\qq\nn
\widehat\cK{}^{\la_{\rm loc}}\ :\ \ggt_{\rm loc}\x\cM\ni(\G,m)\longmapsto\cK{}^{\la_{\rm loc}}_\G(m)\in\cT_m\cM\,.
\qqq 
As we intend to investigate a linearisation of the above structure, we restrict our considerations to group elements of the form $\,\ee^\G\,$ from an infinitesimal (contractible) neighbourhood $\,\cO_e\,$ of the group unit, $\,e\in\txG_{\rm loc}$,\ for which the $\txG_{\rm loc}$-equivariant structure can be presented (locally) as a collection of $(p+1-l)$-cochains $\,\underset{\tx{\ciut{(p+1-l)}}}{\xcP}\hspace{-10pt}{}^{(l)}\in\textrm{\Cv D}^{p+1-l}\bigl(\{\cO_e\}^{\x l}\x\cO_\cM,\cD(p+1)^\bullet\bigr),\ l\in\ovl{1,p+1}\,$ satisfying the coupled equations
\qq\nn
\la_{\rm loc}^*\underset{\tx{\ciut{(p+1)}}}{\xcB}-\pr_2^*\underset{\tx{\ciut{(p+1)}}}{\xcB}+D^{(p)}\underset{\tx{\ciut{(p)}}}{\xcP}{}^{(1)}\equiv d^{(1)\,*}_1\underset{\tx{\ciut{(p+1)}}}{\xcB}-d^{(1)\,*}_0\underset{\tx{\ciut{(p+1)}}}{\xcB}+D^{(p)}\underset{\tx{\ciut{(p)}}}{\xcP}{}^{(1)}&=&0\,,\cr\cr
\sum_{r=0}^{k+1}\,(-1)^{k+1-r}\,d^{(k+1)\,*}_r\underset{\tx{\ciut{(p+1-k)}}}{\xcP}\hspace{-10pt}{}^{(k)}+D^{(p-k)}\underset{\tx{\ciut{(p-k)}}}{\xcP}\hspace{-4pt}{}^{(k+1)}&=&0\,,\qquad k\in\ovl{1,p}\cr\cr
\sum_{s=0}^{p+2}\,(-1)^{p+2-s}\,d^{(p+2)\,*}_s\underset{\tx{\ciut{(0)}}}{\xcP}\hspace{+1pt}{}^{(p+1)}&=&0\,,
\qqq
to be evaluated on $\,(\ee^{\G_1},m),(\ee^{\G_{k+1}},\ee^{\G_k},\ldots,\ee^{\G_1},m)\,$ and $\,(\ee^{\G_{p+2}},\ee^{\G_{p+1}},\ldots,\ee^{\G_1},m)$,\ respectively. In the r\'egime of `small' $\,\G_l$,\ in which we may -- by a mild abuse of the notation -- write $\,\ee^{\G_l}=e+\G_l+\xcO(\G_l^2)$,\ we expand the various cochains appearing in the above equations in the powers of the $\,\G_l\,$ and keep only terms at most linear in (any of) the $\,\G_l$,\ in particular ({\it cp} \Reqref{eq:fundglocup}),
\qq\nn
\bigl(\la_{\rm loc}^*\underset{\tx{\ciut{(p+1)}}}{\xcB}-\pr_2^*\underset{\tx{\ciut{(p+1)}}}{\xcB}\bigr)\bigl(\ee^{\G_1},m\bigr)=\pLie{\widehat\cK{}^{\la_{\rm loc}}}\pr_2^*\underset{\tx{\ciut{(p+1)}}}{\xcB}\bigl(\G_1,m\bigr)+\xcO\bigl(\G_1^2\bigr)
\qqq
(and similarly for the pullbacks of the $\,\underset{\tx{\ciut{(p+1-l)}}}{\xcP}\hspace{-10pt}{}^{(l)}\,$ along the respective $\,d^{(l+1)}_{l+1}$) and
\qq\nn
&\underset{\tx{\ciut{(p+1-l)}}}{\xcP}\hspace{-10pt}{}^{(l)}\bigl(\ee^{\G_l},\ee^{\G_{l-1}},\ldots,\ee^{\G_1},m\bigr)=:\underset{\tx{\ciut{(p+1-l)}}}{\xcP}\hspace{-9pt}{}^{(l),0}(m)+\sum_{r=1}^l\,\underset{\tx{\ciut{(p+1-l)}}}{\xcP}\hspace{-8pt}{}^{(l),1(r)}\bigl(\G_r,m\bigr)+\xcO\bigl(\G_r\,\G_s\bigr)\,,&\cr\cr
&\underset{\tx{\ciut{(p+1-l)}}}{\xcP}\hspace{-8pt}{}^{(l),1(r)}\bigl(\G_r,m\bigr)\equiv\G_r^{\widetilde A}\,\underset{\tx{\ciut{(p+1-l)}}}{\xcP}\hspace{-8pt}{}^{(l),1(r)}_{\ \ \ \widetilde A}(m)\,.&
\qqq
When substituted to the equations, these yield their linearisation (written in the natural coordinates on the $\,\sgt_{\rm loc}^{\x l}\x\cO_\cM\,$ introduced before):
\qq\nn
&&\pLie{\widehat\cK{}^{\la_{\rm loc}}}\pr_2^*\underset{\tx{\ciut{(p+1)}}}{\xcB}(\G_1,m)+D^{(p)}\underset{\tx{\ciut{(p)}}}{\xcP}{}^{(1),0}(m)+D^{(p)}\underset{\tx{\ciut{(p)}}}{\xcP}{}^{(1),1(1)}(\G_1,m)=0\,,\cr\cr\cr
&&\underset{\tx{\ciut{(p)}}}{\xcP}{}^{(1),0}(m)+\pLie{\widehat\cK{}^{\la_{\rm loc}}}\pr_2^*\underset{\tx{\ciut{(p)}}}{\xcP}{}^{(1),0}(\G_1,m)+D^{(p-1)}\underset{\tx{\ciut{(p-1)}}}{\xcP}{}^{(2),0}(m)+\sum_{r\in\{1,2\}}D^{(p-1)}\underset{\tx{\ciut{(p-1)}}}{\xcP}{}^{(2),1(r)}(\G_r,m)=0\,,\cr\cr\cr
&&\underset{\tx{\ciut{(p-1)}}}{\xcP}{}^{(2),1(1)}(\G_1,m)-\underset{\tx{\ciut{(p-1)}}}{\xcP}{}^{(2),1(2)}(\G_3,m)-\pLie{\widehat\cK{}^{\la_{\rm loc}}}\underset{\tx{\ciut{(p-1)}}}{\xcP}{}^{(2),0}(\G_1,m)+D^{(p-2)}\underset{\tx{\ciut{(p-2)}}}{\xcP}{}^{(3),0}(m)\cr\cr
&+&\sum_{r\in\{1,2,3\}}D^{(p-2)}\underset{\tx{\ciut{(p-2)}}}{\xcP}{}^{(3),1(r)}(\G_r,m)=0\,,\cr\cr\cr
&&\cdots\,,\cr\cr\cr
&&\sum_{s=1}^{p+1}\,(-1)^{p+2-s}\,\bigl(\sum_{r=1}^{s-1}\,\underset{\tx{\ciut{(0)}}}{\xcP}{}^{(p+1),1(r)}(\G_r,m)+\sum_{r=s+1}^{p+1}\,\underset{\tx{\ciut{(0)}}}{\xcP}{}^{(p+1),1(r)}(\G_{r+1},m)+\underset{\tx{\ciut{(0)}}}{\xcP}{}^{(p+1),1(s)}(\G_{s+1}+\G_s,m)\bigr)\cr\cr
&+&\sum_{r=1}^{p+1}\,\underset{\tx{\ciut{(0)}}}{\xcP}{}^{(p+1),1(r)}(\G_r,m)+(-1)^{p+2}\,\sum_{r=1}^{p+1}\,\underset{\tx{\ciut{(0)}}}{\xcP}{}^{(p+1),1(r)}(\G_{r+1},m)+\tfrac{1+(-1)^p}{2}\,\underset{\tx{\ciut{(0)}}}{\xcP}{}^{(p+1),0}(m)\cr\cr
&+&(-1)^{p+2}\,\pLie{\widehat\cK{}^{\la_{\rm loc}}}\pr_2^*\underset{\tx{\ciut{(0)}}}{\xcP}{}^{(p+1),0}(\G_1,m)=0
\qqq
We have not written out the structure of the lower-order equations explicitly (although it is straightforward to do so) because they turn out to be completely irrelevant from our point of view. This is so because being valid for any values of the $\,\G_l\,$ (and so, in particular, for $\,\G_l\equiv 0$), the above system of equations splits into a single equation involving the data of the $p$-gerbe,
\qq\nn
\pLie{\widehat\cK{}^{\la_{\rm loc}}}\pr_2^*\underset{\tx{\ciut{(p+1)}}}{\xcB}(\G_1,m)+D^{(p)}\underset{\tx{\ciut{(p)}}}{\xcP}{}^{(1),1(1)}(\G_1,m)=0\,,
\qqq
and a system of coupled equations involving all the $\,\underset{\tx{\ciut{(p+1-l)}}}{\xcP}\hspace{-9pt}{}^{(l),0}\,$ and the $\,\underset{\tx{\ciut{(p+1-l)}}}{\xcP}\hspace{-8pt}{}^{(l),1(r)}\,$ with $\,(l,r)\neq(1,1)$.\ The latter is always solvable\footnote{Indeed, it admits the trivial (zero) solution.}, and so it is but an artifact of our sheaf-theoretic description. Thus, by the end of the day, we are left with a single equation (written in terms of the $p$-cochain $\,\underset{\tx{\ciut{(p)}}}{\widehat\xcP}\equiv\underset{\tx{\ciut{(p)}}}{\xcP}{}^{(1),1(1)}$)
\qq\nn
\pLie{\widehat\cK{}^{\la_{\rm loc}}}\pr_2^*\underset{\tx{\ciut{(p+1)}}}{\xcB}+D^{(p)}\underset{\tx{\ciut{(p)}}}{\widehat\xcP}=0\,.
\qqq
This equation has an obvious \emph{higher-geometric} interpretation which leads us directly to the desired
\bedef\label{def:desclinequivstr}
Let $\,\cM\,$ be a manifold and $\,\txG_{\rm loc}\,$ a Lie group (with Lie algebra $\,\ggt_{\rm loc}$) realised on it by (local) diffeomorphisms, as in Def.\,\ref{def:symmequivstr}, that induce a \textbf{local-symmetry distribution} 
\qq\nn
\cS_{\ggt_{\rm loc}}\subset\cT\cM
\qqq
spanned by the fundamental vector fields  
\qq\nn
\cK{}^{\la_{\rm loc}}_\cdot\ :\ \ggt_{\rm loc}\too\G(\sfT\cM)\ :\ \G\longmapsto\cK_\G\,,
\qqq 
and so giving rise to a $\ggt_{\rm loc}$-linear vector field on $\,\ggt_{\rm loc}\x\cM$,\ 
\qq\label{eq:KglocMeqstr}
\widehat\cK{}^{\la_{\rm loc}}\ :\ \ggt_{\rm loc}\x\cM\ni(\G,m)\longmapsto\cK_\G(m)\in\cT_{(\G,m)}\bigl(\ggt_{\rm loc}\x\cM\bigr)\,.
\qqq
Let, next, $\,\cG^{(p)}\,$ be a $p$-gerbe over $\,\cM\,$ presented by a \Cv ech--Deligne $(p+1)$-cocycle $\,\underset{\tx{\ciut{(p+1)}}}{\xcB}\in{\rm Ker}\,D^{(p+1)}\subset\textrm{\Cv D}^{p+1}(\cO_\cM,\cD(p+1)^\bullet)\,$ associated with an open cover $\,\cO_\cM\,$ of $\,\cM$.\ A (\textbf{descendable}) \textbf{$\ggt_{\rm loc}$-equivariant structure} on $\,\cG^{(p)}\,$ is a $p$-gerbe 1-isomorphism
\qq\nn
\widehat\La_{\ggt_{\rm loc}}\ :\ \pLie{\widehat\cK{}^{\la_{\rm loc}}}\pr_2^*\cG^{(p)}\xrightarrow{\ \cong\ }\cI_0^{(p)}
\qqq
over $\,\ggt_{\rm loc}\x\cM$,\ written for the trivial $p$-gerbe $\,\cI_0^{(p)}\,$ over $\,\ggt_{\rm loc}\x\cM\,$ with a vanishing global curving and a $p$-gerbe $\,\pLie{\widehat\cK{}^{\la_{\rm loc}}}\pr_2^*\cG^{(p)}\,$ over the same base with local (sheaf-cohomological) data over the open cover $\,\{\ggt_{\rm loc}\}\x\cO_\cM\,$ obtained from the pullback of the local data $\,\underset{\tx{\ciut{(p+1)}}}{\xcB}\,$ of $\,\cG^{(p)}\,$ to it along $\,\pr_2\,$ by taking their Lie derivative along $\,\widehat\cK{}^{\la_{\rm loc}}\,$ (component-wise) -- we call this latter $p$-gerbe the \textbf{Lie derivative of the $p$-gerbe $\,\pr_2^*\cG^{(p)}\,$ along the vector field} $\,\widehat\cK{}^{\la_{\rm loc}}$.

We extend the above definition \emph{verbatim} to the category of supermanifolds, whereby the notion of a (descendable) $\ggt_{\rm loc}$-equivariant structure on a $p$-gerbe over a supermanifold (and, in particular, on a super-$p$-gerbe) for $\,\ggt_{\rm loc}\,$ a Lie superalgebra arises.
\exdef
\brem
The only thing that requires a word of justification is the claim that $\,\pLie{\widehat\cK{}^{\la_{\rm loc}}}\pr_2^*\underset{\tx{\ciut{(p+1)}}}{\xcB}\,$ is a \Cv ech--Deligne $(p+1)$-cocycle over $\,\{\ggt_{\rm loc}\}\x\cO_\cM\,$ whenever $\,\underset{\tx{\ciut{(p+1)}}}{\xcB}\,$ is one over $\,\cO_\cM$.\ This follows straightforwardly from Prop.\,\ref{prop:LieD}.

Note also that it is merely the vector-space structure on $\,\ggt_{\rm loc}\,$ that is relevant to the above definition as a algebraic model of the local-symmetry distribution $\,\cS_{\ggt_{\rm loc}}$.\ This is to be kept in mind when generalising the definition to the supergeometric setting.
\erem
\noindent We have
\berop\label{prop:canglocequivstr}
Adopt the notation of Def.\,\ref{def:desclinequivstr} and Prop.\,\ref{prop:LieD}. There exists a canonical descendable $\ggt_{\rm loc}$-equivariant structure on every $p$-gerbe $\,\cG^{(p)}\,$ with a $\cS_{\ggt_{\rm loc}}$-horizontal curvature. The local data $\,\underset{\tx{\ciut{(p)}}}{\widehat\xcP}\in\textrm{\Cv D}^p(\{\ggt_{\rm loc}\}\x\cO_\cM,\cD(p+1)^\bullet)\,$ of the structure associated with the open cover $\,\cO_\cM\,$ of $\,\cM\,$ are determined by those of the $p$-gerbe, $\,\underset{\tx{\ciut{(p+1)}}}{\xcB}$,\ as
\qq\label{eq:canglocequivstr}
\underset{\tx{\ciut{(p)}}}{\widehat\xcP}=-\widehat\cK{}^{\la_{\rm loc}}\con\pr_2^*\underset{\tx{\ciut{(p+1)}}}{\xcB}\,.
\qqq
Consequently, we may write the $p$-gerbe 1-isomorphism of the $\ggt_{\rm loc}$-equivariant structure as
\qq\nn
\widehat\La_{\ggt_{\rm loc}}\equiv-\widehat\cK{}^{\la_{\rm loc}}\con\pr_2^*\cG^{(p)}\,.
\qqq
\eerop
\beroof
For a compact proof, write the Deligne coboundary operator in an obvious shorthand notation as
\qq\nn
D^{(k)}&\equiv&D^{(k)}_{(\bullet_1,\bullet_2)}\,,\cr\cr 
D^{(k)}_{(\bullet_1,\bullet_2)}&\equiv&\bigl(D^{(k)}_{(m,n)}\equiv\sfd^{(n)}+(-1)^{n+1}\,\vd^{(m)}\bigr)_{\substack{(m,n)\in\bN\x(\{-1\}\cup\bN) \\ m+n=k}}\equiv\sfd_{(k)}^{(\bullet_2)}+(-1)^{\bullet_2+1}\,\vd_{(k)}^{(\bullet_1)}
\qqq
{\it cp} \Reqref{eq:Delignecob}. In this notation, we readily compute
\qq\nn
\pLie{\widehat\cK{}^{\la_{\rm loc}}}\pr_2^*\underset{\tx{\ciut{(p+1)}}}{\xcB}&\equiv&\widehat\cK{}^{\la_{\rm loc}}\con\sfd^{(\bullet_2)}_{(p+1)}\pr_2^*\underset{\tx{\ciut{(p+1)}}}{\xcB}+\sfd^{(\bullet_2)}_{(p)}\bigl(\widehat\cK{}^{\la_{\rm loc}}\con\pr_2^*\underset{\tx{\ciut{(p+1)}}}{\xcB}\bigr)\cr\cr
&=&\widehat\cK{}^{\la_{\rm loc}}\con\pr_2^*\bigl(\underset{\tx{\ciut{(p+2)}}}{\chi},D^{(p+1)}\underset{\tx{\ciut{(p+1)}}}{\xcB}\bigr)-\widehat\cK{}^{\la_{\rm loc}}\con\pr_2^*(-1)^{\bullet_2+1}\,\vd_{(p+1)}^{(\bullet_1)}\underset{\tx{\ciut{(p+1)}}}{\xcB}+\sfd^{(\bullet_2)}_{(p)}\bigl(\widehat\cK{}^{\la_{\rm loc}}\con\pr_2^*\underset{\tx{\ciut{(p+1)}}}{\xcB}\bigr)\cr\cr
&=&\sfd^{(\bullet_2)}_{(p)}\bigl(\widehat\cK{}^{\la_{\rm loc}}\con\pr_2^*\underset{\tx{\ciut{(p+1)}}}{\xcB}\bigr)+\widehat\cK{}^{\la_{\rm loc}}\con\pr_2^*(-1)^{\bullet_2}\,\vd_{(p+1)}^{(\bullet_1)}\underset{\tx{\ciut{(p+1)}}}{\xcB}\cr\cr
&\equiv&\sfd^{(\bullet_2)}_{(p)}\bigl(\widehat\cK{}^{\la_{\rm loc}}\con\pr_2^*\underset{\tx{\ciut{(p+1)}}}{\xcB}\bigr)+(-1)^{\bullet_2+1}\,\vd_{(p)}^{(\bullet_1)}\bigl(\widehat\cK{}^{\la_{\rm loc}}\con\pr_2^*\underset{\tx{\ciut{(p+1)}}}{\xcB}\bigr)\equiv D^{(p)}\bigl(\widehat\cK{}^{\la_{\rm loc}}\con\pr_2^*\underset{\tx{\ciut{(p+1)}}}{\xcB}\bigr)\,,
\qqq
whence the thesis of the proposition follows.
\eroof

\brem
Structures of the kind considered above, or -- indeed -- their full-fledged Lie-algebraic extensions induced, through linearisation, from group-equivariant structures on the bi-category of 1-gerbes, appeared for the first time in the context of the gauging of global $\si$-model symmetries in \Rxcite{Sec.\,7.2}{Gawedzki:2010rn}. We shall encounter them again when we take up the issue of (globally linearised-supersymmetric) equivariance with respect to a realisation of the $\k$-symmetry Lie superalgebra in a forthcoming paper \cite{Suszek:2020ksL}.
\erem

\noindent As a corollary to the above, we may state our first physical higher-(super)geometric result:
\berop\label{prop:canequivstr}
Adopt the hitherto notation, in particular that of Def.\,\ref{def:restrextHPpgerb} and Prop.\,\ref{prop:commlocgloblin}, and assume integrability of the vacuum superdistribution $\,{\rm Vac}(\sgt\Bgt^{{\rm (HP)}}_{p,\la_p})\,$ of Def.\,\ref{def:vacHPsdistro}. There exists a canonical $\ggt\sgt_{\rm vac}(\sgt\Bgt^{{\rm (HP)}}_{p,\la_p})$-equivariant structure, in the sense of Def.\,\ref{def:desclinequivstr}, on the vacuum restriction $\,\iota_{\rm vac}^*\widehat\cG{}^{(p)}\,$ of the extended Hughes--Polchinski $p$-gerbe $\,\widehat\cG{}^{(p)}$.\ It takes the form
\qq\nn
\widehat\La_{\ggt\sgt_{\rm vac}(\sgt\Bgt^{{\rm (HP)}}_{p,\la_p})}=-\widehat\cT\con\pr_2^*\iota_{\rm vac}^*\widehat\cG{}^{(p)}
\qqq 
(as introduced in Prop.\,\ref{prop:canglocequivstr}), where $\,\widehat\cT\,$ is the vector field on $\,\ggt\sgt_{\rm vac}(\sgt\Bgt^{{\rm (HP)}}_{p,\la_p})\x\Si^{\rm HP}_{\rm vac}\,$ given by \Reqref{eq:fundglocup}. 
\eerop
\beroof
Follows from an adaptation of Prop.\,\ref{prop:canglocequivstr}.
\eroof

\noindent Next, we investigate compatibility of the above (canonical) $\ggt_{\rm loc}$-equivariant structure on $\,\cG^{(p)}\,$ with its assumed linearised $\txG$-symmetry, to be defined next.

As the first step towards this goal, we adapt the result of our linearisation procedure to the case of a global symmetry, as recapitulated on p.\,\pageref{p:globsym}. A moment's thought about the nature of the difference between the gerbe-theoretic manifestations of a global symmetry and a local one convinces us of the adequacy of the following
\bedef\label{def:linsymstr}
Let $\,\txG\,$ be a Lie group with a Lie algebra $\,\ggt$,\ and let $\,\cM\,$ be a manifold on which $\,\txG\,$ is realised by (local) diffeomorphisms, as in Def.\,\ref{def:symmequivstr}, that induce a global-symmetry subspace $\,S_\txG\subset\G(\cT\cM)\,$ of Def.\,\ref{def:globlinGsdistro}, spanned by the fundamental vector fields 
\qq\nn
\cK^\la_\cdot\ :\ \ggt\too\G(\sfT\cM)\ :\ X\longmapsto\cK^\la_X\,.
\qqq 
A $p$-gerbe $\,\cG^{(p)}\,$ over $\,\cM\,$ is termed a \textbf{$\ggt$-invariant $p$-gerbe} if there exists a family of $p$-gerbe 1-isomorphisms
\qq\nn
\bigl\{\ \ \widetilde\La_X\ :\ \pLie{\cK^\la_X}\cG^{(p)}\xrightarrow{\ \cong\ }\cI_0^{(p)}\ \ \bigr\}_{X\in\ggt}
\qqq
over $\,\cM$,\ written for the trivial $p$-gerbe $\,\cI_0^{(p)}\,$ over $\,\cM\,$ with a vanishing global curving and for the Lie derivative $\,\pLie{\cK^\la_X}\cG^{(p)}\,$ of $\,\cG^{(p)}\,$ along $\,\cK^\la_X$,\ as described in Def.\,\ref{def:desclinequivstr}.

Given two $p$-gerbes $\,\cG^{(p)}_A,\ A\in\{1,2\}\,$ over a common base $\,\cM\,$ that are $\ggt$-symmetric, with the respective families of 1-isomorphisms
\qq\nn
\bigl\{\ \ \widetilde\La_X^A\ :\ \pLie{\cK^\la_X}\cG_A^{(p)}\xrightarrow{\ \cong\ }\cI_0^{(p)}\ \ \bigr\}_{X\in\ggt}\,,
\qqq
and a $p$-gerbe 1-isomorphism 
\qq\nn
\Phi\ :\ \cG^{(p)}_1\xrightarrow{\ \cong\ }\cG^{(p)}_2
\qqq
between them, we call the latter a \textbf{$\ggt$-invariant $p$-gerbe 1-isomorphism} if there exists a family of $p$-gerbe 2-isomorphism 
\qq\nn
\alxydim{@C=2.5cm@R=2.cm}{ \pLie{\cK^\la_X}\cG^{(p)}_1 \ar[r]^{\pLie{\cK^\la_X}\Phi}
\ar[d]_{\widetilde\La_X^1} & \pLie{\cK^\la_X}\cG^{(p)}_2
\ar[d]^{\widetilde\La_X^2} \ar@{=>}[dl]|{\,\psi_X\ } \\
\cI_0^{(p)} \ar@{=}[r]_{{\rm Id}_{\cI_0^{(p)}}} & \cI_0^{(p)}}\,,\qquad X\in\ggt\,,
\qqq
written for the corresponding $p$-gerbe 1-isomorphisms $\,\pLie{\cK^\la_X}\Phi\,$ with local (sheaf-cohomological) data over an open cover $\,\cO_\cM\,$ of $\,\cM\,$ common to all three: $\,\cG^{(p)}_1,\cG^{(p)}_2\,$ and $\,\Phi\,$ obtained from the corresponding local data of $\,\Phi\,$ by taking their Lie derivative along $\,\cK^\la_X\,$ -- we call the $p$-gerbe 1-isomorphism thus formed the \textbf{Lie derivative of the $p$-gerbe 1-isomorphim $\,\Phi\,$ along the vector field} $\,\cK^\la_X$.

We extend the above definition \emph{verbatim} to the category of supermanifolds, whereby the notions of a $\ggt$-invariant structure on a $p$-gerbe over a supermanifold (and, in particular, on a super-$p$-gerbe), as well as that of a $\ggt$-invariant 1-isomorphism between such $p$-gerbes arise for a Lie superalgebra $\,\ggt$.
\exdef
\brem
As in the case of the Lie derivative of a $p$-gerbe, the existence of the Lie derivative $\,\pLie{\cK^\la_X}\Phi\,$ for a given $p$-gerbe 1-isomorphism $\,\Phi\,$ is ensured by the commutativity of the Lie derivative with the Deligne coboundary operator ({\it cp} Prop.\,\ref{prop:LieD}).

Furthermore, in analogy with Def.\,\ref{def:desclinequivstr}, it is only the vector-space structure on $\,\ggt\,$ that enters the definition as an algebraic model for the global-symmetry subspace $\,S_\txG$.
\erem

\noindent Thus, by the assumed global supersymmetry of the GS super-$p$-gerbe $\,\cG^{(p)}\,$ of Def.\,\ref{def:spG} and in view of the manifestly supersymmetric definition \eqref{eq:HPcurv} of the HP super-$(p+1)$-form, we obtain a $\ggt$-indexed family of $p$-gerbe 1-isomorphisms
\qq\nn
\La_X\ :\ \pLie{\cK_X}\widehat\cG{}^{(p)}\xrightarrow{\ \cong\ }\cI^{(p)}_0\,,\qquad X\in\ggt
\qqq
defined for the extended HP $p$-gerbe $\,\cG^{(p)}\,$ of Def.\,\ref{def:restrextHPpgerb}. Those associated with generators of the residual global-supersymmetry subalgebra $\,\sgt_{\rm vac}\subset\ggt\,$ of Prop.\,\ref{prop:resglobsusysub} are readily seen to descend to its vacuum restriction (being induced by vector fields tangent to the vacuum), and so we obtain a family of $p$-gerbe 1-isomorphisms 
\qq\nn
\breve\La_{\breve A}\ :\ \pLie{\cK_{S_{\breve A}}}\iota_{\rm vac}^*\widehat\cG{}^{(p)}\xrightarrow{\ \cong\ }\cI^{(p)}_0\,,\qquad\breve A\in\ovl{1,S_{\rm vac}}\,.
\qqq
These will prove instrumental in understanding $\sgt_{\rm vac}$-invariance of the previously established canonical $\ggt\sgt_{\rm vac}(\sgt\Bgt^{{\rm (HP)}}_{p,\la_p})$-equivariant structure on $\,\iota_{\rm vac}^*\widehat\cG{}^{(p)}\,$ as they form the basis for the application of the following
\berop\label{prop:transinvstrthrLie}
Let $\,\cM,\txG\,$ and $\,\txG_{\rm loc}\,$ all be as in Def.\,\ref{def:symmequivstr}, and let $\,\cG^{(p)}\,$ be a $p$-gerbe over $\,\cM\,$ that we assume $\ggt$-invariant in the sense of Def.\,\ref{def:linsymstr}, with the basis $p$-gerbe 1-isomorphisms
\qq\nn
\widetilde\La_A\ :\ \pLie{\cK^\la_A}\cG^{(p)}\xrightarrow{\ \cong\ }\cI_0^{(p)}\,,\qquad A\in\ovl{1,\dim\,\ggt}\,.
\qqq 
Fix a vector field $\,\widehat\cV\in\G(\cT(\ggt_{\rm loc}\x\cM))$.\ If there exists a lift of the global-symmetry subspace $\,S_\txG\,$ to a $\ggt_{\rm loc}$-linear subspace $\,\widehat S_\txG\subset\G(\cT(\ggt_{\rm loc}\x\cM))$,\ spanned on the lifts $\,\breve\cK{}^\la_A,\ A\in\ovl{1,\dim\,\ggt}\,$ of the respective vector fields $\,\cK^\la_A$ and defined as in Prop.\,\ref{prop:commlocgloblin}, with the property 
\qq\nn
\forall_{A\in\ovl{1,\dim\,\ggt}}\ :\ [\breve\cK{}^\la_A,\widehat\cV]=0\,,
\qqq
then the Lie derivative $\,\pLie{\widehat\cV}\pr_2^*\cG^{(2)}\,$ of the pullback of $\,\cG^{(p)}\,$ to $\,\ggt_{\rm loc}\x\cM\,$ along the canonical projection, understood as in Def.\,\ref{def:desclinequivstr}, is canonically $\ggt$-invariant, with the corresponding $p$-gerbe 1-isomorphisms
\qq\nn
\breve\La{}_A^{\widehat\cV}\ :\ \pLie{\breve\cK{}^\la_A}\bigl(\pLie{\widehat\cV}\pr_2^*\cG^{(p)}\bigr)\xrightarrow{\ \cong\ }\cI_0^{(p)}
\qqq 
given by the formula
\qq\nn
\breve\La{}_A^{\widehat\cV}=\pLie{\widehat\cV}\pr_2^*\widetilde\La_A\,,
\qqq
to be understood in the spirit of Def.\,\ref{def:desclinequivstr} and Prop.\,\ref{prop:LieD}.
\eerop
\beroof
Adopt the previously introduced notation. Let $\,\underset{\tx{\ciut{(p+1)}}}{\xcB}\in{\rm Ker}\,D^{(p+1)}\,$ be a \Cv ech--Deligne $(p+1)$-cocycle presenting $\,\cG^{(p)}\,$ for an open cover $\,\cO_\cM\,$ of $\,\cM\,$ over which -- for every value\footnote{We might have to vary the choice of the open cover as $\,A\,$ ranges over $\,\ovl{1,\dim\,\ggt}$,\ which -- however -- would not affect the validity of our sheaf-cohomological argument, and we might ultimately take their common refinement.\label{page:commonref}} $\,A\in\ovl{1,\dim\,\ggt}\,$ -- there exists, by assumption, a $p$-cochain $\,\underset{\tx{\ciut{(p)}}}{\xcP}\hspace{-2pt}{}_A\in\textrm{\Cv D}^p(\cO_\cM,\cD(p+1)^\bullet)\,$ that represents the $\ggt$-invariant structure,
\qq\nn
\pLie{\cK^\la_A}\underset{\tx{\ciut{(p+1)}}}{\xcB}=-D^{(p)}\underset{\tx{\ciut{(p)}}}{\xcP}\hspace{-2pt}{}_A\,.
\qqq
Invoking the definitions (and the properties that follow therefrom) of the Lie derivatives (of $p$-gerbes) involved and using the defining relation between $\,\cK^\la_A\,$ and (its \emph{lift}) $\,\breve\cK{}^\la_A$,\ as well as the explicit form of the induced cover $\,\{\ggt_{\rm loc}\}\x\cO_\cM\,$ of $\,\ggt_{\rm loc}\x\cM$,\ we readily calculate
\qq\nn
\pLie{\breve\cK{}^\la_A}\pLie{\widehat\cV}\pr_2^*\underset{\tx{\ciut{(p+1)}}}{\xcB}&=&\pLie{[\breve\cK{}^\la_A,\widehat\cV]}\pr_2^*\underset{\tx{\ciut{(p+1)}}}{\xcB}+\pLie{\widehat\cV}\pLie{\breve\cK{}^\la_A}\pr_2^*\underset{\tx{\ciut{(p+1)}}}{\xcB}\equiv\pLie{\widehat\cV}\pr_2^*\pLie{\cK^\la_A}\underset{\tx{\ciut{(p+1)}}}{\xcB}=-\pLie{\widehat\cV}\pr_2^*D^{(p)}\underset{\tx{\ciut{(p)}}}{\xcP}\hspace{-2pt}{}_A\cr\cr
&=&-D^{(p)}\pLie{\widehat\cV}\pr_2^*\underset{\tx{\ciut{(p)}}}{\xcP}\hspace{-2pt}{}_A\,,
\qqq
and thus identify the $p$-cochain
\qq\nn
\pLie{\widehat\cV}\pr_2^*\underset{\tx{\ciut{(p)}}}{\xcP}\hspace{-2pt}{}_A\in\textrm{\Cv D}^p\bigl(\{\ggt_{\rm loc}\}\x\cO_\cM,\cD(p+1)^\bullet\bigr)
\qqq
as a local presentation of the component, associated with the generator $\,\t_A\in\ggt$,\ of the (canonical) $\ggt$-invariant structure on $\,\pLie{\widehat\cV}\pr_2^*\cG^{(2)}\,$ predicted by the proposition.
\eroof
\noindent Again, we obtain a physically relevant corollary
\berop\label{canextinvstr}
Adopt the hitherto notation, in particular that of Def.\,\ref{def:restrextHPpgerb} and Props.\,\ref{prop:resglobsusysub} and \ref{prop:commlocgloblin}, and assume integrability of the vacuum superdistribution $\,{\rm Vac}(\sgt\Bgt^{{\rm (HP)}}_{p,\la_p})\,$ of Def.\,\ref{def:vacHPsdistro}. The Lie derivative $\,\pLie{\widehat\cT}\pr_2^*\iota_{\rm vac}^*\widehat\cG{}^{(p)}\,$ of the pullback of the vacuum restriction of the extended Hughes--Polchinski $p$-gerbe to the supermanifold $\,\ggt\sgt_{\rm vac}(\sgt\Bgt^{{\rm (HP)}}_{p,\la_p})\x\Si^{\rm HP}_{\rm vac}\,$ along the vector field $\,\widehat\cT\,$ on the latter defined in \Reqref{eq:fundglocup} is canonically $\sgt_{\rm vac}$-invariant, with the corresponding $p$-gerbe 1-isomorphisms
\qq\nn
\breve\La{}_{\breve A}\ :\ \pLie{\breve\cK{}_{\breve A}}\bigl(\pLie{\widehat\cT}\pr_2^*\iota_{\rm vac}^*\widehat\cG{}^{(p)}\bigr)\xrightarrow{\ \cong\ }\cI_0^{(p)}
\qqq 
given by the formula
\qq\nn
\breve\La{}_{\breve A}=\pLie{\widehat\cT}\pr_2^*\La{}_{\breve A}
\qqq
in terms of the $p$-gerbe 1-isomorphisms 
\qq\nn
\La_{\breve A}\ :\ \pLie{\cK_{S_{\breve A}}}\iota_{\rm vac}^*\widehat\cG{}^{(p)}\xrightarrow{\ \cong\ }\cI^{(p)}_0\,,\qquad\breve A\in\ovl{1,S_{\rm vac}}
\qqq
whose existence is ensured by $\sgt_{\rm vac}$-invariance of $\,\iota_{\rm vac}^*\widehat\cG{}^{(p)}$.
\eerop
\beroof
Follows from an adaptation of Prop.\,\ref{prop:transinvstrthrLie}, taken in conjunction with Prop.\,\ref{prop:commlocgloblin}.
\eroof

So far, we have been dealing with independent linearisations of a $\txG_{\rm loc}$-equivariant structure on a $p$-gerbe and of a realisation of the global-symmetry group $\,\txG\,$ on it (by $p$-gerbe 1-isomorphisms). We may now finally come to the context of immediate interest in which both linearisations are combined in a cohrent manner, that is to a gerbe-theoretic rendering of the compatibility scenario laid out in Def.\,\ref{def:symmequivstr}. Our goal is to acquire tools that enable us to verify \emph{canonical} $\sgt_{\rm vac}$-invariance of the \emph{canonical} $\ggt\sgt_{\rm vac}(\sgt\Bgt^{{\rm (HP)}}_{p,\la_p})$-equivariant structure on the vacuum restriction of the extended HP $p$-gerbe.

As previously, we work out the requisite intuitions in the Gra\ss mann-even setting first.
\berop\label{prop:commcan}
Adopt the hitherto notation, and in particular that of Defs.\,\ref{def:symmequivstr}, \ref{def:desclinequivstr} and \ref{def:linsymstr}, as well as that of Prop.\,\ref{prop:transinvstrthrLie}. Let $\,\cM,\txG\,$ and $\,\txG_{\rm loc}\,$ be all as in Def.\,\ref{def:symmequivstr}, with $\,S_\txG\,$ the global-symmetry subspace of Def.\,\ref{def:globlinGsdistro} and $\,\cS_{\ggt_{\rm loc}}\subset\cT\cM\,$ the local-symmetry distribution introduced in Def.\,\ref{def:desclinequivstr}, and let $\,\cG^{(p)}\,$ (for $p>0$) be a $p$-gerbe over $\,\cM\,$ that we assume $\ggt$-invariant in the sense of Def.\,\ref{def:linsymstr}, with the basis $p$-gerbe 1-isomorphisms
\qq\nn
\widetilde\La_A\ :\ \pLie{\cK^\la_A}\cG^{(p)}\xrightarrow{\ \cong\ }\cI_0^{(p)}\,,\qquad A\in\ovl{1,\dim\,\ggt}\,,
\qqq 
and suppose that the curvature $\,\underset{\tx{\ciut{(p+2)}}}{\chi}\,$ of $\,\cG^{(p)}\,$ is $\cS_{\ggt_{\rm loc}}$-horizontal, as expressed in \Reqref{eq:curvGphor}. If there exists a lift of $\,S_\txG\,$ to a $\ggt_{\rm loc}$-linear subspace $\,\widehat S_\txG\subset\G(\cT(\ggt_{\rm loc}\x\cM))$,\ spanned on the lifts $\,\breve\cK{}^\la_A,\ A\in\ovl{1,\dim\,\ggt}\,$ of the respective vector fields $\,\cK^\la_A$ and defined as in Prop.\,\ref{prop:commlocgloblin}, with the property
\qq\label{eq:brevecommhat}
\forall_{A\in\ovl{1,\dim\,\ggt}}\ :\ [\breve\cK{}^\la_A,\widehat\cK^{\la_{\rm loc}}]=0\,,
\qqq
expressed in terms of the vector field $\,\widehat\cK^{\la_{\rm loc}}\,$ of \Reqref{eq:KglocMeqstr}, then the canonical $\ggt_{\rm loc}$-equivariant structure \eqref{eq:canglocequivstr} on $\,\cG^{(p)}\,$ is canonically $\ggt$-invariant in the sense of Def.\,\ref{def:linsymstr}. The corresponding $p$-gerbe 2-isomorphisms 
\qq\nn
\alxydim{@C=3.cm@R=2.cm}{ \pLie{\breve\cK{}^\la_A}\bigl(\pLie{\widehat\cK{}^{\la_{\rm loc}}}\pr_2^*\cG^{(p)}\bigr) \ar[r]^{\qquad\qquad\pLie{\breve\cK{}^\la_A}\widehat\La{}_{\ggt_{\rm loc}}}
\ar[d]_{\breve\La{}_A^{\widehat\cK{}^{\la_{\rm loc}}}} & \cI^{(p)}_0
\ar[d]^{\id_{\cI^{(p)}_0}} \ar@{=>}[dl]|{\,\psi_A\ } \\
\cI_0^{(p)} \ar@{=}[r]_{{\rm Id}_{\cI_0^{(p)}}} & \cI_0^{(p)}}\,,\qquad A\in\ovl{1,\dim\,\ggt}
\qqq
can be written in the form
\qq\nn
\psi_A=\widehat\cK{}^{\la_{\rm loc}}\con\pr_2^*\widetilde\La_A\,.
\qqq
\eerop
\beroof
Adopt the previously introduced notation. Let $\,\underset{\tx{\ciut{(p+1)}}}{\xcB}\in{\rm Ker}\,D^{(p+1)}\,$ be a \Cv ech--Deligne $(p+1)$-cocycle presenting $\,\cG^{(p)}\,$ for an open cover $\,\cO_\cM\,$ of $\,\cM\,$ over which there exist 
\bit
\item $\underset{\tx{\ciut{(p)}}}{\xcP}\hspace{-2pt}{}_A\in\textrm{\Cv D}^p(\cO_\cM,\cD(p+1)^\bullet),\ A\in\ovl{1,\dim\,\ggt}\,$ -- a local presentation of the $\ggt$-invariant structure\footnote{{\it Cp} the comment on p.\,\pageref{page:commonref}.} associated with a basis $\,\{\t_A\}_{A\in\ovl{1,\dim\,\ggt}}\,$ of $\,\ggt$,
\qq\nn
\pLie{\cK^\la_A}\underset{\tx{\ciut{(p+1)}}}{\xcB}=-D^{(p)}\underset{\tx{\ciut{(p)}}}{\xcP}\hspace{-2pt}{}_A\,;
\qqq
\item $\underset{\tx{\ciut{(p)}}}{\widehat\xcP}\in\textrm{\Cv D}^p(\{\ggt_{\rm loc}\}\x\cO_\cM,\cD(p+1)^\bullet)\,$ -- a local presentation of the $\ggt_{\rm loc}$-equivariant structure,
\qq\nn
\pLie{\widehat\cK{}^{\la_{\rm loc}}}\pr_2^*\underset{\tx{\ciut{(p+1)}}}{\xcB}=-D^{(p)}\underset{\tx{\ciut{(p)}}}{\widehat\xcP}\,.
\qqq
\eit
We compute, along similar lines as in the proof of Prop.\,\ref{prop:transinvstrthrLie} and using Eqs.\,\eqref{eq:canglocequivstr} and \eqref{eq:brevecommhat}, alongside Prop.\,\ref{prop:LieD}, 
\qq\nn
\pLie{\breve\cK{}^\la_A}\widehat{\underset{\tx{\ciut{(p)}}}{\xcP}}-\pLie{\widehat\cK{}^{\la_{\rm loc}}}\pr_2^*\underset{\tx{\ciut{(p)}}}{\xcP}\hspace{-2pt}{}_A&\equiv&-\pLie{\breve\cK{}^\la_A}\bigl(\widehat\cK{}^{\la_{\rm loc}}\con\pr_2^*\underset{\tx{\ciut{(p+1)}}}{\xcB}\bigr)-\pLie{\widehat\cK{}^{\la_{\rm loc}}}\pr_2^*\underset{\tx{\ciut{(p)}}}{\xcP}\hspace{-2pt}{}_A\cr\cr
&=&-[\breve\cK{}^\la_A,\widehat\cK{}^{\la_{\rm loc}}]\con\pr_2^*\underset{\tx{\ciut{(p+1)}}}{\xcB}-\widehat\cK{}^{\la_{\rm loc}}\con\pr_2^*\bigl(\pLie{\cK{}^\la_A}\underset{\tx{\ciut{(p+1)}}}{\xcB}\bigr)-\pLie{\widehat\cK{}^{\la_{\rm loc}}}\pr_2^*\underset{\tx{\ciut{(p)}}}{\xcP}\hspace{-2pt}{}_A\cr\cr
&=&\widehat\cK{}^{\la_{\rm loc}}\con D^{(p)}\pr_2^*\underset{\tx{\ciut{(p)}}}{\xcP}{}_A-\pLie{\widehat\cK{}^{\la_{\rm loc}}}\pr_2^*\underset{\tx{\ciut{(p)}}}{\xcP}\hspace{-2pt}{}_A=-D^{(p-1)}\bigl(\widehat\cK{}^{\la_{\rm loc}}\con\pr_2^*\underset{\tx{\ciut{(p)}}}{\xcP}\hspace{-2pt}{}_A\bigr)
\qqq
and extract from that computation the definition 
\qq\nn
\widehat\cK{}^{\la_{\rm loc}}\con\pr_2^*\underset{\tx{\ciut{(p)}}}{\xcP}\hspace{-2pt}{}_A\in\textrm{\Cv D}^{p-1}\bigl(\{\ggt_{\rm loc}\}\x\cO_\cM,\cD(p+1)^\bullet\bigr)\,,\qquad A\in\ovl{1,\dim\,\ggt}
\qqq
of a family of $(p-1)$-cochains that compose a local presentation of the (canonical) $\ggt$-invariant 2-isomorphisms
from the thesis of the proposition.
\eroof
\noindent Our hitherto efforts are crowned by the following theorem in which the various results are put together.
\bethe\label{thm:kappagerbified}
Adopt the hitherto notation, and in particular that of Props.\,\ref{prop:canequivstr}, \ref{canextinvstr} and \ref{prop:commcan}. Thus, let $\,\txG\,$ be the supersymmetry group (with the tangent Lie superalgebra\footnote{In the super-Minkowskian setting, we should consider the sub-superalgebra $\,\gt{smink}(d,1\,\vert\,ND_{d,1})\subset\gt{siso}(d,1\,\vert\,ND_{d,1})\,$ as the model for the global-supersymmetry supervector space due to the very nature of the existing constructions ({\it i.e.}, extensions), taking as the point of departure \emph{that} Lie supergroup, and not the super-Poincar\'e group.} $\,\ggt$) and $\,\txH_{\rm vac}\,$ the vacuum isotropy group of a Green--Schwarz super-$\si$-model for the super-$p$-brane in the Hughes--Polchinski formulation, let $\,\Si^{\rm HP}_{\rm vac}\,$ be the vacuum foliation of Def.\,\ref{def:HPvacfol} of the Hughes--Polchinski section $\,\Si^{\rm HP}\,$ of \Reqref{eq:HPsec} determined by an integrable Hughes--Polchinski vacuum superdistribution $\,{\rm Vac}(\sgt\Bgt^{{\rm (HP)}}_{p,\la_p})\,$ of Def.\,\ref{def:vacHPsdistro}, and let $\,\sgt_{\rm vac}\,$ and $\,\sgt_{\rm loc}\equiv\ggt\sgt_{\rm vac}(\sgt\Bgt^{{\rm (HP)}}_{p,\la_p})\,$ be the Lie sub-superalgebras of $\,\ggt\,$ that model -- respectively -- the residual global-supersymmetry supspace of Prop.\,\ref{prop:resglobsusysub} and the limit $\,\k^{-\infty}(\sgt\Bgt^{{\rm (HP)}}_{p,\la_p})\,$ of the weak derived flag of the $\k$-symmetry superdistribution $\,\k(\sgt\Bgt^{{\rm (HP)}}_{p,\la_p})\,$ within $\,{\rm Vac}(\sgt\Bgt^{{\rm (HP)}}_{p,\la_p})$.\ There exists a canonical and canonically $\sgt_{\rm vac}$-invariant $\sgt_{\rm loc}$-equivariant structure on the vacuum restriction $\,\iota_{\rm vac}^*\widehat\cG{}^{(p)}\,$ of the extended Hughes--Polchinski $p$-gerbe $\,\widehat\cG{}^{(p)}\,$ of Def\,\ref{def:restrextHPpgerb}. The canonical structure consists of the $p$-gerbe 1-isomorphisms 
\qq\nn
\widehat\La_{\sgt_{\rm loc}}\equiv-\widehat\cT\con\pr_2^*\iota_{\rm vac}^*\widehat\cG{}^{(p)}\ &:&\ \pLie{\widehat\cT}\pr_2^*\iota_{\rm vac}^*\widehat\cG{}^{(p)}\xrightarrow{\ \cong\ }\cI_0^{(p)}\,,\cr\cr
\breve\La{}_{\breve A}\equiv\pLie{\widehat\cT}\pr_2^*\La{}_{\breve A}\ &:&\ \pLie{\breve\cK{}_{\breve A}}\bigl(\pLie{\widehat\cT}\pr_2^*\iota_{\rm vac}^*\widehat\cG{}^{(p)}\bigr)\xrightarrow{\ \cong\ }\cI_0^{(p)}\,,\qquad\breve A\in\ovl{1,S_{\rm vac}}\,,
\qqq
the latter written in terms of the $p$-gerbe 1-isomorphisms
\qq\nn
\La_{\breve A}\ :\ \pLie{\cK_{S_{\breve A}}}\iota_{\rm vac}^*\widehat\cG{}^{(p)}\xrightarrow{\ \cong\ }\cI^{(p)}_0
\qqq
encoding the assumed $\sgt_{\rm vac}$-invariance of $\,\iota_{\rm vac}^*\widehat\cG{}^{(p)}$,\ and of the $p$-gerbe 2-isomorphisms
\qq\nn
\widehat\cT\con\pr_2^*\La_{\breve A}\equiv\psi_A\qquad\equiv\qquad \alxydim{}{\pLie{\cK_{S_{\breve A}}}\bigl(\pLie{\widehat\cT}\pr_2^*\iota_{\rm vac}^*\widehat\cG{}^{(p)}\bigr)
\ar@/^1.6pc/[rrr]^{\pLie{\breve\cK{}_{\breve A}}\widehat\La_{\sgt_{\rm loc}}}="5"
\ar@/_1.6pc/[rrr]_{\breve\La{}_{\breve A}}="6"
\ar@{=>}"5";"6"|{\psi_A} &&& \cI_0^{(p)}}\,,\qquad\breve A\in\ovl{1,S_{\rm vac}}\,.
\qqq
\ethe
\beroof
Follows directly from Props.\,\ref{prop:canequivstr}, \ref{canextinvstr} and \ref{prop:commcan}. 
\eroof

\brem
Clearly, our results can be repeated {\it verbatim} in the situation described in Example \ref{eg:sqroots1bsMink} if we simply replace the $\k$-symmetry superdistribution appearing above with the \emph{extended} one.
\erem

\noindent The statement of the theorem concludes our study of linearised equivariance of the vacuum restriction of the extended Hughes--Polchinski $p$-gerbe of the Green--Schwarz super-$\si$-model (in the topological Hughes--Polchinski formulation) with respect to the vacuum-generating gauged supersymmetry (aka $\k$-symmetry), compatible with the residual global supersymmetry.\bigskip

\brem\label{rem:vision}
The above symmetry analysis, while far from complete, traces a line of reasoning that we are tempted to pursue speculatively, not least because of the hints as to directions of further development that it may provide. Thus, it is to be noted that the existence of a full-fledged \emph{Lie superalgebra}-equivariant structure on the vacuum restriction $\,\iota_{\rm vac}^*\widehat\cG{}^{(p)}\,$ of the extended Hughes--Polchinski $p$-gerbe would essentially imply -- in virtue of the standard interpretation of an equivariant structure, corroborated in Refs.\,\cite{Gawedzki:2010rn,Suszek:2012ddg,Gawedzki:2012fu,Suszek:2013} in the (1-)gerbe-theoretic context --  that $\,\iota_{\rm vac}^*\widehat\cG{}^{(p)}\,$ descends to the orbispace of the `action' of the symmetry structure, {\it i.e.}, of $\,\k^{-\infty}(\sgt\Bgt^{{\rm (HP)}}_{p,\la_p})$,\ on the vacuum Hughes--Polchinski section. But, then, the vacuum foliation is actually \emph{generated} by the flows of the latter superdistribution, which means -- in the $\cS$-point picture -- that each of its leaves, or vacua, can be retracted to any one of its points. Consequently, $\,\iota_{\rm vac}^*\widehat\cG{}^{(p)}\,$ would be supersymmetrically 1-isomorphic to\ldots the trivial $p$-gerbe with zero curving -- the unique $p$-gerbe over a point. This, in turn, would infer a \emph{vacuum trivialisation}
\qq\nn
\iota_{\rm vac}^*\pi_{\txG/\txH}^*\cG{}^{(p)}\cong\cI^{(p)}_{-\la_p\,\iota_{\rm vac}^*\underset{\tx{\ciut{(p+1)}}}{\b}\hspace{-7pt}{}^{\rm (HP)}}\,.
\qqq
Trivialisations of that kind have long been recognised as hallmarks of the presence of a \emph{defect} in the field theory, (D-)branes and bi-branes of string theory being examples of such structures. In the light of the above, we are led to consider a potential correspondence between vacua of the Green--Schwarz super-$\si$-model for (a homogeneous space of) a given super-Harish--Chandra pair and (homogeneous spaces of) sub-super-Harish--Chandra pairs thereof with a $(p+1)$-dimensional body over which the super-$p$-gerbe of the super-$\si$-model trivisalises in the manner suggested above. We shall certainly contemplate this attractive idea in the future.
\erem

\newpage

\section{Conclusions \& Outlook}\label{ref:CandO}

In the present paper, we have established a simple and complete geometric interpretation of the tangential gauge supersymmetry, also known as $\k$-symmetry, of a large class of Green--Schwarz super-$\si$-models for the super-$p$-brane functorially embedded in a reductive homogeneous space of a Lie supergroup, and lifted the supersymmetry to the higher-geometric structure associated with the topological component of the super-$\si$-model superbackground -- the extended Hughes--Polchinski $p$-gerbe -- in the form of a linearised equivariant structure on the latter, compatible with the global supersymmetry present. The gauge supersymmetry has acquired the interpretation of an odd-generated superdistribution (bracket-)generating -- through its weak derived flag -- the tangent sheaf of the (classical) vacuum of the field theory (resp.\ its chiral component, {\it cp} Example \ref{eg:sqroots1bsMink}) and thus enveloping the vacuum, {\it cp} Sec.\,\ref{sec:susy} and in particular Thm.\,\ref{thm:kdemyst}, whence also the name given to it in the title of the paper -- the \textbf{square root of the vacuum}. Its gerbification as a tangential equivariant structure, consistent with the findings of previous studies on gerbe-theoretic realisations of gauge symmetries in the two-dimensional non-linear bosonic $\si$-model, has been proven to possess a canonical form, canonically invariant with respect to the residual global supersymmetry of the vacuum, {\it cp} Thm.\,\ref{thm:kappagerbified}. These results have been obtained through an in-depth analysis of the universal phenomenon of enhancement of gauge (super)symmetry in the topological Hughes--Polchinski formulation of the Green--Schwarz super-$\si$-model that occurs in the physical correspondence sector thereof, introduced in Sec.\,\ref{sec:physmod} with direct reference to the correspondence superdistribution of Def.\,\ref{def:Corrsdistro}, in which the field theory is (classically) dual to the standard Nambu--Goto formulation of the super-$\si$-model, the conditions for the duality having been worked out in the present paper in greater generality than heretofore, {\it cp} Thm.\,\ref{thm:HpdualNGext}, on the basis of a rigorous supergeometric description of the supertarget(s) using Kostant's approach to the theory of Lie supergroups, {\it cp} Sec.\,\ref{sec:sCart}. Instrumental in the analysis has been the derivation and subsequent geometrisation of the Euler--Lagrange equations of the topological super-$\si$-model, {\it cp} Prop.\,\ref{prop:HPELeqs} and Def.\,\ref{def:vacHPsdistro}, and an exhaustive study of the algebraic conditions of integrability, global supersymmetry and descendability of the vacuum superdistribution determined by these equations in Def.\,\ref{def:vacHPsdistro}. The abstract considerations have been illustrated on a large number of concrete examples of super-$\si$-models with a single topological charge: the Green--Schwarz super-$p$-branes in the super-Minkowskian background and the super-1-branes in $\,{\rm s}({\rm AdS}_n\x\bS^n)\,$ for $\,n\in\{2,3,5\}$.\medskip

While essentially resolving the issue of the geometric nature of $\k$-symmetry, the study reported in the present paper leaves us with several follow-up questions and challenges. The first obvious one is a Lie-superalgebraic classification of super-$\si$-models with an integrable vacuum superdistribution generated by its $\k$-symmetry superdistribution in terms of Lie sub-superalgebras of (physically motivated) Lie superalgebras associated with a pair of projectors defining the vacuum subspace $\,\tgt_{\rm vac}\subset\tgt\,$ in the direct-sum complement of the isotropy subalgebra $\,\hgt\,$ of the mother supersymmetry algebra $\,\ggt$,\ {\it cp} Thm.\,\ref{thm:kdemyst}. This question is intimately related to the issue of existence of minimal spinors in a given metric geometry $\,(|\txG/\txH|,\unl\txg)\,$ of Sec.\,\ref{sec:sCart} ({\it cp}, in particular, Remark \ref{rem:4krule}). Further self-consistency conditions are anticipated to follow from imposition of the requirement of existence of a full-fledged \emph{(residual) supersymmetry-invariant Lie superalgebra}-equivariant structure on the vacuum restriction of the Hughes--Polchinski $p$-gerbe -- its derivation and detailed investigation should, therefore, be undertaken next. In its course, we should most certainly keep in mind and elaborate the attractive idea formulated in Rem.\,\ref{rem:vision}. In this context, we are fortunate to have at our disposal a rich pool of field-theoretic/supergeometric examples with a hands-on Lie-superalgebraic and (local-)coordinate descriptions -- this richness ought to be exploited towards further advancement of our understanding of the intricate nature of the vacuum of the super-$\si$-model in the format of a case-by-case study that is certain to provide us with new insights, just as it has done so far, and to yield concrete results for field theories with a potential application in the modelling of realistic strongly coupled systems of the QCD-type. Thus, in particular, the study reported and motivated herein is hoped to serve -- in the long run -- the outstanding goal of elucidating the still (mathematically) elusive AdS/CFT correspondence. The differential-supergeometric and Lie-superalgebraic discussion of the super-$\si$-model and its vacuum elaborated in the present work does not seal the fate of field theories and field configurations that depart from the neat scenario of a regular ({\it i.e.}, integrable) vacuum superdistribution, and it does not tell the story of super-$\si$-models that do not satisfy the set of constraints imposed in the derivation of the Euler--Lagrange field equations in Sec.\,\ref{sec:vac} -- it would certainly be both interesting and useful to study these departures in greater detail, with view to extending and generalising the results of our work. The example of the super-0-brane in the Zhou superbackground of $\,{\rm s}({\rm AdS}_2\x\bS^2)\,$ treated in the Appendix, taken in conjunction with the supergerbe-theoretic results for this particular superbackground obtained in \Rcite{Suszek:2018ugf}, seems to be a perfect point of departure of a quest thus oriented. Finally, reaching out beyond the compass of the present work, there is the fascinating question of general field-theoretic and geometric consequences of the duality between the \emph{dynamical} Nambu--Goto formulation of the (super-)$\si$-model and the purely \emph{topological} Hughes--Polchinski one. Among other things, one could envisage its application as a new potent tool in the by now fairly advanced study of T-duality -- another symmetry entangling the metric and topological degrees of freedom in the standard Nambu--Goto formulation -- where it is expected to lead to a unified topological (that is gerbe-theoretic) description of this loop-mechanical duality and, in this manner, pave the way to a systematic description and construction of T-folds ({\it via} gerbe-theoretic T-duality gauge defects). We hope to return to these issues in the future.
\newpage

\appendix

\section{The Zhou super-0-brane in $\,{\rm s}({\rm AdS}_2\x\bS^2)$.}
We describe the relevant superbackground along the lines of Examples \ref{eq:s0gsMink}--\ref{eg:MT1}.
\bit
\item[$\bullet$] The mother super-Harish--Chandra pair:
\qq\nn
{\rm SU}(1,1\,|\,2)_2\equiv\bigl({\rm SO}(1,2)\x{\rm SO}(3),\gt{su}(1,1\,|\,2)_2\bigr)\,,
\qqq
as in Example \ref{eg:Zhou1};
\item[$\bullet$] The (infinitesimal) $\sfT_e\Ad_\txH$-invariance of the vacuum splitting: follows by the same argument as in Example \ref{eq:spgsMink} due to the identical structure of and relations between the commutators $\,[\dgt,\tgt^{(0)}_{\rm vac}]\,$ and $\,[\dgt,\egt^{(0)}]$;
\item[$\bullet$] The homogeneous spaces: the NG one
\qq\nn
&{\rm s}\bigl({\rm AdS}_2\x\bS^2\bigr)={\rm SU}(1,1\,|\,2)_2/\bigl({\rm SO}(1,1)\x{\rm SO}(2)\bigr)\,,&\cr\cr
&\txH={\rm SO}(1,1)\x{\rm SO}(2)\,,\qquad\qquad\hgt\equiv\gt{so}(1,1)\oplus\gt{so}(2)=\corr{J_{01}}\oplus\corr{J_{23}}\,,&
\qqq
with the body as in Example \ref{eg:Zhou1} and the HP one
\qq\nn
&{\rm SU}(1,1\,|\,2)_2/{\rm SO}(2)\,,&\cr\cr
&\txH_{\rm vac}={\rm SO}(2)\,,\qquad\qquad\hgt_{\rm vac}=\corr{J_{23}}\,,&\cr\cr
&\dgt=\corr{J_{01}}\,,&\cr\cr
&\tgt_{\rm vac}^{(0)}=\corr{P_0}\,,\qquad\qquad\dgt^{-1}\tgt_{\rm vac}^{(0)}=\corr{P_1}\subsetneq\egt^{(0)}\,;&
\qqq
\item[$\bullet$] The exponential superparametrisation(s):
\qq\nn
\si_0^{\rm vac}\bigl(\theta^{\a'\a''I},x^a,\phi^{01}\bigr)=\ee^{\theta^{\a'\a''I}\ox Q_{\a'\a''I}}\cdot\ee^{x^a\ox P_a}\cdot\ee^{\phi^{01}\ox J_{01}}\,;
\qqq
\item[$\bullet$] The superbackgrounds: the NG one
\qq\nn
\sgt\Bgt_0^{{\rm (NG)}}&=&\bigl({\rm s}\bigl({\rm AdS}_2\x\bS^2\bigr),\eta_{ab}\,\theta^a_{\rm L}\ox\theta_{\rm L}^b,\sfi\,\Si_{\rm L}\wedge\bigl(\unl C\ox\si_2\bigr)\,\Si_{\rm L}+\theta_{\rm L}^0\wedge\theta_{\rm L}^1\equiv\underset{\tx{\ciut{(2)}}}{\chi}\hspace{-1pt}{}^{\rm Zh}\bigr)\,,\cr\cr
&&\theta_{\rm L}=\Si_{\rm L}^{\a'\a''I}\ox Q_{\a'\a''I}+\theta_{\rm L}^a\ox P_a+\theta^{01}_{\rm L}\ox J_{01}+\theta^{23}_{\rm L}\ox J_{23}
\qqq
with a supersymmetric global predecessor of the curving 
\qq\nn
\underset{\tx{\ciut{(1)}}}{\b}\hspace{-1pt}{}^{\rm Zh}=-\theta^{01}_{\rm L}
\qqq
on $\,{\rm SU}(1,1\,|\,2)_2\,$ that does \emph{not} descend to $\,{\rm s}({\rm AdS}_2\x\bS^2)$,\ and the HP one
\qq\nn
\sgt\Bgt_{0,\la_0}^{{\rm (HP)}}&=&\bigl({\rm SU}(1,1\,|\,2)_2/{\rm SO}(2),\underset{\tx{\ciut{(2)}}}{\chi}\hspace{-1pt}{}^{\rm Zh}+\la_0\,\bigl(\Si_{\rm L}\wedge\bigl(\unl C\,\unl\g{}^0\ox\bd1_2\bigr)\,\Si_{\rm L}-\theta_{\rm L}^{01}\wedge\theta_{\rm L}^1\bigr)\equiv\underset{\tx{\ciut{(2)}}}{\widehat\chi}\hspace{-1pt}{}^{\rm Zh}\bigr)\,,\cr\cr
&&\corr{P_0}\perp_\eta\corr{P_1}\,;
\qqq
\item[$\bullet$] The Body-Localisation Constraints: 
\qq\label{eq:Zh0BLCs}
\theta_{\rm L}^{a''}\approx 0\,,\qquad a''\in\{2,3\}\,.
\qqq
\eit

The logarithmic variation of the DF amplitude of the corresponding GS super-$\si$-model in the HP formulation, computed as in Sec.\,\ref{sec:vac}, takes the explicit form 
\qq
&&-\sfi\,\d_{\d\widehat\xi}\log\,\cA_{\rm DF}^{{\rm (HP)},0,\la_0}[\xi]\cr\cr
&=&\sum_{\t\in\Tgt_1}\,\int_\t\,\bigl(\si_{\imath_\t}^{\rm vac}\circ\widehat\xi_\t\bigr)^*\bigl[4\d\theta_{\iota_\t}^{\a'\a''I}\,\bigl(\unl C\,\unl\g^0\ox\bd1_2\bigr)_{\a'\a''I\b'\b''J}\,\left(\tfrac{\la_0\,\bd1_8-\sfi\,\unl\g^0\ox\si_2}{2}\right)^{\b'\b''J}_{\ \ \g'\g''K}\,\Si_{\rm L}^{\g'\g''K}\label{eq:logvarZh0}\\ \cr
&&+\d x^0_{\iota_\t}\,\theta_{\rm L}^1+\d x^1_{\iota_\t}\,\bigl(\la_0\,\theta_{\rm L}^{01}-\theta_{\rm L}^0\bigr)-\la_0\,\d\phi_{\iota_\t}^{01}\,\theta_{\rm L}^1\bigr]\,,\nn
\qqq
from which we read off -- for $\,\la_0\in\{-1,1\}\,$ -- the definition of the possible projectors
\qq\nn
\sfP^{(1)}_{\pm 1}=\tfrac{\bd1_8\pm\sfi\,\unl\g^0\ox\si_2}{2}
\qqq
with properties
\qq\nn
&\bigl(\unl\g^0\ox\bd1_2\bigr)\,\sfP^{(1)}_{\pm 1}=\sfP^{(1)}_{\pm 1}\,\bigl(\unl\g^0\ox\bd1_2\bigr)\,,\qquad\qquad\bigl(\unl\g^{\widehat a}\ox\bd1_2\bigr)\,\sfP^{(1)}_{\pm 1}=\bigl(\bd1_8-\sfP^{(1)}_{\pm 1}\bigr)\,\bigl(\unl\g^{\widehat a}\ox\bd1_2\bigr)\,,&\cr\cr 
&\bigl(\unl C\ox\bd1_2\bigr)\,\sfP^{(1)}\,\bigl(\unl C\ox\bd1_2\bigr)^{-1}=\sfP^{(1)\,{\rm T}}\,,&
\qqq
and so upon choosing
\qq\nn
\sfP^{(1)}\equiv\sfP^{(1)}_{+1}=\tfrac{\bd1_8+\sfi\,\unl\g^0\ox\si_2}{2}\,,
\qqq
also the EL equations \qq\nn
\bigl(\bigl(\bd1_8-\sfP^{(1)}\bigr)\,\Si_{\rm L},\theta_{\rm L}^1,\theta_{\rm L}^0-\theta_{\rm L}^{01}\bigr)\approx 0\,,
\qqq
to be augmented with the BLCs \eqref{eq:Zh0BLCs}. The last of the EL equations indicates that the body of the vacuum is actually `(gauge-)tilted' in the direction of $\,J_{01}\,$ relative to the `canonical' direction $\,P_0$.\ Altogether, we deduce the HP vacuum superdistribution with restrictions
\qq\nn
{\rm Vac}\bigl({\rm SU}(1,1\,|\,2)_2/{\rm SO}(2),\underset{\tx{\ciut{(2)}}}{\widehat\chi}\hspace{-1pt}{}^{\rm Zh}\bigr)\rstr_{\cV_i}=\bigoplus_{\unl\a=1}^4\,\corr{\cT_{\unl\a\,i}}\oplus\corr{\cT_{0\,i}+\cT_{01\,i}}\,.
\qqq 
At this stage, it suffices to compute the supercommutators
\qq\nn
&&\{\sfP^{(1)\,\g'\g''K}_{\ \ \ \ \ \a'\a''I}\,Q_{\g'\g''K},\sfP^{(1)\,\d'\d''L}_{\ \ \ \ \ \b'\b''J}\,Q_{\d'\d''L}\}\cr\cr
&=&2\sfP^{(1)\,\g'\g''K}_{\ \ \ \ \ \b'\b''J}\,\bigl(\bigl(\unl C\,\unl\g^0\ox\bd1_2\bigr)_{\a'\a''I\g'\g''K}\,P_0-\sfi\,\bigl(\unl C\ox\si_2\bigr)_{\a'\a''I\g'\g''K}\,J_{01}\bigr)\cr\cr
&\equiv&2\bigl(\unl C\,\unl\g^0\ox\bd1_2\bigr)_{\a'\a''I\g'\g''K}\,\bigl(\bd1_8\,T_0+2\bigl(\sfP^{(1)}-\bd1_8\bigr)\,J_{01}\bigr)^{\g'\g''K}_{\ \ \d'\d''L}\,\sfP^{(1)\,\d'\d''L}_{\ \ \ \ \ \b'\b''J}\cr\cr
&=&2\bigl(\bigl(\unl C\,\unl\g^0\ox\bd1_2\bigr)\,\sfP^{(1)}\bigr)_{\a'\a''I\b'\b''J}\,T_0
\qqq
and
\qq\nn
&[T_0,T_0]=0=[J_{23},J_{23}]\,,\qquad\qquad[J_{23},T_0]=0\,,&\cr\cr
&[T_0,\sfP^{(1)\,\b'\b''J}_{\ \ \ \ \ \a'\a''I}\,Q_{\b'\b''J}]=\tfrac{1}{2}\,\bigl(\bigl(\sfi\,\widetilde\g_3'\,\unl\g{}_0\ox\si_2+\unl\g{}_{01}\ox\bd1_2\bigr)\cdot\sfP^{(1)}\bigr)^{\b'\b''J}_{\ \ \a'\a''I}\,Q_{\b'\b''J}&\cr\cr
&\hspace{3.85cm}\equiv\bigl(\bigl(\sfi\,\widetilde\g_3'\ox\bd1_2\bigr)\cdot\bigl(\bd1_8-\sfP^{(1)}\bigr)\cdot\sfP^{(1)}\bigr)^{\b'\b''J}_{\ \ \a'\a''I}\,Q_{\b'\b''J}=0\,,&\cr\cr
&[J_{23},\sfP^{(1)\,\b'\b''J}_{\ \ \ \ \ \a'\a''I}\,Q_{\b'\b''J}]=\tfrac{1}{2}\,\bigl(\unl\g{}_{23}\ox\bd1_2\bigr)^{\b'\b''J}_{\ \a'\a''I}\,\sfP^{(1)\,\g'\g''K}_{\ \ \ \ \ \b'\b''J}\,Q_{\g'\g''K}&
\qqq
to conclude that it is an ${\rm SO}(2)$-descendable integrable superdistribution associated with the Lie superalgebra
\qq\nn
\gt{vac}\bigl({\rm SU}(1,1\,|\,2)_2/{\rm SO}(2),\underset{\tx{\ciut{(2)}}}{\widehat\chi}\hspace{-1pt}{}^{\rm Zh}\bigr)=\bigoplus_{\unl\a=1}^4\,\corr{\unl Q{}_{\unl\a}}\oplus\corr{P_0+J_{01}}\oplus\corr{J_{23}}\,.
\qqq

Another look at \Reqref{eq:logvarZh0} readily convinces us that the enhanced gauge-symmetry superdistribution is exceptionally large in this superbackground as it coincides with the HP vacuum superdistribution,
\qq\nn
\cG\cS\bigl({\rm SU}(1,1\,|\,2)_2/{\rm SO}(2),\underset{\tx{\ciut{(2)}}}{\widehat\chi}\hspace{-1pt}{}^{\rm Zh}\bigr)\equiv{\rm Vac}\bigl({\rm SU}(1,1\,|\,2)_2/{\rm SO}(2),\underset{\tx{\ciut{(2)}}}{\widehat\chi}\hspace{-1pt}{}^{\rm Zh}\bigr)\,,
\qqq
and that, in fact, even without any further reduction. Clearly, the (odd-generated) $\k$-symmetry superdistribution with restrictions
\qq\nn
\k^{-\infty}\bigl({\rm SU}(1,1\,|\,2)_2/{\rm SO}(2),\underset{\tx{\ciut{(2)}}}{\widehat\chi}\hspace{-1pt}{}^{\rm Zh}\bigr)=\bigoplus_{\unl\a=1}^4\,\corr{\cT_{\unl\a\,i}}
\qqq
is an ${\rm SO}(2)$-descendable superdistribution with the limit of its weak derived flag with restrictions
\qq\nn
\k^{-\infty}\bigl({\rm SU}(1,1\,|\,2)_2/{\rm SO}(2),\underset{\tx{\ciut{(2)}}}{\widehat\chi}\hspace{-1pt}{}^{\rm Zh}\bigr)\rstr_{\cV_i}=\bigoplus_{\unl\a=1}^4\,\corr{\cT_{\unl\a\,i}}\oplus\corr{\cT_{0\,i}+\cT_{01\,i}}\equiv{\rm Vac}\bigl({\rm SU}(1,1\,|\,2)_2/{\rm SO}(2),\underset{\tx{\ciut{(2)}}}{\widehat\chi}\hspace{-1pt}{}^{\rm Zh}\bigr)\rstr_{\cV_i}\,.
\qqq

\end{document}

%% file: 1armagedef.tex
\newcommand{\alxydim}[2]{\begin{aligned}\xymatrix#1{#2}\end{aligned}}

\newcommand{\brem}{\begin{Rem}}
\newcommand{\erem}{\end{Rem}\medskip}
\newcommand{\beg}{\begin{Eg}}
\newcommand{\eeg}{\end{Eg}}
\newcommand{\bedef}{\begin{Def}}
\newcommand{\exdef}{\begin{flushright}$\diamond$\end{flushright}
\end{Def}\vskip0.1cm}
\newcommand{\berop}{\begin{Prop}}
\newcommand{\eerop}{\end{Prop}}
\newcommand{\belem}{\begin{Lem}}
\newcommand{\elem}{\end{Lem}}
\newcommand{\bethe}{\begin{Thm}}
\newcommand{\ethe}{\end{Thm}}
\newcommand{\becor}{\begin{Cor}}
\newcommand{\ecor}{\end{Cor}}
\newcommand{\beroof}{\noindent\begin{proof}}
\newcommand{\eroof}{\end{proof}}
\newcommand{\becon}{\begin{Conv}}
\newcommand{\econ}{\begin{flushright}$\checkmark$\end{flushright}\end{Conv}}
\newcommand{\befact}{\begin{Fact}}
\newcommand{\efact}{\begin{flushright}$\checkmark$\end{flushright}\end{Fact}}
\newcommand{\bequest}{\begin{Quest}}
\newcommand{\equest}{\end{Quest}}
\newcommand{\brob}{\begin{Prob}}
\newcommand{\erob}{\end{Prob}}

\newcommand{\barr}{\begin{array}}
\newcommand{\earr}{\end{array}}
\newcommand{\ben}{\begin{enumerate}}
\newcommand{\een}{\end{enumerate}}
\newcommand{\bit}{\begin{itemize}}
\newcommand{\eit}{\end{itemize}}

\newcommand{\qq}{\begin{eqnarray}}
\newcommand{\qqq}{\end{eqnarray}}

\newcommand{\nn}{\nonumber}

\newcommand{\ovl}[1]{\overline{#1}}
\newcommand{\unl}[1]{\underline{#1}}

\newcommand{\Reqref}[1]{Eq.\,\eqref{#1}}
\newcommand{\Rcite}[1]{Ref.\,\cite{#1}}
\newcommand{\Rxcite}[2]{Ref.\,\cite[#1]{#2}}

\newcommand\void[1]{}

\newcommand{\tx}[1]{\textrm{#1}} 
\newcommand{\ciut}[1]{\tiny$#1$}

\newcommand{\gt}[1]{\mathfrak{#1}}

\def\cA{\mathcal{A}}
\def\cB{\mathcal{B}}
\def\cC{\mathcal{C}}
\def\cD{\mathcal{D}}

\def\cG{\mathcal{G}}
\def\ceH{\mathcal{H}}
\def\cI{\mathcal{I}}

\def\cK{\mathcal{K}}

\def\cM{\mathcal{M}}

\def\cO{\mathcal{O}}

\def\cS{\mathcal{S}}
\def\cT{\mathcal{T}}
\def\cU{\mathcal{U}}
\def\cV{\mathcal{V}}
\def\cW{\mathcal{W}}

\def\xcA{\mathscr{A}}
\def\xcB{\mathscr{B}}
\def\xcC{\mathscr{C}}
\def\xcD{\mathscr{D}}

\def\xcK{\mathscr{K}}
\def\xcL{\mathscr{L}}

\def\xcO{\mathscr{O}}
\def\xcP{\mathscr{P}}

\def\xcT{\mathscr{T}}

\def\t{\mathbf{t}}

\def\bC{{\mathbb{C}}}

\def\bH{{\mathbb{H}}}

\def\bN{{\mathbb{N}}}

\def\bR{{\mathbb{R}}}
\def\bS{{\mathbb{S}}}

\def\bZ{{\mathbb{Z}}}

\def\a{\alpha}
\def\b{\beta}
\def\g{\gamma}
\def\G{\Gamma}
\def\d{\delta}
\def\D{\Delta}
\def\ep{\epsilon}
\def\vep{\varepsilon}

\def\k{\kappa}

\def\la{\lambda}
\def\La{\Lambda}
\def\om{\omega}
\def\Om{\Omega}

\def\si{\sigma}
\def\Si{\Sigma}

\def\t{\tau}

\def\z{\zeta}

\def\agt{\gt{a}}

\def\Bgt{\gt{B}}

\def\dgt{\gt{d}}

\def\egt{\gt{e}}

\def\fgt{\gt{f}}

\def\ggt{\gt{g}}

\def\hgt{\gt{h}}

\def\kgt{\gt{k}}

\def\lgt{\gt{l}}

\def\mgt{\gt{m}}

\def\gtq{\gt{q}}

\def\sgt{\gt{s}}
\def\Sgt{\gt{S}}
\def\tgt{\gt{t}}
\def\Tgt{\gt{T}}

\def\zgt{\gt{z}}

\newcommand{\sfd}{{\mathsf d}}

\newcommand{\sfF}{{\mathsf F}}

\newcommand{\sfi}{{\mathsf i}}

\newcommand{\sfN}{{\mathsf N}}

\newcommand{\sfP}{{\mathsf P}}

\newcommand{\sfT}{{\mathsf T}}

\newcommand{\sfY}{{\mathsf Y}}

\newcommand{\txB}{{\rm B}}

\newcommand{\ee}{{\rm e}}

\newcommand{\txg}{{\rm g}}
\newcommand{\txG}{{\rm G}}

\newcommand{\txH}{{\rm H}}

\newcommand{\txK}{{\rm K}}

\newcommand{\txm}{{\rm m}}

\newcommand{\txp}{{\rm p}}

\newcommand{\txU}{{\rm U}}

\def\vC{\check{C}}
\def\Cv{\v{C}}

\def\vd{\check{\d}}

\def\id{{\rm id}}
\newcommand{\pr}{{\rm pr}}
\def\sign{{\rm sign}}

\def\too{\longrightarrow}
\def\ev{{\rm ev}}

\def\obj{{\rm Ob}}
\def\mor{{\rm Hom}}

\def\1morf{1{\rm -Mor}}
\def\2morf{2{\rm -Mor}}
\def\dim{{\rm dim}}

\def\End{{\rm End}}

\def\Inv{{\rm Inv}}

\newcommand\Yon{{\rm Yon}}

\newcommand{\Set}{{\rm {\bf Set}}}

\newcommand{\sMan}{{\rm {\bf sMan}}}
\newcommand{\Presh}{{\rm {\bf Presh}}}
\newcommand{\sLieGrp}{{\rm {\bf sLieGrp}}}
\newcommand{\scommAlg}{{\rm {\bf sAlg_{scomm}}}}
\newcommand{\sLieAlg}{{\rm {\bf sLieAlg}}}
\newcommand{\sHCp}{{\rm {\bf sHCp}}}

\newcommand{\pLie}[1]{\,{-\hspace{-8pt}\xcL}_{#1}}
\def\p{\partial}

\def\con{\righthalfcup}

\def\emb{\hookrightarrow}

\def\curv{{\rm curv}}

\def\bd1{{\boldsymbol{1}}}
\def\brd0{{\boldsymbol{0}}}

\def\det{{\rm det}}
\def\tr{{\rm tr}}
\def\diag{\textrm{diag}}

\def\ad{{\rm ad}}
\def\Ad{{\rm Ad}}

\newcommand{\uj}{{\rm U}(1)}

\def\x{\times}
\def\ox{\otimes}

\def\lx{{\hspace{-0.04cm}\ltimes\hspace{-0.05cm}}}
\def\rx{\rtimes}

\def\lact{\vartriangleright}
\def\must{\stackrel{!}{=}}

\def\rstr{\mathord{\restriction}}

\newcommand{\corr}[1]{\left\langle #1 \right\rangle}
